\documentclass[preprint,prd,aps,amsmath,nofootinbib,superscriptaddress,tightenlines]{revtex4}

\usepackage{graphicx}

\usepackage[hypertex]{hyperref}
\usepackage{xspace}

\def\lqcd{\Lambda_{\text{QCD}}}

\def\OMIT#1{}

\newcommand{\nn}{\nonumber \\ } 
\newcommand{\bn}{{\bar n}}
\newcommand{\bnslash}{\bar n\!\!\!\slash}
\newcommand{\nslash}{n\!\!\!\slash}

\newcommand{\bsigma}{\mbox{$\mathbf \sigma$}}

\newcommand{\mcdot}{\!\cdot\!}
\newcommand{\SCETa}{\ensuremath{{\rm SCET}_{\rm I}}\xspace}
\newcommand{\SCETb}{\ensuremath{{\rm SCET}_{\rm II}}\xspace}

\newcommand{\epsU}{\epsilon_{\text{UV}}}
\newcommand{\epsI}{\epsilon_{\text{IR}}}

\newcommand{\bnP}{\bar {\cal P}}
\newcommand{\cP}{{\cal P}}

\newcommand{\muplus}{\mu_+}
\newcommand{\muminus}{\mu_-}
\newcommand{\muplusminus}{\mu_\pm}
\newcommand{\muplustwo}{\mu_+^2}
\newcommand{\muminustwo}{\mu_-^2}

\newcommand{\mupluse}{\mu_+^{\,\epsilon}}
\newcommand{\muminuse}{\mu_-^{\,\epsilon}}
\newcommand{\muplusee}{\mu_+^{2\epsilon}}
\newcommand{\muminusee}{\mu_-^{2\epsilon}}

  \newcount\hour \newcount\minute
  \hour=\time \divide \hour by 60
  \minute=\time
  \newcommand{\mydate}{\ \today \ - \number\hour :\ifnum \minute<10 0\fi 
\number\minute}

\begin{document}
\setlength\baselineskip{17pt}

\preprint{ \vbox{ 
\hbox{MIT-CTP-3726} \hbox{UCSD/PTH 06-04} \hbox{hep-ph/0605001}  } 
}

\title{\phantom{x}\vspace{0.1cm}
The Zero-Bin and Mode Factorization in Quantum Field Theory
\vspace{0.5cm}
}

\author{Aneesh V. Manohar}
\affiliation{Department of Physics, University of California at San Diego,
        La Jolla, CA 92093\footnote{Electronic address: amanohar@ucsd.edu}}

\author{Iain W. Stewart\vspace{0.4cm}}
\affiliation{Center for Theoretical
  Physics, Laboratory for Nuclear Science, Massachusetts \\ 
   Institute of Technology,
  Cambridge, MA 02139\footnote{Electronic address: iains@mit.edu}
  \vspace{0.5cm}}

 
\begin{abstract}
  \vspace{0.3cm}
  
  We study a Lagrangian formalism that avoids double counting in effective field
  theories where distinct fields are used to describe different infrared
  momentum regions for the same particle. The formalism leads to extra
  subtractions in certain diagrams and to a new way of thinking about
  factorization of modes in quantum field theory.  In non-relativistic field
  theories, the subtractions remove unphysical pinch singularities in box type
  diagrams, and give a derivation of the known pull-up mechanism between soft
  and ultrasoft fields which is required by the renormalization group evolution.
  In a field theory for energetic particles, the soft-collinear effective theory
  (SCET), the subtractions allow the theory to be defined with different
  infrared and ultraviolet regulators, remove double counting between soft,
  ultrasoft, and collinear modes, and give results which reproduce the infrared
  divergences of the full theory.  Our analysis shows that convolution
  divergences in factorization formul\ae\ occur due to an overlap of momentum
  regions. We propose a method that avoids this double counting, which helps to
  resolve a long standing puzzle with singularities in collinear factorization
  in QCD.  The analysis gives evidence for a factorization in rapidity space in
  exclusive decays.

\end{abstract}

\maketitle

\tableofcontents
\newpage

\section{Introduction}
\label{sec:intro}

Many problems of interest in quantum field theory have several momentum scales,
and are efficiently treated using effective field theory (EFT) methods. One
constructs a sequence of effective field theories which focus on one scale at a
time. This greatly simplifies the calculations, partly because new symmetries
emerge, and partly because Feynman graphs in each effective theory are much
simpler to evaluate than the multiscale integrals of the full theory.  More
recently, theoretical methods have been developed which allow one to analyze
field theories with several small momentum scales which are coupled by the
dynamics. In these theories, it becomes necessary to treat the coupled momentum
scales simultaneously within a single effective theory, rather than sequentially
in a series of several different effective theories.  Examples of theories with
coupled scales are the soft-collinear effective theory (SCET) for energetic
particles~\cite{Bauer:2000ew,Bauer:2000yr,Bauer:2001ct,Bauer:2001yt}, and any
non-relativistic theory where the kinetic energy is a relevant operator,
examples being non-relativistic QED (NRQED)~\cite{Caswell:1985ui} and
non-relativistic QCD
(NRQCD)~\cite{Bodwin:1994jh,Brambilla:1999xf,Luke:1999kz,Manohar:2000hj,Hoang:2002ae}.

We wish to discuss an issue in the separation of infrared (IR) regions
which appears at first to be a technical subtlety, but turns out to
have important physical ramifications. It results in a tiling theorem
for IR modes in quantum field theory. In the examples we discuss, it
has to do with the proper treatment of soft modes in NRQCD/NRQED and
of collinear and soft modes in SCET.  In NRQED, the photon field of
the fundamental QED theory is replaced by two fields, describing soft
and ultrasoft (usoft) photons with energies of order $mv$ and $mv^2$
respectively, where $v\ll1$ is the typical fermion velocity in the
non-relativistic bound state. The soft and usoft NRQED gauge fields
are $A^\mu_{p}(x)$ and $A^\mu(x)$, where $p$ is a label momentum of
order $mv$, and $k$, the Fourier transform of $x$, is of order
$mv^2$~\cite{Luke:1999kz}. The two fields describe photons with
momenta $p+k$ and $k$, respectively. In the special case that $p=0$
(the zero-bin), the soft photon becomes ultrasoft, and there is a
double-counting of modes. To avoid double counting, the soft sector of
the theory must have the additional constraint that $p\not=0$. This
paper explores the consequences of implementing zero-bin constraints
for soft modes in NRQED/NRQCD, and the analog for collinear modes in
SCET. In loop graphs, for example, the sum over soft intermediate
states should be $\sum_{p\not=0}$ rather than the conventional
$\sum_{p}$. The difference, as we discuss in detail, is that
conventional results have to be modified by zero-bin subtractions.

The zero-bin subtraction solves a number of problems in NRQCD and
SCET, and also resolves the long standing puzzle of divergent
convolutions in QCD factorization formulas. We discuss several
applications:
\begin{enumerate}
  
\item Soft box graphs in NRQCD have unphysical pinch singularities in the energy
  integral, 
  \begin{eqnarray}
   \int \frac{dk^0}{(k^0+i0^+)(-k^0+i0^+)}\, f(k^0)\,,
  \end{eqnarray}
  which make them ill-defined, even in dimensional regularization. In previous
  computations, it has been argued that these pinch singularities should be
  dropped in evaluating box graphs at any order in
  $v$~\cite{Griesshammer:1998tm,Manohar:2000hj,Brambilla:2004jw}. Pinch
  singularities are also a problem for the method of
  regions~\cite{Beneke:1997zp}. A direct application of the method of regions
  for $d^4k$ leads to ill-defined integrals, so it was defined to apply to NRQCD
  only after first doing the energy integrals.  The zero-bin subtraction
  modifies the soft box graphs so that pinch singularities are absent, and the
  graphs are well defined.

\item The zero-bin subtraction automatically implements the previously studied
  pullup mechanism in NRQCD~\cite{Manohar:2000kr,Hoang:2001rr}, which was shown
  to be a necessary part of the definition of this type of theory with multiple
  overlapping low energy modes. Through the pullup, infrared (IR) divergences
  in soft diagrams are converted to ultraviolet (UV) divergences and contribute
  to anomalous dimensions.
  
\item There is a similar pullup mechanism at work in SCET for collinear
  diagrams. The anomalous dimensions of the SCET currents for endpoint $B \to
  X_s\gamma$ and $B\to X_u\ell\bar\nu$ were computed in
  Ref.~\cite{Bauer:2000ew,Bauer:2000yr} from the $1/\epsilon$ and $1/\epsilon^2$
  terms. Some of these terms in the collinear graphs are actually infrared
  divergences. The zero-bin subtraction converts these infrared divergences to
  ultraviolet divergences so that IR-logs in QCD can be resummed as UV-logs in
  the effective theory. This formally justifies the results used for anomalous
  dimensions in these computations, and in subsequent work for other processes
  with similar anomalous dimensions
  eg.~\cite{Fleming:2002sr,Bauer:2003pi,Fleming:2003gt,Manohar:2003vb,Bosch:2004th,Hill:2004if}.
  
\item In high energy inclusive production such as $\gamma^* \to q \bar q g$,
  there is a potential double counting at the edges of the Dalitz plot in
  SCET, which is resolved by properly taking into account the zero-bin in both
  fully differential and partially integrated cross sections.
  
\item As a by-product of our analysis we give definitions for NRQCD and SCET
  that are independent of the UV and IR regulators. We also demonstrate a link
  between power counting and reproducing infrared divergences in the EFT.
  Exploiting this link, we demonstrate that a choice exists for the degrees of
  freedom in NRQCD and SCET which is complete, covering all infrared regions
  for a broad class of physical situations.  For these cases no new modes are
  required at any order in $\alpha_s$ or in the power expansion.

\item In high energy exclusive production, such as $\gamma^* \to \pi
  \rho$ or $\gamma^* \to \pi\pi$, there are unphysical singularities
  in convolution integrals of some hard kernels with the light-cone
  wavefunctions $\phi_\pi(x)$. For example
\begin{eqnarray}   \label{0.1}
\int_0^1 {\rm d }x \ \frac {\phi_\pi(x)}{x^2} \,,
\end{eqnarray}
which is divergent at $x \to 0$ if $\phi_\pi(x)$ vanishes linearly as $x \to 0$.
The same is true for exclusive light meson form factors at large $Q^2$, as well
as processes like $B\to \pi\ell\bar\nu$ and $B\to\pi\pi$ for
$E_\pi\gg\Lambda_{\rm QCD}$.  The zero-bin subtraction removes the singularity,
and induces a corresponding UV divergence. After renormalization we have a
finite convolution with $\phi_\pi(x)$, and the kernel behaves as a distribution
we call \o:
\begin{eqnarray} \label{0.2}
\int_0^1 {\rm d }x \ \frac {\phi_\pi(x)}{x^2} \to 
\int_0^1 {\rm d }x\ \frac{\phi_\pi(x)}{\left( x^2 \right)_{\mbox{\o}}}=
\int_0^1 {\rm d }x\  \frac{ \phi_\pi(x) \!-\! \phi_\pi(0) 
  \!-\! x \phi_\pi'(0)}{x^2} + \tilde \phi_\pi
 < \infty \,.
\end{eqnarray}
The $\tilde \phi_\pi$-type term involves a $\ln(E_\pi)$ and is induced
by UV renormalization in rapidity space. These terms are discussed in
the body of the paper.  Thus using SCET finite amplitudes are obtained
for apparently singular hard scattering kernels.

\item The zero-bin procedure gives insight into factorization
formulas which separate modes in rapidity space rather
  than by scale separation in their invariant mass. For example, two
  hadrons both built of non-perturbative modes with $p^2\sim
  \Lambda_{\rm QCD}^2$ can have their modes factorize by being in
  different corners of phase space or rapidity.  We discuss this by applying the
  zero-bin technique to the formulation of degrees of freedom in
  \SCETb with Wilson lines on the light-cone. The separation between
  soft and collinear regions is controlled by perturbation theory with
 dependence on a rapidity parameter.

\end{enumerate}
Examples of processes where endpoint singularities in convolution integrals have
been encountered include the pion form factor at large $Q^2$ and subleading
twist~\cite{Geshkenbein:1982zs}, the $\rho-\pi$ form
factor~\cite{Chernyak:1983ej}, the $B\to \pi\ell\bar\nu$ form
factor~\cite{Akhoury:1993uw}, form factor terms in $B\to
\pi\pi$~\cite{Szczepaniak:1990dt}, the Pauli nucleon form factor
$F_2$~\cite{Belitsky:2002kj}, color-suppressed $B$-decays involving
light isodoublets, such as $\bar B^0\to D^0 \bar K^0$ and $\bar B^0\to
D_s K^-$~\cite{Mantry:2003uz}, and annihilation contributions in
two-body
$B$-decays~\cite{Keum:2000wi,Keum:2000ph,Beneke:2001ev}. Endpoint
singularities also appear in non-exclusive processes such as
semi-inclusive deep inelastic scattering at low transverse
momentum~\cite{Collins:2003fm,Ji:2004wu}. When the zero-bin procedure
is applied to these cases, individual pieces of the amplitudes exhibit
dependence on a rapidity parameter.

In the work of Collins and Soper~\cite{Collins:1981uk,Collins:1981uw},
factorization formulas involving a rapidity parameter were derived for
fragmentation in $e^+e^-\to A+B+X$ where $A$ and $B$ are hadrons. The separation
of degrees of freedom in \SCETb gives finite amplitudes that appear to indicate
that rapidity dependent factorization also occurs in other two-hadron processes,
including purely exclusive ones. Our finite amplitudes are shown to be a direct
consequence of defining the degrees of freedom in \SCETb carefully. We will not
give a complete derivation of an exclusive rapidity factorization formula here,
because in our analysis we will make the simplifying assumption that the
renormalization of rapidity space effects and invariant mass effects can be
carried out independently.

It is useful to have a physical understanding of why resolving the double
counting issue also resolves the singularity problems. A hint comes from the
fact that neither the pinch singularities or endpoint singularities are present
in the full QCD computations, as pointed out for SCET in~\cite{Bauer:2002aj} in
the context of $B\to\pi\ell\bar\nu$.  For case 1) above, the soft pinch
singularities are removed by the kinetic energy of the quarks, $1/[k^0-{\mathbf
  k}^2/(2m)]$.  In case 7) above, the endpoint singularities are soft limits of
the full theory diagrams and are removed by $\Lambda_{\rm QCD}$, or in
non-relativistic systems~\cite{Bell:2005gw}, by the binding energy. An improper
interpretation of the singularities can occur by not being careful about taking
{\em double limits}.  In 1) the singularity arises from first taking $k^0 \gg
{\mathbf k}^2/(2m)$ and then $k^0\to 0$.  The $k^0 \gg {\mathbf k}^2/(2m)$ limit
gives soft quarks which can not properly describe the potential-quark region
where $k^0 \sim {\mathbf k}^2/(2m)$.  Likewise, in 7) the endpoint singularity
comes from first taking $k^- \gg k_\perp,k^+$ and then taking $k^-\to 0$.  The
collinear particles obtained from $k^- \gg k_\perp,k^+$ do not properly describe
the soft particle region where $k^-\sim k_\perp,k^+$.  To avoid double counting
we must ensure that the soft quarks do not double count the potential region in
case 1), and that the collinear quarks do not double count the soft region in
7).  In the effective theory implementation of the expansion of QCD these
singular limits are properly described by other degrees of freedom. Once we
avoid the double counting, the unphysical singularities never appear because the
potential limit of the soft quarks and the soft limit of the collinear quarks
are rendered harmless. It should be emphasized that the pinch and endpoint
singularities we are discussing are unphysical artifacts of certain
approximations, and are reflected by double counting problem that must be fixed
in the Effective Field Theory (EFT). They are not the same as the classification
of physical IR divergences from the Landau Equations (see
eg.~\cite{Sterman:2004pd,Mueller:1989hs}) which go by similar names. The true
IR structure of the full theory is properly reproduced by contributions from the
full set of EFT degrees of freedom.

It is important to note that the zero-bin subtractions avoids double counting
independent of the choice of UV and IR regulators in effective theory
computations. Our implementation of zero-bin subtractions is unique up to
possible finite scheme dependent contributions, and provides an explicit
connection to methods which introduce hard factorization cutoffs. It also
provides a definition of the modes in these effective theories independent of
dimensional regularization (and with some work could be used for example to take
the cutoff formulation of SCET described below and implement it on the lattice).
Since the proper formulation of an EFT should not depend on the choice of IR
regulator used in perturbative computations, this is not
surprising.\footnote{The choice of IR or UV regulators can make it more
  difficult to perform the power counting, for example by leading to integrals
  that are not homogeneous and require power counting violating counterterms to
  give back a power counting for the renormalized graphs. }  Physical results in
QCD are IR finite with divergences removed by quantities like $\Lambda_{\rm
  QCD}$, binding energies, or cancellations between real and virtual diagrams,
and the same is true in the EFT.

The outline of the paper is as follows. In section~\ref{sec:nrqcdscet} we give a
brief introduction to NRQCD and SCET which serve as our main examples. The
tiling of IR regions with modes is discussed in section~\ref{sec:dblecount}, and
the zero-bin subtraction is formulated in section~\ref{sec:zerobin}. Our
discussion is in the context of NRQCD and SCET, but is general enough to be
readily adapted to other physical situations. In section~\ref{sec:zerobinNRQCD}
we give examples in NRQCD to show that the zero-bin removes pinch
singularities, implies the pullup mechanism, and avoids double counting
problems. In section~\ref{sec:sceta} we give examples in \SCETa which
demonstrate the regulator independence of the zero-bin method, and the removal
of double counting in loop integrals and in inclusive phase space computations.
Finally, in section~\ref{sec:scetb} we give examples in \SCETb which is
formulated with zero-bin subtractions and only soft and collinear modes. The
zero-bin subtraction resolves the endpoint singularity issue in exclusive
processes to leave finite amplitudes, and require the introduction of
a rapidity parameter.  Conclusions are given in section~\ref{sec:conclusion}.


\section{NRQCD and SCET}
\label{sec:nrqcdscet}

NRQCD is an effective theory for non-relativistic quark-antiquark ($Q\bar Q$)
bound states, where the typical relative fermion velocity, $v$, is small, $v \ll
1$. The relevant scales in NRQCD are the quark mass, $m$, momentum $p\sim mv $,
and energy $E \sim mv^2$, with $E \ll p \ll m$.  The energy and momentum are not
independent; they are coupled via the quark equation of motion, $2 m E=p^2$.

SCET describes the interaction of energetic particles; examples include the
inclusive decay $B \to X_s \gamma$ at large $E_\gamma$ via the partonic decay $b
\to s \gamma$, inclusive jet production, or exclusive semileptonic decays such
as $B\to \pi\ell\bar\nu$.  It is convenient to orient the coordinate system in
the direction of the energetic jet or hadron, for example by introducing null
vectors $n=(1,0,0,1)$ and $\bn=(1,0,0,-1)$, and use light-cone coordinates with
$p^+ \equiv n \cdot p$, $p^- \equiv \bn \cdot p$ for any four-vector $p$.
Energetic particles moving near the $n$ direction have momenta $p^- \sim Q$,
$p^+ \sim Q \lambda^2$ and $\mathbf{p}_\perp \sim Q \lambda$, where $\lambda \ll
1$ and $Q \gg \lqcd$. $Q$ is the large energy scale, and is of order $m_b$ for
$B$-decays.  Often the choice $\lambda\sim\sqrt{\Lambda_{\rm QCD}/Q}$ is made
for inclusive processes in \SCETa, but parametrically larger choices for this
small parameter are allowed. For exclusive processes in \SCETb we use $\eta$
rather than $\lambda$ for the expansion parameter to avoid confusion, since here
$\eta\sim \Lambda_{\rm QCD}/Q$.  The $p^+$, $p^-$ and $\mathbf{p}_\perp$ scales
are coupled by the on-shell condition $p^+ p^-=\mathbf{p}_\perp^2$, and despite
the hierarchy $p^-\gg p_\perp \gg p^+$, the effective theory must simultaneously
deal with the low energy scales associated with $p_\perp$ and $p^+$.


\subsection{Comparison with the NRQCD Method of Regions}

One can evaluate Feynman integrals in NRQCD using the method of
regions~\cite{Beneke:1997zp} (also called the threshold expansion),
which divides up an integral into hard, soft, potential and usoft
contributions based on scaling for the loop momentum.  The idea is
that in dimensional regularization the sum of these contributions
exactly reproduces the full theory diagram, so
\begin{eqnarray} \label{Amor}
  A^{\rm full}(p_i) =\prod_j \int \!\! d^dk_j \: F(p_i,k_j) = \sum_{{\rm regions}\ \ell}
\prod_j \int \!\! d^dk_j \: F^{(\ell)}(p_i,k_j) \,.
\end{eqnarray}
Dimensional regularization is required here because there are
cancellations between UV and IR divergences from different regions,
which occur if $\epsilon_{\rm IR}=\epsilon_{\rm UV}=\epsilon$.
Eq.~(\ref{Amor}) does not define an effective field theory, but it is
sometimes taken as a way of defining EFT contributions to amplitudes,
by demanding that in dimensional regularization each mode in the
effective Lagrangian should reproduce a term from a region on the RHS.
Although Eq.~(\ref{Amor}) is quite powerful, a few points must be
treated carefully: i) The division of regions is gauge
dependent.\footnote{A well known example in NRQCD is applying the
method of regions to potential and soft contributions.  Another
example is the division between $(E,{\bf p})\sim (m,m)$ and a new
region $(E,{\bf p})\sim (m,mv)$ that shows up at fourth order in the
$v$ expansion in Coulomb gauge.}  ii) The requirement that scaleless
integrals be set to zero, $1/\epsilon_{\rm UV}-1/\epsilon_{\rm IR}=0$,
does not allow all UV divergences to be treated by counterterms, nor a
verification that every IR divergence has a correspondence with the
full theory.  iii) In Eq.~(\ref{Amor}) one must sum over all possible
momentum routings in loops to determine the relevant regions (or
consider the scaling of all combinations of loop momenta and external
momenta). This is because it is individual propagators in the EFT that
belong to a region rather than the loop momenta. 

In the remainder of this section we will explore how the terms
$(1/\epsilon_{\rm UV} - 1/\epsilon_{\rm IR})$ allow a residual freedom
in associating amplitudes with degrees of freedom beyond that in
Eq.~(\ref{Amor}).  The treatment of these terms effects the
correspondence of the EFT modes with physical regions.  Lets consider
a one-loop graph in NRQCD with contributions from different regions.
The soft contribution depends on external soft scales such as the
momentum transfer $\mathbf{r}$, and has the (schematic) form
\begin{eqnarray}
I_{\text{soft}} &=& \frac{A}{\epsilon_{\text{UV}}} 
  +  \frac{B}{ \epsilon_{\text{IR}}} + f(\mathbf{r},\mu).
\label{2.11}
\end{eqnarray}
As the momentum in the soft graphs vanishes, the graph matches on to an
usoft diagram, with the (schematic) structure
\begin{eqnarray}
I_{\text{usoft}} &=& - \frac{B}{ \epsilon_{\text{UV}}} 
  +  \frac{C}{ \epsilon_{\text{IR}}} + g(E,\mu) \,,
\label{2.12}
\end{eqnarray}
where the coefficient $B$ is the same as in Eq.~(\ref{2.11}).  The usoft
graph depends on external usoft scales such as the energy $E$.  The IR
divergences in the usoft sector are true IR divergences.  They arise if one
is computing an IR divergent quantity such as an on-shell Green's function, but
cancel in measurable quantities such as physical scattering cross-sections and
bound state energies.

\begin{figure}
 \begin{center}
\includegraphics[width=6cm]{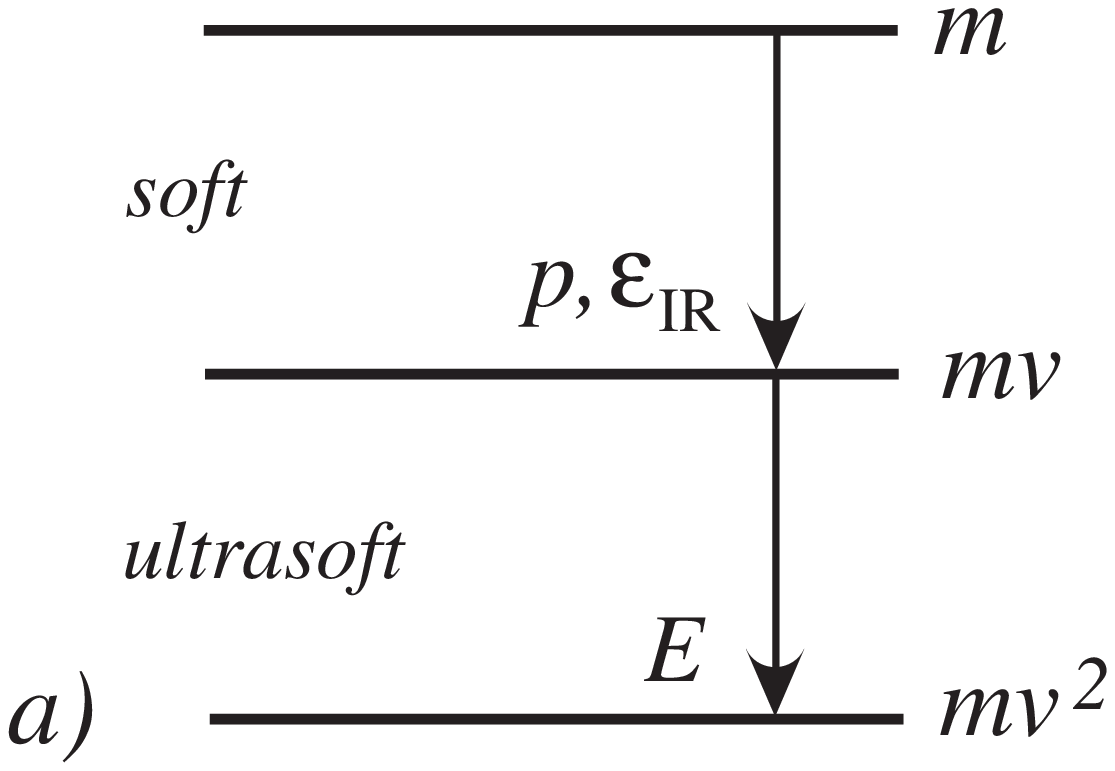}\qquad\quad
\includegraphics[width=6cm]{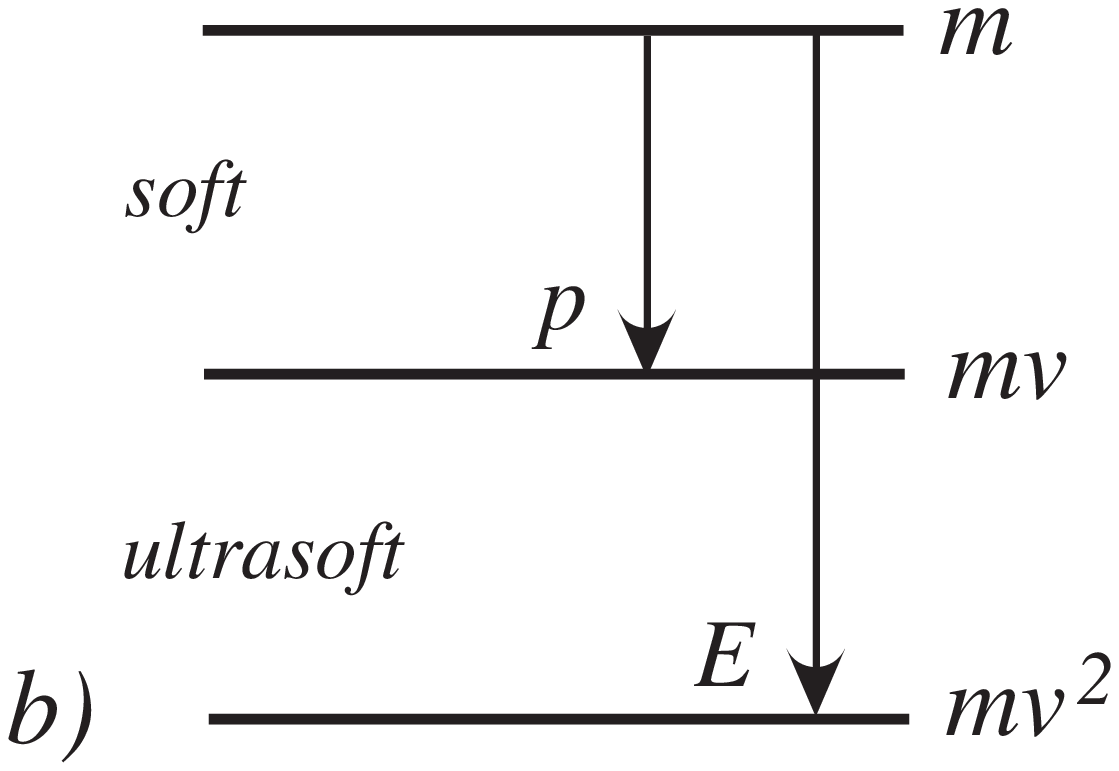}
 \end{center}
\vskip-0.7cm
 \caption{Comparison of two setups for the soft and usoft contributions to 
  an NRQCD Feynman graph. In a) the $1/\epsilon$ divergences at the intermediate
  scale $mv$ cancel between the soft and usoft contributions. In b) there
  are no IR divergences at this intermediate scale.
  \label{fig:pict1}}
\end{figure}

The IR divergence in the soft graphs and the ultraviolet divergence in
the usoft graph are at the intermediate scale $mv$ and cancel each
other; they are not true divergences of the theory.  Since in the
method of regions a rule is applied that scaleless integrals are set
to zero, one is free to consider the $B/\epsilon$ terms canceling in
the sum of Eqs.~(\ref{2.11},\ref{2.12}) to give
\begin{eqnarray} \label{MoRsum}
I_{\text{soft}} + I_{\text{usoft}} &=& \frac{A}{\epsilon_{\text{UV}}} 
  +  \frac{C}{\epsilon_{\text{IR}}} + f(\mathbf{r},\mu) + g(E,\mu).
\label{2.13}
\end{eqnarray}
This interpretation leads to the picture shown in
Fig.~\ref{fig:pict1}a, where the IR soft effects and UV ultrasoft
effects meet at $mv$.
Thus this particular mapping of the method of regions with effective
theory amplitudes does not lead to simultaneously having degrees of
freedom for soft and usoft gluons at a scale $\mu$, but instead in the
renormalization group the soft contribution covers the momentum region
between $m$ and $mv$ and the usoft contribution covers the region
between $mv$ and $mv^2$.

In an effective field theory, the low energy effective Lagrangian is
usually required to reproduce all the IR effects of the original
theory from the start. In NRQCD, this plus a manifest power counting
in $v$ requires that the effective theory include soft and usoft
degrees of freedom at the same time to reproduce the dependence of the
full theory on the momentum transfer $\mathbf{r}
\sim mv$ and the energy $E \sim mv^2$, which are considered IR scales
at $\mu=m$ where the effective Lagrangian is constructed, as found in
Ref.~\cite{Luke:1999kz}. This happens because the power counting links
the scales $mv$ and $mv^2$. A similar result holds for SCET---the SCET
Lagrangian at the hard scale $Q$ must simultaneously include collinear
and usoft degrees of freedom to correctly reproduce the IR behavior of
QCD with a power counting in
$\lambda$~\cite{Bauer:2000ew,Bauer:2000yr}.  In the effective theory,
counterterms must be added for all $1/\epsU$ divergences, including
those from scaleless integrals as is familiar from the study of
HQET~\cite{Manohar:1996cq,Manohar:1997qy,Manohar:2000dt}.  The
counterterm structure can not depend on the choice of IR regulator.

The above line of reasoning gives the picture Fig.~\ref{fig:pict1}b,
rather than Fig.~\ref{fig:pict1}a. In NRQCD this picture has been
implemented in the past by a procedure referred to as a pullup
mechanism~\cite{Manohar:2000kr,Hoang:2001rr}.  One modifies
Fig.~\ref{fig:pict1}a by pulling up the usoft modes to the scale $m$,
and then subtracting the usoft contribution from the soft contribution
in the region between $m$ and $mv$. The picture now looks like
Fig.~\ref{fig:pict1}b.  This modification leaves the usoft
contribution Eq.~(\ref{2.12}) with the same form, but it must now be
included in the theory between the scales $m$ and $mv^2$. The usoft
integral in the region between $m$ and $mv$ has the form
\begin{eqnarray}
I_0 &=& -\frac{B}{ \epsilon_{\text{UV}}} +\frac{B}{ \epsilon_{\text{IR}}} ,
\label{2.15}
\end{eqnarray}
so it corresponds to a $(1/\epsilon_{\rm UV} - 1/\epsilon_{\rm IR})$
that we are free to move between amplitudes from the point of view of
Eq.~(\ref{Amor}).  When Eq.~(\ref{2.15}) is subtracted from
Eq.~(\ref{2.11}) it gives the modified soft contribution
\begin{eqnarray}
I_{\text{soft}} &=& \frac{A}{  \epsilon_{\text{UV}}} 
  +  \frac{B}{ \epsilon_{\text{UV}}} + f(\mathbf{r},\mu).
\label{2.16}
\end{eqnarray}
The IR divergent parts of the soft graphs have been removed. There is
no need for any cancellation of divergences between the soft and usoft
degrees of freedom, and they can be renormalized separately. This is
important, because it is known that the soft and usoft degrees of
freedom have independent coupling constants, and the cancellation of
divergences between the two can be problematic in renormalization
group improved perturbation theory. One of the consequences of the
pullup mechanism is that the anomalous dimensions in the soft sector
are given by Eq.~(\ref{2.16}), and are proportional to $A+B$. As a
short cut one can compute the anomalous dimension from the original
form in Eqs.~(\ref{2.11},\ref{2.12}) by treating the IR divergence as
though it were a UV divergence in Eq.~(\ref{2.11}), and taking the UV
divergence in Eq.~(\ref{2.12}) to be at the hard scale. This is the
procedure that has been followed in previous NRQCD and NRQED
computations~\cite{Luke:1999kz,Manohar:1999xd,Manohar:2000kr,Hoang:2001rr,Manohar:2000rz,Hoang:2002ae,Hoang:2003ns,Hoang:2005dk},
and is known to be necessary to correctly reproduce the high-order
logarithmic terms in Lamb shifts and hyperfine splittings which are
determined from independent fixed order QED computations.  We will see
that the zero-bin subtraction automatically gives the final result
implied by the pullup mechanism, so that the soft integrals with the
zero-bin subtraction have the form in Eq.~(\ref{2.16}), rather than
that in Eq.~(\ref{2.11}). Thus the zero-bin formulation no longer
requires implementing a pullup by hand.

An alternative setup has been explored in Ref.~\cite{Pineda:2000sz,
  Pineda:2001ra, Pineda:2001et,Penin:2004ay}, which keeps Fig.~\ref{fig:pict1}a
but postpones obtaining power counting in $v$. In this setup one starts by
matching onto a purely soft theory with a power counting in $1/m$ and constrains
states to $E=0$ so the IR divergences are properly reproduced. The power
counting in $v$ is not yet manifest in this theory. The soft theory is then
matched at a soft cutoff scale $\Omega_s$ onto an usoft theory, pNRQCD,
where $E\ne 0$ and the velocity power counting is restored.  The key is to
maintain $\Omega_s$ as a variable in the usoft theory so that when required
it can be run down from the scale $m$ including the required correlation with
the usoft scale.  While this setup seems to differ from the pullup, in
several cases it has been shown that both methods give equivalent final answers
for observables with log summation.

Note that so far an analogous setup that would avoid simultaneously
introducing usoft and collinear modes does not exist for ${\rm
SCET}_{\rm I}$.  It is known that one can avoid simultaneously
introducing hard-collinear $p^2\sim Q\Lambda_{\rm QCD}$ and collinear
$p^2\sim \Lambda^2_{\rm QCD}$ modes, by matching \SCETa onto \SCETb as
discussed in Ref.~\cite{Bauer:2002aj}. This postpones obtaining the
final power counting until one matches onto \SCETb, and the
(hard-collinear)--(collinear) setup is similar to the soft-usoft setup
in pNRQCD. Just as in pNRQCD one must in general maintain the matching
scale as a free parameter with this method of matching \SCETa onto
\SCETb. Alternatively, one can match QCD directly onto
\SCETb~\cite{Bauer:2001yt,Hill:2002vw}.


\subsection{Summing Logarithms}
\label{sec:rge}

The dynamical relations $p^2=2mE$ and $p^+p^-=\mathbf{p}_\perp^2$ have
implications for the summation of logarithms using renormalization group
evolution. In NRQCD and NRQED, one must simultaneously run from $m\to p$ and
$m\to E$ in the soft and usoft sectors of the theory~\cite{Luke:1999kz}.
This is implemented in practice by using the velocity renormalization
group~\cite{Luke:1999kz}. Graphs in the theory are evaluated using two different
$\mu$ parameters, $\mu_S$ for soft and potential loops, and $\mu_U$ for
usoft loops. One then sets $\mu_S=m\nu$, $\mu_U=m \nu^2$, and runs from
$\nu=1$ to $\nu=v$. This procedure is also referred to as one-stage or
correlated running~\cite{Manohar:2000mx}, and corresponds to
Fig.~\ref{fig:pict1}b. In NRQED, this correlated running is required to
correctly compute the $\alpha^8\ln^3\alpha$ Lamb shift and $\alpha^7\ln^2\alpha$
hyperfine splittings for Hydrogen and positronium, as well as the
$\alpha^3\ln^2\alpha$ positronium widths~\cite{Manohar:2000rz,Manohar:2000mx}.
It has also been shown to be necessary to properly implement counterterms in
subdivergences at three-loops~\cite{Hoang:2001mm,Hoang:2003ns}.  In some cases,
correlated running is not essential, and one can follow an alternative procedure
called two-stage or uncorrelated running, in which $\mu_S$ is scaled from $m$ to
$mv$, and $\mu_U$ from $mv$ to $mv^2$, corresponding to
Fig.~\ref{fig:pict1}a.\footnote{See Ref.~\cite{Manohar:2000mx} for the precise
  relation between these two methods.  The anomalous dimensions have different
  definitions in the two approaches, so the single $\ln$ terms agree. There is a
  difference only for $\ln^2$ and higher terms if the anomalous dimension does
  not factor into separate soft/potential and usoft pieces.}  The summation
of logarithms for the $1/m^2$ QCD potentials can be done with or without the
correlated running~\cite{Manohar:1999xd,Pineda:2001ra,Hoang:2002yy}; both
methods give the same result.

So far no examples where correlated running is essential have been encountered
in SCET.\footnote{See the discussion in Ref.~\cite{Fleming:2003gt} on the
  equivalence of correlated and uncorrelated running at leading order for $B\to
  X_s\gamma$ and $e^+e^-\to J/\psi X$.} This does not mean that correlated
running is not necessary in SCET. In NRQED, correlated running is first required
for computations of recoil corrections at order $v^3$ ($m_e/m_p$ terms in the
Lamb shift), because it is at this order that the potential and usoft
divergences are tied together. Until this order, both correlated and
uncorrelated running give the same result, and it is the Lamb shift computation
which shows that, in general, one should use correlated running. Thus, it is
very likely that only the running of subleading factorization formulas in SCET
will demonstrate the manner in which correlated running is required.


\section{Infrared Modes in  NRQCD and SCET} \label{sec:dblecount}

In its region of validity, an effective field theory needs to systematically
reproduce the IR structure of the full theory order by order in its power
expansion.  In both NRQCD and SCET, a strict interpretation of this requirement
makes it necessary to include distinct fields for different moment regions of
the same physical particle. These distinct fields have different power counting.
Multiple gluon fields were first introduced in Ref.~\cite{Luke:1997ys} for
potential and usoft gluons in NRQCD, with energy and momentum scaling of
order $(E\sim mv^2,p\sim mv)$ and $(E\sim mv^2,p\sim mv^2)$, respectively. We
will not introduce fields for potential gluons since they are not propagating
degrees of freedom. In NRQCD it is also necessary to introduce soft gluons with
momentum scaling $(E\sim mv,p\sim mv)$
~\cite{Beneke:1997zp,Griesshammer:1997wz}.  We use the NRQCD Lagrangians defined
as in~\cite{Luke:1997ys,Manohar:2000hj} with potential quarks and usoft
gluons/quarks, and also soft quarks. We treat soft gluon vertices with a
soft-HQET effective Lagrangian, rather than integrating out the soft quarks as in
Ref.~\cite{Luke:1997ys}. At two-loops and beyond it is important to keep track
of the $i0^+$ in the soft quark propagators $1/(v\cdot k+i0^+)$ and
leaving them in an action facilitates this. In SCET one requires both collinear
and usoft gluons in a theory often called \SCETa or collinear and soft gluons in
a theory called \hbox{\SCETb\!.} We use the Lagrangians from
Ref.~\cite{Bauer:2001yt} for these theories.

\begin{figure} 
\begin{center}
\includegraphics[height=5cm]{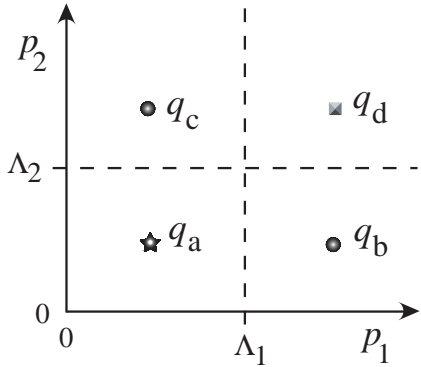} 
\end{center}
\vskip-18pt
\caption[1]{
  Toy model to illustrate the scales captured by an effective theory with
  multiple low energy modes (here $q_a$, $q_b$, and $q_c$).
}\label{fig_UVvsIR}
\end{figure}

One might expect that introducing multiple fields for the same particle would
lead to double-counting problems. In constructing effective theories, one needs
to know not only the power counting for the degrees of freedom, but also
understand the range of scales for which these modes are included in the
effective Lagrangian.  If degrees of freedom overlap in some region of momentum
space, an understanding of how their definitions avoid double counting is
necessary. In the effective theories we study, loop integrals are dominated by
external momenta by construction, so the power counting guarantees that fields
give contributions that can overlap only in UV or IR limits.\footnote{The
  statement actually holds for any renormalization procedure that respects the
  power counting, or at worst requires power counting violating counterterms.}
As a simple toy example consider Fig.~\ref{fig_UVvsIR}.  We imagine that there
are two relevant momentum coordinates $p_1$ and $p_2$, and that physically there
are four interesting sets of momenta labeled $q_a$, $q_b$, $q_c$, and $q_d$
which could be set kinematically or by bound state dynamics. The hard cutoffs
$\Lambda_1$ and $\Lambda_2$ distinguish the momentum regions dominated by these
$q$'s.

Consider first the simplified case where we ignore the $p_2$ axis, and only have
$q_a$ and $q_b$. This situation applies to many physical problems, including
that of integrating out massive particles like the $W$-boson or $b$-quark.  Here
$q_b$ denotes hard fluctuations that are integrated out into Wilson coefficients
$C$, while $q_a$ denotes low energy IR modes. To simplify the renormalization
group evolution and leave symmetries unbroken, it is convenient to trade
$\Lambda_1$ for a scaleless regulator such as dimensional regularization.  The
low energy theory has $C=C(q_b,\mu)$ where $\mu$ is the dimensional
regularization parameter. The effective theory for $q_a$ with a scaleless
regulator takes $\Lambda_1\to \infty$, and thus overlaps with the hard region of
momentum space, but {\em only through ultraviolet effects}. In the effective
theory this double counting is removed by UV counterterms $\propto
1/\epsilon$, as well as through finite terms in the Wilson coefficients.

Now consider both the $p_1$ and $p_2$ axis. Here $q_d$ denotes hard
fluctuations, and $q_a$, $q_b$, and $q_c$ are all low energy modes because they
border a region where one or both of $p_{1,2}$ can be zero. Using dimensional
regularization for each of these modes effectively takes $\Lambda_{1,2}\to
\infty$ for $q_a$; $\Lambda_1\to 0$, $\Lambda_2\to\infty$ for $q_b$; and
$\Lambda_2\to 0$, $\Lambda_1\to\infty$ for $q_c$. Again there is double counting
in the UV which is taken care of by counterterms. However, there is now also a
double counting in the IR. For example, taking $\Lambda_1\to 0$ for $q_b$ runs
into the region for $q_a$. The $\Lambda_1\to 0$ limit is necessary for any
scaleless regulator. In fact this limit is actually important physically,
because we would like to define $q_b$ without reference to $q_a$ in order for it
to be possible that contributions from $q_a$ and $q_b$ can factorize into
independent well defined objects. This double counting in the IR is removed by
the pullup mechanism, which as we will demonstrate, is a consequence of a proper
treatment of the zero-momentum bin, namely $p_2=0$ for $q_c$ and $p_1=0$ for
$q_b$.  In pure dimensional regularization these zero-bin contributions amount to
a correct interpretation of terms $\propto 1/\epsilon_{\rm UV} - 1/\epsilon_{\rm
  IR}$. For example, $1/\epsilon_{\rm UV}$ terms must be canceled by
counterterms while $1/\epsilon_{\rm IR}$ terms match up with IR divergences from
the full theory.

\begin{figure} 
\begin{center}
\includegraphics[height=6.5cm]{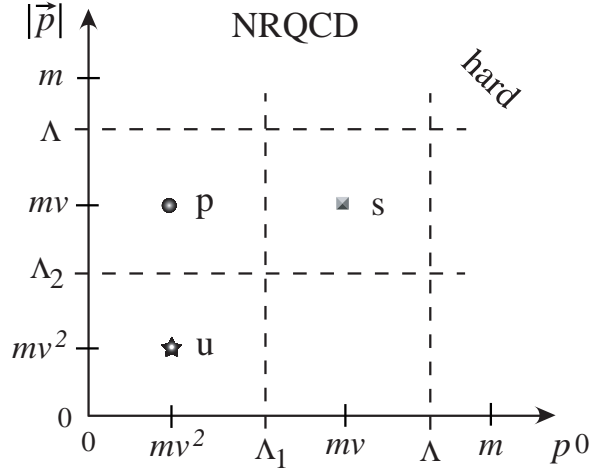}\hspace{0.cm}
\end{center}
\vskip-18pt
\caption[1]{
  a) Scales and momentum modes for nonrelativistic field theories like NRQCD.
  Here s, p, and u denote soft, potential, and usoft respectively. 
}\label{fig_NRQCDUVvsIR}
\end{figure}

We now turn to realistic effective theories and their zero-bin's. For NRQCD the
modes are shown in Fig.~\ref{fig_NRQCDUVvsIR}a. The hard scale is $E \sim m$ or
$\mathbf{p} \sim m$. We have propagating soft and usoft gluons/light quarks with
power counting $E\sim \mathbf{p} \sim mv$ and $E \sim \mathbf{p} \sim mv^2$
respectively, and potential heavy quarks with $E\sim mv^2$ and $\mathbf{p} \sim
mv$, and soft heavy quarks with $E\sim \mathbf{p} \sim mv$.  Hard cutoffs
$\Lambda$, $\Lambda_1$, and $\Lambda_2$ have been introduced to facilitate the
discussion.  Here $\Lambda \sim m $ is an ultraviolet scale below which one uses
the effective theory, and $\Lambda_{1,2}$ divide up the low energy modes.  These
cutoffs will be removed exactly as in the toy example above.
First consider the soft and usoft gluons. In a theory with both present,
the energy/momentum regions we want them to cover are
\begin{eqnarray} \label{regionNRQCD}
 && \mbox{soft gluons:}\qquad\quad\quad \Lambda \gg mv \gtrsim \Omega^s \,, \\
 && \mbox{usoft gluons:}\qquad \Lambda \gg mv^2 \gtrsim \Omega^u \,,
  \nonumber 
\end{eqnarray}
where $\Omega^s$ and $\Omega^u$ are soft and usoft scales denoted by the box and
star in Fig.~\ref{fig_NRQCDUVvsIR}. These scales are usually set by external
variables such as the momentum transfer or energy of the quarks, or by the
nonperturbative scale $\Lambda_{\rm QCD}$. In order to reproduce all possible IR
effects associated with $E\sim mv^2$, the EFT necessarily must have usoft modes
just below the UV scale $\Lambda$. The division in Eq.~(\ref{regionNRQCD}) also
implies that the UV divergences associated with soft and usoft modes contribute
to the same anomalous dimension. This setup corresponds to the result obtained
from the same limits for the cutoffs $\Lambda_{1,2}$ as discussed in the toy
model above. Now double counting must be avoided in the region between $\Lambda$
and $\Omega^s$, between the IR of the soft gluons and the ultraviolet of the
usoft gluons (i.e.\ when virtual momenta for the soft gluons becomes comparable
to virtual momenta for the usoft gluons in loops).

A line of reasoning that gets close to seeing how this is achieved is to
start with soft gluons over the interval $\Lambda \gg mv \sim \Lambda_{1}$ and
usoft gluons over $\Lambda_1 \gg mv^2 \sim \Omega^u$, where we set
$\Lambda_2=\Lambda_1$.  Here $\Lambda_1$ is considered as an intermediate
factorization scale~\cite{Pineda:2000sz, Pineda:2001ra, Pineda:2001et}.  Next
one adds a pullup contribution to the usoft gluons and simultaneously subtracts
the same contribution from the soft gluons as described in
Refs.~\cite{Manohar:2000kr,Hoang:2001rr}. This pulls the upper limit $\Lambda_1$
for the usoft gluons all the way up to the scale $\Lambda$, while at the same
time avoiding double counting between $m$ and $mv$. The pre-pullup setup is
shown in Fig.~\ref{fig:pict1}a and post-pullup in Fig.~\ref{fig:pict1}b.  Taking
into account the zero-bin gives the post-pullup setup. With the pullup, the
contributions from soft and usoft gluons are now as in Eq.~(\ref{regionNRQCD}).
If the cutoffs had been swapped for a scaleless regulator, this would amount to
a proper interpretation of $1/\epsilon$ poles as discussed in
Ref.~\cite{Hoang:2001rr}.

In this paper we show that Eq.~(\ref{regionNRQCD}) is obtained automatically by
carefully considering the zero-bin in the NRQCD effective Lagrangian.  The usoft
gluons are defined all the way up to $\Lambda$, and the subtraction for the soft
gluons is associated with properly removing their zero momentum bin. Thus the
effective Lagrangian gives the full procedure for the evaluation of Feynman
graphs and is decoupled from the additional choice of which regulators to use.
With the zero-bin taken into account there is no longer a need to implement a
separate pullup.

A second type of division between potential and soft IR modes in NRQCD is also
shown in Fig.~\ref{fig_NRQCDUVvsIR}a.  In this case the distinction is solely in
the energy variable $p^0$. Potential gluons are not propagating degrees of
freedom since they have $p^0\sim\!\!\!\!\!\slash\:\, \mathbf{p} $, and so
potential gluon fields should not be introduced (they would have problems with
gauge invariance for example).  The matching of soft gluons onto four-quark
operators with potential coefficients can be thought of in a similar manner to
integrating out a massive particle~\cite{Pineda:1997bj}. On the other hand a
zero-bin subtraction is necessary to distinguish soft and potential quarks.
Their momentum regions and propagators are
\begin{eqnarray}
  && \mbox{soft quarks:}\qquad\quad
    (p^0:\ \ \Lambda > p^0\!\sim\! mv > \Omega^s) \ ;\quad\ \ 
   \mbox{propagator:}\ \ \  \frac{i}{[p^0 +i0^+]}
    \,,
  \\[5pt]
 && \mbox{potential quarks:}\quad 
 (p^0:\ \ \Lambda > p^0 \!\sim\! mv^2 > \Omega^u) \ ;\quad\: 
  \mbox{propagator:}\ \ \  \frac{i}{[p^0 -\frac{{\mathbf{p} }^{\,2}}{2m}
   +i0^+]}
    \,.  \nonumber 
\end{eqnarray}
Without the zero-bin, double counting occurs when $p^0\to 0$ in the soft
propagator, and this reveals itself through the presence of pinch singularities
in soft loop diagrams with quarks and antiquarks, which have the form
\begin{eqnarray}
\int {\rm d}p^0   \frac{1}{[p^0+i0^+] [-p^0 +i0^+]  \cdots } \,.
\end{eqnarray}
Here we will show that the zero-bin subtraction removes all pinch singularities
in the soft regime. At the same time it avoids double counting of soft and
potential contributions.  It is useful to recall that the same is not true in
the method of regions~\cite{Beneke:1997zp}, where the method used to avoid the
pinch singularities is to consider the expansion only after doing the $p^0$
integral by contours. 

Note that in NRQCD there is no issue of a possible potential-usoft overlap
for quarks since propagating usoft quarks are light quarks and therefore of
a different flavor from the potential quarks. 

\begin{figure} 
\begin{center}
\includegraphics[height=6.4cm]{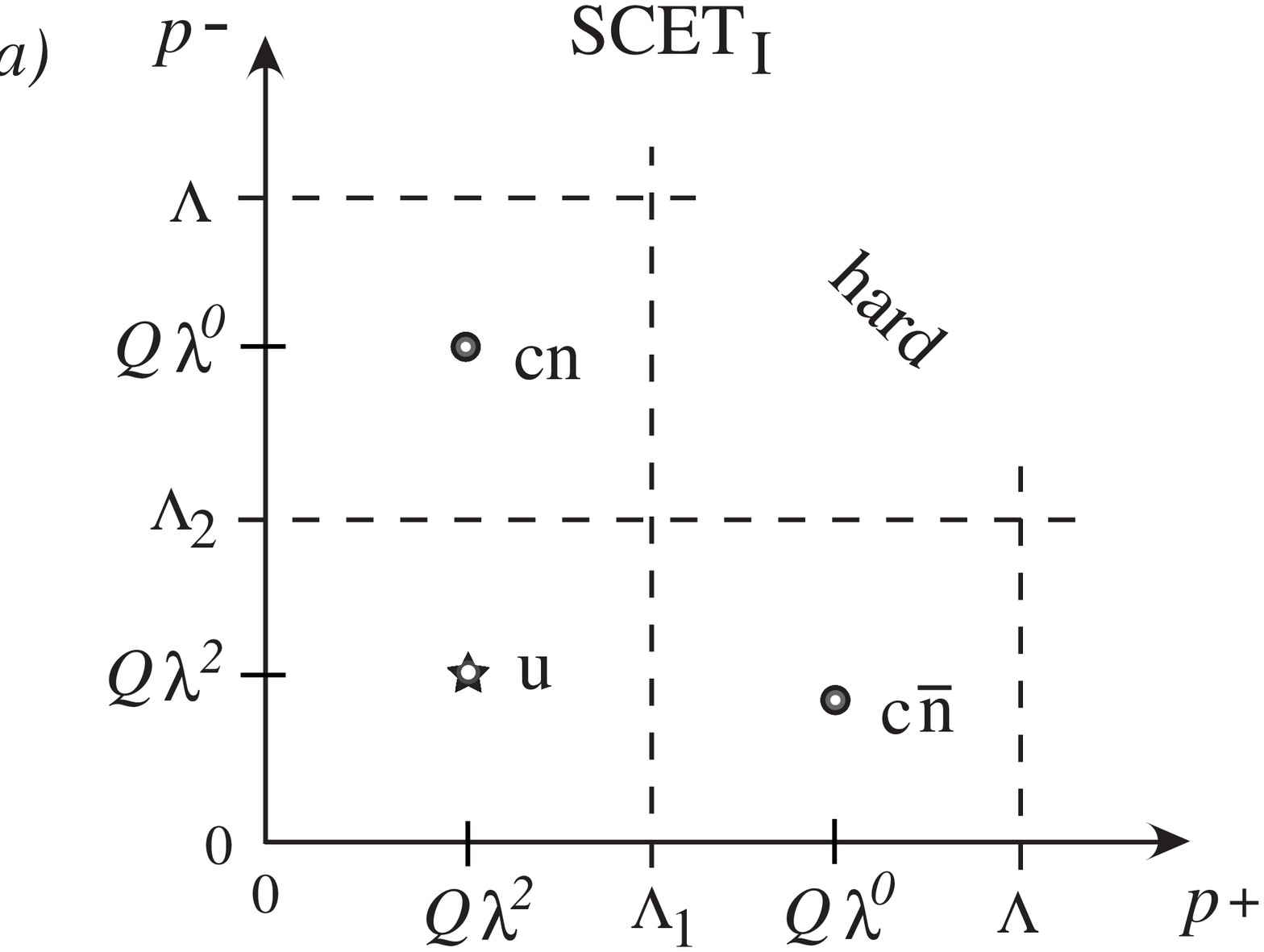}\hspace{-0.1cm}
\includegraphics[height=6.4cm]{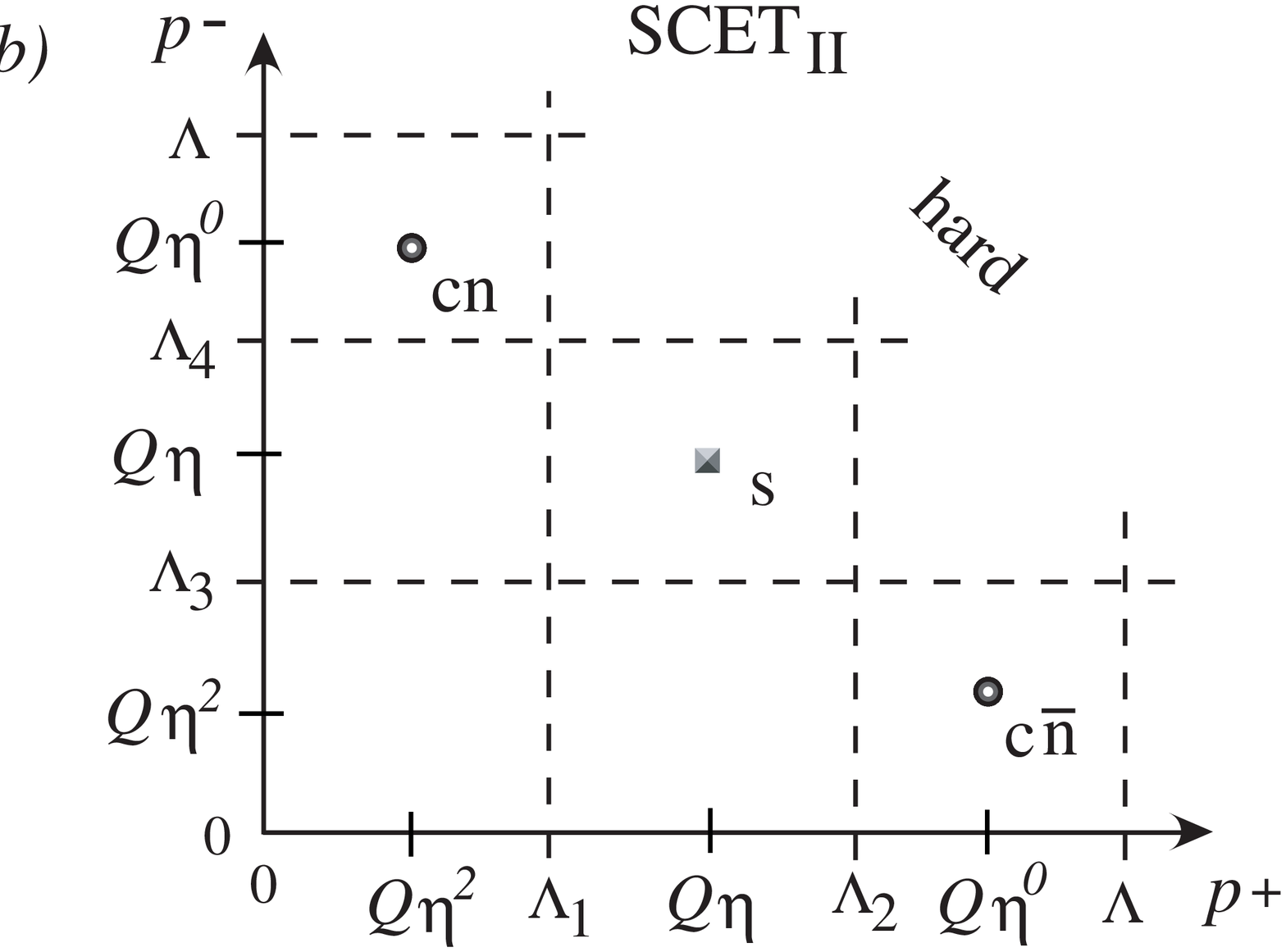} 
\end{center}
\vskip-18pt
\caption[1]{
  a) Scales and momentum modes for \SCETa. Here $cn$, $c\bn$, and $u$ denote
  collinear-$n$, collinear-$\bn$, and usoft modes respectively and
  $\lambda\sim \sqrt{\Lambda_{\rm QCD}/Q}$.  b) Scales and momentum modes for
  \SCETb. Here $cn$, $c\bn$, and $s$ denote collinear-$n$, collinear-$\bn$, and
  soft modes respectively and $\eta\sim \Lambda_{\rm QCD}/Q$.
}\label{fig_SCETUVvsIR}
\end{figure}

So far the discussion was for NRQCD, but similar logic holds for \SCETa. In this
case the hard scale is set by $Q$ and the expansion parameter is $\lambda$.
Below the scale $Q$ we have usoft and collinear gluons with power
counting $(p^+,p^-,p^\perp)\sim Q(\lambda^2,\lambda^2,\lambda^2)$ and
$Q(\lambda^2,1,\lambda)$ respectively. These gluons cover the regions
\begin{eqnarray} \label{regionSCET}
 && \mbox{usoft gluons:}\qquad
    (p^-:\ \ \Lambda \gg Q\lambda^2 \gtrsim \Omega^u) \ ;\quad
     (p^\perp: \ \ \Lambda\gg Q\lambda^2 \gtrsim \Omega^u) \,,
  \\
 && \mbox{collinear gluons:}\qquad 
 (p^-:\ \ \Lambda > Q\lambda^0 \gtrsim \Omega^c_-) \ ;\quad\: 
     (p^\perp: \ \ \Lambda\gg Q\lambda \gtrsim \Omega^c_\perp) \nonumber \,,
\end{eqnarray}
and a common region for $p^+$: $\Lambda \gg Q\lambda^2 \gtrsim \Omega^u$. These
regions are denoted by a circle and star in Fig.~\ref{fig_SCETUVvsIR}a, where we
have also included a second collinear region c$\bn$ for later convenience. In
Eq.~(\ref{regionSCET}) we have used a common UV scale $\Lambda$ and a common
usoft IR scale $\Omega^u$ in the $-$ and $\perp$ components which have power
counting $\sim \lambda^2$. In \SCETa we see that we must avoid double counting
in the region $\Lambda$ to $\Omega_-^c$ for the $p^-$ momenta, and the region
$\Lambda$ to $\Omega_\perp^c$ for the $p^\perp$ momenta.\footnote{For
  typographical convenience we use $\Omega_-$ although a superscript,
  $\Omega^-$, would be more appropriate.}  As in NRQCD, this is achieved by a
proper treatment of the zero momentum bin for collinear fields (which implements
a pullup in SCET).

Finally, we can consider the theory \SCETb which has soft and collinear gluons
with power counting $(p^+,p^-,p^\perp)\sim Q(\eta,\eta,\eta)$ and
$Q(\eta^2,1,\eta)$ respectively, for $\eta\sim \Lambda_{\rm QCD}/Q$.  Here
the regions are
\begin{eqnarray} \label{regionSCET2}
 && \mbox{soft gluons:}\qquad\quad\quad
    (p^-:\ \ \Lambda \gg Q\eta \gtrsim \Omega^s) \ ;\quad\ \ 
     (p^+: \ \ \Lambda\gg Q\eta \gtrsim \Omega^s) \,,
  \\
 && \mbox{collinear gluons:}\qquad 
 (p^-:\ \ \Lambda > Q\eta^0 \gtrsim \Omega^c_-) \ ;\quad\: 
     (p^+: \ \ \Lambda\gg Q\eta^2 \gtrsim \Omega^c_+) \nonumber \,,
\end{eqnarray}
and for convenience a common region for $p^\perp$: $\Lambda \gg Q\eta \gtrsim
\Omega^s$.  Both soft and collinear modes describe non-perturbative fluctuations
close to the mass shell, $p^2\sim Q^2\eta^2\sim \Lambda_{\rm QCD}^2$.
Interactions between these modes are offshell~\cite{Bauer:2001yt} by an amount,
$p_{hc}^2 \sim Q^2 \eta$, and dependence on this momentum is integrated out,
appearing in the coefficient functions for mixed soft-collinear operators. Here
double counting in \SCETb occurs when a collinear momentum overlaps the soft
region, and when a soft momentum overlaps the collinear region. This case
differs from our discussion of NRQCD and \SCETa because here the overlapping
modes have the same $p^2$, but differ in their rapidity $y$, or more
conveniently their value of
\begin{eqnarray} \label{zetap}
  \zeta_p = e^{2y} = \frac{p^-}{p^+} \,.
\end{eqnarray}
The $n$-collinear modes have $\zeta_p\gg 1$, the soft modes have $\zeta_p\sim
1$, and the $\bn$-collinear modes have $\zeta_p \ll 1$.  Consider a process for
which $\bn$-collinear modes are irrelevant. Double counting is avoided by a
proper treatment of the zero-bins: the ``$p^-=0$ bin'' for $n$-collinear modes and
the ``$p^+=0$ bin'' for soft modes. However, here double counting of a physical IR
region in QCD requires a correlated change in the $+$ and $-$ momenta: $p^-$
gets small while $p^+$ gets big for collinear, and $p^-$ gets big while $p^+$
gets small for soft. The implementation of zero-bin's in \SCETb is discussed
further in section~\ref{sec:scetb}.

The treatment of the zero-bin ensure that the double counting is removed in the
infrared, and that the overlap in the ultraviolet is properly handled by
renormalization in the effective theory irrespective of the choice of regulator.
If dimensional regularization is used to regulate both the IR and UV, then the
added contributions are scaleless loop integrals that appear to be zero.  In
logarithmically divergent integrals, this occurs because the integral is the
difference of UV and IR divergences, $0=1/\epsU-1/\epsI$, since there is only
one $\epsilon$. If these added contributions are ignored, then one must be
careful to properly interpret the divergences as UV or IR.  This conversion of
IR to UV divergences has been used implicitly in much of the NRQCD and SCET
literature.  However, if one wants to fully understand the physical significance
of certain divergences or use another regulator in the UV or IR then explicitly
including the subtractions discussed here is necessary.


\section{Zero-Bin Subtractions} \label{sec:zerobin}

\begin{figure} 
\begin{center}
\includegraphics[height=5cm]{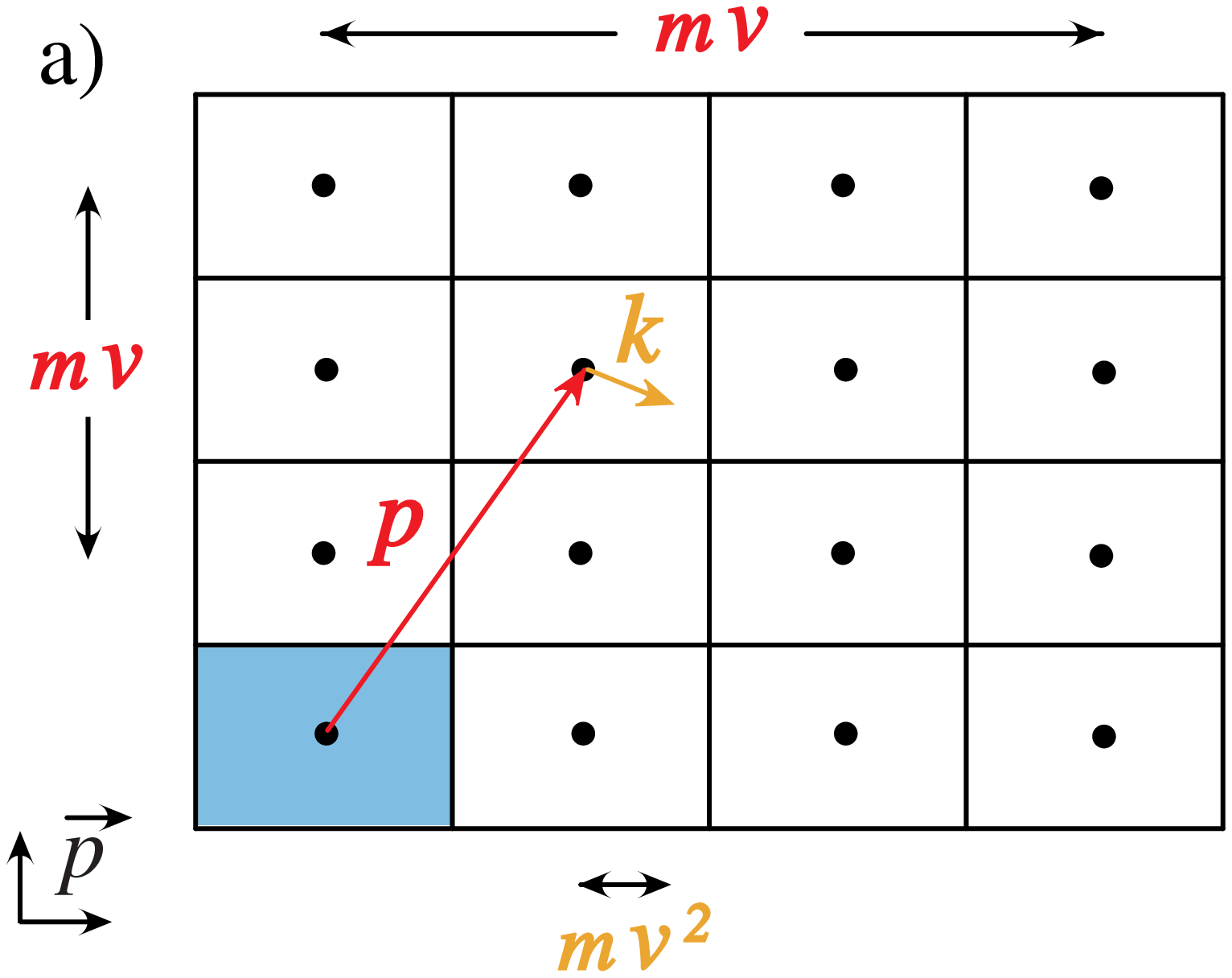}\qquad\qquad
\includegraphics[height=5cm]{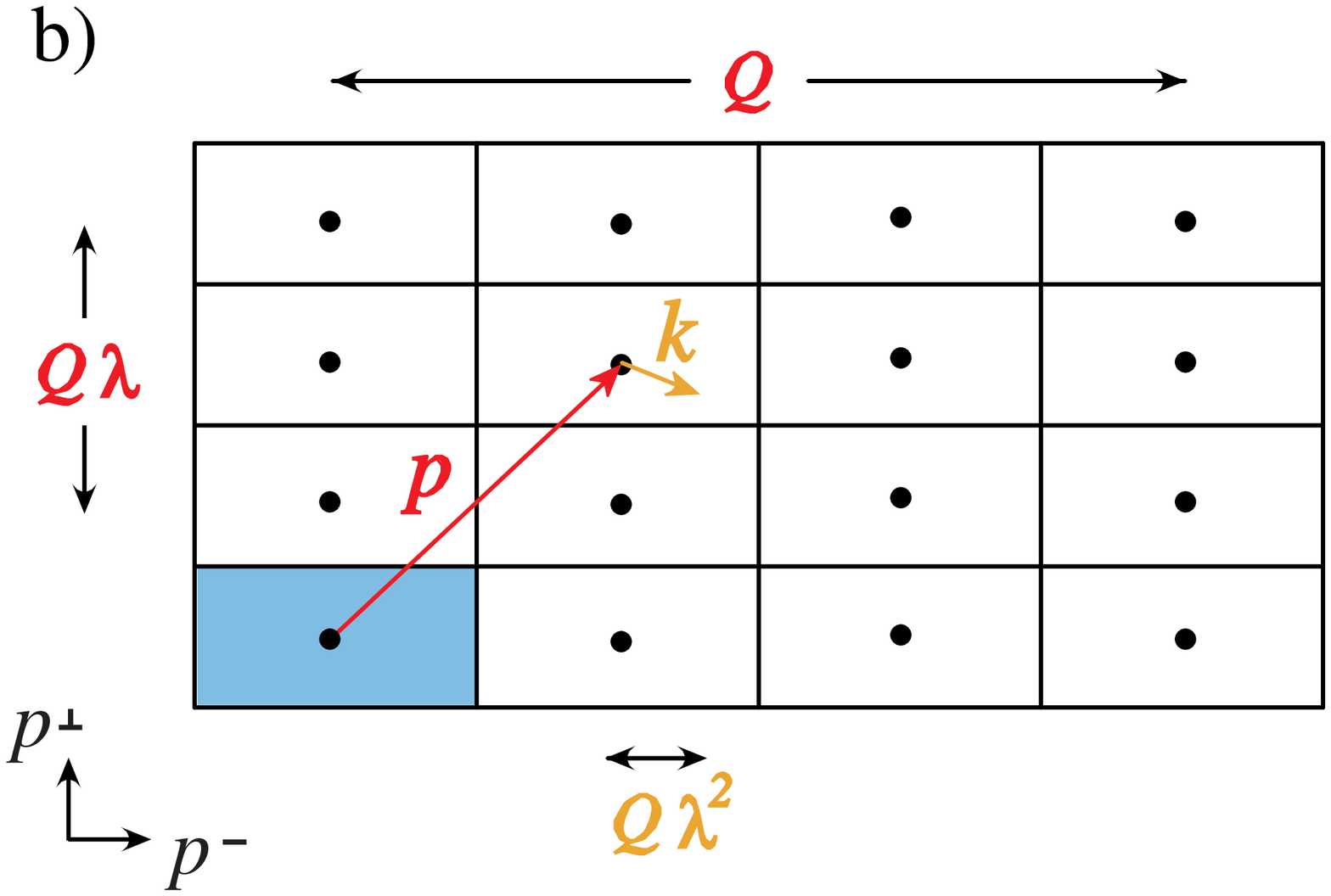}
\end{center}
\vskip-18pt
\caption[1]{
  Label and Residual momenta for a) NRQCD quarks and b) \SCETa. In both cases $p$
  denotes a large momentum, and labels a particular box, whereas $k$ is a small
  momentum, and gives the final momentum location relative to the reference
  momentum point in the box labeled by $p$.
\label{fig:grids}}
\end{figure}

We start by reviewing how the relevant momentum scales are separated in the
effective theory using labeled fields.  In NRQCD, one first removes the large
mass $m$ of the quark (and antiquark) from the problem just as in heavy quark
effective theory (HQET)~\cite{Georgi:1990um}.  The total momentum of the quark,
$P^\mu$, is written as the sum $P^\mu = m {\rm v}^\mu + q^\mu$, where ${\rm
  v}^\mu \equiv (1,0,0,0)$. This subtracts $m$ from all the energies.  The
residual momentum $q^\mu$ is much smaller than $m$, and contains the
non-relativistic energy $E \sim mv^2$ and momentum $p \sim mv$ of the particle,
where $v$ is a scaling parameter of order the typical relative velocity between
the heavy quarks.  This mixes different powers of $v$. As shown in
Ref.~\cite{Luke:1999kz}, it is useful to make a further division of $q^\mu$,
$q^\mu = p^\mu + k^\mu$, where $p^\mu$ is of order $mv$, and $k^\mu$ is of order
$mv^2$. This second separation allows the power counting in $v$ to be manifest
in the effective theory.  The break-up of $q^\mu$ is shown schematically in
Fig.~\ref{fig:grids}a.
One breaks momentum space into a \emph{discrete} variable $\mathbf{p}$ of order
$mv$, and a continuous variable $k^\mu$ of order $mv^2$. Often $k$ is referred
to as the residual momentum. The discrete label $\mathbf{p}$ does not have a
time-component for quarks, since the energy is of order $mv^2$. The entire
$q^\mu$ momentum space is covered by integrating over $k^\mu$ and summing over
the labels $\mathbf{p}$. Quarks are described by fields $\psi_{\mathbf{p}}(x)$,
with an explicit label $\mathbf{p}$, and the momentum $k^\mu$ is the Fourier
transform of $x$. Similarly, soft gluons with energy and momentum of order $mv$
are described by gauge fields $A^\mu_{p}(x)$ with a four-vector label $p^\mu$.
Ultrasoft gluons with energy and momentum of order $mv^2$ are described by gauge
fields $A^\mu(x)$. Those unfamiliar with how the field theory with label and
residual momenta works are referred to Ref.~\cite{Luke:1997ys} or the example in
the next section.

An analogous procedure was applied to SCET in
Ref.~\cite{Bauer:2000yr,Bauer:2001ct}. In SCET, one breaks up the collinear
momentum into a label and residual momentum, $P^\mu = p^\mu + k^\mu$.  The label
$p$ contains the $Q$ and $Q \lambda$ pieces of the momentum, and the Fourier
transform of the coordinate $x$ is the $Q \lambda^2$ part of the momentum $k$.
Unlike NRQCD, the collinear interactions can still change the large label
momentum $\sim Q$.  In a theory referred to as \SCETa, one
has collinear quark and gluon fields $\xi_{n,p}(x)$ and $A_{n,p}^\mu(x)$, where
the label $p$ has minus and $\perp$ components, and usoft quark fields
$A^\mu(x)$ which have energy and momentum of order $Q\lambda^2$. The \SCETa
decomposition is shown schematically in Fig.~\ref{fig:grids}b. In a theory
called \SCETb one has collinear quarks and gluons, $\xi_{n,p}$, $A^\mu_{n,p}$
with $p^-$ label momenta, and soft quarks and gluons, $q_{s,k}$, $A^\mu_{s,k}$
with $k^+$ label momenta. Here soft fields have plus-momenta much bigger than
their collinear counterparts, and collinear fields have minus-momenta larger than
their soft counterparts.

The quark and soft gluon fields in NRQCD and the collinear quark and gluon
fields in SCET will be referred to as labeled fields. As is clear from
Fig.~\ref{fig:grids}, labeled fields must have a non-zero value for their label,
i.e.\ they must be outside the zero-bin. Otherwise, they cover the same momentum
region as the usoft fields. \emph{This is implemented at the level of the
  effective theory Lagrangian by requiring that all labeled fields in the
  Lagrangian have a non-zero value for the label.}  Terms in the effective
theory Lagrangian have sums over the field labels, and all such sums are over
non-zero values of the labels. One then has to carefully derive the rules for
effective theory graphs including this constraint on the labels.

In loop integrals, one finds expressions which involve a
sum over labels and an integral over the residual momentum, which can be
converted to an integral over the entire momentum space~\cite{Luke:1999kz},
\begin{equation}
\sum_{p} \int {\rm d} k \longrightarrow \int {\rm d} p \:,
\label{1.01}
\end{equation}
as is clear from Fig.~\ref{fig:grids}. Here the label $p$ denotes a generic
label, such as $\mathbf{p}$ for NRQCD quarks, $(\bar n \cdot
p,\mathbf{p}_\perp)$ for collinear SCET quarks, etc.  One residual momentum is
removed for each sum over a label momentum.\footnote{From the reparameterization
  invariance~\cite{Luke:1992cs} in splitting $p+k$ it is equivalent to think of
  this as first fixing the lattice of $p$'s and adding the integrals over $k$,
  or as using the freedom in the choice of $p$'s to fix $k$ and then extending
  the sum over $p$'s to an integral.} Any remaining residual momenta appear as
integrals in their own right, providing a proper implementation of the multipole
expansion.\footnote{See~\cite{Grinstein:1997gv} for why this is relevant to
  power counting in NRQCD.} In using equations such as Eq.~(\ref{1.01}), one
should imagine using Fig.~\ref{fig:grids} with explicit bins (i.e.\ hard
cutoffs) for the $k$ integrals. Once the final expressions are derived, one can
evaluate the integrals in dimensional regularization, treating the bin sizes as
infinite as discussed in section~\ref{sec:dblecount}. Eq.~(\ref{1.01}) is only
true if we sum over all $p$, including $p=0$.  The replacement in
Eq.~(\ref{1.01}) is what we will call the result for the naive integral,
generically denoted with a tilde, $\tilde I$.

We can finally formulate the zero-bin subtractions mentioned at the
beginning of this article.  The sum on $p$ in Eq.~(\ref{1.01}) is over $p \not =
0$ because the effective Lagrangian terms are a sum over $p \not = 0$.  The
restriction $p \not =0$ in Eq.~(\ref{1.01}) modifies the right hand side. In a
Feynman graph, let $F\left(\left\{p_i\right\},\left\{k_i\right\}\right)$ be the
integrand, including all the momentum conserving $\delta$-functions and label
preserving Kronecker-$\delta$'s at the vertices, where $i$ runs over all the
internal propagators.  When integrating over a function $F$, the correct form of
Eq.~(\ref{1.01}) is actually
\begin{equation}
\sum_{ \vee \{p_i \not=0 \} }  \int \prod_i {\rm d} k_i\:  
  F\left(\left\{p_i\right\},\left\{k_i\right\}\right) \longrightarrow 
   \int  \prod_i {\rm d} p_i\: \Big[  F\left(\left\{p_i\right\}\right) 
   - \sum_{j \in U}    F^{\rm sub}_j\left(\left\{p_i\right\}\right) \Big]
  \,.
\label{1.02}
\end{equation}
On the l.h.s.\ the sum is over all label momenta avoiding the zero bins, $p_i=0$.
On the r.h.s.\ we integrate $p_i$ over all of momentum space and the second term
subtracts the contributions from regions $j\in U$ where one or more $p_i$
vanish. The set of such regions $U$ can be broken up into $U_i$ where $p_i=0$,
$U_{ij}$ where $p_i=0$ and $p_j=0$, etc. The subtractions over $U$ are defined
iteratively by first subtracting over each $U_i$, then adding back $U_{ij}$,
subtracting $U_{ijk}$, etc. Note that since the $p_i\not=0$ constraint comes
from the fields in the Lagrangian, it is implemented at the level of propagators
in a graph, i.e.\ for each internal line, not for each loop momentum. Once the
momentum conserving delta functions are accounted for the subtractions are
implemented at the level of the full integrand.\footnote{If there are less sums
  over label momenta $p_i$ than integrals $dk_i$ then these extra integrals over
  $k$'s will appear on the RHS with corresponding dependence in $F$. These extra
  $k$'s are momenta that are truly small for the physical process, see
  Ref.~\cite{Luke:1997ys}. For simplicity this complication was suppressed in
  writing Eq.~(\ref{1.02}), since it is not the most important aspect.  }

Each of the $p_i\ne 0$ terms in Eq.~(\ref{1.02}) represent the full label on a
field, which for example will be a four-vector for soft gluons in NRQCD,
$q^\mu$, and the $q^\mu= q^- n^\mu/2 + q_\perp^\mu$ components for a collinear
quark in \SCETa. The proper subtraction integrand $F^{\rm sub}$ is obtained from
$F$ by assigning a scaling to all the $p_i$ appropriate for the zero-bin region,
and expanding in powers of the momentum which vanishes in the zero-bin (as we
scale towards this region with the power counting parameter in the sense of a
standard OPE).  For example, if we sum over $q^\mu\ne 0$ for a soft momentum
(i.e.\ order $mv$) in NRQCD, then we define $F^{\rm sub}$ by an expansion of the
integrand $F$ in powers of $q^\mu$ by assuming that $q^\mu\sim v^2$, i.e.\ by
assuming, for the purposes of the expansion, that $q^\mu$ is usoft. This
subtraction is done at the level of the integrand.  The expansion is done to
high enough order that the resulting integrand $F- F^{\rm sub}$ vanishes in the
zero-bin, i.e. vanishes as $q^\mu \to 0$.

Note that there is a freedom to define a scheme which leaves a finite integrand
in the scaling limit since this just moves finite pieces around between the
matching and matrix elements. We will use the scheme where all singularities are
removed but not finite pieces. Also note that in a given subtraction some terms
will be power divergences which in dimensional regularization are set to zero.
This implies that if we had subtracted additional polynomial pieces they would
not change the result of loop graphs in dimensional regularization since they
integrate to zero.

To show that the RHS of Eq.~(\ref{1.02}) provides the proper implementation of
the zero-bin independent of imposing hard momentum cutoffs we use a logic
similar to section~\ref{sec:dblecount} in discussing
Figs.~\ref{fig_UVvsIR}--\ref{fig_SCETUVvsIR}. The $\sum_{p\ne 0}$ can be turned
into a full integral if we add the $p=0$ bin, but we must subtract it again. The
subtraction term has only an integral over residual momentum, but when we send
the hard cutoffs on the sides of the zero-bin to $\infty$ the subtraction is
also integrated over all momentum. Thus we end up subtracting terms derived from
the scaling limit of the original integrand integrated over all of momentum
space. The full integrand on the r.h.s.\ of Eq.~(\ref{1.02}) ensures that we do
not double count the zero-bin because the integrand vanishes when the loop
momentum is sent towards the zero-bin momenta. This ensures there is no double
counting in the IR. Any double counting in the UV is taken care of by Wilson
coefficients and renormalization as usual.

It is worth emphasizing that the result in Eq.~(\ref{1.02}) applies equally well
to the use of scaleless regulators like dimensional regularization, and to the
case where hard Wilsonian cutoffs are applied to distinguish modes. For the
Wilsonian case, consider the cutoffs as $\theta$-functions multiplying the
integrand. In this situation the regulator ensures that the integrand is zero in
the scaling limit so the subtraction terms all turn out to be zero, and the
naive replacement in Eq.~(\ref{1.01}) with the cutoffs gives the correct answer.

The result in Eq.~(\ref{1.02}) applies to NRQCD or SCET or any other quantum
field theory of this type.  It is necessary to avoid double counting the
zero-bin momenta which correspond to different degrees of freedom in the
effective theory. It provides a means for tiling the infrared regions of a
quantum field theory with different degrees of freedom while avoiding double
counting. In the following sections we explore the difference between
Eqs.~(\ref{1.01}) and (\ref{1.02}), and its consequences, with the help of
several examples.

Before leaving the general discussion it is worth emphasizing that
Eq.~(\ref{1.02}) together with the definition of the NRQCD and SCET degrees of
freedom given in Figs.~\ref{fig_NRQCDUVvsIR} and~\ref{fig_SCETUVvsIR} give
complete coverage of all momentum regions where IR divergences can occur.  We
used multiple degrees of freedom to cover these IR regions because this
facilitates setting up the proper EFT power counting expansion. In general one
can look at combining regions together to describe a larger region with only a
single degree of freedom. Doing so comes at the expense of making the power
counting expansion difficult to formulate. In many cases it is actually unknown
how to formulate the EFT expansion when regions are combined, thus necessitating
multiple modes.  We see that in general, the concepts of i) an EFT having a
complete set of degrees of freedom to reproduce all IR divergences, and ii) the
EFT having a valid power counting expansion, are tied together. We use this
freedom to {\em define} NRQCD and SCET to cover the IR regions with our chosen
degrees of freedom, so they reproduce the IR divergences. In this case proving
that these EFT's are complete is equivalent to proving that their power counting
expansions do not break down at any order.  Demonstrating this is easier, since
the power counting can only break down if we have missed a relevant operator at
leading order in the expansion of some observable.  All subleading operators are
treated as insertions and do not upset the power counting.  Thus, we see that
constructing a complete EFT is equivalent to identifying the proper physical
degrees of freedom in the leading order action, which is related to identifying
the set of physical processes for which the EFT applies.


\section{Zero-bin Subtractions in NRQCD (Non-relativistic Processes)} 
\label{sec:zerobinNRQCD}

In this section we consider examples of the use of Eq.~(\ref{1.02}) for
non-relativistic field theories. The results are quite general, applying whether
the non-relativistic particles are quarks, nucleons, ions, or quasi-particles.
The fields generating the potential are different in these cases, but the same
momentum regions are important.  To be definite we use a gauge theory, and so
take our examples from non-relativistic QCD (NRQCD).
 

\subsection{NRQCD Soft Crossed-Box Graph}

\begin{figure}
\includegraphics[width=7cm]{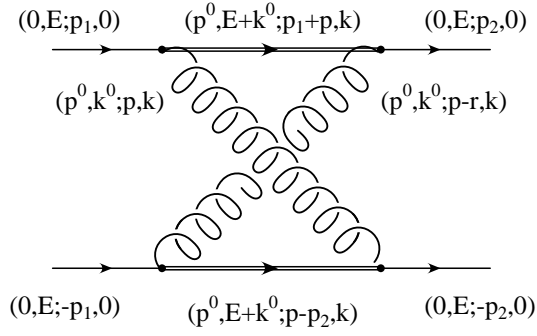}
\vskip-10pt
\caption{
  Soft crossed-box graph in the effective theory. The zigzag lines are soft
  gluons, the double lines are soft quarks, and the single lines are potential
  quarks. For each line we show (label energy, residual energy; label momentum,
  residual momentum).
\label{fig:cbox}}
\end{figure}

It is helpful to consider a concrete example to study the consequences of
Eq.~(\ref{1.02})---we will start with the crossed-box graph in NRQCD. The full
theory integral contains hard, soft, and usoft contributions, and we examine
the soft crossed-box graph in the effective theory shown in Fig.~\ref{fig:cbox}.
Here $\mathbf{p}_1$ and $\mathbf{p}_2$ are the external momenta, with
$\mathbf{r}=\mathbf{p}_2-\mathbf{p}_1$ the momentum transfer, and $2m E =
\mathbf{p}_1^2=\mathbf{p}_2^2$.
The momentum of all the particles has been denoted as (label energy, residual
energy; label momentum, residual momentum), and the external particles have been
chosen to have zero residual momentum.

In Feynman gauge the propagator for a soft gluon with momentum
$(p^0,k^0;\mathbf{p},\mathbf{k})$ is $1/[(p^0)^2-\mathbf{p}^2+i0^+]$, and
for a soft quark with momentum $(p^0,k^0;\mathbf{p},\mathbf{k})$ is $1/[p^0+i
\epsilon]$.  We neglect overall factors such as color Casimirs and
coupling constants, and focus on the integral for this graph:
\begin{eqnarray}
I_S^{\rm cross} &=&\hspace{-0.35cm}
 \sum_{p^0\not=0,p^\mu\not=0,p^\mu\not=(0,\mathbf{r})}
\int\!\! \frac{d^D k}{ (2 \pi)^D} \frac{1}{ p^0 \!+\! i0^+}\  
 \frac{1}{ p^0 \!+\! i0^+} \
 \frac{1}{ (p^0)^2 \!-\! \mathbf{p}^2 \!+\! i0^+}\ 
 \frac{1}{ (p^0)^2 \!-\! (\mathbf{p-r})^2 \!+\! i0^+} \,.
\label{2.02sum}
\end{eqnarray}
These soft propagators do not depend on the residual momentum components and so
the propagator takes different values at each grid site in
Fig.~\ref{fig_NRQCDUVvsIR}, but the same constant value for all the points in
each box. Thus there is no change to the integrand when the label and
residual momenta are combined into a continuous integration using
Eq.~(\ref{1.01}), which gives
\begin{eqnarray}
\tilde I_S^{\rm cross} &=& \int \frac{d^D p}{ (2 \pi)^D} 
 \frac{1}{ p^0 + i0^+}\  \frac{1}{ p^0 + i0^+} \
 \frac{1}{ (p^0)^2 - \mathbf{p}^2 + i0^+}\ 
 \frac{1}{ (p^0)^2 - (\mathbf{p-r})^2 + i0^+} \,.
\label{2.02}
\end{eqnarray}
However, Eq.~(\ref{2.02}) still includes the zero-bin contribution, which must
be subtracted out. A more careful analysis taking account of the zero-bin for
each soft particle propagator, and instead using Eq.~(\ref{1.02}) implies that
the value of the soft crossed-box graph is not Eq.~(\ref{2.02}) but rather
\begin{eqnarray}
I_S^{\rm cross} &=& \tilde I_S^{\rm cross}- I_1^{\rm cross} - I_2^{\rm cross}
\label{2.03}
\end{eqnarray}
where the subtractions are: (a) $I_1$ from the region $(p^0=0,\mathbf{p}=0)$ and
(b) $I_2$ from the region $(p^0=0,\mathbf{p}=\mathbf{r})$:
\begin{eqnarray}
I_1^{\rm cross} &=& \int \frac{d^D p}{ (2 \pi)^D} \frac{1}{ p^0 + i0^+}\  
 \frac{1}{ p^0 + i0^+} \
 \frac{1}{ (p^0)^2 - \mathbf{p}^2 + i0^+}\ \frac{1}{ - (\mathbf{r})^2 + i
  \epsilon} \,, \nn
I_2^{\rm cross} &=& \int \frac{d^D p}{ (2 \pi)^D} \frac{1}{ p^0 + i0^+}\  
 \frac{1}{ p^0 + i0^+} \
 \frac{1}{ - \mathbf{r}^2 + i0^+}\ \frac{1}{ (p^0)^2 - (\mathbf{p-r})^2 
  + i0^+}. 
\label{2.04}
\end{eqnarray}
By the shift symmetry in $\mathbf{p}$ we have $I_1^{\rm cross}=I_2^{\rm cross}$.

The $I_1^{\rm cross}$ subtraction comes from the region where the
$(p^0,k^0;\mathbf{p},\mathbf{k})$ gluon is usoft. Similarly, the $I_2^{\rm
  cross}$ subtraction comes from the region where the other gluon becomes
usoft.  Subtractions from the region where the quark is potential ($p^0= 0$)
vanish, as do the double-subtractions where the regions for $I_{1,2}$ overlap
with $p^0=0$. This is because for the crossed box all the $p^0$ poles are on the
same side of the contour of integration.

\begin{figure}[tp]
\begin{center}
\includegraphics[width=8cm]{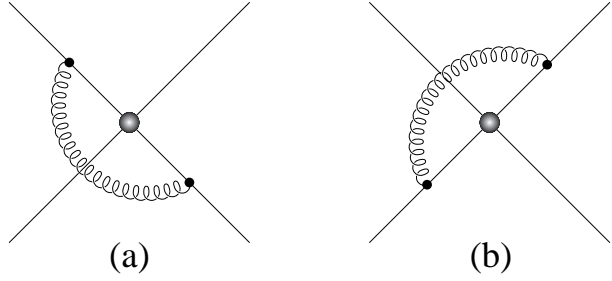}
\end{center}
\vskip-0.8cm
\caption{Ultrasoft corrections to potential scattering related to the crossed 
box graph. \label{fig:cusoft}}
\end{figure}

The subtraction $I_1^{\rm cross}$ avoids double counting the usoft graph
shown in Fig.~\ref{fig:cusoft}(a) and similarly $I_2^{\rm cross}$ avoids double
counting the usoft graph Fig.~\ref{fig:cusoft}(b).  The usoft graphs
depend on external usoft variables such as the energy.  If these are set to
zero, then the integral Fig.~\ref{fig:cusoft}(a) is $I_1^{\rm cross}$, including
the omitted color factors. The reason that $I_1^{\rm cross}$ does not depend on
external usoft variables, but Fig.~\ref{fig:cusoft}(a) could, is because
$I_1^{\rm cross}$ is obtained by considering a soft graph, and then taking its
usoft limit. The effective field theory Feynman rules require that all
usoft momentum be expanded out while considering soft diagrams. Thus
$I_1^{\rm cross}$ is the same as Fig.~\ref{fig:cusoft}(a) with the external
usoft variables expanded out.

Prior to the subtraction the soft crossed-box integral in dimensional
regularization is
\begin{eqnarray}
\tilde I_S^{\rm cross} &=&  -\frac{i}{4\pi^2 {\bf r}^2} \bigg[
  \frac{1 }{ \epsilon_{\text{IR}}} +  \ln\Big(\frac{\mu^2}{\bf r^2}\Big) \bigg].
\label{2.08}
\end{eqnarray}
The total subtraction $I_1^{\rm cross}+I_2^{\rm cross}$ gives
\begin{eqnarray}
I_1^{\rm cross}+I_2^{\rm cross} =  -\frac{i}{4\pi^2 {\bf r}^2} \bigg[
  \frac{1}{\epsilon_{\text{IR}}} - \frac{ 1}{ \epsilon_{\text{UV}}} \bigg],
\label{2.09}
\end{eqnarray}
so the final result for the crossed-box integral is
\begin{eqnarray}
I_S^{\rm cross} = \tilde I_S^{\rm cross}-I_1^{\rm cross}-I_2^{\rm cross}  
 &=&  -\frac{i}{4\pi^2 {\bf r}^2} \bigg[
  \frac{1}{ \epsilon_{\text{UV}}} +  \ln\Big(\frac{\mu^2}{\bf r^2}\Big) \bigg].
\label{2.10}
\end{eqnarray}
The subtractions have converted the $1/\epsilon_{\text{IR}}$ divergence in
$\tilde I_S^{\rm cross}$ into a $1/\epsilon_{\text{UV}}$ ultraviolet divergence.
In Ref.~\cite{Manohar:2000kr,Hoang:2001mm}, it was argued that
$1/\epsilon_{\text{IR}}$ divergences in soft graphs should be converted to
ultraviolet divergences by a pullup mechanism and included in the computation
of anomalous dimensions.  We see that the zero-bin subtraction automatically
implements this conversion.  An important feature for NRQCD is that the soft
graph defined with the zero-bin subtraction is infrared finite, and has a
well-defined renormalized value independent of any cancellation with usoft
graphs.

A similar conversion from infrared to ultraviolet divergences also occurs for
collinear graphs in \SCETa as we show in section~\ref{sec:sceta}.


\subsection{NRQCD Box Graph}

\begin{figure}
\includegraphics[width=7cm]{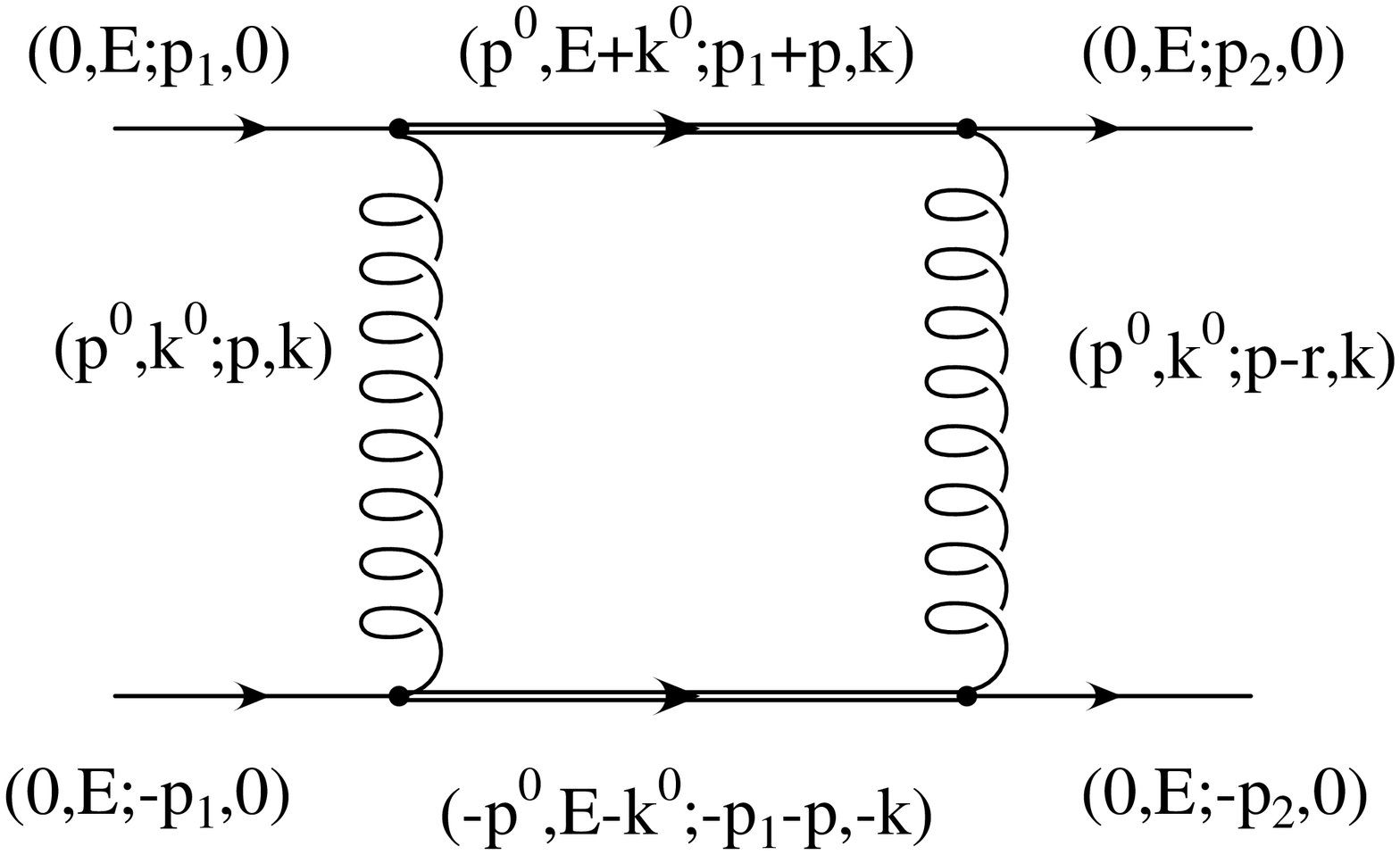}
\caption{
  Soft box graph in the effective theory. The zigzag lines are soft gluons, the
  double lines are soft quarks, and the single lines are potential quarks. For
  each line we show (label energy, residual energy; label momentum, residual
  momentum).
\label{fig:box}}
\end{figure}

The zero-bin subtraction has another important consequence ---it gets rid
of pinch singularities. Consider the soft box graph in NRQCD, shown in
Fig.~\ref{fig:box}, and follow the same procedure as for the crossed-box graph.
The only difference from the crossed-box is the replacement
\begin{eqnarray}
\frac{1}{ p^0 + i0^+}\ \frac{1}{ p^0 + i0^+} \to 
  \frac{1}{ p^0 + i0^+}\ \frac{1}{ -p^0 + i0^+}
\end{eqnarray}
in Eqs.~(\ref{2.02sum}--\ref{2.04}) due to the change in momentum routing
through the antiquark line. The integrals analogous to those in
Eqs.~(\ref{2.02},\ref{2.04}) are
\begin{eqnarray}
\tilde I_S^{\rm box} &=& \int \frac{d^D p}{(2 \pi)^D} 
 \frac{1}{ p^0 + i0^+}\  \frac{1}{ -p^0 + i0^+} \
 \frac{1}{ (p^0)^2 - \mathbf{p}^2 + i0^+}\ 
 \frac{1}{ (p^0)^2 - (\mathbf{p-r})^2 + i0^+},\nn
I_1^{\rm box} &=& \int \frac{d^D p}{ (2 \pi)^D} \frac{1}{ p^0 + i0^+}\  
 \frac{1}{ -p^0 + i0^+} \
 \frac{1}{ (p^0)^2 - \mathbf{p}^2 + i0^+}\ \frac{1}{- (\mathbf{r})^2 + i
  \epsilon} \,, \nn
I_2^{\rm box} &=& \int \frac{d^D p}{ (2 \pi)^D} \frac{1}{ p^0 + i0^+}\  
 \frac{1}{ -p^0 + i0^+} \
 \frac{1}{ - \mathbf{r}^2 + i0^+}\ \frac{1}{ (p^0)^2 - (\mathbf{p-r})^2 
  + i0^+}. 
\label{2.04b}
\end{eqnarray}
where the usoft subtractions $I_{1,2}^{\rm box}$ are for $(p^0=0,{\bf
  p}=0)$, $(p^0=0,{\bf p}={\bf r})$ respectively.
In addition one also has a potential subtraction for $p^0=0$,
\begin{eqnarray}
I_3^{\rm box} &=& \int \frac{d^D p}{ (2 \pi)^D} \frac{1}{ p^0 + i0^+}\  
 \frac{1}{ -p^0 + i0^+} \
 \frac{1}{  - \mathbf{p}^2 + i0^+}\
 \frac{1}{ - (\mathbf{p-r})^2 + i0^+} \,. 
\label{3.04}
\end{eqnarray}
Now this $p^0=0$ subtraction overlaps with the usoft subtractions, so we have to
add back the double subtractions, the ($p^0=0$) limit of $I_{1,2}^{\rm box}$:
\begin{eqnarray}
I_4^{\rm box} &=& \int \frac{d^D p}{(2 \pi)^D} \frac{1}{p^0 + i0^+}\ 
  \frac{1}{ -p^0 + i0^+} \
  \frac{1}{  - \mathbf{p}^2 + i0^+}\ \frac{1}{ - \mathbf{r}^2 
  + i0^+}, \nn
I_5^{\rm box} &=& \int \frac{d^D p }{ (2 \pi)^D} \frac{1 }{p^0 + i0^+}\  
 \frac{1 }{ -p^0 + i0^+} \
 \frac{1 }{  - \mathbf{r}^2 + i0^+}\ \frac{1 }{ - (\mathbf{p-r})^2 
  + i0^+} \,.
\label{3.05}
\end{eqnarray}
The complete expression for the soft box graph is
\begin{eqnarray}
  I_S^{\rm box} &=& \tilde I_S^{\rm box} - I_1^{\rm box} - I_2^{\rm box}
    - I_3^{\rm box} + I_4^{\rm box} + I_5^{\rm box} \,.
\end{eqnarray}

Both $\tilde I_S^{\rm box}$ and $I_3^{\rm box}$ have pinch singularities in the
$p^0$ integral, from the poles at $p^0=\pm i0^+$, and are ill-defined.
However, for the result in the effective theory, we don't need the separate
integrals, but only the difference $\tilde I_S^{\rm box}-I_3^{\rm box}$, which
has no pinch. We have
\begin{eqnarray}
\tilde I_S^{\rm box}-I_3^{\rm box} &=&
 \int \frac{d^D p }{ (2 \pi)^D} \frac{1 }{p^0 + i0^+}\ 
 \frac{1 }{ -p^0 + i0^+} \times \nn
&& \left\{
\frac{1 }{ (p^0)^2 - \mathbf{p}^2 + i0^+}\ \frac{1 }{ (p^0)^2 
 - (\mathbf{p-r})^2 + i0^+}
 -\frac{1 }{ \mathbf{p}^2 + i0^+}\ \frac{1 }{ (\mathbf{p-r})^2 
 + i0^+}\right\} .
\label{10.07}
\end{eqnarray}
One can evaluate the $p^0$ integral in Eq.~(\ref{10.07}) using contour
integration. The result is the same as doing Eq.~(\ref{2.02}) by contours and
dropping the pinch pole at $p^0=0$, since the integrand of Eq.~(\ref{10.07}) has
no $p^0$ pole and the subtraction term does not introduce  new poles in
$p^0$. This prescription for the soft box graph is what was used in
Refs.~\cite{Manohar:1999xd,Griesshammer:1998tm,Manohar:2000hj,Brambilla:2004jw,Penin:2004ay,Kniehl:2004rk},
but we now see how the effective theory automatically gives this
result.\footnote{At one-loop there are many prescriptions that lead to the same
  result, that the pinch is dropped in the contour integration. Examples include
  split dimensional regularization and the principal value prescription. The
  formula in Eq.~(\ref{1.02}) can be applied at any order, and makes adopting a
  prescription moot. }

The double subtractions $I_4^{\rm box}$ and $I_5^{\rm box}$ remove the pinch
poles at $p^0=0$ for the subtractions $I_1^{\rm box}$ and $I_2^{\rm box}$
respectively, so that $I_1^{\rm box}-I_4^{\rm box}$ and $I_2^{\rm box}-I_5^{\rm
  box}$ are free of pinch singularities. This justifies ignoring the $p^0=0$
pole in the calculation of these integrals. The $I_{1,2}^{\rm box}$ usoft
subtractions convert the infrared divergences in $\tilde I_S^{\rm box}$ into
ultraviolet divergences, just as they did for the crossed-box. 
Prior to the subtraction the soft box integral in dimensional
regularization is
\begin{eqnarray}
\tilde I_S^{\rm box} - I_3^{\rm box} &=&  \frac{i}{4\pi^2 {\bf r}^2} \bigg[
  \frac{1 }{ \epsilon_{\text{IR}}} +  \ln\Big(\frac{\mu^2}{\bf r^2}\Big) \bigg].
\label{2.08a}
\end{eqnarray}
The total subtraction is
\begin{eqnarray}
I_1^{\rm box} - I_4^{\rm box}+I_2^{\rm box} - I_5^{\rm box} 
  =  \frac{i}{4\pi^2 {\bf r}^2} \bigg[
  \frac{1}{\epsilon_{\text{IR}}} - \frac{ 1}{ \epsilon_{\text{UV}}} \bigg],
\label{2.09a}
\end{eqnarray}
so the final result for the box integral is
\begin{eqnarray}
I_S^{\rm box} = \tilde I_S^{\rm box}-I_1^{\rm box}-I_2^{\rm box}  
  - I_3^{\rm box} + I_4^{\rm box} + I_5^{\rm box}
 &=&  \frac{i}{4\pi^2 {\bf r}^2} \bigg[
  \frac{1}{ \epsilon_{\text{UV}}} +  \ln\Big(\frac{\mu^2}{\bf r^2}\Big) \bigg].
\label{2.10a}
\end{eqnarray}
The subtractions have converted the $1/\epsilon_{\text{IR}}$ divergence in
$\tilde I_S^{\rm box}$ into a $1/\epsilon_{\text{UV}}$ ultraviolet divergence
just like the cross-box. The zero-bin subtraction has removed the infrared
divergences and the pinch singularities, since these regions are properly taken
care of by usoft and potential graphs, respectively. The properly defined
soft box graph is infrared finite, and has no pinch singularity.

The use of Eq.~(\ref{1.02}) with the zero-bin subtraction works at higher orders
as well. One can check explicitly that the subtractions remove the pinch
singularities in the double box. If the loop momenta for the two single-box
subgraphs are called $p$ and $\ell$, the subtracted double-box is given by
subtracting the region where $p^0=0$ and where $\ell^0=0$, and adding back the
region where $p^0=\ell^0=0$. This gives an expression for the double box which
is free of both single and double pinch singularities. We have checked that the
subtracted three gluon exchange graphs give the correct contribution to the
two-loop static potential~\cite{Schroder:1998vy} in Feynman gauge.

The standard computation of the Coulomb potential at ${\cal O}(\alpha_s^2)$ in
QCD is free of IR singularities for a different reason, because the IR
divergences in the box and crossed-box cancel against the vertex and
wavefunction diagrams.  With the zero-bin subtractions, the scaleless soft
vertex and wavefunction diagrams are set to zero, and the box and cross-box
together with the non-Abelian vacuum polarization and Y-graphs give the complete
IR finite answer. At this order there is also no overall usoft contribution
to this four point function, and the same is true for the Coulomb potential at
two loop order~\cite{Peter:1997me,Schroder:1998vy}. At three loops the zero-bin
subtraction removes the ADM singularity in the Coulomb
potential~\cite{Appelquist:1977tw,Brambilla:1999qa,Hoang:2001rr}. A one-loop
example at ${\cal O}(v^2)$ where the subtractions do not cancel in the sum of
diagrams is discussed in the next section.


\subsection{Results for the Box and Crossed-Box at order $\mathcal{O}(v^2)$}

In this section, we study the soft box and crossed-box graphs to second order in
the $v$ expansion, i.e.\ to the same order as the spin-orbit, Darwin and
tensor-force contributions to the $Q \bar Q$ potential. At order $v^2$, the
naive soft box and crossed-box have IR divergences and there are also
non-trivial contributions from usoft diagrams.  We will summarize results for
these graphs to illustrate how the zero-bin subtractions work. This example also
illustrates a case where $F^{\rm sub}_j$ in Eq.~(\ref{1.02}) involves a series
of terms.

Prior to any subtractions, the necessary diagrams are simply given by all
quark-antiquark scattering diagrams that are derived using the HQET Lagrangian
up to $1/m^2$. The full integrands are lengthy and we refer the reader
to Refs.~\cite{Manohar:1999xd,Manohar:2000hj}.  After using standard tricks to
trivialize the numerator momenta we are left with the basic
integrals
\begin{align}
  J(\alpha,\beta) &=
 \sum_{p \in Z^c}
\int \frac{d^D k }{ (2 \pi)^D} \frac{1 }{ (p^0 \!+\! i0^+)^\alpha 
 (-p^0 \!+\! i0^+)^\beta} \: \frac{1}{  
 [ (p^0)^2 \!-\! \mathbf{p}^2 \!+\! i0^+ ]\:
 [ (p^0)^2 \!-\! (\mathbf{p\!-\!r})^2 \!+\! i0^+ ] },
\end{align}
where $Z^c=\{ p^0\ne 0, p^\mu\ne 0, p^\mu\ne (0,{\bf r})\}$.  The subtractions
that account for $p^0\ne 0$ and remove the pinch singularity from the first two
denominators involve $\delta$ derivatives of the second two denominators where
$\delta$ is the nearest integer $\le (\alpha+\beta)$. For this particular
computation this is equivalent to ignoring these poles in the contour integral.

As in the previous section the removal of the remaining constraints is similar,
with or without the pinches, so we will consider the case $\beta=0$ for
simplicity. The naive integral and its subtractions for $\{ p^\mu\ne (0,{\bf
  r}), p^\mu\ne 0\}$ respectively are
\begin{eqnarray}\label{15.19}
 \tilde J(\alpha,0) &=& \int \frac{d^D p }{ (2 \pi)^D} 
  \frac{1 }{ (p^0 + i0^+)^\alpha \:  
 [ (p^0)^2 \!-\! \mathbf{p}^2 + i0^+ ]\:
 [ (p^0)^2 \!-\! (\mathbf{p-r})^2 + i0^+ ] }\,, \nn
 J_1(\alpha,0) &=& \int \frac{d^D p }{ (2 \pi)^D} 
  \frac{1 }{ (p^0 + i0^+)^\alpha \:  
 [ (p^0)^2 \!-\! (\mathbf{p-r})^2 + i0^+ ]  } 
  \sum_{k=0}^{\alpha-2}
   \frac{[-2\, \mathbf{(p-r)\cdot r}]^k }{[-\mathbf{r}^2]^{k+1}} 
  \:\,, \nn
 J_2(\alpha,0) &=& \int \frac{d^D p }{ (2 \pi)^D} 
  \frac{1 }{ (p^0 + i0^+)^\alpha \:  
 [ (p^0)^2 \!-\! \mathbf{p}^2 + i0^+ ] } \:
 \sum_{k=0}^{\alpha-2}
   \frac{   [-2\,\mathbf{p\cdot r}]^k }{[ -\mathbf{r}^2 ]^{k+1} }\,.
\end{eqnarray}
where in $J_1$ and $J_2$ we have dropped terms that are obviously zero.  Note
that here removing the zero-bin requires a series of subtractions obtained from
expanding the naive integrand about the zero-bin values. By translation
invariance in dimensional regularization the two subtraction integrals are
equal, $J_1(\alpha,0)=J_2(\alpha,0)$.

For $\alpha=1$, the subtractions are zero, $J_1=J_2=0$, and $\tilde J(1,0)$ is
finite.  For any other odd $\alpha$, both the
naive integral and subtractions give zero. For even $\alpha \le 0$, $\tilde
J(\alpha,0)$ is UV divergent and the subtractions give zero. Finally for even
$\alpha \ge 2$ the base integral is IR divergent and the subtractions convert
this to a UV divergence for $J(\alpha,0)=\tilde J(\alpha,0) -
J_1(\alpha,0)-J_2(\alpha,0)$.  As an example, consider $\alpha=4$. The naive
integral is
\begin{align}
  \tilde J(4,0) &= \frac{i}{16\pi^2}\: \Big(- \frac{16}{3 \mathbf{r}^4} \Big)
  \Big[\frac{1}{\epsilon_{\rm IR}} +  \ln\Big(\frac{\mu^2}{\mathbf{r}^2}\Big) 
   + 2 \Big]\,,
\end{align}
and the subtractions $J_1=J_2$ are given by
\begin{align}
  J_2(4,0) & = \int\!\! \frac{d^D p }{(2 \pi)^D} 
  \frac{1 }{(p^0 + i0^+)^4 \:  
 [ (p^0)^2 \!-\! \mathbf{p}^2 + i0^+ ] } \:
 \bigg[ -\frac{1}{\mathbf{r}^2} - \frac{2\mathbf{p\mcdot r}}{ \mathbf{r}^4}
  - \frac{4 (\mathbf{p\mcdot r})^2 }{ \mathbf{r}^6} \bigg] \nn
  & = 0 + 0 +  \frac{i}{16\pi^2}\: \Big( -\frac{16}{3\mathbf{r}^4} \Big)
  \Big[\frac{1}{\epsilon_{\rm IR}} - \frac{1}{\epsilon_{\rm UV}} \Big] \,.
\end{align}
Thus the full integral is $J(4,0)= \tilde J(4,0)-2 J_2(4,0)$ giving
\begin{align}
  \ J(4,0) &= \frac{i}{16\pi^2}\: \Big(- \frac{16}{3 \mathbf{r}^4} \Big)
  \Big[\frac{1}{\epsilon_{\rm UV}} +  \ln\Big(\frac{\mu^2}{\mathbf{r}^2}\Big) 
   + 2 \Big]\,,
\end{align}
which does not have an IR pole. The same is true for all even $\alpha\ge 2$.

Lets consider the sum of all order $v^2$ NRQCD soft exchange diagrams (boxes,
cross-boxes, and triple gluon graphs), and the vertex and wavefunction graphs,
all computed in Feynman gauge with equal mass quarks and antiquarks. Using the
naive $\tilde J$ integrands we find\footnote{For the purpose of this example we
  have set Wilson coefficients in the HQET Lagrangian to their tree level
  values. For the summation of logs in
  Refs.~\cite{Manohar:1999xd,Pineda:2001ra,Hoang:2002yy}, their renormalization
  group evolution of course had to be kept.}
\begin{align}
  \tilde S_{\rm exchange} &= \frac{i\alpha_s^2}{m^2} \bigg\{
  \Big[C_1 (1\otimes 1) -\frac{C_d}{4} (T^A\otimes T^A) \Big]  
   \Big(\frac{1}{\epsilon_{\rm UV}} + \frac{4}{3\epsilon_{\rm IR}} \Big)
  \\[5pt]
  &\quad
   + C_A (T^A\otimes T^A) \bigg[ \Big(\frac{13}{4\epsilon_{\rm UV}} 
      + \frac{1}{3\epsilon_{\rm IR}} \Big) 
      + \frac{({\mathbf p}^2+{\mathbf p}^{\prime\,2})}{2 {\mathbf r}^2} 
      \Big(-\frac{5}{3\epsilon_{\rm UV}} - \frac{14}{3\epsilon_{\rm IR}} \Big)
     \Big)\nn
  &\quad 
   + {\Lambda}
     \Big(\frac{1}{2\epsilon_{\rm UV}} + \frac{3}{\epsilon_{\rm IR}} \Big)
   + {\mathbf S^2} \Big(-\frac{11}{18\epsilon_{\rm UV}} + \frac{2}{3\epsilon_{\rm
       IR}} \Big)
   + {\rm T} \Big(- \frac{1}{36\epsilon_{\rm UV}} + \frac{1}{6\epsilon_{\rm
       IR}} \Big) \bigg] \bigg\} \,,
 \nonumber \\[5pt]
 \tilde S_{\rm vertex+w.fn.} &= \frac{i\alpha_s^2}{m^2}  (T^A\otimes T^A) 
  \bigg\{ \frac{4}{3}\Big(C_F \!-\!\frac{C_A}{2}\Big) 
 - 2 C_A \Big( \frac{{\mathbf p}^2 + {\mathbf p'}^2}{2{\mathbf r}^2}
  - \frac{\mathbf S^2}{3} -\frac{3{ \Lambda}}{2} -\frac{\rm T}{12}
  \Big) \bigg\}\nn
 &\quad
  \times
  \Big(  \frac{1}{\epsilon_{\rm UV}} - \frac{1}{\epsilon_{\rm IR}} \Big)
  \nonumber
\end{align}
for the pole structure.  We used the notation in Ref.~\cite{Manohar:2000hj}
where in $SU(N_c)$ the color coefficients $C_1=(N_c^2-1)/(4N_c^2)$,
$C_d=N_c-4/N_c$, $C_A=N_C$, $C_F=(N_c^2-1)/(2N_c)$, there are two color
structures $(1\otimes 1)$ and $(T^A\otimes T^A)$, and $\Lambda$, ${\mathbf
  S}^2$, and ${\rm T}$ are spin and momentum dependent structures,
\begin{eqnarray}
\mathbf S &=& \frac{ {\mathbf \bsigma_1 + \bsigma_2} }{ 2},
 \qquad \Lambda = -i \frac{\mathbf S \cdot ( p^\prime
 \times p) }{  {\mathbf r}^2 },\qquad
 T = {\mathbf \bsigma_1 \cdot \bsigma_2} - \frac{3\, {\mathbf r
 \cdot \bsigma_1}\,  {\mathbf r \cdot \bsigma_2} }{ {\mathbf r}^2} \,.
\end{eqnarray}
The sum of diagrams with naive integrands is
\begin{align}
  \tilde S  &= \frac{i\alpha_s^2}{m^2} \bigg\{
  C_1 (1\otimes 1) \Big(\frac{1}{\epsilon_{\rm UV}} + \frac{4}{3\epsilon_{\rm
      IR}} \Big)
+ (T^A\otimes T^A) \bigg[ \frac{4 C_F}{3}
   \Big(  \frac{1}{\epsilon_{\rm UV}} - \frac{1}{\epsilon_{\rm IR}} \Big)
    -C_d\Big(\frac{1}{4\epsilon_{\rm UV}} + \frac{1}{3\epsilon_{\rm
      IR}} \Big) \bigg]
  \nonumber \\[5pt]
  &\qquad
   + C_A (T^A\otimes T^A) \bigg[ \Big(\frac{31}{12\epsilon_{\rm UV}} 
      + \frac{1}{\epsilon_{\rm IR}} \Big) 
      + \frac{({\mathbf p}^2+{\mathbf p}^{\prime\,2})}{2 {\mathbf r}^2} 
      \Big(-\frac{11}{3\epsilon_{\rm UV}} - \frac{8}{3\epsilon_{\rm IR}} \Big)
     \Big)\nn
  &\qquad 
   + {\Lambda}
     \Big(\frac{7}{2\epsilon_{\rm UV}}  \Big)
   + {\mathbf S^2} \Big(\frac{1}{18\epsilon_{\rm UV}}  \Big)
   + {\rm T} \Big( \frac{5}{36\epsilon_{\rm UV}}  \Big) \bigg] \bigg\}  \,.
\end{align}
After subtracting the zero-bin contributions to get the proper integrals
$J(\alpha,\beta)$ we find
\begin{align}
  S  &= \frac{i\alpha_s^2}{m^2} \bigg\{
  C_1 (1\otimes 1) \Big(\frac{7}{3\epsilon_{\rm UV}}  \Big)
 -C_d (T^A\otimes T^A) \Big(\frac{7}{12\epsilon_{\rm UV}}  \Big) 
   + C_A (T^A\otimes T^A) \bigg[ \Big(\frac{43}{12\epsilon_{\rm UV}} 
       \Big) \nonumber \\[5pt]
  &\qquad
      + \frac{({\mathbf p}^2+{\mathbf p}^{\prime\,2})}{2 {\mathbf r}^2} 
      \Big(\frac{-19}{3\epsilon_{\rm UV}} \Big)
   + {\Lambda}
     \Big(\frac{7}{2\epsilon_{\rm UV}} \Big)
   + {\mathbf S^2} \Big(\frac{1}{18\epsilon_{\rm UV}}  \Big)
   + {\rm T} \Big( \frac{5}{36\epsilon_{\rm UV}}  \Big) \bigg] \bigg\}  \,.
\end{align}
At ${\cal O}(v^2)$ there are also UV divergences from the usoft graphs which
can be found from Ref.~\cite{Manohar:1999xd}. Using dimensional regularization
for both the UV and the IR they give:
\begin{align} \label{IusUV}
  U &= \frac{i\alpha_s^2}{m^2} \Big(  \frac{1}{\epsilon_{\rm UV}} -
  \frac{1}{\epsilon_{\rm IR}} \Big)
\bigg\{
  \frac{4C_1}{3} (1\otimes 1) 
+ \Big[ C_A \!-\! \frac{4 C_F}{3} \!-\! \frac{C_d}{3} 
   \!-\! \frac{8C_A}{3} 
       \frac{({\mathbf p}^2+{\mathbf p}^{\prime\,2})}{2 {\mathbf r}^2} 
      \Big] (T^A\otimes T^A)  \bigg\}  \,.
\end{align}
We see explicitly that the usoft graphs have UV divergences which match up
with the fake IR divergences from the unsubtracted soft computation in $\tilde
S$. The true soft computation gives an IR finite result and the usoft
contribution exactly matches the IR divergences in the full theory computation,
see Ref.~\cite{Manohar:2000hj}. To interpret the UV divergence in
Eq.~(\ref{IusUV}) as occurring at the hard scale, it is crucial to make the
zero-bin subtractions to avoid double counting in the soft region.


\section{Zero-bin Subtractions in SCET$_{\rm I}$ (Inclusive Processes) } 
\label{sec:sceta}

SCET is another theory with correlated scales, and with multiple fields for the
same particle. As a result, one expects the zero-bin subtraction to also apply
in this theory (yielding a pullup here too). In this part, we consider examples
of the application of Eq.~(\ref{1.02}) to \SCETa which has collinear fields
appropriate for the description of perturbative energetic jets, and
non-perturbative usoft fields.\footnote{These collinear modes are sometimes
  called hard-collinear, and the usoft modes are then called soft.} The new
feature of \SCETa is the appearance of double-logarithmic divergences at
one-loop order. In the following discussion we show how the zero-bin subtraction
works in this case.

We first consider the heavy to light vertex diagram in
Fig.~\ref{fig:heavylight}, which appears in processes such as the $b \to s
\gamma$ transition magnetic moment operator needed to compute inclusive $B \to
X_s \gamma$ decays. In section~\ref{sect_sceta1} we work on-shell with finite
cutoffs and demonstrate that the infrared divergences of the full theory are
reproduced in \SCETa . The subtraction from Eq.~(\ref{1.02}) in the collinear
diagram is required for this to be true. To demonstrate that Eq.~(\ref{1.02}) is
independent of the choice of ultraviolet and infrared regulator, in
section~\ref{sect_sceta2} we consider the more standard choice of dimensional
regularization with an offshellness infrared regulator, and explain how the
zero-bin subtraction works for this case.  In section~\ref{sect_sceta3} we treat
the example of the current relevant for inclusive two-jet production where we
have collinear fields in two directions, $n$ and $\bn$, and usoft fields.
The $\sum_{p\ne 0}$ are also important at tree level and for phase space
integration as demonstrated by the $\gamma^*\to q\bar q g$ example that we take up in
section~\ref{sect_sceta4}.


\subsection{ Onshell Integrals and a Cutoff Regulator for $B\to X_s\gamma$}
\label{sect_sceta1}

For $b\to s\gamma$ at lowest order in $\Lambda_{\rm QCD}/E_\gamma$ we have the
SCET current~\cite{Bauer:2000yr}
\begin{eqnarray} \label{J01}
  J^{(0)} &= C(\omega)  [ (\bar\xi_n W)_\omega \Gamma h_v ] \,.
\end{eqnarray}
After making the decoupling field redefinition~\cite{Bauer:2001yt} on the
collinear fields this becomes
\begin{eqnarray} \label{J02}
  J^{\prime(0)} &= C(\omega)  [ (\bar\xi_n W)_\omega \Gamma (Y^\dagger h_v) ] \,.
\end{eqnarray}
We start by making use of the current in Eq.~(\ref{J01}) and will discuss the
equivalence of using (\ref{J02}) at the end of this section.
\begin{figure}
 \begin{center}
\includegraphics[width=4cm]{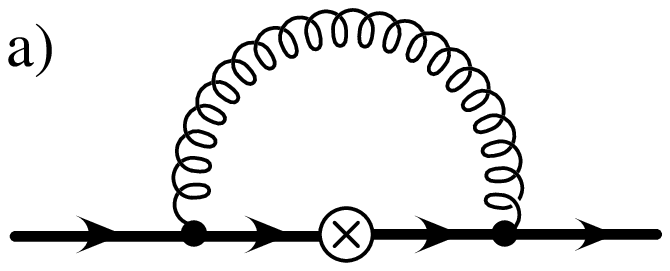}\qquad
\includegraphics[width=4cm]{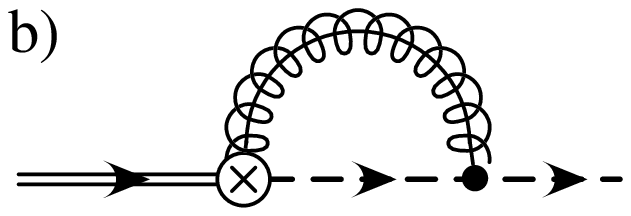}\qquad
\includegraphics[width=4cm]{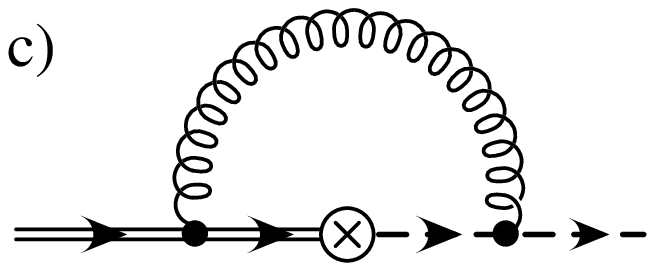}
 \end{center}
 \caption{Heavy-to-light vertex diagrams. a)
     full theory, b) \SCETa collinear graph, and c) \SCETa usoft graph.
\label{fig:heavylight}}
\end{figure}
The incoming heavy quark momentum is $p_b^\mu=m\, v^\mu$ with $v^2=1$ and the
outgoing light quark momentum is $p^\mu=p_- n^\mu/2$ where $n^2=0$ and
$p_-=\bn\mcdot p$. Both the incoming and outgoing quarks are taken onshell. 

In this section we first demonstrate the effect of the pullup on collinear
diagrams, prior to making a specific choice of regulator. We then use infrared
cutoffs $\Omega_\perp$ and $\Omega_-$ on $p_\perp$ and $p_-$ in both the full
and effective theories, so that the loop momenta are restricted to the region
$q_\perp^2\ge \Omega_\perp^2$ and $q_-^2 \ge \Omega_-^2$. For the usoft
graphs, we use a ultraviolet cutoff $\Lambda_-$, while for the collinear
graphs, we use an ultraviolet cutoff $\Lambda_\perp$.  Hard cutoffs make the
computation of anomalous dimensions more difficult and in more generic diagrams
would require gauge violating (and power counting violating) counterterms to
restore these symmetries. Our focus is on showing how the IR divergences are
reproduced for a particular example where these problems do not occur, so for the
purpose of this computation these issues are not a concern.

The part of the full theory diagram in Fig.~\ref{fig:heavylight}a with the
double logarithmic infrared divergence involves the integral
\begin{eqnarray}
 I_{\rm full}^{b\to s\gamma} &=& 
  \int \!\!\frac{d^Dq}{(2\pi)^D}  \frac{4 p_b \cdot p}
 {(q^2+i0^+)(q^2+2p_b\cdot q+i0^+)(q^2+2p\cdot q+i0^+)}  \,,
\end{eqnarray}
which is ultraviolet finite. Taking $p_b=m_b (n+\bn)/2$ and $0 < p^- < m_b$, we
use the identity $d^D q = d(n\cdot q) d (\bn \cdot q) d^n q_\perp/2$ to write
the measure in light-cone variables, where here the exponent $n=D-2$ is the
dimension of the $\perp$-space, not to be confused with the light-like vector
$n^\mu$.  Next we perform the $n\mcdot q$ integral by contours. There are poles
for
\begin{align}
 n\mcdot q &= -\frac{q_\perp^2}{\bn\mcdot q} 
  -i0^+\: {\rm sign}(\bn\mcdot q)\,, 
 &n\mcdot q &= -\frac{(q_\perp^2 \!+\! m_b\bn\mcdot q)}{\bn\mcdot q\!+\! m_b}
    -i0^+\: {\rm sign}(\bn\mcdot q \!+\! m_b)\,, \nn
 n\mcdot q &= -\frac{q_\perp^2 }{\bn\mcdot q\!+\! \bn\mcdot p}
    -i0^+\: {\rm sign}(\bn\mcdot q\!+\! \bn\mcdot p)\,, 
\end{align} 
which gives three poles above the axis for $q^- < -m_b$, one moving below in the
region $-m_b < q^- < -p_-$, two below and one above for $-p^- < q^- < 0$, and
all below for $q^->0$. Thus only the middle two regions contribute. We will
drop the integral over the interval $-m < \bn\mcdot q < -p_-$, since it is finite
in the UV and IR. This gives
\begin{eqnarray}
 I_{\rm full}^{b\to s\gamma} 
 &=& \frac{i}{2\pi} \int_{-p_-}^0 \!\!\!\!\! d\bn\mcdot q \:
   \frac{d^nq_\perp}{(2\pi)^n} \frac{\bn\mcdot q}{[q_\perp^2+(\bn\mcdot q)^2]
   (q_\perp^2)} 
   + \mbox{finite} \,,
\end{eqnarray}

The usoft graph in \SCETa is shown in Fig.~\ref{fig:heavylight}c and just
has an integral over residual momentum. It is important to recall that the
leading order SCET Lagrangian involves a momentum space multipole expansion for
the residual momentum~\cite{Bauer:2000yr}, so that only the residual $n\cdot k$
momentum appears in the collinear quark propagator.  The integral is
\begin{align} \label{Ius}
 I_{\rm us}^{b\to s\gamma} &=
  \int \!\!\frac{d^Dk}{(2\pi)^D}  \frac{1}
 {(k^2+i0^+)(v\mcdot k +i0^+)(n\mcdot k +i0^+)} 
 = \frac{i}{2\pi} \int_{-\infty}^0 \!\!\!\!\! d\bn\mcdot k \:
   \frac{d^nk_\perp}{(2\pi)^n} \frac{\bn\mcdot k}{[k_\perp^2+(\bn\mcdot
     k)^2](k_\perp^2)}\,. 
\end{align}
For the collinear graph, Fig.~\ref{fig:heavylight}b, we have the label loop
momentum $\tilde q^\mu=\bn\mcdot qn^\mu/2 + q_\perp^\mu$ and residual loop
momentum $q_r^\mu$, and we will denote $q^\mu =\tilde q^\mu + n\mcdot q_r\,
\bn^\mu/2$.  The original integral is
\begin{eqnarray}
 I_{\rm C}^{b\to s\gamma} \!\!&=& \!\!
 \sum_{\tilde q\ne 0,\: \tilde q\ne -\tilde p} 
  \ \int \!\!\frac{d^Dq_r}{(2\pi)^D}  \frac{2 \bn\mcdot (q+p)}
 {(\bn\mcdot q +i0^+)(q^2+2p\mcdot q + i0^+)(q^2 +i0^+)} \,.
\end{eqnarray}
Eq.~(\ref{1.02}) is used to take into account the subtractions from the
zero-bins. For $\tilde q\ne 0$ we examine the scaling $\tilde q^\mu\sim
\lambda^2$, for which case the loop-measure scales as $[\lambda^8]$ and combines
with the integrand to give: $[\lambda^8]/[(\lambda^2)(\lambda^2)(\lambda^4)]\sim
\lambda^0$, so there is a non-trivial subtraction for this region. For $\tilde
q\ne -\tilde p$ we examine $q^-+p^- \sim \lambda^2$, $q^\perp\sim\lambda^2$ and
have: $[\lambda^8]/[(\lambda^0)(\lambda^4)(\lambda^2)]\sim \lambda^2$, so this
zero-bin can be ignored. When we combine the sum over label momentum with the
integral over residual momentum we get the naive result $\tilde I_C$ and a
subtraction $I_0$ from Eq.~(\ref{1.02}):
\begin{eqnarray}
 \tilde I_{\rm C}^{b\to s\gamma} \!\!&=& \!\!
  \int \!\!\frac{d^Dq}{(2\pi)^D}  \frac{2 \bn\mcdot (q+p)}
 {(\bn\mcdot q +i0^+)(q^2+2p\mcdot q + i0^+)(q^2 +i0^+)} 
 = \frac{i}{2\pi} \int_{-p^-}^0 \!\!\!\!\! d\bn\mcdot q \:
   \frac{d^nq_\perp}{(2\pi)^n} \frac{\bn\mcdot (q+p)}{(\bn\mcdot q)(\bn\mcdot p)
   (q_\perp^2)}
   \,,\nn
 I_{\rm 0}^{b\to s\gamma} \!\! &=&\!\!
  \int \!\!\frac{d^Dq}{(2\pi)^D}  \frac{2 \bn\mcdot p}
 {(\bn\mcdot q +i0^+)(n\mcdot q\, \bn\mcdot p + i0^+)(q^2 +i0^+)} 
 = \frac{i}{2\pi} \int_{-\infty}^0 \!\!\!\!\! d\bn\mcdot q \:
   \frac{d^nq_\perp}{(2\pi)^n} \frac{\bn\mcdot p}{(\bn\mcdot q)(\bn\mcdot p)
   (q_\perp^2)}
   \,. \ \ \ \phantom{x} 
\end{eqnarray}
Here the zero-bin subtraction $I_{\rm 0}^{b\to s\gamma}$ is obtained from the
$q^\mu\to Q\lambda^2$ scaling limit of the $I_C^{b\to s\gamma}$ integrand
and avoids double counting for the usoft region of momentum space.  To double
logarithmic accuracy the $\bn\cdot q$ in the numerator of $\tilde I_C^{b\to
  s\gamma}$ can be dropped so
\begin{eqnarray}
 I_{\rm C}^{b\to s\gamma} = \tilde I_{\rm C}^{b\to s\gamma}-I_0^{b\to s\gamma}
   &=& \frac{i}{2\pi} \int_{-p^-}^{-\infty} \!\!\!\!\! d\bn\mcdot q \:
   \frac{d^nq_\perp}{(2\pi)^n} \frac{1}{(\bn\mcdot q)
   (q_\perp^2)} + \ldots 
   \,.
\end{eqnarray}
We see that the subtraction integral changes an infrared divergence in $I_C$ at
$\bn\mcdot q=0$ into a ultraviolet divergence for $\bn\mcdot q\to -\infty$.
Since we can see this at the level of the integrand it is obviously independent
of the choice of ultraviolet and infrared regulators.

With the prescribed cutoff regulators and $\perp$-spacetime dimension $n=2$,
these integrals can be evaluated to give
\begin{eqnarray}
 I_{\rm full}^{b\to s\gamma} &=&
  \frac{i}{8\pi^2} \bigg[ {\rm Li}_2\Big(\frac{-\Omega_\perp^2}{\Omega_-^2}\Big)
  + \ln\Big(\frac{\Omega_-}{p^-}\Big) 
    \ln\Big(\frac{\Omega_- p^-}{\Omega_\perp^2}\Big) 
   \bigg] +\ldots \,,\nn
 I_{\rm us}^{b\to s\gamma} &=&
  \frac{i}{8\pi^2} \bigg[ {\rm Li}_2\Big(\frac{-\Omega_\perp^2}{\Omega_-^2}\Big)
  + \ln\Big(\frac{\Omega_-}{\Lambda_-}\Big) 
    \ln\Big(\frac{\Omega_- \Lambda_-}{\Omega_\perp^2}\Big)
   \bigg] \,,\nn
 \tilde I_{\rm C}^{b\to s\gamma} &=&
  \frac{i}{8\pi^2} \bigg[
  -  \ln\Big(\frac{\Omega_\perp^2}{\Lambda_\perp^2}\Big)
     \ln\Big(\frac{\Omega_-}{p^-}\Big)
   \bigg] +\ldots\,,\nn
 I_{\rm 0}^{b\to s\gamma} &=&
  \frac{i}{8\pi^2} \bigg[
   -  \ln\Big(\frac{\Omega_\perp^2}{\Lambda_\perp^2}\Big)
     \ln\Big(\frac{\Omega_-}{\Lambda_-}\Big)
   \bigg] \,.
\end{eqnarray}
The full result for the collinear graph is therefore
\begin{eqnarray} \label{ICbsg}
  I_{\rm C}^{b\to s\gamma} 
  = \tilde I_{\rm C}^{b\to s\gamma} -I_{\rm 0}^{b\to s\gamma}
  &=& \frac{i}{8\pi^2} \bigg[
  -  \ln\Big(\frac{\Omega_\perp^2}{\Lambda_\perp^2}\Big)
     \ln\Big(\frac{\Lambda_-}{p^-}\Big)
   \bigg] +\ldots\,,
\end{eqnarray}
and we see that the zero-bin subtraction $I_0^{b\to s\gamma}$ has converted an IR
divergence $\ln(\Omega_-)$ for the $q^-$ variable in $\tilde I_C^{b\to s\gamma}$
into a UV divergence, $\ln(\Lambda_-)$. The sum of the \SCETa effective theory
contributions gives
\begin{eqnarray} \label{scetIsum}
  I_{\rm us}^{b\to s\gamma}  + I_{\rm C}^{b\to s\gamma} 
      &=& 
  \frac{i}{8\pi^2} \bigg[ {\rm Li}_2\Big(\frac{-\Omega_\perp^2}{\Omega_-^2}\Big)
  + \ln\Big(\frac{\Omega_-}{p^-}\Big) 
   \ln\Big(\frac{\Omega_- p^-}{\Omega_\perp^2}\Big) 
  +\ln^2\Big(\frac{\Lambda_\perp}{p^-}\Big)
  \!-\! \ln^2\Big(\frac{\Lambda_\perp}{\Lambda_-}\Big)
   \bigg] \nn
  && + \ldots \,.
\end{eqnarray}
The first two terms on the r.h.s.\ contain the infrared divergences and exactly
reproduce these divergences in the full theory result $I_{\rm full}^{b\to
  s\gamma} $.  Furthermore, the last two terms in Eq.~(\ref{scetIsum}) depend
only on the ultraviolet cutoffs and the large label momentum $p^-$ and can be
compensated by a counterterm for the current in \SCETa.  If $I_0^{b\to s\gamma}
$ in Eq.~(\ref{ICbsg}) had been left out, then we would not properly reproduce
the IR divergences in the full theory result. Furthermore, without $I_0^{b\to
  s\gamma}$, the ultraviolet cutoff dependent term would have cross terms
$\ln(\Lambda_-)\ln(\Omega_\perp^2)$ and $\ln(\Lambda_\perp^2)\ln(\Omega_-)$ and
it would not be possible to cancel the cutoff dependence by a counterterm
independent of the IR regulator.

The above calculation was performed for the current $J^{(0)}$ in
Eq.~(\ref{J01}). Since our regulator leaves all external lines onshell we obtain
exactly the same results if we had started with the current $J^{\prime (0)}$ in
Eq.~(\ref{J02}), which is obtained after making a field redefinition involving
the Wilson line $Y$.  Since we work onshell the two forms of the current are
equivalent, and the Feynman rule from the Wilson line $Y$ give exactly the same
integral in Eq.~(\ref{Ius}). Thus our implementation of a cutoff IR regulator
does not destroy the eikonal factorization embodied by the field redefinitions
involving the Wilson line $Y$. This property of the field theory is not
maintained with the offshellness IR regulator which we consider in the next
section. This should be considered as a fault of this IR regulator as pointed
out in Ref.~\cite{Bauer:2003td}. In Ref.~\cite{Bauer:2003td} an energy dependent
gluon mass regulator was studied which also preserves the field
redefinition.\footnote{Ref.~\cite{Bauer:2003td} also argued that the $\bn\mcdot
  k\to 0$ divergence must be treated as a UV in the EFT since it comes from
  angles opposite to the collinear direction. The renormalizability properties
  of field theory only appear for large momenta, and the zero-bin turns this
  divergence into a true UV divergence. One must be careful about the
  distinction between angles for particle and antiparticle poles when
  determining that the $\bn\mcdot k\to 0$ divergence is IR.}


\subsection{Offshell Regulator with Dimensional Regularization 
for $B\to X_s\gamma$}
\label{sect_sceta2}

We now repeat the calculation of the effective theory diagrams in the previous
section but keep $p^2\ne 0$ to regulate the infrared and use dimensional
regularization for the ultraviolet, $D=4-2\epsilon$. The full theory integral
is
\begin{eqnarray}
  I_{\rm full}^{b\to s\gamma}
   &=& \int \!\!\frac{d^Dq}{(2\pi)^D}  \frac{4 p_b \cdot p}
 {(q^2+i0^+)(q^2+2p_b\cdot q+i0^+)[(q+p)^2+i0^+]} \,.
\end{eqnarray}
The SCET integrals are
\begin{eqnarray} \label{drbsg}
 I_{\rm us}^{b\to s\gamma}
   &=& \int \!\!\frac{d^Dk}{(2\pi)^D}  \frac{1}
 {(k^2+i0^+)(v\mcdot k +i0^+)(n\mcdot k + p^2/\bn\mcdot p + i0^+)} 
 \,, \nn
 \tilde I_{\rm C}^{b\to s\gamma}
   &=& \int \!\!\frac{d^Dq}{(2\pi)^D}  \frac{2 \bn\mcdot (q+p)}
 {(\bn\mcdot q +i0^+)[(q+p)^2+ i0^+](q^2 +i0^+)} 
  \,,\nn
 I_{\rm 0}^{b\to s\gamma}
   &=& \int \!\!\frac{d^Dq}{(2\pi)^D}  
 \frac{2 \bn\mcdot p}
 {(\bn\mcdot q +i0^+)(n\mcdot q\, \bn\mcdot p + p^2 +i0^+)
  (q^2 +i0^+)}
   \,. \ \ \ \phantom{x} 
\end{eqnarray}
Again, one can see that as $\bn\cdot q\to 0$ the difference $\tilde I^{b\to
  s\gamma}_C-I^{b\to s\gamma}_0$ does not have an infrared divergence from this
region.  However in $I_C$ alone, there is an
infrared divergence from this region that is not regulated by $p^2\ne 0$.  It is
regulated by dimensional regularization, and so contributes to the $1/\epsilon$
singular terms.  Evaluating the above integrals we find,
\begin{eqnarray}
 I_{\rm full}^{b\to s\gamma} 
  &=& -\frac{i}{16\pi^2} \bigg[
   \ln^2 \Big(\frac{-p^2}{[\bn\mcdot
   p]^2}\Big) \bigg] + \ldots \,,\nn
 I_{\rm us}^{b\to s\gamma} 
  &=&  -\frac{i}{16\pi^2} \bigg[ \frac{1}{\epsilon_{\rm UV}^2} 
   +\frac{2}{\epsilon_{\rm UV}}\ln\Big(\frac{\mu \bn\mcdot p}{-p^2}\Big) 
   + 2 \ln^2\Big(\frac{\mu\bn\mcdot p}{-p^2}\Big) \bigg] + \ldots
 \,, \nn
 \tilde I_{\rm C}^{b\to s\gamma}
   &=& -\frac{i}{16\pi^2} \bigg[ -\frac{2}{\epsilon_{\rm IR}\epsilon_{\rm UV}} 
   -\frac{2}{\epsilon_{\rm IR}}\ln\Big(\frac{\mu^2}{-p^2}\Big) 
   - \ln^2\Big(\frac{\mu^2}{-p^2}\Big) + \Big(\frac{2}{\epsilon_{\rm IR}}
   -\frac{2}{\epsilon_{\rm UV}}\Big)\ln\Big(\frac{\mu}{\bn\mcdot p}\Big)\bigg] + \ldots
  \,,\nn
 I_{\rm 0}^{b\to s\gamma}
   &=& -\frac{i}{16\pi^2} \bigg[ \Big(\frac{2}{\epsilon_{\rm UV}} -
   \frac{2}{\epsilon_{\rm IR}}
   \Big) \Big\{\frac{1}{\epsilon_{\rm UV}} +
     \ln\Big(\frac{\mu^2}{-p^2}\Big) 
      -  \ln\Big(\frac{\mu}{\bn\mcdot p}\Big) \Big\} \bigg] 
   \,, \ \ \ \phantom{x} 
\end{eqnarray}
where we have distinguished between ultraviolet and infrared divergences. Here
we see that the zero-bin contribution $I_0^{b\to s\gamma}$ is responsible for
canceling IR divergences in $\tilde I_{\rm C}^{b\to s\gamma}$ that were not
regulated by the offshellness,
\begin{eqnarray} \label{ICbsgdr}
  I_{\rm C}^{b\to s\gamma} 
   = \tilde  I_{\rm C}^{b\to s\gamma} -  I_{\rm 0}^{b\to s\gamma}
  = -\frac{i}{16\pi^2} \bigg[ -\frac{2}{\epsilon_{\rm UV}^2} 
   -\frac{2}{\epsilon_{\rm UV}}\ln\Big(\frac{\mu^2}{-p^2}\Big) 
   - \ln^2\Big(\frac{\mu^2}{-p^2}\Big) \bigg] + \ldots
  \,.
\end{eqnarray}
  Therefore the sum of the \SCETa contributions gives
\begin{eqnarray} \label{Iusc_DR}
 I_{\rm us}^{b\to s\gamma} \!+\! I_{\rm C}^{b\to s\gamma}   &=& 
   -\frac{i}{16\pi^2} \bigg[-\frac{1}{\epsilon_{\rm UV}^2} 
   \!-\!\frac{2}{\epsilon_{\rm UV}}\ln\Big(\frac{\mu}{\bn\mcdot p}\Big) 
   \!-\!2\ln^2 \Big(\frac{\mu}{\bn\mcdot
   p}\Big)  \!+\! \ln^2 \Big(\frac{-p^2}{[\bn\mcdot
   p]^2}\Big) \bigg] + \ldots
  \,.
\end{eqnarray}
The last term reproduces the infrared structure of the full theory result. The
first two terms are canceled by a counterterm. The third term contributes a
finite contribution to the hard Wilson coefficient of the heavy-to-light current
in matching onto the full theory. Again we see that the contribution from
$I^{b\to s\gamma}_{\rm 0}$ is necessary in order for the infrared divergences in
the full and effective theories to match up. The ellipses in Eq.~(\ref{Iusc_DR})
denote $1/\epsilon$, single log, and finite terms that we have not bothered to
display in the quoted results, but which have the same desired properties.

The computation of SCET anomalous dimensions in Ref.~\cite{Bauer:2000yr} and all
subsequent papers used the entire $1/\epsilon$ divergent terms in $\tilde
I_C^{b\to s\gamma}$ to compute the anomalous dimension. As we have shown above,
some of these divergences are, in fact, infrared divergences.  The pullup
mechanism is needed to convert these into ultraviolet divergences which can then
be canceled by local counterterms in the effective theory, and so properly
contribute to the anomalous dimension. 

One can see the problem with having $\tilde I_C^{b\to s\gamma}$ and no
subtraction term in another way.  Consider adding a very small gluon mass to the
collinear calculation, where $m^2 \ll p^2$. The gluon mass can only affect the
IR, and we should obtain the same form for the $1/\epsilon_{\rm UV}$ divergences
if we expand in $m^2$ before or after the integration. If we consider expanding
after the integration, then performing the $q^+$ integral by contours followed by the $q_\perp$
integration gives
\begin{eqnarray} \label{Ic0mp}
  \tilde I_{\rm C}^{b\to s\gamma} &=& \frac{2 i \Gamma(\epsilon)\mu^{2\epsilon}}{16\pi^2}
   \int_{-p^-}^0\!\! \frac{dq^-}{q^-} 
      \Big(m^2+ \frac{p^2q^-}{p^-}\Big)^{-\epsilon} 
      \Big(1+\frac{q^-}{p^-}\Big)^{-\epsilon}
   \,,\nn
   I_{0}^{b\to s\gamma} &=& \frac{2 i \Gamma(\epsilon)\mu^{2\epsilon}}{16\pi^2}
   \int_{-\infty}^0\!\! \frac{dq^-}{q^-} 
      \Big(m^2+ \frac{p^2q^-}{p^-}\Big)^{-\epsilon} \,.
\end{eqnarray}
Thus with $m^2\ne 0$ the IR singularity at $q^-\to 0$ is no longer regulated by
dimensional regularization in $\tilde I_C^{b\to s\gamma}$ or $I_{0}^{b\to
  s\gamma}$, however the difference $I_C^{b\to s\gamma}=\tilde I_C^{b\to
  s\gamma}-I_{0}^{b\to s\gamma}$ remains well defined. Here the $1/\epsilon$
terms in $\tilde I_{\rm C}^{b\to s\gamma}$ do not give the correct counterterm
structure even if we set $\epsilon_{\rm IR}=\epsilon_{\rm UV}$.  Letting $q^-=
-xp^-$ gives
\begin{eqnarray}
   I_{\rm C}^{b\to s\gamma} &=& \frac{-2 i
     \Gamma(\epsilon)\mu^{2\epsilon}}{16\pi^2}
    \bigg[ 
     \int_0^1  \frac{dx}{x} (1-x)^{-\epsilon} (m^2-x p^2)^{-\epsilon} - 
     \int_0^\infty \frac{dx}{x} (m^2-x p^2)^{-\epsilon} \bigg] 
   \nn
    &=& \frac{-2 i \Gamma(\epsilon)\mu^{2\epsilon}}{16\pi^2} \bigg[ 
      \frac{\pi^2\epsilon}{6} - 
     \int_1^\infty \frac{dx}{x} (m^2-x p^2)^{-\epsilon} \bigg]
   \nn
   &=& = -\frac{i}{16\pi^2} \bigg[ -\frac{2}{\epsilon_{\rm UV}^2} 
   -\frac{2}{\epsilon_{\rm UV}}\ln\Big(\frac{\mu^2}{-p^2}\Big) 
   - \ln^2\Big(\frac{\mu^2}{-p^2}\Big)  - 2 {\rm Li}_2\Big(\frac{-m^2}{p^2}\Big)
  +\ldots \bigg],
\end{eqnarray}
and in the limit $m^2\ll p^2$ this reproduces Eq.~(\ref{ICbsgdr}). Thus with the
zerobin subtractions the $1/\epsilon_{\rm UV}$ divergences in the effective
theory are independent of the choice of IR regulator.

Note that in this section it was crucial to use the current $J^{(0)}$ in order
that taking $p^2\ne 0$ provides the same IR regulator in the full and effective
theories.  For the SCET current $J^{\prime (0)}$ this is no longer possible.
Working with this current, only onshell IR regulators should be considered. This
happens because the field redefinitions involving $Y$'s modify the LO collinear
Lagrangian, rather than just subleading terms, and are therefore sensitive to
regulation of the propagator.


\subsection{Production of $n$-$\bn$ jets}
\label{sect_sceta3}

As another example of the zero-bin subtractions in \SCETa we consider the
one-loop diagrams contributing to two jet production, $\gamma^* \to q\bar q$.
The degrees of freedom required in \SCETa are those pictured in
Fig.~\ref{fig_SCETUVvsIR}a. In this case we have subtractions for both the
$n$-collinear and $\bn$-collinear fields, which ensure that they do not overlap
with the usoft region. The leading order \SCETa current is~\cite{Bauer:2002nz}
\begin{align} \label{J0prod}
  J^{(0)} =  C(\omega,\omega')  (\bar\xi_n W_n)_\omega \gamma^\mu
    (W_\bn^\dagger \xi_\bn)_{\omega'} \,.
\end{align}
If we make the decoupling field redefinition~\cite{Bauer:2001yt} which encodes
the eikonal coupling to all collinear quarks and gluons then $J^{(0)}$ becomes
$J^{\prime(0)}=C(\omega,\omega') (\bar\xi_n W_n)_\omega Y_n^\dagger Y_\bn \gamma^\mu
(W_\bn^\dagger \xi_\bn)_{\omega'}$.  Below we work with the current $J^{(0)}$
since we will use an offshellness IR regulator.

\begin{figure}
\begin{center}
\includegraphics[width=2.8cm]{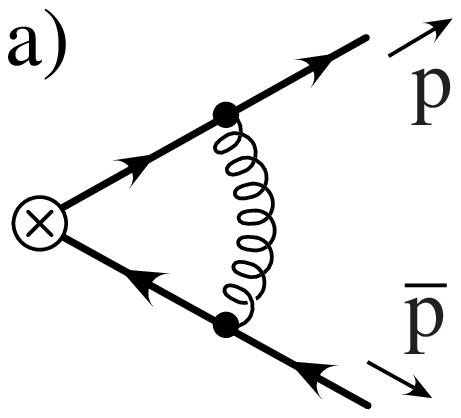}\qquad\qquad
\includegraphics[width=6.5cm]{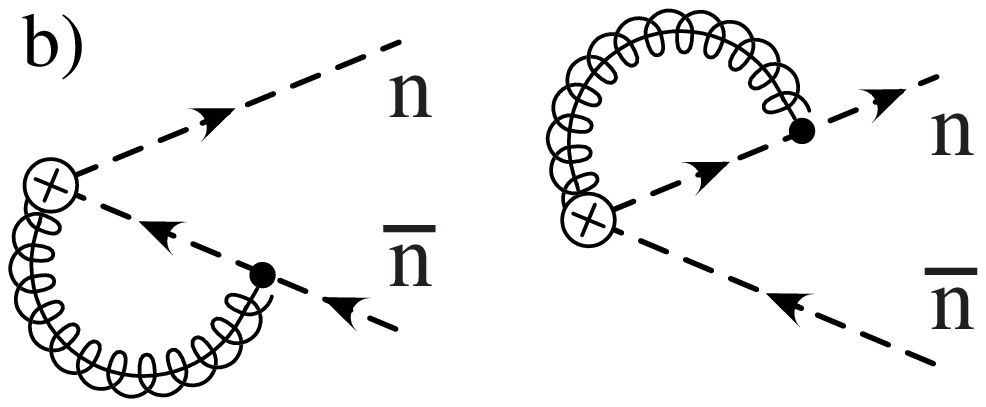}\qquad\qquad
\includegraphics[width=2.5cm]{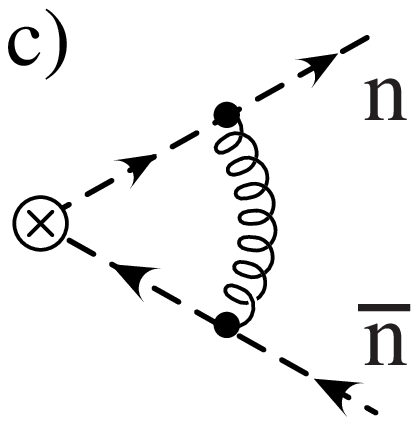}
\end{center}
\vskip-0.4cm
\caption{Vertex diagrams for $n$--$\bn$ production. a)
     full theory, b) \SCETa collinear graphs, and c) \SCETa usoft graph.
\label{fig:sudakov}}
\end{figure}
The one-loop vertex graphs are shown in Fig.~\ref{fig:sudakov}. Wavefunction
graphs are not shown, but in Feynman gauge the collinear gluon wavefunction
renormalization for a collinear quark is equal to the full theory result and the
usoft gluon contribution vanishes. Working to double logarithmic order and
breaking up the collinear terms into the naive result and subtractions the
relevant integrals are
\begin{eqnarray}
  I_{\rm full}^{\rm prod}
   &=& \int \!\!\frac{d^Dq}{(2\pi)^D}  \frac{-4\, p \cdot \bar p}
{[q^2+i0^+][(q-\bar p)^2+i0^+)[(q+p)^2+i0^+]}, \nn
I_{\rm usoft}^{\rm prod}
   &=& \int \!\!\frac{d^Dq}{(2\pi)^D}  \frac{2 (\bn\mcdot p)(-n\mcdot\bar p)}
{[q^2+i0^+][\bn\mcdot p\, n\mcdot q + p^2 +i0^+]
  [-n\mcdot \bar p\, \bn\mcdot q + \bar p^2 + i0^+)}
\,, \nn
\tilde I_{n}^{\rm prod}
   &=& \int \!\!\frac{d^Dq}{(2\pi)^D}  \frac{2 \bn\mcdot p}
{[\bn\mcdot q +i0^+][(q+p)^2+ i0^+][q^2 +i0^+ ]}
  \,,\nn
I_{n\rm 0}^{\rm prod}
   &=& \int \!\!\frac{d^Dq}{(2\pi)^D}
\frac{2 \bn\mcdot p}
{[\bn\mcdot q +i0^+][\bn\mcdot p\, n\mcdot q + p^2 +i0^+]
  [q^2 +i0^+]}
  \,,\nn
\tilde I_{\bn}^{\rm prod}
   &=& \int \!\!\frac{d^Dq}{(2\pi)^D}  \frac{2 (-n\mcdot \bar p)}
{[n\mcdot q +i0^+][(q-\bar p)^2+ i0^+][q^2 +i0^+]}
  \,,\nn
I_{\bn\rm 0}^{\rm prod}
   &=& \int \!\!\frac{d^Dq}{(2\pi)^D}
\frac{2 (-n\mcdot \bar p)}
{[n\mcdot q +i0^+][- n\mcdot \bar p\,\bn\mcdot q + \bar p^2 +i0^+]
  [q^2 +i0^+]}
   \,. \ \ \ \phantom{x}
\end{eqnarray}
We have kept offshellnesses, $p^2\ne 0$ and $\bar p^2\ne 0$ to regulate the IR
and will use dimensional regularization for the UV. Much like the heavy-to-light
computation, this does not regulate all the IR divergences in the naive
collinear integrands, $\tilde I_n^{\rm prod}$ and $\tilde I_\bn^{\rm prod}$.
Note that we took $\bn\mcdot (p+q)\to \bn\mcdot p$ in the numerator of the
collinear graphs since we only examine the double logarithms and have made a
corresponding approximation in $I_{\rm full}^{\rm prod}$.  Evaluating the full
theory integral we find\footnote{In the computation of the cross-section in the
  full theory the $\ln(p^2)$ IR divergences are canceled by analogous IR
  divergences in the bremsstrahlung graphs.}
\begin{eqnarray}
  I_{\rm full}^{\rm prod} = -\frac{i}{8\pi^2} \ \ln\Big(\frac{p^2}{Q^2}\Big)\:
    \ln\Big(\frac{\bar p^2}{Q^2}\Big) + \ldots \,.
\end{eqnarray}
The usoft loop graph in the effective theory gives
\begin{eqnarray}
  I_{\rm usoft}^{\rm prod} =- \frac{i}{8\pi^2} \bigg\{ \frac{1}{\epsilon_{\rm UV}^2}
   \!+\! \frac{1}{\epsilon_{\rm UV}} \ln\Big( \frac{\mu n\mcdot \bar p}{\bar
     p^2} \Big)
   \!+\! \frac{1}{\epsilon_{\rm UV}} \ln \Big( \frac{\mu \bn\mcdot p}{-p^2} \Big)
   \!+\! \frac{1}{2} \ln^2 \!\bigg[ \frac{\bar p^2 (-p^2)}
     {(\mu\, \bn\mcdot p)(\mu\, n\mcdot\bar p) } \bigg]
   \!+\! \ldots \bigg\} \,.
\end{eqnarray}
For the $n$-collinear naive integral and subtraction we find
\begin{eqnarray}
\tilde I_{n}^{\rm prod}
   &=& -\frac{i}{16\pi^2} \bigg[- \frac{2}{\epsilon_{\rm IR}\epsilon_{\rm UV}}
   -\frac{2}{\epsilon_{\rm IR}}\ln\Big(\frac{\mu^2}{-p^2}\Big)
   \!-\! \ln^2\Big(\frac{\mu^2}{-p^2}\Big) \!+\! \Big(\frac{2}{\epsilon_{\rm IR}}
   -\frac{2}{\epsilon_{\rm UV}}\Big)\ln\Big(\frac{\mu}{\bn\mcdot p}\Big)\bigg] + \ldots
  \,,\nn
I_{n\rm 0}^{\rm prod}
   &=& -\frac{i}{16\pi^2} \bigg[ \Big(\frac{2}{\epsilon_{\rm UV}} -
   \frac{2}{\epsilon_{\rm IR}}
   \Big) \Big\{\frac{1}{\epsilon_{\rm UV}} +
     \ln\Big(\frac{\mu^2}{-p^2}\Big)
      -  \ln\Big(\frac{\mu}{\bn\mcdot p}\Big) \Big\} \bigg]
   \,,
\end{eqnarray}
so the full $n$-collinear result is
\begin{eqnarray}
I_{n}^{\rm prod}
   &=& \tilde I_n^{\rm prod} - I_{n\rm 0}^{\rm prod} =
   -\frac{i}{8\pi^2} \bigg[ -\frac{1}{\epsilon_{\rm UV}^2}
   -\frac{1}{\epsilon_{\rm UV}}\ln\Big(\frac{\mu^2}{-p^2}\Big)
   - \frac{1}{2} \ln^2\Big(\frac{\mu^2}{-p^2}\Big) \bigg] + \ldots
   \,. \ \ \ \phantom{x}
\end{eqnarray}
Just as in the $b\to s\gamma$ example the subtraction terms remove the
$1/\epsilon_{\rm IR}$ poles, and the IR in the complete collinear integral is
regulated by the offshellness. The ellipses denote $1/\epsilon$, single log, and
finite terms that we have not bothered to display in the quoted results. The
results for the $\bn$-collinear terms are similar
\begin{eqnarray}
\tilde I_{\bn}^{\rm prod}
   &=& -\frac{i}{16\pi^2} \bigg[ -\frac{2}{\epsilon_{\rm IR}\epsilon_{\rm UV}}
   -\frac{2}{\epsilon_{\rm IR}}\ln\Big(\frac{\mu^2}{-\bar p^2}\Big)
   \!-\!
   \ln^2\Big(\frac{\mu^2}{-\bar p^2}\Big) \!+\! \Big(\frac{2}{\epsilon_{\rm IR}}
   -\frac{2}{\epsilon_{\rm UV}}\Big)\ln\Big(\frac{\mu}{n\mcdot \bar
     p}\Big)\bigg]
   + \ldots
  \,,\nn
I_{\bn\rm 0}^{\rm prod}
   &=& -\frac{i}{16\pi^2} \bigg[ \Big(\frac{2}{\epsilon_{\rm UV}} -
   \frac{2}{\epsilon_{\rm IR}}
   \Big) \Big\{\frac{1}{\epsilon_{\rm UV}} +
     \ln\Big(\frac{\mu^2}{-\bar p^2}\Big)
      -  \ln\Big(\frac{\mu}{n\mcdot \bar p}\Big) \Big\} \bigg]
   \,,\nn
I_{\bn}^{\rm prod}
   &=& \tilde I_\bn^{\rm prod} - I_{\bn\rm 0}^{\rm prod} =
   -\frac{i}{8\pi^2} \bigg[ -\frac{1}{\epsilon_{\rm UV}^2}
   -\frac{1}{\epsilon_{\rm UV}}\ln\Big(\frac{\mu^2}{-\bar p^2}\Big)
   - \frac{1}{2} \ln^2\Big(\frac{\mu^2}{-\bar p^2}\Big) \bigg] + \ldots
   \,. \ \ \ \phantom{x}
\end{eqnarray}
Adding up the \SCETa integrals, $I_{\rm scet}^{\rm prod} = I_{\rm usoft}^{\rm prod} +
I_{\bn}^{\rm prod} + I_{n}^{\rm prod}$,  we find
\begin{align}
I_{\rm scet}^{\rm prod}
    &=  -\frac{i}{8\pi^2} \bigg\{ \!- \frac{1}{\epsilon_{\rm UV}^2}
   \!-\!\frac{1}{\epsilon_{\rm UV}}\ln\Big(\frac{\mu^2}{-Q^2}\Big)
   \!-\! \frac{1}{2} \ln^2\Big(\frac{\mu^2}{-\bar p^2}\Big)
   \!-\! \frac{1}{2} \ln^2\Big(\frac{\mu^2}{-p^2}\Big)
   \!+\! \frac{1}{2} \ln^2 \!\Big( \frac{- \bar p^2 p^2}
     {\mu^2 Q^2} \Big)
   \bigg\} \nonumber \\
  &=-\frac{i}{8\pi^2} \bigg[ -\frac{1}{\epsilon_{\rm UV}^2}
   -\frac{1}{\epsilon_{\rm UV}}\ln\Big(\frac{\mu^2}{-Q^2}\Big)
   - \frac12 \ln^2 \Big(\frac{\mu^2}{-Q^2}\Big)
    +  \ln \! \Big(\frac{\bar p^2}{Q^2}\Big) \ln \Big(\frac{p^2}{Q^2}\Big)
   \bigg].
\end{align}
In the last line, the first two terms are removed by a counterterm in $\overline
{\rm MS}$ and the third term contributes to the Wilson coefficient $C^{\rm
  prod}$ in the one-loop matching, see Ref.~\cite{Manohar:2003vb}. The fourth
term exactly reproduces the IR divergences in the full theory result.

In the above computation there was an interplay between the usoft loop and the
$n$ and $\bn$ collinear loops which combine to reproduce the IR of the full
theory. The exact way in which these IR divergences combine depends on the
choice of IR regulator as we saw in the $b\to s\gamma$ example.  Again
we see that the zero-bin subtractions are important to correctly reproduce
the IR divergences once we distinguish between $\epsilon_{\rm UV}$ and
$\epsilon_{\rm IR}$.


\subsection{$\gamma^* \to q\bar q g$\,: The Zero-bin at Tree Level and for Phase
Space Integrals} \label{sect_sceta4}

In this section we show how the zero-bin is kept track of in tree level
computations. One has to consider it even at tree level because in SCET multiple
fields are present for the same physical particle.  We will demonstrate how
double counting is avoided in fully differential cross sections and how the
zero-bin subtraction affects phase space integrations. In the context of proving
factorization in Drell-Yan, subtractions which avoid overcounting in phase
space regions have been considered in Ref.~\cite{Collins:1982wa}.

Consider the high energy process $\gamma^*(q) \to q(p_1) \bar q(p_2) g(p_3)$ in
the rest frame of the $\gamma^*$ with $q^2=Q^2 \gg \Lambda_{\rm QCD}^2$.  This
is a basic ingredient in two-jet and three-jet production, which were considered
in SCET in Refs.~\cite{Bauer:2002ie,Bauer:2003di,Lee:2006fn}. We take the full
theory production current, $J=\bar\psi \gamma^\mu \psi$. The external lines have
$p_i^2=0$ and we define dimensionless momentum fractions $x_i = 2 q\cdot p_i
/q^2$ so that momentum conservation reads $2=x_1+x_2+x_3$. Computing the phase
space integrals in $(4\!-\!2\epsilon)$-dimensions, the standard full theory
cross section from the bremsstrahlung graphs is
\begin{align} \label{fullS12}
 \frac{1}{\sigma_0}\:
  \frac{d\sigma_{\rm full}}{dx_1\,dx_2} &= \frac{C_F \alpha_s}{2\pi}\:
   \frac{\mu^{2\epsilon}}{q^{2\epsilon}}\:
  \frac{1}{\Gamma(1\!-\!\epsilon)(x_1\!+\! x_2 \!-\! 1)^\epsilon }
      \frac{x_1^2+x_2^2 -\epsilon (2\!-\!x_1\!-\! x_2)^2}
      {(1\!-\! x_1)^{1+\epsilon}(1\!-\! x_2)^{1+\epsilon}}
\end{align}
where $\sigma_0$ is the Born cross section in dimensional regularization,
$\sigma_0=(4\pi\alpha^2/Q^2) \sum_f e_f^2 +{\cal O}(\epsilon)$ with a sum over
the quark charges $e_f=2/3$ or $-1/3$.

\begin{figure}[!t]
\begin{center}
 \includegraphics[width=16cm]{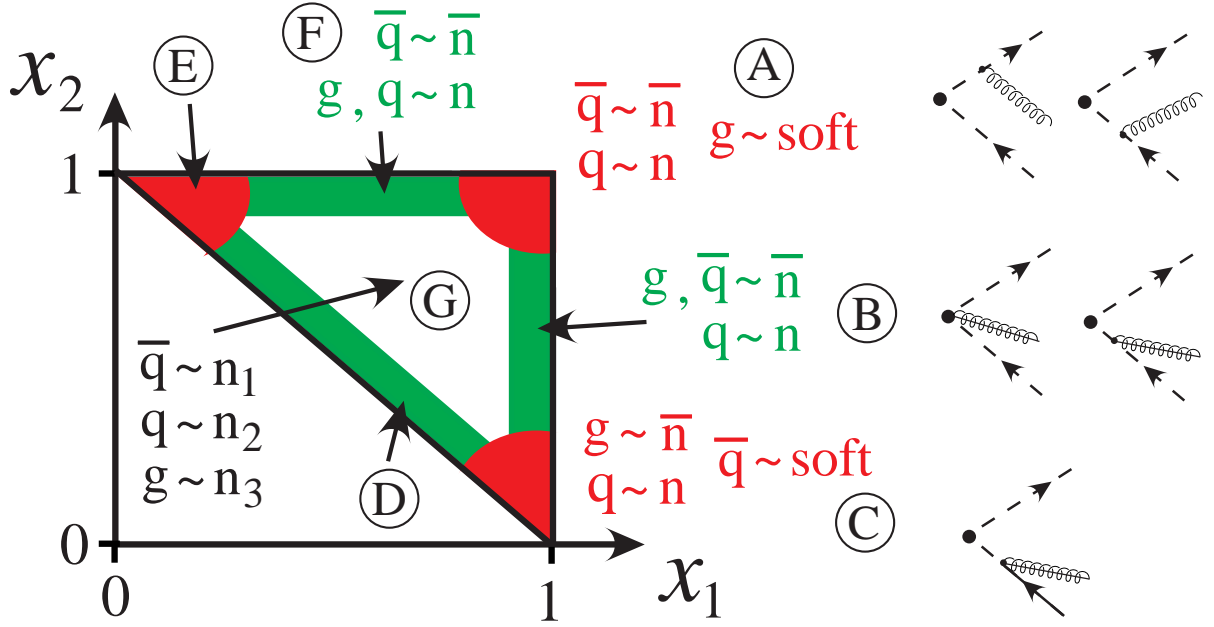}
\end{center}
 \vskip -.3cm 
\caption[1]{ 
  Regions R=A--G for $\gamma^*\to q\,\bar q\,g$ where the particles become
  usoft and collinear. For regions A,B,C the graphs in SCET are shown. The
  soft regions are slightly exaggerated for visibility.
}\label{fig:qqg} 
\end{figure}
Taking $x_1$ and $x_2$ as the independent variables we have the phase space
shown in Fig.~\ref{fig:qqg}.  In SCET the different regions of the phase space
plot are described by distinct EFT diagrams with the particles being created by
either a collinear or usoft field. The relevant limits are
\begin{align} \label{qqglimit}
  {\rm A}:\ \  & x_{1,2}\to 1 \,,
  & {\rm B}:\ \  & x_{1}\to 1, x_2\sim x \,,
  & {\rm C}:\ \  & x_{1}\to 1, x_2\to 0 \,, \nonumber \\
  {\rm D}:\ \  &  x_{1}\!+\!x_2\to 1, x_{1,2}\sim x \,,
  & {\rm E}:\ \  & x_{1}\to 0, x_2\to 1 \,,
  & {\rm F}:\ \  & x_{2}\to 1, x_1\sim x \,, \nonumber \\
  {\rm G}: \ \ & x_1, x_2, x_1+x_2 \sim x \,,
\end{align}
where $x$ denotes generic values not near the two ends.  In
Fig.~\ref{fig:qqg} the SCET graphs for regions A,B,C are shown. We will compute
the sum of the square of SCET diagrams for each region and compare them with the
full theory result for the double differential cross section and a single
differential cross section. In the $\gamma^*$ rest frame with $2q^\mu = Qn^\mu +
Q\bn^\mu$ there is still a rotational freedom in the $\perp$-plane which we can
fix in performing the calculations. For A,B,C we take $p_1^\perp=0$, while for
E,F it is more convenient to take $p_2^\perp =0$.

Computing these SCET graphs with the phase space integrals in dimensional
regularization we find cross sections in each of the regions $R$
\begin{align} \label{qqgAmps}
  \frac{d\sigma_R}{\sigma_0\: dx_1\,dx_2} =\frac{C_F\alpha_s}{2\pi}\  
   \frac{1}{\Gamma(1\!-\!\epsilon)(1\!-\! x_1)^\epsilon (1\!-\! x_2)^\epsilon 
     (x_1\!+\! x_2 \!-\! 1)^\epsilon }\ 
    \big| {\cal A}_R(x_1,x_2)\big|^2 \,.
\end{align} 
Here ${\cal A}_R$ is the amplitude in region $R$ divided by
$Z=2g\sqrt{2(1\!-\!\epsilon)}$.  From the tree level diagrams we find
\begin{align}
  |{\cal A}_A|^2
  \Big|_{\parbox{1.35cm}{\scriptsize  $x_{1}\ne 0$\\ $x_2\ne 1\!-\!x_1$}} 
  &= \frac{1}{Z^2} \: 
   \sum_{\rm spins} \bigg|  
   \bar u_n(p_1) \Big\{ \frac{ -g n\mcdot \varepsilon}{n\mcdot p_3}
   + \frac{ g \bn\mcdot \varepsilon}{\bn\mcdot p_3}  \Big\}
    \gamma^\mu_\perp T^A v_{\bn}(p_2) 
   \bigg|^2
  \nonumber \\
  & = \frac{2}{(1\!-\! x_1)(1\!-\! x_2)} \,,\nonumber \\[5pt]
 |{\cal A}_B|^2 
  \Big|_{\parbox{1.35cm}{\scriptsize $x_{1}\ne 0$\\  $x_{2}\ne 1$\\ 
    $x_2\ne 1\!-\! x_1$}}
  &= \frac{1}{Z^2} \: 
   \sum_{\rm spins} \bigg|  
  \bar u_n(p_1) \Big\{ \frac{-g n\mcdot \varepsilon}{n\mcdot p_3}
     \gamma^\mu_\perp \!+\! \frac{g n\mcdot(p_3\!-\!p_2)}{(p_3\!-\!p_2)^2}
  \gamma_\perp^\mu  \Big( \bn\mcdot\varepsilon \!+\! 
   \frac{\slash\!\!\!\varepsilon_\perp\,
    \slash\!\!\! p_2^\perp}{n\mcdot p_2}\Big)  \Big\} T^A v_{\bn}(p_2) \bigg|^2
  \nn
 & =  \frac{1+x_2^2}{(1\!-\! x_1)(1\!-\! x_2)} 
   -  \frac{\,\epsilon\,(1\!-\! x_2)}{(1\!-\! x_1)}
  \,,\nonumber\\[5pt]
|{\cal A}_C|^2 
  \Big|_{\parbox{1.35cm}{\scriptsize $x_1\ne 0$\\ $x_2\ne 1$}}
  &= \frac{1}{Z^2} \: 
   \sum_{\rm spins}  \bigg|  
    \bar u_n(p_1) \Big\{ \Big(
   \frac{g\bnslash\, \slash\!\!\!\varepsilon_\perp}{2\,\bn\mcdot p_1}
    \Big)\gamma_\perp^\mu -
   \gamma_\perp^\mu \Big(
   \frac{g\bnslash\, \slash\!\!\!\varepsilon_\perp}{2\,\bn\mcdot p_1}  
    \Big) 
    \Big\} T^A v(p_2) \bigg|^2
  \nn
 &= \frac{1-\epsilon}{(1-x_1)} 
 \,,
\end{align}
and 
\begin{align}
 |{\cal A}_D|^2 
  \Big|_{\parbox{1.4cm}{\scriptsize $x_1\!+\!x_2\ne 2$\\ $x_{1,2}\ne 0$}}
  &=  \frac{1+(x_1-x_2)^2}{2 x_1 x_2} +{\cal O}(\epsilon)
  \,,\nonumber\\[5pt]
 |{\cal A}_E|^2 
 \Big|_{\parbox{1.35cm}{\scriptsize $x_1\ne 1\!-\! x_2$\\ $x_2\ne 0$}}
  &=  \frac{1-\epsilon}{(1-x_2)}
 \,, \nonumber\\[5pt]
 |{\cal A}_F|^2 
   \Big|_{\parbox{1.35cm}{\scriptsize $x_1\ne 1\!-\! x_2$\\ $x_1\ne 1$\\ 
     $x_2\ne 0$}}
  &=   \frac{1+x_1^2}{(1\!-\! x_1)(1\!-\! x_2)}
    -  \frac{\,\epsilon\,(1\!-\! x_1)}{(1\!-\! x_2)}
  \,,\nonumber\\[5pt]
 |{\cal A}_G|^2 
  \Big|_{\parbox{1.4cm}{\scriptsize $x_1\!+\!x_2\ne 1$\\ $x_{1,2}\ne 1$}}
  &=  \frac{x_1^2+x_2^2}{(1\!-\! x_1)(1\!-\! x_2)} +{\cal O}(\epsilon)
  \,.
\end{align}
Here $u_n$, $u_\bn$, and $u$ are an $n$-collinear spinor, an $\bn$-collinear
spinor, and an usoft spinor respectively, all with relativistic
normalization.  For the regions A, B, C, we explicitly show the amplitudes that
follow from the SCET Feynman diagrams in Fig.~\ref{fig:qqg} and include
$\epsilon$-dependent terms in the results.  The amplitudes for A and B follow
from the LO Lagrangians and LO SCET production current in Eq.~(\ref{J0prod}).
The result for C requires an insertion of the subleading Lagrangian ${\cal
  L}_{\xi q}^{(1)}= \bar q_{us} W_\bn^\dagger ig\, \slash\!\!\!\!{ B}_\perp^\bn
\xi_\bn +{\rm h.c.}$~\cite{Beneke:2002ph}, where the field strength
\mbox{$ig\,\slash\!\!\!\!{B}_\perp^\bn = [i\bn\mcdot D_\bn^c,i\slash\!\!\!\!
  D_\perp^\bn]$}.  In Eq.~(\ref{qqgAmps}) we have translated the zero-bin
restrictions on the large momenta of collinear particles to restrictions on
$x_1$ and $x_2$ as shown on the RHS of the equations.

It should be obvious from the form of ${\cal A}_A$--${\cal A}_G$ that one can
not simply add the SCET diagrams to reproduce the doubly differential cross
section in Eq.~(\ref{fullS12}).  The point is that the effective theory results
do not overlap, as made explicit by the sums which exclude the zero-bins, and
constrain the valid region of phase space.  Given an $x_1$ and $x_2$, only one
of the effective theory expressions is relevant. It is straightforward to
determine which one once we specify parametric definitions of the scaling limits
in Eq.~(\ref{qqglimit}), and pick values for $x_1$ and $x_2$.\footnote{The
  particles in the final state can be treated as \emph{observed}, by a
  measurement of the final state. For each final state particle, one can assign
  a label $p$ and residual momentum $k$ as given by the binning of momentum
  space, and classify particles as collinear or usoft depending on whether
  $p\not=0$. } The SCET diagrams in this region reproduce the full theory double
differential cross section order by order in the expansion. Thus, it is crucial
to take the zero-bin into account even at tree level in order to avoid double
counting.

Often we would like to deal with a less differential cross section which
involves integrating over kinematic variables. In this case we should implement
the zero-bin subtractions in the phase space integrals using Eq.~(\ref{1.02}) to
avoid double counting when combining regions. 
As an example of the zero-bin subtractions in phase space integrals we consider
the $\gamma^*\to q\bar q g$ cross section $d\sigma/dx_1$ for {\em fixed} $x_1 =
1 - \delta$ with $\delta \sim \lambda^2$ and nonzero. In the full theory the
single differential cross section for $x_1\to 1$ is obtained by integrating
Eq.~(\ref{fullS12}) over $1-x_1 < x_2 < 1$ and expanding about $\delta$.
\begin{align} \label{Kfull}
 \frac{1}{\sigma_0}\:
  \frac{d\sigma_{\rm brem}^{\rm full}}{dx_1} \bigg|_{x_1\to 1} 
  &= \frac{C_F \alpha_s}{\pi}\:
    \frac{\mu^{2\epsilon}}{q^{2\epsilon}}\:
  \frac{1}{(1\!-\! x_1)^{1+\epsilon}\, \Gamma(1\!-\!\epsilon)}
   K^{\rm full} \,, \nonumber\\
  K^{\rm full} &=
  \Big[ -\frac{1}{\epsilon_{\rm IR}} - \frac{3}{4} + {\cal O}(\delta) \Big] \,,
\end{align}
where the IR divergence was regulated by dimensional regularization.  To
reproduce this result from the SCET computation requires adding contributions
from regions A, B, and C. By using Eq.~(\ref{1.01}) we can add these
contributions and still integrate over the full phase space in $x_2$. For each
region we can also expand the prefactor in Eq.~(\ref{qqgAmps}) in the
appropriate manner without effecting the LO results.  Since $x_1$ is fixed we
need not worry about subtractions involving this variable. Overlap occurs from
$|{\cal A}_B|^2$ with region C if $x_2=0$ and with region A if $x_2=1$, so we
find two zero-bin subtractions for these contributions.  We multiply by
$(1-x_1)^{1+\epsilon}$ to give the same normalization as Eq.~(\ref{Kfull}). The
regions with a soft particle are unsubtracted and give
\begin{align}
  K^A &= (1\!-\! x_1)^{1+\epsilon} \int_{1-x_1}^1 \!\!\!\!\! dx_2  \ \frac{2}{
    (1\!-\! x_1)^{1+\epsilon} (1\!-\! x_2)^{1+\epsilon} }  
  = -\frac{1}{\epsilon_{\rm IR}} +{\cal O}(\delta) \,,\nonumber \\
  K^C &= (1\!-\! x_1)^{1+\epsilon} \int_{1-x_1}^1 \!\!\!\!\! dx_2  \
   \frac{1-\epsilon}{
    (1\!-\! x_1)^{1+\epsilon} (x_1\!+\!x_2\!-\! 1)^{1+\epsilon} }  = \frac{1}{2}
  +{\cal O}(\delta) \,.
\end{align}
For region B with three collinear particles the naive contribution and its two
subtractions $(x_2=1, x_2=0)$ are
\begin{align}
  \tilde K^B &= (1\!-\! x_1)^{1+\epsilon} \int_{1-x_1}^1 \!\!\!\!\! dx_2 \ 
  \frac{[(1\!-\!x_2)^2(1\!-\!\epsilon)\!+\! 2 x_2]}{ (1\!-\! x_1)^{1+\epsilon}
    (1\!-\! x_2)^{1+\epsilon} \: x_2^\epsilon} 
   = -\frac{1}{\epsilon_{\rm IR}} - \frac{3}{4}
  +{\cal O}(\delta) \,,\nonumber \\
  K^B_1 &= (1\!-\! x_1)^{1+\epsilon} \int_{1-x_1}^1 \!\!\!\!\! dx_2  \
   \frac{2}{
    (1\!-\! x_1)^{1+\epsilon} (1\!-\!x_2)^{1+\epsilon} }  
   = -\frac{1}{\epsilon_{\rm IR}}
  +{\cal O}(\delta) \,,\nonumber \\
  K^B_2 &= (1\!-\! x_1)^{1+\epsilon} \int_{1-x_1}^1 \!\!\!\!\! dx_2  \
   \frac{1-\epsilon}{
    (1\!-\! x_1)^{1+\epsilon}\: x_2^{\epsilon} }  = \frac{1}{2}
  +{\cal O}(\delta) \,,
\end{align}
For the collinear integral we therefore find
\begin{align}
  K^B &= \tilde K^B - K^B_1 -K^B_2 =  - \frac{5}{4}  +{\cal O}(\delta) \,.
\end{align}
This result is IR finite as expected from the fact that the IR divergence comes
from the soft region A in Fig.~\ref{fig:qqg} and not from the collinear region
B.

Adding the contributions from the three regions we find
\begin{align}
  K^A + K^B + K^C = -\frac{1}{\epsilon_{\rm IR}} - \frac{3}{4} \,,
\end{align}
in agreement with the result for $K^{\rm full}$ at this order. Thus with the
zero-bin subtractions the sum of SCET diagrams reproduces the expected result
for the $1/(1-x_1)$ bremsstrahlung term in the cross section $d\sigma/d x_1$ as
$x_1\to 1$.

\begin{figure}[!t]
\begin{center}
 \includegraphics[width=14cm]{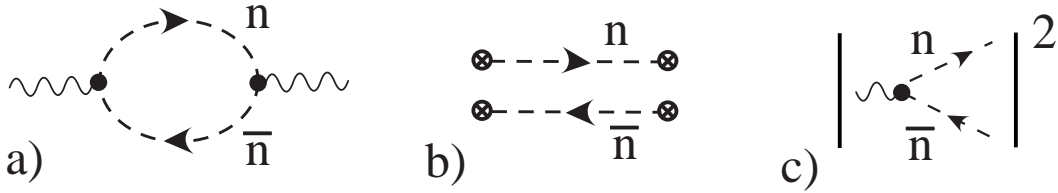}
\end{center}
 \vskip -.5cm 
\caption[1]{ 
  Contraction of fields in the time ordered product of LO SCET currents,
  $T\{J^{(0)},J^{(0)\dagger}\}$.
}\label{fig:qqborn} 
\end{figure}
Finally, we briefly remark as to whether we could have turned the phase space
computation into the imaginary part of a loop graph that we already know how to
deal with from zero-bin examples in previous sections. One might think that the
total cross section can be obtained by computing a forward scattering loop
diagram in SCET and taking the imaginary part. However in some cases the optical
theorem must be applied with care due to the momentum scaling of different types
of SCET fields.  A simple example is the Born cross section for $\gamma^*\to
q\bar q$ which we can consider computing from ${\rm Im} [i\int\!\! d^4x
\exp(-iq\mcdot x) \langle 0| J(0) J^\dagger(x) | 0 \rangle ]$.  In the full
theory there is a contribution with hard loop momentum and the imaginary part
contributes to the total-$\sigma$. In the SCET we are focusing on corners of
phase space like back-to-back jets in the $n$ and $\bn$ directions.  The product
of LO currents $J^{(0)}$ allows for a loop with usoft momentum, shown in
Fig.~\ref{fig:qqborn}a, without violating momentum conservation. However this
graph evaluates to zero due to the multipole expansion on collinear lines,
\begin{align}
  \int\!\! \frac{d^Dk}{(2\pi)^D} \frac{1}{(n\mcdot (k+p))(\bn\mcdot (k+p)} =0
   \,.
\end{align}
The product of currents does give a nonzero contribution, just not from the
imaginary part of this usoft loop in SCET. Instead, the SCET fields give the
imaginary part of two propagators as in Fig.~\ref{fig:qqborn}b, yielding an
integrand for the phase space integral that is accurate in the desired phase
space region for each line.  Because the matrix element factorizes into a
product of two matrix elements, Fig.~\ref{fig:qqborn}b is not a disconnected
contribution that can be discarded.  This reduces the problem back to squaring
the current, as depicted in Fig.~\ref{fig:qqborn}c. It also gives a hint as to
why the derivation of factorization formulas from SCET is more predictive than
requiring a strict OPE in QCD, much as for diagrammatic
factorization~\cite{Ellis:1978ty,Mueller:1989hs,Collins:1989gx,Sterman:1995fz}.


\section{Zero-bin Subtractions in SCET$_{\rm II}$ (Exclusive Processes) } 
\label{sec:scetb}

\begin{figure}[t!]  
 \begin{center}
\includegraphics[height=9cm]{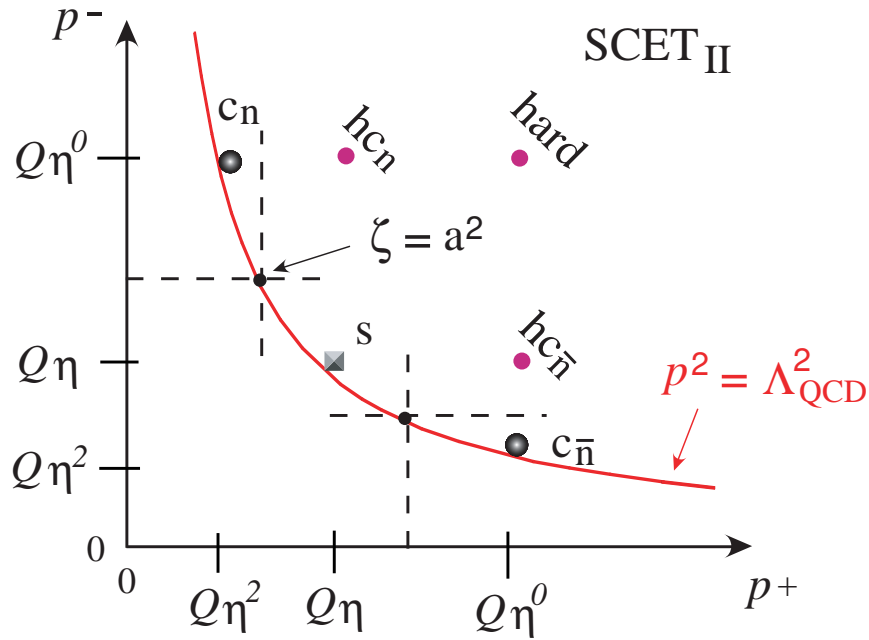}
 \end{center}
\vskip-0.7cm
 \caption{
   Degrees of freedom and momentum regions for \SCETb which describe
   non-perturbative fluctuations with $p^2\sim \Lambda_{\rm QCD}^2$. The modes
   include $n$--collinear ($c_n$), soft ($s$), and when applicable also
   $\bn$--collinear ($c_\bn$). In perturbation theory these modes extend all the
   way in to zero-momenta. For QCD the solid (red) curve represents the region where
   infrared divergences are rendered finite by $\Lambda_{\rm QCD}$. Also shown
   (in pink) are three regions of perturbative momenta, two with hard-collinear
   momenta ($hc_n$, $hc_\bn$) and one where the momenta are hard.
  \label{fig:scet2}}
\end{figure}

In this section, we consider an SCET with degrees of freedom which are suitable
for describing exclusive QCD processes with both energetic and soft hadrons.
This theory is usually called \SCETb, and contains fields that describe
nonperturbative collinear and soft momenta as pictured in the $p^+$--$p^-$ plane
shown in Fig.~\ref{fig:scet2}.\footnote{We do not need messenger or
  soft-collinear modes~\cite{Becher:2003qh} for the reasons discussed in
  section~\ref{sect:scalars}.}  It is also necessary to include these momentum regions
when considering mixed inclusive and exclusive processes. We begin with a
discussion of the ways in which \SCETb differs from the \SCETa and NRQCD
examples discussed previously.

In perturbation theory with massless particles, physical IR divergences occur as
$p^2\to 0$ either with collinear scaling $(p^+,p^-)\sim Q(\eta^2, 1)$ or soft
scaling $(p^+,p^-) \sim Q(\eta,\eta)$ for small dimensionless power counting
parameters $\eta$ and a large momentum scale $Q$. This is well known from
the study of the Landau equations and use of the Coleman-Norton
theorem~\cite{Sterman:1994ce,Sterman:1995fz}. In QCD, IR divergences either
cancel between diagrams or are cutoff by the nonperturbative effects that
generate confinement at a scale $p^2 \sim \Lambda_{\rm QCD}^2$. In
Fig.~\ref{fig:scet2} we show the confinement scale by a red solid line.  To
formulate the power counting for these non-perturbative momenta, we take $\eta\sim
\Lambda_{\rm QCD}/Q$. The collinear and soft fields represent distinct IR
sectors as given by their momentum scaling, and together cover all approaches to
the solid (red) curve in the $p^+$--$p^-$ plane.\footnote{In this paper we do not
  consider processes which have important contributions from potential momenta
  for forward $n$-$\bn$ scattering, $k^+k^- \ll k_\perp^2$, which are sometimes
  referred to as Glauber modes. } The different sectors are separated by
perturbative rapidity gaps.

The distinction between soft and collinear regions can be made with the variable
\begin{align}
  \zeta_p = \frac{p^-}{p^+} \,,
\end{align}
which provides a measure along the solid red curve in Fig.~\ref{fig:scet2}.  For
the different regions we have:
\begin{align} \label{rapidity}
  \mbox{$n$-collinear: } &\qquad \zeta_p \sim \eta^{-2} \gg 1 
     \,,\\
  \mbox{soft: } &\qquad \zeta_p \sim \eta^{0\phantom{-}} \sim 1 
     \,,\nonumber\\
  \mbox{$\bn$-collinear: } &\qquad \zeta_p \sim \eta^{2\phantom{-}} \ll 1 
     \,. \nonumber
\end{align}  
Thus to avoid double counting, we must make sure that in the variable $\zeta_p$
the $n$-collinear mode does not double count the soft mode and vice-versa, and
also that the $\bn$-collinear mode does not double count the soft mode and
vice-versa, etc. Note that the variable $\zeta_p$ provides a way of
distinguishing the modes and at the same time allows us to maintain the
boost-inversion symmetry~\cite{Becher:2003qh,Becher:2003kh}. The boost-inversion
symmetry allows one to swap the soft and $n$-collinear fields, etc.\ when setting
up the modes for the description of a physical process.

As discussed in Ref.~\cite{Bauer:2001yt}, momentum conservation strongly
constrains the form of soft-collinear interactions in \SCETb. Adding a soft mode
$p_s\sim Q (\eta,\eta,\eta)$ to a collinear mode $p_c \sim Q(\eta^2,1,\eta)$,
which both have $p^2\sim Q^2\eta^2$, gives an offshell hard-collinear momentum
$p_{hc} = p_s+p_c$ with $p^2_{hc}\sim Q^2\eta$. Thus all physical interaction
Lagrangians and operators in \SCETb will have $\ge 2$ soft fields and $\ge 2$
collinear fields.  As long as double counting (and divergent convolutions) are
avoided, we can group like fields together in gauge invariant products to obtain
factorized amplitudes at any order in the power expansion in $\eta$. In some
cases, one can more directly prove that the convolution integrals
converge~\cite{Beneke:2003pa}, and for these cases it is less important to be
careful about the zero-bins.

We will show that avoiding double counting in \SCETb involves zero-bin
subtractions similar to the previous sections, with the added complication
associated with ensuring that regions in $\zeta_p$ are treated correctly.  Due
to UV divergences in rapidity, this requires a regularization method. It also
requires extra renormalization parameters for the insertion of any operator that
connects soft and collinear fields, which we denote by $\muplus$ and
$\muminus$.\footnote{For typographical convenience we use $\mu_-$ and $\mu_+$
  although superscripts would be more appropriate.}  In dimensional
regularization, the parts of the \SCETb action that are purely soft or purely
collinear have the standard $\mu^{2\epsilon}$ multiplying couplings, so all
factors of $\alpha_s$ are $\alpha_s(\mu)$. The factors of $\muplus$ and
$\muminus$ only occur from mixed soft-collinear operators. We will show below
how $\muplusminus$ appear in dimensional regularization, and also with a cutoff
regulator.  The $\muplus$ and $\muminus$ parameters are tied together with the
usual $\mu$ by the dynamics of factorization, which, independent of the UV and
IR regulators, gives
\begin{align} \label{mupm}
  \mu^2 = \muplus\, \muminus \ .
\end{align}

Under an RPI-III transformation on the basis vectors, $n\to e^\alpha n$ and
$\bn\to e^{-\alpha} \bn$~\cite{Manohar:2002fd} (a longitudinal boost on
coordinates and fields), $\muplus$ behaves like a momentum $p^+=n\mcdot p$ and
$\muminus$ behaves like $p^-=\bn\cdot p$.  Furthermore $\zeta_p$ scales under a
RPI-III transformation. These boosts correspond to a universal shift of all
degrees of freedom along the solid red curve in Fig.~\ref{fig:scet2}, and thus
do not change the fact that having distinguished between modes using $\zeta_p$
in one frame we also avoid double counting in any other frame. For a process
with only soft and $n$-collinear modes, the boost-inversion symmetry allows us
to interchange the role of these modes~\cite{Becher:2003qh}. In
Fig.~\ref{fig:scet2}, we boost to lower $p^-$ and increase $p^+$, so that the
$c_n$ overlaps the $s$, and the $s$ overlaps the $c_\bn$. We then switch our
definition of plus and minus, $p^+ \leftrightarrow p^-$, with the overall
outcome that $c_n \leftrightarrow s$. Differentiating between modes using the
variable $\zeta_p$ keeps them distinct throughout this process. 

The basic structure that we have in mind for a factorization formula in \SCETb is
\begin{eqnarray}  \label{newFACT}
  \int\!\! dk^+ dk^- dp^- dp^+\: J(k^\pm,p^\pm, \muplusminus, \muplusminus')\: 
  \phi_n(p^-,\muminus,\mu^2)\:
  \phi_s(k^+,k^-,\muplus,\muminus') \:  \phi_\bn(p^+,\muplus',\mu^2),
\end{eqnarray}
where $J$ contains perturbative contributions from both hard-collinear and
hard momenta (as shown by the solid pink dots in Fig.~\ref{fig:scet2}).
For cases where only the $n$-collinear and soft modes are relevant, we have
the slightly simpler form
\begin{eqnarray}  \label{newFACT1}
  \int\!\! dk^+ dp^- \: J(k^+,p^-, \muplus,\muminus)\: 
  \phi_n(p^-,\muminus,\mu^2)\:
  \phi_s(k^+,\muplus,\mu^2) \,,
\end{eqnarray}
with $J$ purely hard-collinear.  If only the
$n$-collinear and $\bn$-collinear modes are relevant, we have
\begin{eqnarray}  \label{newFACT2}
  \int\!\! dk^+ dp^- \: J(p^+,p^-, \muminus,\muplus')\: 
  \phi_n(p^-,\muminus,\mu^2)\:
  \phi_\bn(p^+,\muplus',\mu^2) \,,
\end{eqnarray}
with $J$ having hard momenta. The idea is that due to the separation of degrees
of freedom in rapidity space the distributions can depend on $\mu_\pm$. This
dependence is similar to that for fragmentation functions in
Ref.~\cite{Collins:1981uk,Collins:1981uw}.  The meaning of the $\mu_+$ and
$\mu_-$ variables in the distribution functions is described further below in
section~\ref{sect:kernel} below Eq.~(\ref{Apif}). The presence of the $\mu_\pm$
parameters allows us to formulate the non-perturbative matrix elements that give
$\phi_n$ and $\phi_s$ as boost invariant objects. This evades an argument made
in Ref.~\cite{Becher:2003kh} that no IR regulator will allow a boost invariant
factorization of soft and collinear modes in \hbox{\SCETb\!.}  Our proposed
factorization formula differs from the conclusion of non-factorization in
Refs.~\cite{Beneke:2003pa,Becher:2003kh,Lange:2003pk}. The effects due to
$\muminus$ and $\muplus$ are actually not IR sensitive: they denote a choice we
have to distinguish the IR regions. They behave like the dimensional
regularization parameter $\mu$ in that we can compute the dependence on these
parameters in perturbation theory because of the large rapidity gaps.

The soft and collinear modes in \SCETb generate the physical hadron states in
the effective theory, with each mode generating the physical states in its
sector. The two sectors are separated by a perturbatively large rapidity gap, so
we do not need to consider hadrons made of both modes. Since near the mass-shell
$p^+ \sim {p_\perp^2}/p^-$, the rapidity scaling in Eq.~(\ref{rapidity}) gives
gaps in $p_\perp/E$ of spacing $\Lambda_{\rm QCD}/Q$. If we try to generate a
hadron with an interpolating field built from soft and collinear fields, such as
one collinear antiquark and one soft quark, then there are no physical
non-perturbative poles by momentum conservation. Thus, the Hilbert space of
states in the soft and collinear sectors are individually complete.

In section~\ref{sect:scalars} we begin by discussing a loop integral
in \SCETb taking into account the zero-bins. Our first one-loop
example uses a hard cutoff regulator, in
subsection~\ref{sect:cutoffscalar}. In section~\ref{sect:dimregII} we
formulate the separation of soft and collinear modes using dimensional
regularization, and repeat the one-loop example in
subsection~\ref{sect:dimregscalar}. In section~\ref{sect:kernel} we
give a general discussion on how the zero-bin subtractions work on
singular hard kernels to give what we call \o--distributions (the
complete definition can be found in this section). In
section~\ref{sect:rhopi} we apply this formalism to obtain a result
for the $\gamma^*\rho\to\pi$ form factor at large $Q^2$ which is free
from convolution endpoint singularities. In section~\ref{sect:Bpi} we
discuss the so-called ``soft'' form factor for $B\to\pi$ transitions,
$\zeta^{B\pi}(E)$, to argue that \SCETb yields a result in terms
of individual $B$ and $\pi$ distribution functions.


\subsection{A \SCETb Loop Integral with Subtractions} \label{sect:scalars}

As our first \SCETb example, we consider a one-loop integral for the process
``$B\to \gamma\ell\bar\nu$'' with $E_\gamma \gg\Lambda_{\rm QCD}$ but using
scalar quarks and gluons. The LO factorization formula for the full QCD
process was considered in
Refs.~\cite{Korchemsky:1999qb,Lunghi:2002ju,Descotes-Genon:2002ja,Bosch:2003fc}
using SCET. It involves $n$-hard-collinear fields and soft fields, but does not
suffer from the subtleties in \SCETb we wish to address.  The toy example with
scalars was considered in Ref.~\cite{Beneke:2003pa}, where it was pointed out
that this process with scalar quarks does not factor (naively) into a product of
scalar and collinear terms, due to endpoint divergences which connect the soft
and collinear matrix elements.  This issue only shows up at subleading order for
fermions.

In Ref.~\cite{Becher:2003kh,Lange:2003pk} it was independently concluded that
the convolution divergences encountered in these situations spoil factorization.
The analysis was based on a different IR regulator, implemented with an
offshellness, and adding to \SCETb a so-called messenger or soft-collinear IR
regulator mode which has $p^2\sim \Lambda_{\rm QCD}^3/Q$. Any long distance
colored interaction in \SCETb would violate confinement in QCD and therefore be
forbidden, but in perturbation theory one is free to introduce modes with $p^2$
below $\Lambda_{\rm QCD}^2$ if they facilitate the regulation of IR
divergences.\footnote{In particular, any mode that would leave both soft and
  collinear modes onshell must have $p^2 \ll \Lambda_{\rm QCD}^2$ by a
  parametric amount~\cite{Becher:2003qh}. In QCD a physical IR cutoff is
  provided by confinement which eliminates any such modes~\cite{Manohar:2005az}
  as they would physically correspond to colored degrees of freedom propagating
  between color singlet bound states that have already hadronized. They have
  been termed hyper-confining modes by Rothstein~\cite{IraPC}.  The
  soft-collinear messenger mode considered in
  Refs.~\cite{Becher:2003qh,Becher:2003kh,Lange:2003pk} are in the
  hyper-confining category.  } The fact that messenger modes should be
considered as part of the IR regulator was discussed in
Ref.~\cite{Bauer:2003td}, where it was shown that they are absent with an energy
dependent gluon mass IR regulator, but that a common dependence on this
regulator still appears in the soft and collinear matrix elements.\footnote{Note
  that in general matching computations between QCD and \SCETb can be a bit
  tricky because one must be sure that ones choice of IR regulator is treating
  the Hilbert space of full QCD in exactly the same way as the Hilbert space of
  \hbox{\SCETb\!.}  An example of this type that we encountered in \SCETa was
  our discussion of the field redefined current $J^{\prime (0)}$ in $b\to
  s\gamma$ in sections~\ref{sect_sceta1} and \ref{sect_sceta2}. For \SCETb
  further discussion of this point is given in Appendix~\ref{AppB}.} Messenger modes
  were also absent with the calculations using analytic regulators in
  Ref.~\cite{Beneke:2003pa}.

With our definition of modes in \SCETb in Fig.~\ref{fig:scet2},
messenger modes are not needed. If, on the other hand, one were to
take the picture for \SCETa in Fig.~\ref{fig_SCETUVvsIR}a and
translate it into modes for \SCETb in a one-to-one correspondence,
then the usoft mode in \SCETa becomes a messenger mode for
\SCETb~\cite{Becher:2003qh}. So the \SCETa mode decomposition seems to
want a messenger mode in \SCETb.  The fault here is with the
translation, which required both a boost and a scale
transformation. There is nothing wrong with the boost, but QCD is not
scale invariant. Due to the presence of $\Lambda_{\rm QCD}$,
translating the IR tiling of modes in \SCETa into a tiling of IR modes
for \SCETb, gives modes which hide the physical situation, in
particular the perturbative split of the hadronic physics in rapidity
space. In our definition of \SCETb the IR regions that show up in
perturbation theory and were described by the messenger mode in
Ref.~\cite{Becher:2003qh} are absorbed into the soft and collinear
fields.

\begin{figure}[!t]
\begin{center}
 \includegraphics[width=14cm]{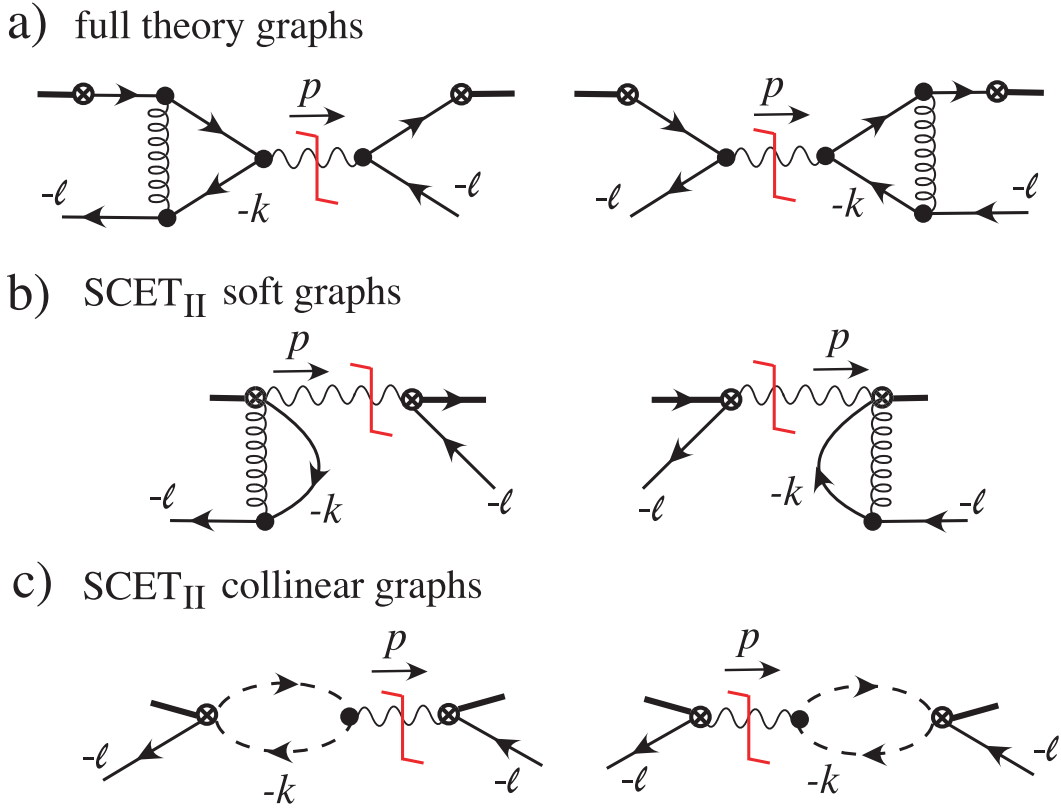}
\end{center}
 \vskip -.5cm 
\caption[1]{ 
  One loop cut-graphs for the $\gamma$-fragmentation contribution to $B\to
  \gamma\ell\bar\nu$ in the full and effective theories.  The $\otimes$'s denote
  an insertion of the weak current and the leptons are not shown. The graphs in
  a) are in full QCD, while the graphs in b) and c) are in \hbox{\SCETb\!.}  For
  b) we have soft propagators, while for c) the propagators are collinear
  (dashed).
}\label{fig:Bgamfrag} 
\end{figure}

In considering scalar ``$B\to \gamma\ell\bar\nu$'' we are really treating a
subleading contribution to the physical process with fermions where the photon
comes from fragmentation of $q\bar q \to \gamma$ or from a subleading
contribution to the direct $\gamma$-production. In the fragmentation case we
will have a non-perturbative soft distribution associated with the initial state
$B$, $\phi_s(k^+)$, and a non-perturbative collinear distribution for the
$\gamma$ associated with the fragmentation, $\phi_n(p^-)$.  To factorize this
physical process we must consider the imaginary part of the forward scattering,
since due to the probabilistic interpretation for the fragmentation function,
there will be a factorization formula for the decay rate but not for the
amplitude.  Examples of one loop diagrams in the full theory and \SCETb are
shown in Fig.~\ref{fig:Bgamfrag}. The \SCETb graphs in Fig.~\ref{fig:Bgamfrag}b
correspond to direct production, while those in Fig.~\ref{fig:Bgamfrag}c are
fragmentation. The \SCETb factorization allows these two effects to be
distinguished. 

In the full scalar theory we have a charged scalar ``b-quark'' with field
$\phi^b$, charged light scalar ``u-quark'', $\phi$, and neutral ``gluon'' and
``photon'' fields $\phi^g$ and $\phi^\gamma$ respectively. The ``weak'' current
and interaction terms are
\begin{align}
  J^{\rm weak}_{\rm full} &= G \phi^\dagger \phi^b \,,
  & {\cal L}^{\rm int}_{\rm full} &= g \phi^g \phi^\dagger \phi 
   + e \phi^\gamma \phi^\dagger\phi \,,
\end{align}
with standard kinetic terms for $\phi_b$ with mass $m_b$, and for the massless
charged and neutral scalars. Here the coupling $G$ tracks the current, and
$g$ and $e$ are coupling constants of mass dimension one. In \SCETb the leading
order Lagrangian is split into soft and collinear fields,
\begin{align}
  {\cal L}^{(0)}_{\rm II} &=
     2m_b\: \phi_{v}^{b\dagger}\, iv\mcdot\partial\, \phi_v^b 
     + \phi_s^\dagger\, (i\partial)^2 \,\phi_s 
     + \frac{1}{2}\: \phi^{g}_s \, (i\partial)^2 \, \phi^{g}_s 
     + g\,  \big[\phi^g_{s,\ell-k} \, \phi_{s,\ell}^\dagger\, \phi_{s,k}\big]
    \\
 & \ + \phi_n^\dagger\, (i\partial)^2 \,\phi_n 
     \!+\! \frac{1}{2}\: \phi^{\gamma}_n \, (i\partial)^2 \, \phi^{\gamma}_n
     \!+\! \frac{1}{2}\: \phi^{g}_n \, (i\partial)^2 \, \phi^{g}_n 
      + g \big[\phi^g_{n,p-q} \, \phi_{n,p}^\dagger\, \phi_{n,q}\big]
     + e  \big[\phi^\gamma_{n,p-q} \, \phi_{n,p}^\dagger\, \phi_{n,q}\big] 
   \nonumber ,
\end{align}
where $\phi_v^b$ is a scalar HQET field~\cite{Luke:1992cs}, the $\phi_s$ and
$\phi_n$ are soft and collinear massless ``quarks'', and $\phi_s^g$, $\phi_n^g$,
and $\phi_n^\gamma$ are scalars for the ``gauge'' fields. The collinear fields
have label momenta $p^-\sim \eta^0$ and residual momenta $p_r^\perp\sim \eta$
and $p_r^+\sim \eta^2$, and the soft fields have label momenta $k^+\sim \eta$
and residual momenta $k_r^\perp\sim \eta$ and $k_r^-\sim\eta$. Writing out the
label and residual terms explicitly in the kinetic terms we would have
$(i\partial^\mu) \phi_s \to (\bn^\mu \cP/2 + i\partial^\mu_r) \phi_{s,\ell}$,
$(i\partial^\mu) \phi_n \to (n^\mu \bnP/2 + i\partial^\mu_r) \phi_{n,p}$ etc.,
with the standard treatment of leading and subleading terms.  Demanding ${\cal
  L}^{(0)}\sim \eta^0$ the power counting in $\eta$ is
\begin{align}
 & \phi_v^b \sim \eta^{3/2} \,,
 & \phi_s \sim \phi_n \sim \eta & \,,
 & \phi_n^\gamma \sim \eta &\,,
 & \phi_s^g \sim \phi_n^g \sim \eta \,.
\end{align}
The leading order currents we will need are
\begin{align} \label{LOcurr}
  O^{(0a)}_{\rm II} &= \sum_{p^-,\ell^+\ne 0} 
    \frac{J^{(0a)}}{(\bn\mcdot p\, n\mcdot \ell \!-\! i0^+)}\:
    \big[ \phi^\dagger_{s,-\ell}\, \phi^b_{v}\, \phi_{n,p}^\gamma \big], \\
  O^{(0b)}_{\rm II} &= \sum_{p^-,\ell^+,k^+\ne 0}  \frac{J^{(0b)}}
 {(\bn\mcdot p\, n\mcdot \ell \!-\! i0^+)(\bn\mcdot p\, n\mcdot k \!-\!
   i0^+)} \: \big[ \phi^\dagger_{s,-k}\, \phi^b_{v}\, \phi_{n,p}^\gamma\,
    \phi_{s,\ell-k}^g \big] ,
   \nn
  O^{(0c)}_{\rm II} &=   \sum_{p^-,\ell^+,q^-\ne 0} \frac{J^{(0c)}}
 {(\bn\mcdot p\, n\mcdot \ell \!-\! i0^+)(\bn\mcdot q\, n\mcdot \ell \!-\!
   i0^+)} \: \big[ \phi^\dagger_{s,-\ell}\, \phi^b_{v}\, \phi_{n,p-q}^\dagger\,
    \phi_{n,-q} \big] ,   
   \nonumber
\end{align}
where $\bn\mcdot p\sim \eta^0$ while $n\mcdot \ell\sim n\mcdot k\sim
\eta$.  Note that the operators $O_{\rm II}^{(0b)}$ and $O_{\rm
II}^{(0c)}$ include restrictions on the label sums with $n\mcdot
\ell\ne 0$ and $\bn\mcdot p\ne 0$ respectively. In these bins the
$1/(\bn\mcdot p\,n\mcdot \ell)$ factor would be a collinear or soft
propagator in \SCETb, and these bins are taken into account by
time-ordered products with subleading
\SCETb Lagrangians as discussed in Ref.~\cite{Mantry:2003uz}.

Note that in scalar \SCETb, we can construct leading order operators with
additional scalar fields since the extra powers of $\eta$ are compensated by
$1/(\bn\cdot p\, n\cdot \ell)$ factors and the mass dimension is compensated by
the dimension of the couplings. In our perturbative example only the currents
shown are needed (plus counterterm operators). A non-perturbative treatment
would require additional terms.  In the scalar theory, matching tree level
graphs gives $J^{(0a)}=e G$, $J^{(0b)}= e g G$, and $J^{(0c)}= g^2 G$. In
dimensional regularization, the currents are modified in a manner described in
section~\ref{sect:dimregII}.

In the gauge theory with fermions, the full set of operators for $B\to \gamma
e\bar\nu$ would be determined by gauge invariance and leading order matching. The
current analogous to $O_{\rm II}^{(0a)}$ is LO, while the currents analogous to
$O_{\rm II}^{(0b,0c)}$ are suppressed by one power of $\eta$. The
non-perturbative treatment is simpler in gauge theories since it is constrained
by more symmetries. See for example, the construction of soft-collinear \SCETb
operators in Refs.~\cite{Beneke:2003pa,Becher:2005fg,Hill:2005ju}.

A goal of this section is to demonstrate that although there is not a
simple factorization for the scalar $B\to \gamma e\bar\nu$ process of
the form $J(\mu)\otimes \phi_n(\mu)\otimes \phi_s(\mu)$, there appears
to be a more involved factorization which contains terms of the form
\begin{align} \label{newFACT3}
  \int\!\! dk^+ dp^-\: J(k^+,p^-, \muplus, \muminus)\: \phi_n(p^-,\muminus,\mu^2)\:
  \phi_s(k^+,\muplus,\mu^2) \,.
\end{align}
Here $J$ is a perturbative jet function, and the $\phi$'s are
non-perturbative, with $\phi_n$ given by a matrix element of collinear fields
and $\phi_s$ given by a matrix element of soft fields.  In Eq.~(\ref{newFACT3})
we have two additional scale parameters $\muplus$ and $\muminus$ in \SCETb (or
two factorization scales in a more traditional language in full QCD). We will
demonstrate that these scales are connected in a specific way to the standard
renormalization scale 
\begin{align}
  \muminus\: \muplus = \mu^2  \,.
\end{align}
Although our conjecture about the existence of a factorization formula valid to
all orders in $\alpha_s$ differs from
Refs.~\cite{Beneke:2003pa,Becher:2003kh,Lange:2003pk}, the structure of the
result is also different from standard factorization formulas in the literature.
Because we do not have the identity $\muplus = \muminus = \mu$, there is not a
simple factorization for the amplitude for this process, in agreement with the
conclusions in Refs.~\cite{Beneke:2003pa,Becher:2003kh,Lange:2003pk}.  The
perturbative formula that we find for the observable process,
Eq.~(\ref{newFACT1}), has all the desired properties of a factorization formula,
including correctly reproducing IR divergences (from the zero-bin subtractions),
finite convolution integrals (from the zero-bin and renormalization), and
distinct matrix elements for the soft and collinear objects.

In the context of the analytic IR regulator used
in~\cite{Beneke:2003pa}, it was pointed out that there was an
interesting cancellation between IR divergences in soft and collinear
diagrams.  Our observation is that this cancellation has to do with
avoiding double counting just like the zero-bin subtractions, rather
than having to do with reproducing IR divergences in QCD.  For
physical observables, like the forward scattering graphs, the method
for avoiding the double counting is computable, and can be handled in
perturbation theory. It results in regularization parameters $\muplus$
and $\muminus$ which encode the coupling between soft and collinear
modes with a simple correlation.  Our results turn situations which
were previously plagued by the unphysical convolution endpoint
singularities, into manageable finite amplitudes, which one can then
try to arrange into a predictive factorization formula.  Since the
divergent effects are computable they do not spoil many of the nice
features obtained in simpler QCD factorization formulas.


\subsubsection{Soft-Collinear Division with a Hard Cutoff Regulator} 
\label{sect:cutoffscalar}

We begin by considering a hard cutoff, $a$, between the soft and collinear
modes, as indicated in Fig.~\ref{fig:scet2}. For a loop momentum $k^\mu$ we
define $\zeta_k=\bn\cdot k/n\cdot k=k^-/k^+$ as discussed near
Eq.~(\ref{zetap}). Only the magnitude of the rapidity variable, $|\zeta_k|$, is
relevant for distinguishing the soft and collinear modes.  Switching variables
from $\{k^+,k^-\}$ to $\{k^+,\zeta_k\}$ gives $dk^-=|k^+| d\zeta_k$ when we
integrate over $-\infty < k^+ < \infty$ and $-\infty < \zeta_k < \infty$, so the
loop integral is no longer analytic in $k^+$ but remains analytic in $\zeta_k$,
and likewise if we switch to $\{k^-,\zeta_k\}$.  When imposing hard cutoffs we
need to avoid the physical poles, which can be accomplished using cutoffs in
Euclidean space after Wick rotation. For variables $\{k^+,\zeta_k\}$ the Wick
rotation $k^-\to ik^-$ is equivalent to $\zeta_k=i\zeta'_k$, while for
$\{k^-,\zeta_k\}$ the Wick rotation $k^+\to ik^+$ gives $\zeta_k=-i\zeta'_k$. In
our examples Wick rotation about the origin suffices, and the poles in complex
$\zeta_k'$ occur along the imaginary axis in the first and third quadrants. We
take cutoffs
\begin{align} \label{harda}
  \mbox{ soft: } & \qquad  -a^2 \le \zeta_k' \le a^2 \,,\\
  \mbox{ collinear: } &  \qquad  -a^2 \ge \zeta_k'  \ \ \ \mbox{or}\ \ \ 
   \zeta_k' \ge a^2 \,.\nonumber
\end{align}
As mentioned above we only require $n$-collinear and soft fields in \SCETb for
the example in this section and so are free to include the entire
$\zeta_k'\lesssim 1$ region in the soft modes. (For more complicated problems
the region $\zeta_k'\sim \eta^2$ would need to be disentangled for the
$\bn$--collinear modes.) We can take $a^2\sim \eta$, and note that under an
RPI-III transformation on $n$ and $\bn$~\cite{Manohar:2002fd} (a longitudinal
boost) that $a^2$ behaves like a $(p^-)^2$ momentum just like $\zeta_k$ does.
Also note that the power counting scaling which fixes the soft and collinear
components only depends on $|\zeta_k|$ or $|\zeta_k'|$ and so does not care
about the Wick rotation.

\begin{figure}[!t]
\begin{center}
 \includegraphics[width=14cm]{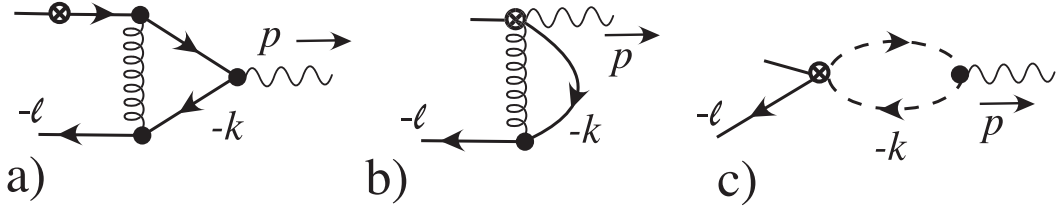}
\end{center}
 \vskip -.5cm 
\caption[1]{ 
  Graphs with scalar propagators as a toy model for the $\gamma$-fragmentation
  contribution to $B\to \gamma\ell\bar\nu$ with fermions.  The $\otimes$ denotes
  the weak current and the leptons are not shown. Graph a) is in full QCD, graph
  b) has an insertion of $O_{\rm II}^{(0b)}$ and a loop with soft fields in
  \hbox{\SCETb\!,} and graph c) has an insertion of $O_{\rm II}^{(0c)}$ and a
  collinear loop (with dashed propagators).
}\label{fig:scalargamma} 
\end{figure}
For simplicity we consider the same diagram as discussed in
Ref.~\cite{Beneke:2003pa} which is shown in Fig.~\ref{fig:scalargamma}a.  Unlike
Ref.~\cite{Beneke:2003pa}, we do not analyze this graph with the method of
regions. Instead we consider the diagram in full scalar field theory
(Fig.~\ref{fig:scalargamma}a) and the corresponding diagrams in scalar
\hbox{\SCETb} (Fig.~\ref{fig:scalargamma}b,c). The difference of the two results
gives a matching contribution, and allows us to check that the full theory IR
divergences are correctly reproduced with zero-bin subtractions implemented in
\mbox{\SCETb\!.}  It also allows us to discuss the factorization formula in
Eq.~(\ref{newFACT1}).  

For simplicity we will leave off the prefactor $i e g^2 G/(p^-\ell^+)$ in
quoting results for graphs in this subsection. For the full theory diagram, we
have the integral
\begin{align} \label{sfullint}
 I_{\rm full}^{\rm scalar} &= \int  
  \!\!\frac{d^Dk}{(2\pi)^D}  \frac{1}
{[(k-\ell)^2+i0^+][k^2  + i0^+][(k- p)^2 + i0^+]} \,. 
\end{align}
Evaluating this onshell with $p^\mu=p^- n^\mu/2$, $\ell^\mu=\ell^+ \bn^\mu/2$,
so that $p^2=\ell^2=0$, we have
\begin{align} \label{sfull}
  I_{\rm full}^{\rm scalar} &= \frac{-i}{16\pi^2 (p^-\ell^+)} 
   \bigg[ \frac{1}{\epsilon_{\rm
      IR}^2} - \frac{1}{\epsilon_{\rm IR}} \ln\Big(\frac{p^-\ell^+}{\mu^2} \Big)
  + \frac{1}{2} \ln^2\Big(\frac{p^-\ell^+}{\mu^2} \Big) - \frac{\pi^2}{12}
  \bigg] \,.
\end{align}
Here the IR divergences are regulated by dimensional regularization. For the
soft and collinear graphs in Fig.~\ref{fig:scalargamma}b,c we find
\begin{align} \label{startsoftcollin}
 I_{\rm soft}^{\rm scalar} &= \sum_{k^+\ne 0} \int  
  \!\!\frac{d^Dk_r}{(2\pi)^D}  \frac{1}
 {[k^2\!-\!n\mcdot\ell\,\bn\mcdot k+i0^+]
 [k^2 \!+\! i0^+][-\bn\mcdot p\, n\mcdot k + i0^+]} \,, \\
 I_{\rm nc}^{\rm scalar} &= \sum_{k^-\ne 0} \int  
  \!\!\frac{d^Dk_r'}{(2\pi)^D}  \frac{1}
 {[-n\mcdot\ell\,\bn\mcdot k+i0^+]
 [k^2 \!+\! i0^+][k^2\!-\!\bn\mcdot p\, n\mcdot k + i0^+]} \,,
  \nonumber
\end{align}
where in both cases the first two terms are the displayed propagators, and the
last factor comes from the non-local vertex which emits the scalar soft or
collinear fields in \SCETb.

To compute the EFT graphs we implement the hard cutoff in Eq.~(\ref{harda}) to
regulate UV effects in the effective theory diagrams. With this regulator the
zero-bin subtractions are automatically zero since they are outside the region
of integration. The hard cutoffs are theta functions in the integrand so they
give identically zero for the integrand evaluated in the subtraction regions.
Therefore with this regulator the full integrals are given by the naive
replacement in Eq.~(\ref{1.01}).  We discuss in detail the calculation of the
\SCETb diagrams in Appendix~\ref{AppA}. For the soft graph the result is
\begin{align} \label{ssrslt}
  I_{\rm soft}^{\rm scalar} =\tilde I_{\rm soft}^{\rm scalar} 
  &= \frac{-i}{16\pi^2(p^-\ell^+)} 
  \bigg[ \frac{1}{2\epsilon_{\rm IR}^2} 
   - \frac{1}{\epsilon_{\rm IR}} \ln\Big(\frac{\ell^+ a }{\mu} \Big)
  +  \ln^2\Big(\frac{\ell^+ a }{\mu} \Big) - \frac{\pi^2}{16}
  \bigg] \nonumber \\
   &= \frac{-i}{16\pi^2(p^-\ell^+)} 
    \bigg[ \frac{1}{2\epsilon_{\rm IR}^2} 
   - \frac{1}{\epsilon_{\rm IR}} \ln\Big(\frac{\ell^+ }{\muplus} \Big)
  +  \ln^2\Big(\frac{\ell^+}{\muplus} \Big) - \frac{\pi^2}{16}
  \bigg] \,,
\end{align}
where we defined $\muplus = \mu/a$. Note that since $a$ boosts like a
minus-momentum, $\muplus$ behaves like a plus-momentum, and the result in
Eq.~(\ref{ssrslt}) is RPI-III invariant. For the collinear graph the
result is
\begin{align} \label{scrslt}
 I_{\rm cn }^{\rm scalar}= \tilde I_{\rm cn }^{\rm scalar}
 &= \frac{-i}{16\pi^2(p^-\ell^+)} \bigg[
  \frac{1}{2\epsilon_{\rm IR}^2} 
  - \frac{1}{\epsilon_{\rm IR}} \ln\Big(\frac{p^- }{a\mu} \Big)
  +  \ln^2\Big(\frac{p^-}{a\mu } \Big) - \frac{\pi^2}{16}
  \bigg] \,, \nonumber \\
 &= \frac{-i}{16\pi^2(p^-\ell^+)} \bigg[
  \frac{1}{2\epsilon_{\rm IR}^2} 
  - \frac{1}{\epsilon_{\rm IR}} \ln\Big(\frac{p^- }{ \muminus} \Big)
  +  \ln^2\Big(\frac{p^- }{ \muminus} \Big) - \frac{\pi^2}{16}
  \bigg] \,, 
\end{align}
where we defined $\muminus = a \mu$. Here $\muminus$ behaves like a minus-momentum and
the result in Eq.~(\ref{scrslt}) is also RPI-III invariant. The soft and
collinear regularization parameters $\mu_\pm$ defined in the computation of
Eqs.~(\ref{ssrslt}) and (\ref{scrslt}) obey the anticipated relation,
\begin{align}
  \muplus \, \muminus = \mu^2 \,,
\end{align}
where the $a$ dependence cancels out in this product.  Moreover, with $a^2\sim
\eta$ we find that with $\mu^2$ at the matching scale, $\mu^2\sim Q\Lambda_{\rm
  QCD}$, one can still take $\muplus\gtrsim \Lambda_{\rm QCD}$ and $\muminus\sim
Q$. Thus we can simultaneously minimize the logarithms in the \SCETb matrix
elements, which in our perturbative computation are represented by $I_{\rm
  soft}^{\rm scalar}$ and $I_{\rm cn}^{\rm scalar}$.

Adding the two \SCETb graphs, $I_{\rm s+cn}^{\rm scalar}= I_{\rm s }^{\rm
  scalar}+ I_{\rm cn }^{\rm scalar}$, we find
\begin{align} \label{sctotrslt}
  I_{\rm s+cn}^{\rm scalar} &= \frac{-i}{16\pi^2(p^-\ell^+)} \bigg[
  \frac{1}{\epsilon_{\rm IR}^2} 
  - \frac{1}{\epsilon_{\rm IR}} \ln\Big(\frac{p^-\ell^+}{\muminus\muplus} \Big)
    +  \ln^2\Big(\frac{p^-}{\muminus} \Big) 
  +  \ln^2\Big(\frac{\ell^+}{\muplus} \Big) - \frac{\pi^2}{8}
  \bigg] \nn
 &= \frac{-i}{16\pi^2(p^-\ell^+)} \bigg[
  \frac{1}{\epsilon_{\rm IR}^2} 
  - \frac{1}{\epsilon_{\rm IR}} \ln\Big(\frac{p^-\ell^+}{\mu^2} \Big)
  +  \ln^2\Big(\frac{p^-}{\muminus  } \Big) 
  +  \ln^2\Big(\frac{\ell^+  }{\muplus} \Big) - \frac{\pi^2}{8}
  \bigg] \,.
\end{align}
We see that with the relation $\mu^2=\muplus\muminus$, the $1/\epsilon_{\rm IR}$ poles
agree exactly with the full theory expression in Eq.~(\ref{sfull}) as required.
To match the full and effective calculations we set $\mu^2 = \muplus \muminus$ and
subtract to find
\begin{align} \label{smatch}
  I_{\rm matching}^{\rm scalar} 
 &= \frac{-i}{16\pi^2(p^-\ell^+)} \bigg[
   + \frac{1}{2} \ln^2\Big(\frac{p^-\ell^+}{\muminus \muplus} \Big)
  -  \ln^2\Big(\frac{p^-}{\muminus} \Big) 
  -  \ln^2\Big(\frac{\ell^+}{\muplus} \Big) + \frac{\pi^2}{24}
  \bigg] \nn
 &= \frac{-i}{16\pi^2(p^-\ell^+)} \bigg[
   - \frac{1}{2} 
   \ln^2\Big(\frac{p^- \muplus  }{\muminus\,\ell^+} \Big) 
   + \frac{\pi^2}{24}
  \bigg]
 \,.
\end{align}
Here the $\ln^2(p^-\mu_+/\ell^+\mu_-)$ contributes $p^-$ and $\ell^+$ dependence
to the jet function $J(\ell^+,p^-,\mu_+,\mu_-)$ at one loop.  To minimize the
large logarithms in the matching calculation in Eq.~(\ref{smatch}) we take
$\mu^2=\mu_-\mu_+$ to be of order the hard-collinear scale, and take
$\muplus/\muminus\sim \eta$. Since the matching result in Eq.~(\ref{smatch})
depends on $\muplus/\muminus$ it depends on $a$, which is not surprising.  Here
$a$ ensures there is no double counting between the soft and collinear modes in
the IR, but $a$ also changes the behavior of the collinear and soft modes in the
ultraviolet. This change is compensated by the perturbative Wilson coefficient,
and in perturbation theory the sum of these contributions reproduce the full
theory result.

The result in Eq.~(\ref{smatch}) is shown for illustration only, since a
complete matching calculation for scalar $B\to \gamma\ell\bar\nu$ requires a
computation of all diagrams, not just the one diagram that we considered. For
example, one should also compute graphs with the scalar gluon attached to the
$b$-quark line, and wavefunction renormalization type diagrams in both the full
and effective theories.


\subsection{Dimensional Regularization Division in \SCETb: General Discussion}
\label{sect:dimregII}

In this section, we discuss the use of dimensional regularization for
the UV divergences and the separation of soft and collinear modes.
This regulator makes higher order computations more feasible and
preserves gauge symmetry. We also expect that it will make it easier
to compute anomalous dimensions and sum logarithms using
renormalization group techniques, although we do not address these
features here. Finally, it is useful to consider dimensional
regularization in order to compare how the separation of modes in
rapidity space appears with a different regulator.

Since the standard application of dimensional regularization is boost invariant,
it does not provide the ability to distinguish modes in rapidity space. This
also means that in general, divergences in the rapidity will not be regulated by
standard dimensional regularization.  For an insertion of a mixed soft-collinear
operator, we can regulate the rapidity space in dimensional regularization by
scaling out factors of the label operators from the Wilson coefficients. To implement
Fig.~\ref{fig:scet2} in dimensional regularization, the correct form of the
operators are
\begin{align}   \label{Op1}
  &  J(p_j^-,k_j^+) 
   \big[(\bar q_s S)_{k_1^{+}} \Gamma_s ( S^\dagger q_s)_{k_2^+} \big]
  \big[ (\bar \xi_n W)_{p_1^{-}} \Gamma_n (W^\dagger \xi_n)_{p_2^-} \big]
  \\[5pt]
  & \stackrel{\rm dim.reg.}{\longrightarrow} J(p_j^-,k_j^+,\mu_\pm)\:
  \mu^{2\epsilon} \bigg[ (\bar q_s S)_{k_1^{+}} \,
  \frac{|\cP^\dagger|^\epsilon}{\mupluse} \, \Gamma_s \,
  \frac{|\cP|^\epsilon}{\mupluse} \, ( S^\dagger q_s)_{k_2^+} \bigg]
  \bigg[(\bar \xi_n W)_{p_1^{-}}\, 
    \frac{|\bnP^\dagger|^\epsilon}{\muminuse} \,\Gamma_n \, 
    \frac{|\bnP|^\epsilon}{\muminuse}\, (W^\dagger \xi_n)_{p_2^-} \bigg]
  \nonumber\\[5pt]
  & \ \ \ \ \ \ 
   = J(p_j^-,k_j^+,\mu_\pm,\mu^2)\: \mu^{2\epsilon} \bigg[
  \frac{|k_1^+k_2^+|^\epsilon}{\muplusee} 
  (\bar q_s S)_{k_1^{+}} \Gamma_s ( S^\dagger q_s)_{k_2^+} \bigg] 
  \bigg[ \frac{|p_1^-p_2^-|^\epsilon}{\muminusee} 
   (\bar \xi_n W)_{p_1^{-}} \Gamma_n
   (W^\dagger \xi_n)_{p_2^-}\bigg] \,. \nonumber
\end{align}
Here the label operator $\cP$ gives the plus momentum from soft fields, and the
label operator $\bar \cP$ gives the minus momentum from collinear fields.  The
momenta subscripts occur for products of quark fields and Wilson lines,
$(S^\dagger q_s)_{k^+}= \delta(k^+\!-\!\cP)(S^\dagger q_s)$, $(W^\dagger
\xi_{n})_{p^-}= \delta(p^-\!-\!\bnP)(W^\dagger \xi_{n})$, which ensures that the
momenta are gauge invariant and that the gauge symmetry is not spoiled by the
factors of $|\bnP|^\epsilon$, $|\cP|^\epsilon$, etc. The absolute values ensure
that we raise a positive physical momentum to the $\epsilon$ power, and thus do
not modify the cut structure of matrix elements.\footnote{Recall that the labels are
positive for particles, and negative for antiparticles~\cite{Bauer:2001ct}. Combining both particles and antiparticles into a single field distinguished by the sign of the label simplifies the formulation of the effective theory. One could instead have used separate fields for the particles and antiparticles, in which case the antiparticle field could also be chosen to have a positive label. The absolute values in $|\cP|^\epsilon$ mean that we are using the momentum of the particle, which is unambiguous, rather than the label on the field, which is convention dependent. Due to the zero-bin conditions $p_i^-\ne 0$ and $k^+_i\ne 0$, there is no problem at the origin.} $J$ is the Wilson coefficient jet function.

This rescaling will allow us to properly distinguish the soft and collinear
modes in dimensional regularization without imposing a hard cutoff to implement
the division in Fig.~\ref{fig:scet2}.  This modification of the current is not
done to solve a problem in the IR -- it is the zero-bin subtractions for the
soft and collinear fields which will ensure that there is no IR double counting.
The zero-bin subtraction terms are integrated over all space, which introduces
new UV divergences in rapidity space, and in Eq.~(\ref{Op1}) the factors of
$|\bnP|^\epsilon$ etc. are necessary to regulate these UV divergences. If one
thinks of splitting the full loop integral $I$ into a naive part $\tilde I$ and
a subtraction part $I_0$, then $\tilde I$ has an IR rapidity divergence, while
$I_0$ has both UV and IR divergences. The IR divergences cancel in $I=\tilde
I-I_0$, so the rapidity divergence in $I$ is pure UV. We will see that these
remaining UV divergences can be removed by counterterms.

Before giving the rules for constructing Eq.~(\ref{Op1}), let us consider how it
should be used.  When we do a collinear loop involving an insertion of this
operator we expand in $(p^-/\muminus)^\epsilon$, but DO NOT expand the
$(k^+/\muplus)^\epsilon$ factors, and we do the opposite for a soft loop.  This
dimensional regularization rule is forced on us in any field theory with a
multipole expansion, and \SCETb has a multipole expansion between components of
the soft and collinear momenta.  The rule was discussed in
Ref.~\cite{Hoang:2002yy} for NRQCD in examples involving mixed usoft-soft loops.
In general one does not expand matrix elements of the soft fields in
$D$-dimensions when doing the collinear loops and one does not expand matrix
elements of the collinear fields when doing soft loops. The factors of
$(p^-/\muminus)^\epsilon$ and $(k^+/\muplus)^\epsilon$ should be thought of as being
associated with the renormalized coupling function $J$, just like a factor of
$\mu^\epsilon$ is associated to the strong coupling $g(\mu)$.  For purely
collinear or purely soft operators we apply dimensional regularization in the
usual manner.  The only place that $\mu_\pm$ appear is in the insertion of a
mixed soft-collinear operator.  All purely soft operators and purely collinear
operators only have $\mu^\epsilon$ factors, and so all couplings are
$\alpha_s(\mu)$. The couplings do not dependent explicitly on $\mu_\pm$. In
multiloop diagrams one can carry out the standard renormalization procedure
first, and leave to the end the rapidity renormalization for the final loop
involving the soft-collinear vertex.

Lets consider how we determined the powers of $\muplus$, $\muminus$ and $\mu$
for Eq.~(\ref{Op1}). In dimensional regularization, factors of $\mu^{\epsilon}$
appear from ensuring that coupling constants in the action are dimensionless.
Demanding that this is the case for $J$ gives the $\mu^{2\epsilon}$ factor.  To
determine the factors of $\muplusminus$ we must examine the scaling of fields in
the operator along the solid red curve in Fig.~\ref{fig:scet2}. An RPI-III
transformation scales all modes by a common amount and separate invariance under
this transformation demands we introduce a $1/\muplus$ to compensate each $\cP$,
and a $1/\muminus$ for each $\bnP$.  Charge conjugation requires the same power
for quarks and antiquarks, and the boost-inversion
symmetry~\cite{Becher:2003qh,Becher:2003kh} requires the same power be used for
the soft and collinear fields.  We can also demand rapidity invariance under
small individual scalings of the soft and collinear sectors. This will determine
the power of the label parameters, $|\bnP|^\epsilon$ and $|\cP|^\epsilon$. The
power is related to the space-time dimension because in doing this rapidity
scaling we demand that the invariant mass $p^2$ remains homogeneous.  As an
example consider a scaling by $\beta>0$: $\chi_{n,p^-}(0)\to \chi_{n,\beta\,
  p^-}(0)= \beta^{-\epsilon}\chi_{n,p^-}(0)$, where here $x=0$ as in the
soft-collinear operator. To derive the $\beta^{-\epsilon}$ factor write the
quark field $\chi_{n,\beta p^-}(0)= \delta(\beta p^- -\bnP)\chi_n(0) =\int
d^dk\: \delta(\beta p^-\!-\!k^-)\delta(k^2) \theta(k^-) a_k$, and then shift
$k^-\to \beta k^-$ and $k_\perp^2\to \beta k_\perp^2$ (to keep $k^2$
homogeneous). This $\beta^{-\epsilon}$ factor from the transformation of
$\chi_{n,p^-}$ is exactly canceled by the $|\bnP|^\epsilon \to |\bnP|^\epsilon
\beta^\epsilon$ factor acting on this field. The symmetry of the problem
dictates that we need one such factor for each collinear field.  Repeating these
arguments with a scaling parameter $\beta'$ in the soft-sector determines the
$|\cP|^\epsilon$ terms.

These arguments determine the proper operator for the dimensional regularization
computations in \SCETb, with an example shown in Eq.~(\ref{Op1}).  The Wilson
coefficient $J$ has non-trivial $\muminus$ and $\muplus$ dependence which cancels the
dependence on these parameters in the matrix element of the effective theory
operator order by order in $\alpha_s(\mu)$.  Thus we see that factorizing the
soft and collinear modes in \SCETb also requires introducing $\muplus$ and $\muminus$, just like with our cutoff regulator.

It is interesting to compare the regulator introduced in Eq.~(\ref{Op1}) with
the use of analytic regulators used for collinear computations in
Refs.~\cite{Smirnov:1998vk,Smirnov:2002pj,Beneke:2003pa}. Much like an analytic
regulator, the result in Eq.~(\ref{Op1}) modifies the power of a momentum
dependent factor in the integrand. It is used to regulate divergences that are
not handled by dimensional regularization, which is also the motivation for
introducing an analytic regulator. However, unlike the use of analytic
regulators, Eq.~(\ref{Op1}) is gauge invariant, defines the modification at the
operator level, and does not modify the power of the propagators in the EFT.
Furthermore, as already emphasized, with our zero-bin setup this regulator is
needed for divergences in the UV, rather than the IR. Since these divergences
arise due to the separation of momentum fractions in hard scattering kernels and
collinear operators, we anticipate that the addition of a power of the momentum
fraction for each labeled field will regulate UV rapidity divergences in a general
situation.


\subsubsection{Dimensional Regularization for the one-loop example} 
\label{sect:dimregscalar}

We now repeat the computation in the last section with dimensional
regularization for the UV. The loop integral in Eq.~(\ref{sfullint}) has IR
divergences for $k\to 0$, $k\to \ell^+$, and $k\to p^-$.  We take one of the
propagator lines to have an infinitesimal mass $m^2$ to regulate these IR
divergences.  For simplicity we will leave off the prefactor $i e g^2
G/(p^-\ell^+)$ in quoting results.  For the full theory diagram we have
\begin{align} \label{scetIIfull}
 I_{\rm full}^{\rm scalar} &= \int  
  \!\!\frac{d^Dk}{(2\pi)^D}  \frac{1}
{[(k-\ell)^2+i0^+][k^2 - m^2+ i0^+][(k- p)^2  + i0^+]} 
 \nn
   &= \frac{-i}{16\pi^2 (p^-\ell^+)} 
   \bigg[ \frac12\, 
   \ln^2\Big(\frac{m^2}{p^-\ell^+} \Big) \, + \, \frac{\pi^2}{3} \,
  \bigg] .
\end{align}
Here $m^2$ regulates the IR divergences in a manner similar to the solid red
curve in Fig.~\ref{fig:scet2}. Other choices of IR regulator can be made, and in
Appendix~\ref{AppB} we repeat the computations done in this section with i)
factors $m_2^2$, $m_1^2$, and $m_3^2$ in the three propagators in
Eq.~(\ref{scetIIfull}), and ii) taking $p^2\ne 0$ and $\ell^2\ne 0$ in
Eq.~(\ref{scetIIfull}). The choice $m_1=0$ is also discussed in
Appendix~\ref{AppB}, but makes the matching more complicated.

The LO mixed soft-collinear \SCETb currents in dimensional regularization
include the UV rapidity regulation factors, and are
\begin{align} \label{OIIdr}
  O^{(0a)}_{\rm II} &= \frac{J^{(0a)}}
  {(\bn\mcdot p\, n\mcdot \ell \!-\! i0^+)}\:
  \bigg[ \phi^\dagger_{s,-\ell}\, \phi^b_{v}\
  \frac{|\ell^+|^\epsilon}{\mupluse}
  \bigg] \:
  \bigg[ \phi_{n,p}^\gamma \, \frac{|p^-|^\epsilon}{\muminuse} \bigg]
  \  , \\
  O^{(0b)}_{\rm II} &=  \frac{J^{(0b)}}
 {(\bn\mcdot p\, n\mcdot \ell \!-\! i0^+)(\bn\mcdot p\, n\mcdot k \!-\!
   i0^+)} \: \bigg[ \phi^\dagger_{s,-k}\, \phi^b_{v}\,
    \phi_{s,\ell-k}^g \frac{|\ell^+\!-\!k^+|^\epsilon |k^+|^\epsilon }
   {\muplusee} 
   \bigg]\ 
   \bigg[ \phi_{n,p}^\gamma \frac{|p^-|^\epsilon}{\muminuse}\bigg] ,
   \nn
  O^{(0c)}_{\rm II} &=  \frac{J^{(0c)}}
 {(\bn\mcdot p\, n\mcdot \ell \!-\! i0^+)(\bn\mcdot q\, n\mcdot \ell \!-\!
   i0^+)} \: \bigg[ \phi^\dagger_{s,-\ell}\, \phi^b_{v} 
    \frac{|\ell^+|^\epsilon}{\mupluse}\bigg]\, 
   \bigg[ \phi_{n,p-q}^\dagger\, \phi_{n,-q} 
   \frac{|p^-\!-\!q^-|^\epsilon |q^-|^\epsilon }
   {\muminusee} \bigg] ,
   \nonumber
\end{align} 
where we have suppressed the sums over label momenta shown in
Eq.~(\ref{LOcurr}), and in general the $J^{(i)}$ are functions of the
label momenta ($p^-$, $\ell^+$, $k^+$, $\ldots$).  Using the currents
$O_{\rm II}^{(0b)}$ and $O_{\rm II}^{(0c)}$ for the soft and collinear
graphs in Figs.~\ref{fig:scalargamma}b,c respectively we have
\begin{align} \label{IsIcstart}
 I_{\rm soft}^{\rm scalar} &= \sum_{k^+\ne 0} \int  
  \!\!\frac{d^Dk_r}{(2\pi)^D}  \frac{\mu^{2\epsilon}}
 {[k^2\!-\!\ell^+\,k^-  \!+\! i 0^+]
 [k^2 \!-\!m^2\!+\! i 0^+][-p^-\,  k^+ \!+\! i0^+]}
\ \frac{|k^+|^\epsilon |k^+\!-\!\ell^+|^\epsilon}{\muplusee} \,, \\
 I_{\rm cn}^{\rm scalar} &= \sum_{k^-\ne 0} \int  
  \!\!\frac{d^Dk_r'}{(2\pi)^D}  \frac{\mu^{2\epsilon}}
 {[-\ell^+\,k^- \!+\! i 0^+]
 [k^2 \!-\!m^2\!+\! i 0^+][k^2\!-\! p^- \, k^+  \!+\! i 0^+]} 
 \ \frac{|k^-|^\epsilon |k^-\!-\!p^-|^\epsilon}{\muminusee} \,.
  \nonumber
\end{align}
Here the $k^+\ne 0$ and $k^-\ne 0$ conditions denote the overlap regions where
the soft integration variable becomes collinear and the collinear integration
variable becomes soft, as in Fig.~\ref{fig:scet2}.  The sums over $k^+\ne 0$ and
$k^-\ne 0$ ensure that the $[-p^-k^+]$ and $[-\ell^+k^-]$ propagators never get
small. By examining the scaling, we find that no subtraction is necessary for
$k^+\ne \ell^+$ and $k^-\ne p^-$ here, so though present, these restrictions
were not shown.  Eq.~(\ref{1.02}) tells us that unlike the \SCETa computations
and the \SCETb cutoff computation, here we have zero-bin subtractions for both
the soft and collinear diagrams. These will ensure that we do not get spurious
singularities from the $[-p^-k^+]$ and $[-\ell^+k^-]$ propagators. The naive
integrals and subtraction integrals are
\begin{align}
 \tilde I_{\rm soft}^{\rm scalar} &=  \int  
  \!\!\frac{d^Dk}{(2\pi)^D}  \frac{\mu^{2\epsilon}}
 {[k^2\!-\!\ell^+\,k^- \!+\! i 0^+]
 [k^2 \!-\!m^2\!+\! i 0^+][-p^-\,  k^+ \!+\! i0^+]}
\ \frac{|k^+|^\epsilon |k^+\!-\!\ell^+|^\epsilon}{\muplusee} \,, \\
  I_{\rm 0soft}^{\rm scalar} &=  \int  
  \!\!\frac{d^Dk}{(2\pi)^D}  \frac{\mu^{2\epsilon}}
 {[-\!\ell^+\,k^- \!+\! i 0^+]
 [k^2 \!-\!m^2\!+\! i 0^+][-p^-\,  k^+ \!+\! i0^+]}
\ \frac{|k^+|^\epsilon |k^+\!-\!\ell^+|^\epsilon}{\muplusee} \,, \nn
 \tilde I_{\rm cn}^{\rm scalar} &=  \int  
  \!\!\frac{d^Dk}{(2\pi)^D}  \frac{\mu^{2\epsilon}}
 {[-\ell^+\,k^- \!+\! i 0^+]
 [k^2 \!-\!m^2\!+\! i 0^+][k^2\!-\! p^- \, k^+ \!+\! i 0^+]} 
 \ \frac{|k^-|^\epsilon |k^-\!-\!p^-|^\epsilon}{\muminusee} \,,\nn
 I_{\rm 0cn}^{\rm scalar} &=  \int  
  \!\!\frac{d^Dk}{(2\pi)^D}  \frac{\mu^{2\epsilon}}
 {[-\ell^+\,k^- \!+\! i 0^+]
 [k^2 \!-\!m^2 \!+\! i 0^+][-\! p^- \, k^+ \!+\! i 0^+]} 
 \ \frac{|k^-|^\epsilon |k^-\!-\!p^-|^\epsilon}{\muminusee} \,.\nonumber
\end{align}
Note that we must keep the $m^2$ dependence in the subtraction integrals to
properly avoid double counting the zero-bin regions in the differences $\tilde
I_{\rm soft}^{\rm scalar}- I_{\rm 0soft}^{\rm scalar}$ and $\tilde I_{\rm
  cn}^{\rm scalar}- I_{\rm 0cn}^{\rm scalar}$, which from Eq.~(\ref{1.02}) 
give the result for $I_{\rm soft}^{\rm scalar}$ and $I_{\rm cn}^{\rm scalar}$
respectively.

For the soft graph we do the $k^-$ integral by contours. Due to the pole
structure this restricts the $k^+$-integration to the region $0< k^+<\ell^+$.
The $k_\perp$ integral is then done. For the soft subtraction integral we follow
the same procedure which this time leaves the integration region $0<
k^+<\infty$. We find
\begin{align} 
  \tilde I_{\rm soft}^{\rm scalar} 
  &= \frac{-i\: \Gamma(\epsilon)\, \mu^{2\epsilon} }{16\pi^2(p^-\ell^+)}  
   \int_0^{\ell^+} \! \frac{dk^+}{k^+} \, \left[ \frac{(\ell^+\!-\! k^+)
       m^2}{\ell^+}\right]^{-\epsilon} 
   \ \bigg|  \frac{k^+ (k^+\!-\!\ell^+)}
   {\muplustwo} \bigg|^\epsilon
    \\
 &= \frac{-i\: \Gamma(\epsilon) }{16\pi^2(p^-\ell^+)}  
   \Big(\frac{m^2}{\mu^2}\Big)^{-\epsilon}
   \Big(\frac{\ell^+}{\muplus}\Big)^{2\epsilon}
   \: \frac{1}{\epsilon_{\rm IR}} \,,
  \nonumber \\[5pt]
  I_{\rm 0soft}^{\rm scalar} 
  &= \frac{-i\: \Gamma(\epsilon)\, \mu^{2\epsilon} }{16\pi^2(p^-\ell^+)}  
   \int_0^{\infty} \! \frac{dk^+}{k^+} \, \left(  m^2\right)^{-\epsilon} 
   \: \bigg|  \frac{k^+ (k^+\!-\!\ell^+)}
   {\muplustwo} \bigg|^\epsilon  \nn
&= \frac{-i\: \Gamma(\epsilon) }{16\pi^2(p^-\ell^+)}  
   \Big(\frac{m^2}{\mu^2}\Big)^{-\epsilon}
   \Big(\frac{\ell^+}{\muplus}\Big)^{2\epsilon}
   \: \left( \Big\{ \frac{1}{\epsilon_{\rm IR}} - \frac{\pi^2\epsilon}{6}\Big\}
   + \Big\{ -\frac{1}{2\epsilon_{\rm UV}} - \frac{\pi^2\epsilon}{6}\Big\}
   \right). \nonumber
\end{align}
In the last line the first $\{\cdots\}$ factor comes from the integral over $0<
k^+< \ell^+$, and the second from $\ell^+ < k^+ <\infty$.  Computing the full
soft integral in Eq.~(\ref{IsIcstart}), $I_{\rm soft}^{\rm scalar} = \tilde I_{\rm
  soft}^{\rm scalar} - I_{\rm 0soft}^{\rm scalar}$,
\begin{align} \label{ssrslt2}
 I_{\rm soft}^{\rm scalar} 
 &= \frac{-i\: \Gamma(\epsilon) }{16\pi^2(p^-\ell^+)}  
   \Big(\frac{m^2}{\mu^2}\Big)^{-\epsilon}
   \Big(\frac{\ell^+}{\muplus}\Big)^{2\epsilon}
   \: \left(\frac{1}{2\epsilon_{\rm UV}} +\frac{\pi^2\epsilon}{3} \right)
  \nonumber \\
 &= \frac{-i\: }{16\pi^2(p^-\ell^+)}  
   \bigg[ \frac{1}{2\epsilon_{\rm
      UV}^2} 
   \!+\! \frac{1}{\epsilon_{\rm UV}} \ln\Big(\frac{\ell^+  }{\muplus} \Big)
   \!-\! \frac{1}{2\epsilon_{\rm UV}} \ln\Big(\frac{m^2 }{\mu^2} \Big)
  +  \ln^2\Big(\frac{\ell^+}{\muplus} \Big) + \frac{3\pi^2}{8} \nn
 & \qquad\qquad\qquad\qquad
  \!+\! \frac{1}{4} \ln^2\Big(\frac{m^2}{\mu^2} \Big) 
   -  \ln\Big(\frac{m^2}{\mu^2} \Big) \ln\Big(\frac{\ell^+}{\muplus}\Big)
  \bigg] \,.
\end{align}
Much like the examples in \SCETa the zero-bin subtraction integral $I_{\rm
  0soft}^{\rm scalar}$ cancels the IR singularity in the $k^+$ integration in
$\tilde I_{\rm soft}^{\rm scalar}$ and replaces it by a UV divergence. 

For the collinear integrals, we do the contour integration in $k^+$ which
restricts the remaining integration region in $k^-$. For the naive and
subtraction integrals we find
\begin{align}  \label{Icns}
  \tilde I_{\rm cn}^{\rm scalar} 
  &= \frac{-i\: \Gamma(\epsilon)\, \mu^{2\epsilon} }{16\pi^2(p^-\ell^+)}  
   \int_0^{p^-} \! \frac{dk^-}{k^-} \, \left[ \frac{(p^-\!-\!k^-)
       m^2}{p^-}\right]^{-\epsilon} 
   \ \ \bigg|  \frac{k^- (k^-\!-\!p^-)}
   {\muminustwo} \bigg|^\epsilon
   \nonumber \\[5pt]
  I_{\rm 0cn}^{\rm scalar} 
  &= \frac{-i\: \Gamma(\epsilon)\, \mu^{2\epsilon} }{16\pi^2(p^-\ell^+)}  
   \int_0^{\infty} \! \frac{dk^-}{k^-} \, \left( m^2\right)^{-\epsilon} 
   \ \ \bigg|  \frac{k^- (k^-\!-\!p^-)}
   {\muminustwo} \bigg|^\epsilon \,,
\end{align}
which for our example, are the same integrals as for the soft loops but with
$\ell^+\to p^-$ and $\muplus\to \muminus$.  Thus for the complete collinear result in
Eq.~(\ref{IsIcstart}), $I_{\rm cn}^{\rm scalar}= \tilde I_{\rm cn}^{\rm
  scalar}-I_{\rm 0cn}^{\rm scalar}$, we find
\begin{align}  \label{scrslt2}
  I_{\rm cn}^{\rm scalar} 
  &= \frac{-i\: \Gamma(\epsilon) }{16\pi^2(p^-\ell^+)}  
   \Big(\frac{m^2}{\mu^2}\Big)^{-\epsilon}
   \Big(\frac{p^-}{\muminus}\Big)^{2\epsilon}
   \: \left(\frac{1}{2\epsilon_{\rm UV}} +\frac{\pi^2\epsilon}{3} \right)
  \nonumber \\
 &= \frac{-i\: }{16\pi^2(p^-\ell^+)}  
   \bigg[ \frac{1}{2\epsilon_{\rm
      UV}^2} 
   \!+\! \frac{1}{\epsilon_{\rm UV}} \ln\Big(\frac{p^-  }{\muminus} \Big)
   \!-\! \frac{1}{2\epsilon_{\rm UV}} \ln\Big(\frac{m^2 }{\mu^2} \Big)
  + \ln^2\Big(\frac{p^-}{\muminus} \Big) + \frac{3\pi^2}{8} \nn
 & \qquad\qquad\qquad\qquad
  \!+\! \frac{1}{4} \ln^2\Big(\frac{m^2}{\mu^2} \Big) 
   -  \ln\Big(\frac{m^2}{\mu^2} \Big) \ln\Big(\frac{p^-}{\muminus}\Big)
  \bigg] \,.
\end{align}

The results in Eqs.~(\ref{ssrslt2}) and (\ref{scrslt2}) have $1/\epsilon_{\rm
  UV} \ln(m^2)$ divergences, terms that did not appear in our example with a
rapidity cutoff, and are simply artifacts of the dimensional regularization
setup. These divergences arise from the fact that the UV collinear
divergences induced by the zero-bin subtraction are multiplicative over all
loops and propagators. They are canceled by a special type of counterterm that
is proportional to the renormalized distribution function at the origin,
$\phi(0,\mu)/\epsilon_{\rm UV}$. The presence of these operators is discussed
further in section~\ref{sect:kernel} below. Here we have one such counterterm
current for the soft loop and one for the collinear loop
\begin{align} \label{J0d}
  O^{(0d)}_{\rm II} &=  \sum_{p^-,\ell^+\ne 0} \frac{J^{(0d)}}
 {[(\bn\mcdot p\, n\mcdot \ell \!-\! i0^+) \bn \mcdot p]} \: 
   \Big[ \phi^\dagger_{s,-k}\, \phi^b_{v}\, \phi_{s,\ell}^g  \Big]\ 
   \big[ \phi_{n,p}^\gamma \big] \Big|_{k^+\to 0}  \,, \nn
   O^{(0e)}_{\rm II} &=  \sum_{p^-,\ell^+\ne 0} 
    \frac{J^{(0e)}} {[(\bn\mcdot p\, n\mcdot \ell \!-\!
     i0^+)n \mcdot \ell]} \: 
  \big[ \phi^\dagger_{s,-\ell}\, \phi^b_{v} \big]\ 
  \big[ \phi^\dagger_{n,p} \phi_{n,-q} \big]  \Big|_{q^- \to 0} \,.
\end{align}
These correspond to counterterms for operators which give a $\phi(0,\mu)$,
corresponding to the initial $B$ meson, and a $\phi(0,\mu)$ for the quark part
of the final photon wavefunction. Note that the limit $k^+\to 0$ and $q^-\to 0$
is done at the end.

The necessary counterterm coefficients for the results in Eqs.~(\ref{ssrslt2}) and
(\ref{scrslt2}) are
\begin{align} \label{C0d0e}
 \delta J^{(0d)} =  \frac{e g G}{2\epsilon_{\rm UV}} \,, \qquad\qquad
  \delta J^{(0e)} =  \frac{e g G}{2\epsilon_{\rm UV}} \,. 
\end{align}
At one loop these operator generate graphs similar to the diagrams in
Fig.~\ref{fig:scalargamma}b,c. Using the same IR mass regulator, pulling out the
same prefactor as the other diagrams, and performing the standard UV
renormalization of the operators in ${\overline {\rm MS}}$ prior to multiplying by
the rapidity counterterm gives
\begin{align}
 I_{\rm ct\, 0d}^{\rm scalar} = \Big(\frac{1}{2\epsilon_{\rm UV}} \Big)
   \frac{i\: }{16\pi^2(p^-\ell^+)}  
   \bigg[  -  \ln\Big(\frac{m^2}{\mu^2} \Big) 
  \bigg]\,,\nn
 I_{\rm ct\, 0e}^{\rm scalar} = \Big(\frac{1}{2\epsilon_{\rm UV}} \Big)
   \frac{i\: }{16\pi^2(p^-\ell^+)}  
   \bigg[  -  \ln\Big(\frac{m^2}{\mu^2} \Big) 
    \bigg]\,.
\end{align}
Due to our choice of $\delta J^{(0d,0e)}$ these counterterm diagrams exactly
cancel the $1/\epsilon_{\rm UV} \ln(m^2)$ terms in the collinear and soft loops.
In dimensional regularization, this type of counterterm operator is quite
important to the rapidity renormalization, as we discuss in the next section. In
particular in going beyond perturbation theory, these same type of counterterms
are required to cancel UV divergences in the convolution over the
non-perturbative matrix element.

Adding the soft, collinear, and $I_{\rm ct\, 0d}^{\rm scalar}$, $I_{\rm ct\,
  0e}^{\rm scalar}$ counterterms we find the \SCETb result
\begin{align} \label{scetIIDRrslt0}
 I_{\rm soft+cn}^{\rm scalar} 
 &= \frac{-i\: }{16\pi^2(p^-\ell^+)}  
   \bigg[  \frac12\, \ln^2\Big(\frac{m^2}{\mu^2} \Big) 
   -  \ln\Big(\frac{m^2}{\mu^2} \Big) \ln\Big(\frac{p^-}{\muminus}\Big)
  -  \ln\Big(\frac{m^2}{\mu^2} \Big) \ln\Big(\frac{\ell^+}{\muplus}\Big)
  \\
 & \qquad
  +\frac{1}{\epsilon_{\rm UV}^2} 
   \!+\! \frac{1}{\epsilon_{\rm UV}} \ln\Big(\frac{p^- \ell^+ }{\muminus\muplus} \Big)
   +  \ln^2\Big(\frac{p^-}{\muminus} \Big) +  \ln^2\Big(\frac{\ell^+}{\muplus} \Big)
  + \frac{3\pi^2}{4}
  \bigg] \nonumber \\[5pt]
&= \frac{-i\: }{16\pi^2(p^-\ell^+)}  
   \bigg[  \frac12\, \ln^2\Big(\frac{m^2}{p^-\ell^+} \Big)
   - \ln\Big(\frac{m^2}{\mu^2} \Big) \ln\Big(\frac{\mu^2}{\muminus\muplus}\Big)
  \nn
 & \qquad
  +\frac{1}{\epsilon_{\rm UV}^2} 
   \!+\! \frac{1}{\epsilon_{\rm UV}} \ln\Big(\frac{p^- \ell^+ }{\muminus\muplus} \Big)
  + \ln^2\Big(\frac{p^-}{\muminus} \Big) +  \ln^2\Big(\frac{\ell^+}{\muplus} \Big)
    - \frac12\, \ln^2\Big(\frac{p^-\ell^+}{\mu^2} \Big) 
   + \frac{3\pi^2}{4}  \bigg]
 \,.\nonumber
\end{align}
The effective theory still has UV divergences shown on the second line of
Eq.~(\ref{scetIIDRrslt0}).  These divergences occur because of the separation of
$\zeta_p$ momenta.  The remaining UV divergences are canceled by a counterterm
for the jet function coefficient $J^{(0a)}$. Putting back the prefactor, we find
the counterterm
\begin{align} \label{delJrslt}
   \delta J^{(0a)} &= \frac{e g^2 G }{16\pi^2(p^-\ell^+)} \bigg[
   - \frac{1}{\epsilon_{\rm UV}^2} 
   - \frac{1}{\epsilon_{\rm UV}}  \ln\Big(\frac{p^- \ell^+ }{\muminus\muplus} \Big) 
   \bigg].
\end{align}
As is familiar from \SCETa, the counterterm depends on logs involving $\mu$'s.
Thus, finally, the renormalized EFT result is
\begin{align}
  \label{scetIIDRrslt}
 I_{\rm soft+cn}^{\rm scalar} 
 &=  \frac{-i\: }{16\pi^2(p^-\ell^+)}  
   \bigg[ \frac{1}{2}\, \ln^2\Big(\frac{m^2}{p^-\ell^+} \Big)
   -  \ln\Big(\frac{m^2}{\mu^2} \Big) \ln\Big(\frac{\mu^2}{\muminus\muplus}\Big)
  \\
 & \qquad \qquad\qquad
  + \ln^2\Big(\frac{p^-}{\muminus} \Big) +  \ln^2\Big(\frac{\ell^+}{\muplus} \Big)
    - \frac12 \ln^2\Big(\frac{p^-\ell^+}{\mu^2} \Big) 
   + \frac{3\pi^2}{4}  \bigg]
 \,.\nonumber
\end{align}

Comparing the first two terms in Eq.~(\ref{scetIIDRrslt}) with the full theory
result in Eq.~(\ref{scetIIfull}) we see that the IR divergences are exactly
reproduced if and only if
\begin{eqnarray}
  \mu^2 = \muplus\: \muminus \,.
\end{eqnarray}
Thus again this condition follows from the dynamics.  It is interesting to note
that the $\ln(p^-\ell^+)\ln(m^2)$ divergence is reproduced independent of the
power of $|\bnP|$ and $|\cP|$ used in Eq.~(\ref{Op1}), but that the $\ln^2(m^2)$
term is only reproduced for the power $\epsilon$, which was derived in
section~\ref{sect:dimregII}.  Using $\mu^2=\muplus\muminus$, the difference of the
remaining finite terms gives a contribution to the one-loop matching
\begin{align} \label{matchrslt}
   I_{\rm match}^{\rm scalar} 
  &= \frac{-i\: }{16\pi^2(p^-\ell^+)}  
   \bigg[ \frac12\, \ln^2\Big(\frac{p^-\ell^+}{\muminus\muplus}\Big) 
  -  \ln^2\Big(\frac{p^-}{\muminus} \Big) -  \ln^2\Big(\frac{\ell^+}{\muplus} \Big)
  - \frac{5\pi^2}{12}
  \bigg] \nn
&= \frac{-i\: }{16\pi^2(p^-\ell^+)}  
   \bigg[ - \frac12\, \ln^2\Big(\frac{p^-\muplus}{\muminus\ell^+} \Big)
  - \frac{5\pi^2}{12}
  \bigg]
 \,.
\end{align}
From Eq.~(\ref{matchrslt}) we see that the jet function will be a non-trivial
function of $\muplus$ and $\muminus$ (and thus $\mu^2=\muplus\muminus$) whose
$\mu$-dependences will cancel against dependence on these variables in the
\SCETb matrix elements.  The matching coefficient arises from integrating out
perturbative effects associated with $p^- l^+$ as well as perturbative effects
responsible for the rapidity gap between the soft and collinear modes, and
therefore has a different structure than what would be obtained if only the
former were integrated out.  In Appendix~\ref{AppB} we verify that the same
result for $I_{\rm match}^{\rm scalar}$ is obtained when we regulate the IR
divergences in the full and effective theories with three non-equal masses,
$m_{1,2,3}$, or when we keep the $\ell^2$ and $p^2$ offshell. The result should
be the same because the matching only depends on the UV regulator which we keep
the same in these computations. Note that in Eq.~(\ref{smatch}) we used a
different UV regulator than in Eq.~(\ref{matchrslt}), which explains why the
$\pi^2$ terms differ.

The results in Eq.~(\ref{delJrslt}) and (\ref{matchrslt}) are shown for
illustration only, since of course, the complete anomalous dimension and
matching calculations require a computation of all diagrams, not just the one
diagram that we considered here for illustration.


\subsection{Zero-Bin Subtractions in  Convolutions: General Discussion}
\label{sect:kernel}

The remaining application of subtractions in \SCETb will be for
factorization formulas which appear to suffer from singular
convolutions at the level of tree level matching.  Much like the
example discussed in \SCETa, in \SCETb we must avoid the zero-bin in
hard scattering kernels defined by tree level matching.  Doing so
removes the double counting problem and renders singular convolutions
finite. Here we only deal with the rapidity renormalization, so we
make the simplifying assumption that the standard UV divergences have
already been taken care of, and do not interfere with the steps carried out here.

We will use dimensional regularization to separate the modes in rapidity space
as in Eq.~(\ref{Op1}). To see why the convolution integrals are always
finite, let us consider the vacuum to pion matrix element of a hard scattering
kernel $J(p_i^-,\muminus,\mu^2)$ and a twist-2 collinear operator in \SCETb that
gives the light-cone distribution function $\phi_\pi(x,\mu)$. For simplicity we
will not write the $\mu$-dependences for $J$ and $\phi_\pi$
below.\footnote{Note that integer powers of the $p_i^-$ can be moved
  from $J$ to the collinear operator by inserting powers of $\bnP$, but that our
  analysis is independent of this freedom.}   This leaves
the matrix element
\begin{align} \label{Api0}
 A_\pi & =  \sum_{p_{1,2}^-\ne 0}  
\int\!\! dp_{1r}^- dp_{2r}^{-}\ J(p_1^-,p_2^{-})\ 
  \big\langle \pi_n(p_\pi) \big| (\bar\xi_n
  W)_{p_1^-} \bnslash\gamma_5 (W^\dagger \xi_n)_{-p_2^{-}} \big| 0 \big\rangle \
  \bigg|  \frac{p_1^{-} p_2^{-}}{\muminustwo} \bigg|^\epsilon  \\
 & = -if_\pi 
  \sum_{p_{1,2}^-\ne 0 }  
\int\!\! dp_{1r}^- dp_{2r}^{-}\ J(p_1^-,p_2^{-})\ 
  \delta(\bn\mcdot p_\pi\!-\!p_1^-\!-\!p_2^{-})\: \phi_\pi(x_1,x_2) \
  \bigg|  \frac{p_1^{-} p_2^{-}}{\muminustwo} \bigg|^\epsilon \nn
  & = -if_\pi \, \bn\mcdot p_\pi \Big( \frac{\bn\mcdot
    p_\pi}{\muminus}\Big)^{2\epsilon} \sum_{x_{1,2}\ne 0 } \int\!\! dx_{1r}
  dx_{2r}\ J(x_1,x_2)\: 
  \delta(1\!-\!x_1\!-\!x_2)\: \phi_\pi(x_1,x_2)\ \big| x_1 x_2 \big|^\epsilon ,
   \nonumber
\end{align}
where we switched to dimensionless variables $x_{1,2}$ via $p_1^- =
x_1\,\bn\mcdot p_\pi$ and $p_2^{-}=x_2\, \bn\mcdot p_\pi$, and in the second
line we inserted the standard definition of the twist-2 distribution function
\begin{align} \label{pimelt}
 \big\langle \pi_n^+(p_\pi) \big| \bar u_{n,p_1^{-}} {\bnslash\gamma_5}
   \, 
  d_{n,-p_2^{-}} \big| 0 \big\rangle 
 &= {-i f_\pi}\: 
    \delta(\bn\mcdot p_\pi\!-\!p_1^-\!-\!p_2^{-})\: \phi_\pi(x_1,x_2,\mu) \,,
\end{align}
where the $\delta$ function gives conservation of momentum.  Now suppose that we
computed $J$ at tree level (by a matching computation) and found that
$J(x_1,x_2) = 1/(p_1^-)^2 = 1/[(\bn\mcdot p_\pi)^2\, x_1^2]$.  If we were not
careful about the $x_1\ne 0$ condition, this would lead to a singular
convolution integral as in Eq.~(\ref{0.1}). The zero-bin subtraction formula in
Eq.~(\ref{1.02}) tells us to impose the momentum conserving $\delta$-functions
carrying through all zero-bin constraints. Since the $x_2$-integration is not
singular, there are no zero-bin subtractions for $x_2\ne 0$ and we can combine
the sum over label $x_2$ momenta and integral over residual $x_{2r}$ momenta
back into a integral over all $x_2$ momenta using Eq.~(\ref{1.01}):
\begin{align} \label{Apisame}
 A_\pi &=  -i \frac{f_\pi}{ \bn\mcdot p_\pi} \Big( \frac{\bn\mcdot
    p_\pi}{\muminus}\Big)^{2\epsilon} \sum_{x_{1}\ne 0 } \int\!\! dx_{1r}\,
  dx_{2}\ \frac{1}{(x_1)^2}\: 
  \delta(1\!-\!x_1\!-\!x_2)\: \phi_\pi(x_1,x_2)\ \big| x_1 x_2 \big|^\epsilon
   \nn
 &=  -i\frac{f_\pi}{ \bn\mcdot p_\pi} \Big( \frac{\bn\mcdot
    p_\pi}{\muminus}\Big)^{2\epsilon} \sum_{x_{1}\ne 0 } \int\!\! dx_{1r}
   \: \frac{1}{(x_1)^2}\: 
  \theta(1\!-\!x_1)\theta(x_1)\: \hat \phi_\pi(x_1) 
  \ \big| x_1 (1\!-\!x_1) \big|^\epsilon\,.
\end{align}
where $\bar x_1=1-x_1$.  If there had been zero-bin subtractions for $x_2$ they
would carry through as additional zero-bin subtractions at $x_1\ne 1$ after
removing the $\delta$-function.  In the last line we set $\phi_\pi(x_1,\bar
x_1)=\theta(1\!-\!x_1)\theta(x_1)\: \hat \phi_\pi(x_1)$ to make the support of
the non-perturbative distribution function explicit. To turn the final sum over
labels and integral over residual momenta into an integral over $x_1$, there will
be zero-bin subtractions from Eq.~(\ref{1.02}).  The subtraction acts on the
integrand including the $\theta$-functions, but just as in our perturbative
analysis, it does not act on the $|x_1(1\!-\! x_1)|^\epsilon$ factor.  The
expansion for $x_1\ne 0$ is from the right, about $x_1=0^+$, since this is how
the variable scales towards the zero-bin region:
\begin{align} \label{piexpn}
  \theta(1\!-\! x_1) \theta(x_1) \hat\phi_\pi(x_1)
   & = \theta(x_1) \Big[ 
    \hat \phi_\pi(0) + x_1 \, \hat \phi_\pi'(0) + \frac{
    x_1^2}{2} \, \hat \phi_\pi''(0) + \ldots \Big] \nn
 & \ \ + \theta(x_1) \hat\phi_\pi(0) 
   \big[ \delta(1\!-\!x_1) \!+\! \ldots \big] 
  \,. 
\end{align}
In the set of terms obtained on the first line, the $\theta(1- x_1)$ disappears
in the series so the support of the $x_1$ integration for the subtraction terms
differs from that for the naive integral. This is the same as what we saw in our
perturbation theory example in Eq.~(\ref{Icns}), where the naive integral was
integrated over $k^- \in [0,p^-]$, ie. $x_1\in [0,1]$, but the the subtraction
integral was integrated over $k^- \in [0,\infty]$, ie. $x_1\in [0,\infty]$. In
the last line in Eq.~(\ref{piexpn}), the terms are all zero (or finite
subtractions) for the cases considered here, and therefore these terms do not
contribute for our choice of zero-bin scheme as discussed in
section~\ref{sec:zerobin}.

Lets make the standard assumption for the twist-2 distribution that
$\phi_\pi(0)=0$. Then using Eq.~(\ref{1.02}), the result for $A_\pi$ is
\begin{align}
 A_\pi &=   \frac{-if_\pi}{ \bn\mcdot p_\pi} 
  \Big( \frac{p_\pi^-}{\muminus}\Big)^{2\epsilon}  \!\! \int\!\! 
  {dx_{1}}\: 
  \frac{\theta(x_1)}{(x_1)^2}\: 
  \Big[ \theta(1\!-\!x_1)\:  \hat \phi_\pi(x_1) 
   \!-\!   x_1 \hat\phi'_\pi(0)\Big] 
  \big| x_1 (1\!-\! x_1) \big|^{-\epsilon} \,,
\end{align}
where as usual only the subtraction needed to remove the singular term was kept.
Next we split the integration into a finite integral $x_1\in [0,1]$ where the
factor of $|x_1(1\!-\! x_1)|^\epsilon$ can be set to $1$, and the integral of
the subtraction term over $x_1\in [1,\infty]$ where the $\epsilon$ dependent
term is needed
\begin{align} \label{Api1}
  A_\pi &= -i \frac{f_\pi}{ \bn\mcdot p_\pi} \Big( \frac{\bn\mcdot
    p_\pi}{\muminus}\Big)^{2\epsilon} \bigg\{ \int_0^1\!\! dx_{1}\, \frac{
    \phi_\pi(x_1) - x_1 \phi'_\pi(0)}{(x_1)^2}\ - \int_1^\infty\!\!\! dx_1\,
  \frac{x_1^\epsilon (x_1\!-\!1)^\epsilon}{(x_1)^2} \: \big[ x_1 \phi'_\pi(0)
  \big] \bigg\} \nn
  &= -i \frac{f_\pi}{ \bn\mcdot p_\pi} \Big( \frac{\bn\mcdot
    p_\pi}{\muminus}\Big)^{2\epsilon} \bigg\{ \int_0^1\!\! dx_{1}\,
  \frac{ \phi_\pi(x_1) - x_1 \phi'_\pi(0)}{(x_1)^2}\ 
   + \frac{1}{2\epsilon_{\rm UV}}\, 
   \phi'_\pi(0) 
  \bigg\} \,.
\end{align}
Here terms of ${\cal O}(\epsilon)$ have been dropped.

Eq.~(\ref{Api1}) is UV divergent, but we must still add to it the pion matrix
element of the counterterm operator. This operator is determined by the UV
counterterms that are necessary to renormalize our original operator, and can be
derived in perturbation theory with any desired external states. Carrying out
a one-loop computation with external quark states and using our perturbative
kernel $J=1/(p_1^-)^2$ we find the counterterm operator
\begin{align} \label{Oct}
  O_{ct}^{[1]} &= C_{ct}^{[1]}  
\int\!\! dp_{2}^- \: 
  \Big[\frac{\partial}{\partial p_1^-} - \frac{\partial}{\partial (p_1^-\!+\!p_2^-)}\Big] (\bar\xi_n
  W)_{p_1^-} \bnslash\gamma_5 (W^\dagger \xi_n)_{-p_2^{-}} \bigg|_{p_1^-\to 0} \,,
\end{align} 
with a counterterm coefficient $\delta C_{ct}^{[1]}=-1/(2\epsilon_{\rm UV})$.  The derivative
with respect to $(p_1^-+p_2^-)$ removes surface terms. In the vacuum
to pion matrix element, they would result from a $d/dp_1^-$ of the
$\delta$-function in Eq.~(\ref{pimelt}) if we had left out the
$d/d(p_1^-+p_2^-)$.  At tree level with quarks the matrix element of
this operator vanishes --- one obtains $\delta'(p^-)$ factors and the
quark states have non-zero $p^-$ momenta. The vacuum to pion matrix
element of $O_{ct}^{[1]}$ gives
\begin{align}
  A_\pi^{ct1} &= -\frac{1}{2\epsilon_{\rm UV}} 
\int\!\! dp_{2}^- \: 
  \Big[\frac{d}{dp_1^-} + \frac{d}{dp_\pi^-}\Big]
 \langle 
 \pi_n(p_\pi) \big| (\bar\xi_n
  W)_{p_1^-} \bnslash\gamma_5 (W^\dagger \xi_n)_{-p_2^{-}} \big| 0 \big\rangle
  \Big|_{p_1^-\to 0}
  \nn
  &= \frac{i f_\pi}{2\epsilon_{\rm UV}\,p_\pi^-} 
\int\!\! dp_{2}^- \: 
  \delta(\bn\mcdot p_\pi\!-\!p_1^-\!-\!p_2^{-})\: 
    \phi_\pi^{(1,0)}(x_1,x_2,\mu)
  \Big|_{p_1^-\to 0}
  \nn
  &= i \frac{f_\pi}{ \bn\mcdot p_\pi} \ \frac{1}{2\epsilon_{\rm UV}}\:
  \phi_\pi'(0) \,,
\end{align}
where the superscript $(1,0)$ indicates a derivative with respect to the first
argument.  We should also include the matrix element of the finite
part of the counterterm operator in Eq.~(\ref{Oct})~\cite{LMS}, which gives
\begin{align} \label{ct2}
   A_\pi^{ct2} &=  -i \frac{f_\pi}{ \bn\mcdot p_\pi} C_{ct}^{[1]}(\mu_-)
  \phi_\pi'(0)    \,.
\end{align}
Adding the $A_\pi^{ct1}$ term to Eq.~(\ref{Api1}), the UV divergence cancels, and
sending $\epsilon\to 0$ we obtain the finite result
\begin{align} \label{Apif}
 A_\pi+ A_\pi^{ct1}   &= -i \frac{f_\pi }{ \bn\mcdot p_\pi}
   \bigg\{ \int_0^1\!\! dx_{1}\,
  \frac{ \phi_\pi(x_1,\mu) - x_1 \phi'_\pi(0,\mu)}{(x_1)^2}\ 
   +  \phi'_\pi(0,\mu)\:  \ln\Big( \frac{\bn\mcdot
    p_\pi}{\muminus}\Big)
  \bigg\} \nn
  &\equiv -i  \frac{f_\pi }{ \bn\mcdot p_\pi} \: \int_0^1\!\! dx_{1}\:
  \frac{ \phi_\pi(x_1,\mu,\mu_-) }{(x_1^2)_{\mbox{\o}} } \,.
\end{align}
As indicated, performing the steps outlined from Eq.~(\ref{Api0}) to
(\ref{Apif}) defines the \o-distribution in dimensional regularization
with our renormalization scheme. The $\mu_-$ in the distribution,
$\phi(x_1,\mu,\mu_-)$ is a short hand for the dependence on the
$\ln(\mu_-)$ in the first line of Eq.~(\ref{Apif}).  Once again, in
Eq.~(\ref{Apif}) the zero-bin subtraction has converted an IR
divergence into a UV divergence -- the naive IR divergence in the
convolution has been converted into a UV divergence for the operator
in Eq.~(\ref{pimelt}), which is canceled by the operator
renormalization counterterm in Eq.~(\ref{Oct}).  Essentially the
\o-distribution notation on a variable, $(x)_{\mbox{\o}}$ indicates
that we have a sum over labels $x\ne 0$, and do an integral over
residuals $dx_r$, together with applying the rapidity renormalization
procedure outlined above for the UV divergences. The $\mu_-$
dependence in Eq.~(\ref{Apif}) is canceled by
$C_{ct}^{[1]}(\mu_-)$~\cite{LMS} in Eq.~(\ref{ct2}).

For other cases, the steps in determining the result for the \o-distribution are
the same as in our example; however it should be clear that the final result
will depend on how singular the perturbative kernel is, as well as the endpoint
properties of the non-perturbative function that the \o-distribution is acting
on. In particular, if the starting kernel was not singular there would be no
zero-bin subtractions and we would obtain the naive result for the convolution
that one finds without the \o-distribution. Note that if we had implemented a
hard cutoff as in section~\ref{sect:cutoffscalar} rather than dimensional
regularization, then lower limits, like $x\ge \delta$, would be induced on the
convolution integrals, together with compensating $\delta$ dependence in the jet
functions.

For illustration, we consider a few other cases in dimensional
regularization that are quite common and which appear in the examples
in the next section. First consider a distribution
$\phi^p_{\pi}(x_1,x_2,\mu)$ that does not vanish at its endpoints,
integrated against a kernel $J= 1/(\bn\mcdot p_\pi\, p_1^-)$.  For the
analog of Eq.~(\ref{pimelt}) we take the matrix element of the
operator to give $-i f_{\pi}\mu_\pi\, \delta(\bn\mcdot p_\pi
-p_1^--p_2^-)\,\phi^p_{\pi}(x_1,x_2,\mu)$ where
$\mu_\pi=m_\pi^2/(m_u\!+\!m_d)$, using Eq.~(\ref{2bdy}) below with
$n\leftrightarrow \bn$.  The steps leading up to Eq.~(\ref{Apisame})
are very similar, giving
\begin{align} \label{Bpi}
 B_\pi & =  -i\frac{f_\pi \mu_\pi}{ \bn\mcdot p_\pi} \Big( \frac{\bn\mcdot
    p_\pi}{\muminus}\Big)^{2\epsilon} \sum_{x_{1}\ne 0 } \int\!\! dx_{1r}\,
  \frac{1}{(x_1)}\:
  \theta(1\!-\!x_1)\theta(x_1)\: \hat \phi^p_\pi(x_1) 
  \ \big| x_1 (1\!-\!x_1)\big|^\epsilon \nn
 &=  -i \frac{f_\pi \mu_\pi}{ \bn\mcdot p_\pi}  \Big( \frac{\bn\mcdot
    p_\pi}{\muminus}\Big)^{2\epsilon}  \! \int\!\! dx_{1}\,
  \frac{\theta(x_1)}{(x_1)}\: 
  \Big[ \theta(1\!-\!x_1)\: 
   \hat \phi^p_\pi(x_1) - \hat \phi^p_\pi(0) \Big] \ 
    \big| x_1 (1\!-\!x_1) \big|^\epsilon\nn
  &= -i \frac{f_\pi \mu_\pi}{ \bn\mcdot p_\pi} \Big( \frac{\bn\mcdot
    p_\pi}{\muminus}\Big)^{2\epsilon} \bigg\{ \int_0^1\!\! dx_{1}\, \frac{
    \phi^p_\pi(x_1) - \phi^p_\pi(0)}{x_1}\ - \int_1^\infty\!\!\! dx_1\,
  \frac{x_1^\epsilon (x_1\!-\!1)^\epsilon}{x_1} \ \phi^p_\pi(0)
   \bigg\} \nn
  &= -i \frac{f_\pi \mu_\pi}{ \bn\mcdot p_\pi} \Big( \frac{\bn\mcdot
    p_\pi}{\muminus}\Big)^{2\epsilon} \bigg\{ \int_0^1\!\! dx_{1}\,
  \frac{ \phi^p_\pi(x_1) -  \phi^p_\pi(0)}{x_1}\ 
   + \frac{1}{2\epsilon_{\rm UV}}\,
  \phi^p_\pi(0) 
  \bigg\} 
  \,.
\end{align}
For the zero-bin subtraction in the second line we kept the first term in the
analog of the expansion in Eq.~(\ref{piexpn}).  Here the counterterm operator is
\begin{align} \label{Oct0}
  O_{ct}^{[0]} &=   C_{ct}^{[0]}
\int\!\! dp_{2}^- \:  (\bar\xi_n W)_{p_1^-} 
  \frac{\bnslash\gamma_5}{2} \Big\{
  \frac{1}{\bnP}(i\,\slash\!\!\!\!{\cal D}_n^\perp) \!-\!
  (i\,\slash\!\!\!\!{\cal D}_n^\perp)^\dagger \frac{1}{\bnP^\dagger} \Big\}
 (W\xi_n)_{-p_2^{-}}\frac{1}{\bnP^\dagger}\ 
   \bigg|_{p_1^-\to 0} \,,
\end{align}
with $\delta C_{ct} = - 1/(2\epsilon_{\rm UV})$. The tree level quark
matrix element of this operator vanishes.  Eq.~(\ref{Oct0}) with
$\delta C_{ct}$ gives a vacuum to pion matrix element
\begin{align}
  B_\pi^{ct1} &= \frac{if_\pi \mu_\pi}{2\epsilon_{\rm UV} p_\pi^-} \int\!\! dp_{2}^- \: 
  \big[\delta(\bn\mcdot p_\pi\!-\!p_1^-\!-\!p_2^{-})\: 
    \phi^p_\pi(x_1,x_2,\mu) \big]
  \Big|_{p_1^-\to 0}
  \nn
  &=  \frac{if_\pi \mu_\pi}{ \bn\mcdot p_\pi} \ \frac{1}{2\epsilon_{\rm UV}}\:
  \phi^p_\pi(0) \,.
\end{align}
This term cancels the UV divergence in Eq.~(\ref{Bpi}) to leave the finite
result
\begin{align} \label{Bpif}
 B_\pi + B_\pi^{ct1} &=  -i \frac{f_\pi \mu_\pi}{ \bn\mcdot p_\pi}
   \bigg\{ \int_0^1\!\! dx_{1}\,
  \frac{ \phi^p_\pi(x_1,\mu) -  \phi^p_\pi(0,\mu)}{x_1}\ 
   + \ln\Big(\frac{\bn\mcdot p_\pi}{\muminus}\Big) \phi^p_\pi(0,\mu) 
   \bigg\} \nn
   &\equiv  -i  \frac{f_\pi \mu_\pi }{ \bn\mcdot p_\pi} \: \int_0^1\!\! dx_{1}\:
  \frac{ \phi^p_\pi(x_1,\mu,\mu_-) }{(x_1)_{\mbox{\o}} } 
  \,.
\end{align}
As indicated the result in the first line defines the \o-distribution for this
case. Again we should add to this the matrix element of $O_{ct}^{[0]}$
with the  $C_{ct}^{[0]}(\mu_-)$ coefficient.

We also will need results for distributions like $\phi^p_\pi$ and $\phi_\pi$
but with zero-bin subtractions at both ends of the integration regions
\begin{align} \label{Op3}
  & \hspace{-0.4cm}
   \int\!\! dx\, dy\  \frac{\phi^p_\pi(x,y)}{x_{\mbox{\o}}\, y_{\mbox{\o} }} \
    \delta(1\!-\!x\!-\!y) \\
  &= \Big( \frac{\bn\mcdot
    p_\pi}{\muminus}\Big)^{2\epsilon} \sum_{x\ne 0,\, y\ne 0\,
   } 
  \int\!\! dx_r\, dy_r \ \frac{\phi^p_\pi(x,y)}{x\, y} \:
    \delta(1\!-\!x\!-\!y)\ \big|x\, y\big|^\epsilon  +\mbox{c.t.}\nn
 & = \Big( \frac{\bn\mcdot
    p_\pi}{\muminus}\Big)^{2\epsilon}
   \sum_{x\ne 0,1} \int\!\! dx_r\,\frac{\hat \phi^p_\pi(x)}
    {x\, (1\!-\! x)} \:
    \theta(x)\theta(1\!-\!x)\ \big|x(1\!-\!x)\big|^\epsilon +\mbox{c.t.} \nn
 &  =   \int_0^1 \!\!\! dx\: \bigg\{ \frac{\hat \phi^p_\pi(x)}{x(1\!-\!x)}
    \!-\! \frac{\hat \phi^p_\pi(0)}{x} \!-\!
   \frac{\hat \phi^p_\pi(1)}{1\!-\!x} \bigg\} \nn
 & \ \ \ \ \ \!-\! \Big( \frac{\bn\mcdot
    p_\pi}{\muminus}\Big)^{2\epsilon} \bigg\{
  \int_1^\infty \!\!\!\! dx\: 
   \frac{\hat \phi^p_\pi(0) x^\epsilon (x\!-\!1)^\epsilon}{x} + 
  \int_{-\infty}^0 \!\!\!\! dx\: 
   \frac{\hat \phi^p_\pi(1) x^\epsilon (1\!-\!x)^\epsilon}{1-x} \bigg\}
   +\mbox{c.t.} 
  \nn
 & =  \int_0^1 \!\!\! dx\: \bigg\{ \frac{\phi^p_\pi(x)}{x(1\!-\!x)}
    \!-\! \frac{\phi^p_\pi(0)}{x} \!-\!
   \frac{\phi^p_\pi(1)}{1\!-\!x} \bigg\} 
   \!+ \ln\Big(\frac{\bn\mcdot p_\pi}{\muminus}\Big) 
   \big[ \phi^p_\pi(0) \!+\! \phi^p_\pi(1) \big] 
 , \nonumber
\end{align}
and 
\begin{align} \label{Op4}
  & \hspace{-0.4cm}
   \int\!\! dx\, dy\  \frac{\phi_{\pi}(x,y)}{(x^2)_{\mbox{\o}}\, 
   (y^2)_{\mbox{\o} }} \
    \delta(1\!-\!x\!-\!y) \\
 & = \Big( \frac{\bn\mcdot
    p_\pi}{\muminus}\Big)^{2\epsilon}
   \sum_{x\ne 0,1} \int\!\! dx_r\,\frac{\hat \phi_{\pi}(x)}
    {x^2\, (1\!-\! x)^2} \:
    \theta(x)\theta(1\!-\!x)\ \big|x(1\!-\!x)\big|^\epsilon +\mbox{c.t.} \nn
 &  =   \int_0^1 \!\!\! dx\: \bigg\{ \frac{\hat \phi^p_\pi(x)}{x^2(1\!-\!x)^2}
    \!-\! \frac{\hat \phi'_{\pi}(0)}{x} \!+\!
   \frac{\hat \phi_{\pi}'(1)}{1\!-\!x} \bigg\} \nn
 & \ \ \ \ \ \!-\! \Big( \frac{\bn\mcdot
    p_\pi}{\muminus}\Big)^{2\epsilon} \bigg\{
  \int_1^\infty \!\!\!\! dx\,
   \frac{\hat \phi'_{\pi}(0) x^\epsilon (x\!-\!1)^\epsilon}{x} - 
  \int_{-\infty}^0 \!\!\!\! dx\: 
   \frac{\hat \phi'_{\pi}(1) x^\epsilon (1\!-\!x)^\epsilon}{1-x} \bigg\}
   +\mbox{c.t.} 
  \nn
 & =  \int_0^1 \!\!\! dx\, \bigg\{ \frac{\phi_{\pi}(x)}{x^2(1\!-\!x)^2}
    \!-\! \frac{\phi_{\pi}'(0)}{x} \!+\!
   \frac{\phi_{\pi}'(1)}{1\!-\!x} \bigg\} 
   \!+\! \ln\Big(\frac{\bn\mcdot p_\pi}{\muminus}\Big) 
   \big[ \phi'_{\pi}(0) \!-\! \phi'_{\pi}(1) \big] \!
  . \nonumber
\end{align}
In the last line the sign for $\phi'_\pi(1)$ appears because we
differentiate with respect to $x$ rather than $1- x$. The notation
$+\mbox{c.t.}$ indicates the matrix element of the counterterm
operators that cancel the $1/\epsilon_{\rm UV}$ divergences leaving
only the logarithm. The dependence of the results on $\mu$ was
suppressed, and terms with counterterm coefficients,
$C_{ct}^{[0]}(\mu_-)[\phi^p_\pi(0) \!+\!
\phi^p_\pi(1)]$ and $C_{ct}^{[1]}(\mu_-)[\phi'_{\pi}(0) \!-\!
\phi'_{\pi}(1)]$ should be added to these amplitudes~\cite{LMS}.
 
The final matrix elements in Eqs.~(\ref{Apif},\ref{Bpif},\ref{Op3},\ref{Op4})
have a linear $\ln(\muminus)$ dependence which comes from the action of the
\o-distribution. The coefficient of these logs is independent of the power taken
for the $x^\epsilon$ type factors, though the analysis in
section~\ref{sect:dimregII} dictates that the $\epsilon$--power should be used.
The $\ln(\muminus)$ dependence will be canceled order by order in
$\alpha_s(\mu)$ by $\ln(\muminus)$ dependence in the perturbative kernel
$J(\muminus)$ including the coefficients $C_{ct}(\mu_-)$ of the counterterm
operators.  Just as in our scalar loop example in
sections~\ref{sect:cutoffscalar} and~\ref{sect:dimregscalar}, when we consider
the resulting factorization formula at the matching scale the $\mu_+$ and
$\mu_-$ dependence will cancel out between logs in the coefficient functions and
those in the $\mu_\pm$ dependent hadronic distributions, where
$\mu^2=\mu_+\mu_-$.

\subsection{The $\rho$--$\pi$ Form Factor at Large $Q^2$}  \label{sect:rhopi}

In this section we consider $\gamma^*\rho \to \pi$ as an example of a process
with convolution integrals that appear to be divergent, but are tamed in our
formulation of \SCETb using the procedure in section~\ref{sect:kernel}.  The QCD dynamics
are described by the $\rho$--$\pi$ form factor
\begin{eqnarray} \label{ffrhopi}
  \langle \pi^+(p') | \bar q \gamma^\nu q | \rho^+(p,\varepsilon^\perp) \rangle 
  &=& i \epsilon^{\nu\alpha\beta\lambda} p_\alpha p'_\beta 
   \varepsilon^\perp_\lambda\ F_{\rho\pi}(q^2) \,,
\end{eqnarray}
with $\varepsilon^\perp$ the transverse rho polarization vector.  At large
$Q^2=-q^2$ the form factor $F_{\rho\pi}(q^2)$ was studied in detail in
Ref.~\cite{Chernyak:1983ej}, and also was discussed in~\cite{Lepage:1980fj}.
The pion form factor $F_{\pi\pi}(q^2)$ and leading proton form factor
$F_{pp}(q^2)$ both scale as $1/Q^2$ at large $Q^2$, however the form factor
$F_{\rho\pi}(q^2)$ scales as $1/Q^4$. This results from an additional $1/Q$
suppression of the QCD matrix element shown in Eq.~(\ref{ffrhopi}) relative to
the pion and proton cases.

In SCET, the two LO currents $J^{(0)}_j$ that mediate any $\gamma^*M_1 \to M_2$
transition with light quarks were derived in Ref.~\cite{Bauer:2002nz}. Using
isospin and working in the Breit frame with incoming momentum transfer
$q^\mu=Q\bn^\mu/2 - Qn^\mu/2$, the incoming partons are $n$-collinear and
outgoing partons are $\bn$-collinear. The LO matching of the QCD vector current
onto SCET is
\begin{align} \label{J12f}
 &  J^\nu \to \frac{n^\nu\!+\!\bn^\nu}{Q^3} 
  \int\!\! d\omega_j \Big[ C_1(\mu,\omega_j)
  J^{(0)}_1(\mu,\omega_j) + C_2(\mu,\omega_j)
  J^{(0)}_2(\mu,\omega_j) \Big]\,, \nn
 & J^{(0)}_j = \big[\bar \chi_{n,\omega_1} \Gamma_j
   \chi_{n,\omega_2}\big] \big[ \bar \chi_{\bn,\omega_3} \Gamma_j^\prime 
   \chi_{\bn,\omega_4} \big]   \,,
\end{align}
and involves $n$-collinear isodoublet $\chi_n$ and $\bn$-collinear isodoublet
$\chi_\bn$ fields in \SCETb.  Since the $\rho$ and $\pi$ have opposite charge
conjugation, the only relevant matrix from Ref.~\cite{Bauer:2002nz} for our
two isotriplet mesons is
\begin{eqnarray} \label{G12}
   \Gamma_2 \otimes \Gamma_2^\prime &=& \frac14 \Big[
   (Q_u+Q_d) (\tau^a \otimes \tau^a)    \big[ \bnslash\otimes \nslash
    + \bnslash\gamma_5\otimes \nslash\gamma^5 \big] \,.
\end{eqnarray} 
From Eq.~(\ref{G12}) we see that there is no $\gamma_\perp$, so at this order
the $\rho_\perp$--$\pi$ transition is forbidden. This is a reflection of the
helicity structure of the
factorization~\cite{Lepage:1980fj,Lepage:1980fj,Chernyak:1983ej} and occurs
despite the existence of matrix elements at leading power for the relevant
hadronic states:
\begin{align} \label{twist2}
 \Big\langle 0 \Big| \bar d_{n,-y}\,  \,
    {\bnslash\gamma_\perp^\mu} 
   \, 
  u_{n,x} \Big| \rho_n^+(p,\varepsilon_\perp) \Big\rangle 
 &=  { f_\rho^T}\: \bn\mcdot p\, \varepsilon_\perp^\mu\,
   \delta(1\!-\!x\!-\!y)\: \phi_{\rho_\perp}\!(\mu,x,y) \,, \\
 \Big\langle \pi_\bn^+(p') \Big| \bar u_{\bn,u} {\nslash\gamma_5}
   \, 
  d_{\bn,-v} \Big| 0 \Big\rangle 
 &= {-i f_\pi}\: n\mcdot p'\: 
    \delta(1\!-\!u\!-\!v)\: \phi_\pi(\mu,u,v) \,. \nonumber
\end{align}
These twist-2 matrix elements will be useful below. The notation $u_{n,x}$
denotes a gauge invariant product of a Wilson line and quark with momentum
fraction $x$ along the quark arrow, $u_{n,x}=\delta(x-\bnP/\bn\cdot p) W^\dagger
\xi_n^{(u)}$.  Similarly, $\bar d_{n,-y} = \bar\xi_n^{(d)} W
\delta(y+\bnP^\dagger/\bn\cdot p)$ etc.  Finally, the $\delta(1\!-\!x\!-\!y)$
factors on the r.h.s.\ of Eq.~(\ref{twist2}) comes from momentum conservation in
the matrix element. When implementing the \o-distribution we start with
expressions containing $\delta$-functions, like those shown in
Eq.~(\ref{twist2}). This is useful for cases where there are simultaneous
zero-bin subtractions in variables appearing in the $\delta$-function.

Once operators suppressed by $1/Q$ are considered, the $\rho_\perp^+\to \pi^+$
transition is allowed~\cite{Chernyak:1983ej}. In \SCETb the relevant operators
can be constructed from products of the bilinear ${\cal O}(\eta^3)$ operators
from Ref.~\cite{Hardmeier:2003ig}
\begin{align} \label{pirhoOPS}
   P_\pm (u,v) &= \Big[\bar u_{\bn}\,  \frac{\nslash}{2} 
     (i \slash\!\!\!\! {\cal D}_\bn^{\perp} )^\dagger\Big]_{u} \,
      \frac{\gamma_5}{\cP^\dagger} \, d_{\bn,-v} 
    \pm \bar u_{\bn,u} \, \frac{\gamma_5}{\cP}\, \Big[
    (i \slash\!\!\!\!{\cal D}_\bn^{\perp} )\frac{\nslash}{2}  \,  d_\bn
    \Big]_{-v} 
  \,,
   \\[4pt]
   V_{1\pm}^\mu(x,y) &=  \Big[\bar d_n\,  \frac{\bnslash}{2} 
     (i \slash\!\!\!\! {\cal D}_\bn^{\perp} )^\dagger\Big]_{-y} \,
    \frac{\gamma^\mu}{\bnP^\dagger} \, u_{n,x} 
    \pm \bar d_{n,-y} \, \frac{\gamma^\mu}{\bnP}\, \Big[ 
    (i \slash\!\!\!\!{\cal D}_\bn^{\perp} )\frac{\bnslash}{2}  \,  u_n 
    \Big]_{x} 
    \,,
  \nonumber \\[4pt]
   V_{2\pm}^\mu(x,y) &= \Big[ \bar d_n\, \frac{\bnslash}{2} 
    (i {\cal D}_\bn^{\perp\mu} )^\dagger \Big]_{-y}
   \frac{1}{\bnP^\dagger} \, u_{n,x} \pm   \bar d_{n,-y} \,
   \frac{1}{\bnP}\, \Big[ (i {\cal D}_\bn^{\perp\mu}
   )\frac{\bnslash}{2} \, d_n \Big]_{x}
   \,, \nonumber \\[5pt]
     T_3^{\mu\nu}(u,v,w) &=  \bar u_{\bn,u}
     \frac{\nslash \gamma_\perp^\mu}{2}
    \, (ig{\cal B}_{\bn\perp}^\nu)_{-w} \,  
    d_{\bn,-v}
   \,, \nonumber \\[5pt]
   V_3^{\mu}(x,y,z) &=   \bar d_{n,-y} \frac{\bnslash}{2}
    \, (ig{\cal B}_{n\perp}^\mu)_{z} \,
    u_{n,x}
   \,, \nonumber \\[5pt]
    A_3^{\mu}(x,y,z) &=    \bar d_{n,-y} \frac{\bnslash\gamma_5}{2}
    \, (ig{\cal B}_{n\perp}^\mu)_{z} 
    \, u_{n,x}
   \,, 
\end{align}
where $\bnP=\bn_\alpha \cP^\alpha$, $\cP = n_\alpha \cP^\alpha$, $ig {\cal
  B}_{n\perp}^\mu = \big[ 1/\bnP\, W_n^\dagger [i\bn\mcdot D_n, iD_{n\perp}^\mu
] W_n \big]$, $i{\cal D}_n^{\perp\mu} = {\cP}_{\perp}^\mu + ig {\cal
  B}_{n\perp}^\mu$, and $(ig{\cal B}_{n\perp}^\mu)_{z}= \delta(z-\bnP/n\cdot
p)ig{\cal B}_{n\perp}^\mu $.  As indicated, in this section we use a rescaling
with respect to the momentum carried by the state in order to make the delta
functions acting on the fields dimensionless. This rescaling hides the process
independence of the operators, but makes the results simpler to present.  Note
that we have used slightly different notation for the operators than
Ref.~\cite{Hardmeier:2003ig}, with relations $P_+(u,v) ={\cal P}^{\rm
  LPW}(\vec\omega)$, $P_-(u,v)=\tilde {\cal P}^{\rm LPW}(\vec\omega)$, etc. We
also included one less power of $1/\bnP$ in the three-body operators.

For the two-body operators we have matrix elements~\cite{Hardmeier:2003ig}
\begin{align} \label{2bdy}
 \big\langle \pi_\bn^+ \big| P_+(u,v) \big| 0 \big\rangle 
 &=  {-i f_\pi \mu_\pi}\: 
    \delta(1\!-\!u\!-\!v)\: \phi_\pi^p(\mu,u,v)  \,, 
  \nonumber\\[5pt]
 \big\langle \pi_\bn^+ \big| P_-(u,v) \big| 0 \big\rangle 
 &=  \frac{- i f_\pi \mu_\pi}{6}\: 
    \delta(1\!-\!u\!-\!v)\: \phi_\pi^{\prime\,\sigma}(\mu,u,v) \,, 
   \nonumber\\[5pt]
  \big\langle 0 \big| V_{1+}^\mu(x,y) \big| \rho^+_{n\perp}(\varepsilon)
   \big\rangle 
 &=  {f_\rho m_\rho  \varepsilon_\perp^\mu}\,
    \delta(1\!-\!x\!-\!y)\: g_{\rho_\perp}^{(v)}(\mu,x,y) \,, 
  \nonumber\\[5pt]
 \big\langle 0\big| V_{1-}^\mu(x,y) \big|\rho^+_{n\perp}(\varepsilon)
   \big\rangle 
 &=  \frac{f_\rho m_\rho  \varepsilon_\perp^\mu}{4}\,
    \delta(1\!-\!x\!-\!y)\: g_{\rho_\perp}^{(a)\,\prime}(\mu,x,y) \nn
 &=  \frac{f_\rho m_\rho  \varepsilon_\perp^\mu}{2(y\!-\! x)}\,
    \delta(1\!-\!x\!-\!y)\: g_{\rho_\perp}^{(A)}(\mu,x,y)
   \,, 
\end{align}
where $\mu_\pi = m_\pi^2/(m_u+m_d)$ and for later convenience we switch from the
distribution $g_{\rho_\perp}^{(a)\,\prime}$ to a distribution
$g_{\rho_\perp}^{(A)}$.  Other matrix elements are related to these by simple
operator relations in SCET.  
\begin{figure}[!t]
\begin{center}
 \includegraphics[width=16cm]{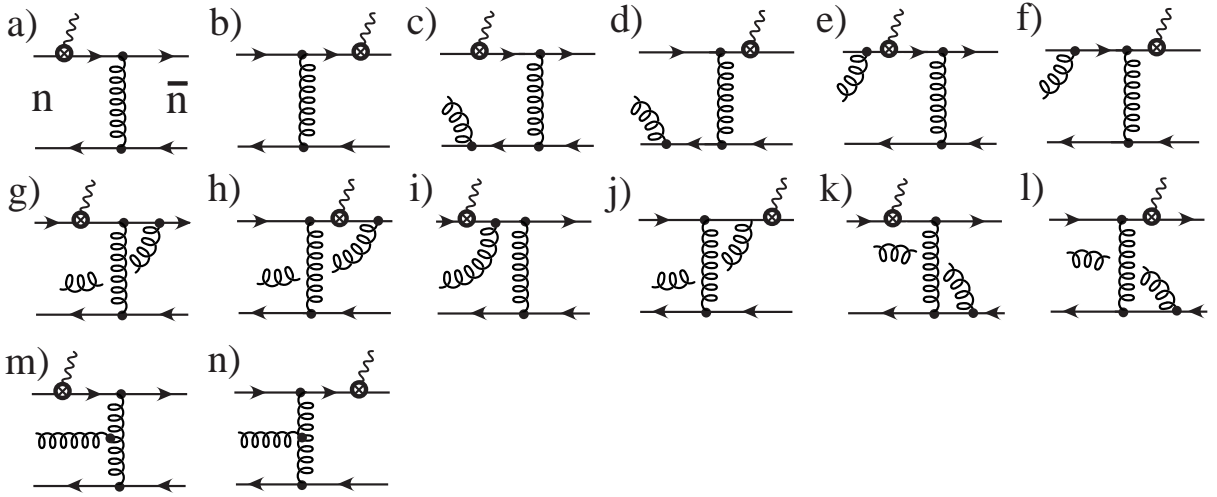}
\end{center}
 \vskip -.5cm 
\caption[1]{  
  Graphs for matching onto the $\rho$--$\pi$ electromagnetic form factor. To
  obtain the full set of diagrams one must add graphs with the gluon exiting to
  the right (the left-right mirror images of c) through n) not flipping the
  direction of the arrows). Then to this entire set one must add the graphs with
  the current insertion on the other quark line (which can be obtained by
  charge conjugation).
}\label{fig:rhopi} 
\end{figure}
For the three-body operators the matrix elements
are~\cite{Hardmeier:2003ig}
\begin{align} \label{3bdy}
 \big\langle \pi_\bn^+(p') \big| T_3^{\mu\nu}(u,v,w) \big| 0 \big\rangle 
 &=  n\mcdot p'\: f_{3\pi}\, \epsilon_\perp^{\mu\nu}\:
    \frac{\delta(1\!-\!u\!-\!v\!-\!w)}{2w} \,
    \: {\phi_{3\pi}(\mu,u,v,w)} \,, 
    \\[5pt]
  \big\langle 0 \big| V_3^\mu(x,y,z) \big| \rho^+_{n\perp}(p,\varepsilon) 
   \big\rangle 
 &= - \bn\mcdot p\: f_{3\rho}^{V}  \varepsilon_\perp^{\mu}\: 
    \frac{\delta(1\!-\!x\!-\!y\!-\!z)}{2z} \: 
    \, 
      \phi_{3\rho}^{V}(\mu,x,y,z) 
   \,, 
    \nonumber \\[5pt]
  \big\langle 0 \big| A_3^\mu(x,y,z) \big| \rho^+_{n\perp}(p,\varepsilon) 
    \big\rangle 
 &=  i\, \bn\mcdot p\: f_{3\rho}^{A}  \epsilon^{\mu\nu}_\perp 
   \varepsilon^\perp_{\nu}\:
   \frac{\delta(1\!-\!x\!-\!y\!-\!z)}{2z} \,\: 
  \phi_{3\rho}^{A}(\mu,x,y,z) 
    \,, \nonumber
\end{align}
where $\epsilon^{\mu\nu}_\perp = \epsilon^{\mu\nu\alpha\beta}\bn_\alpha
n_\beta/2$ (which switches sign under $n\leftrightarrow \bn$).  In comparing to
Ref.~\cite{Hardmeier:2003ig} note that we took $(f_V m_V {\cal V})^{\rm LPW} =
-f_\rho^{3V} \phi_\rho^{3V}$ and $(f_V m_V {\cal A})^{\rm LPW} = f_{3\rho}^{A}
\phi_{3\rho}^{A}$ which agrees with the notation in Ref.~\cite{Chernyak:1983ej}.
To compare with the other notation in Ref.~\cite{Chernyak:1983ej}, note that
$(\varphi_\pi^A)^{\rm CZ} = \phi_\pi/2$, $(\varphi_\pi^p)^{\rm CZ}=\phi_\pi^p/2$,
$(\varphi_{3\pi})^{\rm CZ}=\phi_{3\pi}$, $(\varphi_\rho^T)^{\rm
  CZ}=\phi_{\rho_\perp}/2$, $(\varphi_{\rho}^{V,\perp})^{\rm CZ}=g_{\rho_\perp}^{(v)}/2$,
$(\varphi_{\rho}^{A,\perp})^{\rm CZ}=(1\!-\!2z) g_{\rho_\perp}^{(a)\,\prime}(z)/4 =
g_{\rho_\perp}^{(A)}(z)/2$.

To connect the operators in Eq.~(\ref{pirhoOPS}) to the process $\gamma^*
\rho_\perp^+ \to \pi^+$ we must match the full theory diagrams shown in
Fig.~\ref{fig:rhopi} onto \hbox{\SCETb\!.} The graphs in Fig.~\ref{fig:rhopi}a-b
must be expanded to NLO. The graphs in Fig.~\ref{fig:rhopi}c-f can be obtained
from graphs a) and b) by using the tree level relation between the QCD and
\SCETb fields in place of the lowest order spinor in the LO part of these
diagrams:
\begin{align}
  \psi = W \Big[ 1 + \frac{1}{\bnP} {i\, \slash\!\!\!\! {\cal D}_{n\perp} }
  \frac{\bnslash}{2} \Big] (W^\dagger \xi_n) \,.
\end{align}
Graphs g) through n) require separate computations.

In matching these graphs onto \SCETb we can make different assumptions for the
scaling of the external lines.  We work in the Breit frame and take the
interpolating field for the incoming $\rho$ and outgoing $\pi$ to be built
purely out of $n$ and $\bn$ collinear fields respectively.  The contribution
that matches onto the operators in Eqs.~(\ref{twist2},\ref{pirhoOPS}) will have
two $n$-collinear quark fields and two $\bn$-collinear quark fields with or
without an extra $n$ or $\bn$ collinear gluon. In addition \SCETb has graphs
where one or more of the above fields simultaneously become soft. Although we
can formulate operators with soft fields in \SCETb, they do not contribute to
the $\rho_\perp$--$\pi$ form factor in this frame.  The zero-bin subtractions
ensures that we will not double count the region of momentum space that these
other operators correctly describe. Following Ref.~\cite{Bauer:2002aj} we note
that we do not need to consider interpolating fields for hadrons built out of
mixed soft and collinear components. These interpolating fields do not have
non-perturbative poles as discussed earlier.  Furthermore, in the Breit frame,
an interpolating field that is purely soft would only be needed for a different
physical process than the one we are considering (and would correspondingly
require different current operators).

To simplify the presentation we define
\begin{align}
  \delta_{xy} = \delta(1\!-\!x\!-\!y) \,, \qquad
   \delta_{xyz} = \delta(1\!-\!x\!-\!y\!-\!z) \,.
\end{align}
Using the computation of the tree level graphs done in
Ref.~\cite{Chernyak:1983ej}, but including the $\sum_{p^\pm\ne0}$ terms with
\o-distributions as described in the previous section, gives\footnote{Note that
  we have not independently verified the calculations in
  Ref.~\cite{Chernyak:1983ej}.}
\begin{align} \label{rhopifinal}
  F_{\rho\pi}(Q^2) & = \frac{4\pi\alpha_s(\mu)}{27 Q^4}
   \int\!\! dx\! \int\!\! dy\! \int\!\! dz
  \! \int\!\! du\! \int\!\! dv\! \int\!\! dw\,
  \Bigg\{ 
   4f_\rho^T f_\pi \mu_\pi \: 
     \frac{\delta_{xy}\delta_{uv}\,\phi_{\rho_\perp}(x,y) \phi_\pi^p(u,v)}
         {(y^2)_{\mbox{\o}}v_{\mbox{\o}} } 
    \\[5pt]
 & \hspace{-1.3cm}
   +\! f_\rho^V m_\rho f_\pi \, \delta_{xy}\delta_{uv}\, \bigg[ 
    \frac{g^{(v)}_{\rho_\perp}(x,y) \phi_\pi(u,v)}
         {x_{\mbox{\o}}y_{\mbox{\o}}(v^2)_{\mbox{\o}} } 
   \!+\!
      \frac{g^{(A)}_{\rho_\perp}(x,y) \phi_\pi(u,v)}
  {4x_{\mbox{\o}}y_{\mbox{\o}}(u^2)_{\mbox{\o}}(v^2)_{\mbox{\o}}}
   \bigg]  
  \!+\! \frac{f_{\rho}^{3A} f_\pi}{4} \delta_{uv}\delta_{xyz} 
     \: {\phi_\pi(u,v)\phi_{3\rho}(x,y,z)}
   \nonumber\\[4pt]
 & \hspace{-1.3cm}
   \times \bigg[
    \frac{8}{(\bar y^2 x)_{\mbox{\o}} v_{\mbox{\o}} } 
   \!+\! \frac{2}{  (\bar z z y)_{\mbox{\o}} v_{\mbox{\o}} }
   \!-\! \frac{9}{  (\bar y^2 x)_{\mbox{\o}} (v^2)_{\mbox{\o}}  }
   \!-\! \frac{1}{(\bar z z x)_{\mbox{\o}}(v^2)_{\mbox{\o}} }
   \!-\!  \frac{1}{ (z\, \bar y^2)_{\mbox{\o}}(v^2)_{\mbox{\o}}  }
     \bigg]
   \nonumber\\[5pt]
 & \hspace{-1.3cm}
   - 
  f_{\rho}^T f_{3\pi}  \: 
  \phi_{3\pi}(u,v,w)\phi_{\rho_\perp}\!(x,y)\, \delta_{uvw}\delta_{xy} 
  \bigg[
  \frac{9}{2 (\bar u^2  v)_{\mbox{\o}} (y^2)_{\mbox{\o}}  }
  \!+\! \frac{1}{2(\bar u^2 w)_{\mbox{\o}} (y^2)_{\mbox{\o}}   }
  \!+\! \frac{1}{(\bar u  v w)_{\mbox{\o}}  y_{\mbox{\o}}
     } \bigg] \nn
 & \hspace{-1.3cm} + \mbox{D-terms}\Bigg\} \,,  \nonumber
\end{align}
where $\phi_{3\rho}=\phi_{\rho}^{3A}- f_{\rho}^{3V}/f_\rho^{3A}
\phi_{\rho}^{3V}$. The ``+ D-terms'' factor indicates that at this order we must
also include the renormalized coefficients $D_i(\mu_+,\mu_-)$ just like the
coefficient $C_{ct}^{[1]}(\mu_-)$ in Eq.~(\ref{ct2}). Here they multiply terms with
$\phi_{\rho_\perp}^\prime(1)$, $g_{\rho_\perp}^{(A)}(0)$, $g_{\rho_\perp
}^{(A)}(1)$, $g_{\rho_\perp }^{(v)}(0)$, $g_{\rho_\perp }^{(v)}(1)$,
$\phi_{\pi}^\prime(0)$, $\phi_{\pi}^\prime(1)$, and $\phi_{\pi}^p(1)$. The
$\mu$-dependence of the distributions is suppressed for brevity. The range for
the integrations are determined by the theta functions in the non-perturbative
distributions, which have support from $[0,1]$ in their respective momentum
fraction variables. Any variable denoted with a bar is one minus itself, $\bar
x=1-x$, etc.  For the three-body distributions we will have two convolution
integrals left after using the $\delta$-functions, and the \o-distribution must
in general be treated as two-dimensional. We indicated this in
Eq.~(\ref{rhopifinal}) by having the \o\ subscript act on the product of
three-body momentum fractions. The action of \o\ in these cases can be
determined by the same steps used in section~\ref{sect:kernel}. Due to the
\o-distributions the result in Eq.~(\ref{rhopifinal}) is finite, independent of
assumptions about the non-perturbative distribution functions.

It is possible to study the $\mu_\pm$ dependence of the result in
Eq.~(\ref{rhopifinal}).  To do so one adopts some endpoint behavior for the
distributions, and can make the action of the \o-distributions from the tree
level jet functions explicit. A common assumption for the scaling behavior of
the above distribution functions near their endpoints is $\phi_\pi(x) \sim x
\bar x$, $\phi_{\rho_\perp}(u)\sim u\bar u$, $g_{\rho_\perp}^{(v)}(x)\sim 1$,
$g_{\rho_\perp}^{(a)}(x)\sim x\bar x$ (so $g_{\rho_\perp}^{(A)}(x)\sim 1$),
$\phi_\pi^p(u)\sim 1$, $\phi_{3\rho}(x,y)\sim x y z^2$, $\phi_{3\pi}(u,v)\sim u
v w^2$. With this scaling behavior all integrals over three body distributions
converge without zero-bin subtractions, and we can evaluate the two-body
\o-distributions explicitly using the formulas worked out in
section~\ref{sect:kernel}. For the rapidity logarithms in the distributions we
then get $\ln(\bn\cdot p_\rho/\muminus) =\ln(Q/\muminus)$ and $\ln(n\cdot
p_\pi/\muplus)=\ln(Q/\muplus)$. The $\mu_+$ and $\mu_-$ dependence in these
logarithms is canceled by the $\mu_\pm$ dependence of the $D_i(\mu_\pm)$ Wilson
coefficients.

In this section we showed that the \SCETb zero-bin subtractions together with UV
renormalization yield a finite answer for the $\rho$--$\pi$ form factor at large
$Q^2$, given in Eq.~(\ref{rhopifinal}). Due to the separation in rapidity, the
result has additional dependence on $\ln(Q)$ beyond that in the hard scattering
kernel. The appearance of these logarithms is controlled by the powers of
momentum fractions in the hard scattering kernel and the endpoint behavior of
the distributions. At the matching scale $\muplus\gtrsim \Lambda_{\rm QCD}$ and
$\muminus\sim Q$ and there are no large logarithms.  As we scale $\mu^2$ towards
$\Lambda_{\rm QCD}^2$ keeping $\mu^2=\muplus\muminus$, large logarithms may be
generated. The dependence on this large log is computable, up to its
normalization which is fixed by non-perturbative parameters. These parameters
are determined once we adopt a model for the light-cone distribution functions.


\subsection{Factorization of $\zeta^{B\pi}(E)$ appearing in $B\to
  \pi\ell\bar\nu$ and $B\to \pi\pi$}   \label{sect:Bpi}
  
In this section, we study the implications of our results for $B\to
\pi\ell\bar\nu$ and the related process $B\to\pi\pi$. For small pion energy $E$,
this process is dominated by the $B^*$ pole and can be studied in chiral
perturbation theory~\cite{Wise:1992hn,Burdman:1992gh}. For $E \lesssim 1\,{\rm
  GeV}$ this process is also amenable to HQET and lattice QCD
simulations~\cite{Gulez:2006dt,Okamoto:2004xg}. For large $E$, namely $E^2 \gg
E\Lambda_{\rm QCD} \gg\Lambda^2_{\rm QCD}$, the process is expected to factor
into the convolution of perturbative and non-perturbative contributions.

The exploration of factorization in QCD for the process $B\to \pi\ell\bar\nu$
has a rich history.  In Ref.~\cite{Szczepaniak:1990dt}, exclusive QCD
factorization techniques were applied to the $B\to \pi$ and $B \to \pi\pi$
transitions to examine the hard scattering contributions.  Endpoint divergences
as in Eq.~(\ref{0.1}) were encountered.  Ref.~\cite{Burdman:1992hg} argued that
the soft endpoint contributions dominate, and obey heavy quark symmetry
relations.  This was extended to a bigger class of large energy symmetry
relations in Ref.~\cite{Charles:1998dr} in the context of an EFT known as
LEET~\cite{Dugan:1990de} that was later shown to be inconsistent. Factorization
of soft and collinear regions was studied in Ref.~\cite{Akhoury:1993uw},
including Sudakov suppression of soft contributions, and a result for the hard
parts was obtained in terms of leading twist $B$ and $\pi$ distributions.  $B\to
\pi\ell\bar\nu$ decays were studied with light-cone sum rules in
Refs.~\cite{Belyaev:1993wp,Khodjamirian:1997ub,Bagan:1997bp}, and in
Ref.~\cite{Bagan:1997bp}, a definition of the soft and hard contributions was
given using the so-called local duality approximation. The soft part dominated
numerically. In Ref.~\cite{Beneke:2000wa}, it was shown that terms with endpoint
divergences could be absorbed into a soft form factor without spoiling the large
energy symmetry relations, and $\alpha_s$ contributions which spoil the
relations were evaluated. A dominant non-factorizable soft form factor is an
important ingredient in the power counting used by BBNS in discussing
factorization for $B\to\pi\pi$~\cite{Beneke:1999br}.\footnote{A soft form factor
  does appear in the $B\to D\pi$ process~\cite{Politzer:1991au,Bauer:2001cu}.}
In formulating SCET in Ref.~\cite{Bauer:2000yr}, the leading order low energy
current operators for the soft $B\to \pi$ form factor were derived, and their
hard Wilson coefficients were computed at ${\cal O}(\alpha_s)$.
Ref.~\cite{Kurimoto:2001zj} studied $B\to\pi$ form factors with threshold and
$k_\perp$ resummations, argued that soft contributions are Sudakov suppressed in
this framework, pointed out the importance of the $p^2\sim m_b\Lambda_{\rm QCD}$
scale, and obtained a factorized result given by individual $B$ and $\pi$
wavefunctions with $k_\perp$'s. The ability to completely factorize the
amplitude into individual $B$ and $\pi$ distributions depending on $k_\perp$ is
an important ingredient in the pQCD approach to $B\to\pi\pi$~\cite{Keum:2000ph}.

More recently, \SCETa and \SCETb have been used to analyze $B\to\pi\ell\bar\nu$.
Refs.~\cite{Bauer:2002aj,Pirjol:2002km} pointed out that the interpolating field
for the pion is purely collinear, and that the LO terms therefore necessarily
involve time-ordered products of currents and subleading Lagrangians in \SCETa.
Thus, in the complete set of operator contributions, leading and subleading
currents both contribute to the LO form factor. The hard-scattering and soft
form factor contributions can be defined as subsets of these time-ordered
products with $J^{(1)}$ and $J^{(0)}$, and it was shown that the factorization
of the hard-scattering contributions $J^{(1)}$ involve a hard function for the
scale $m_b^2$, and also a distinct jet function for the scale $m_b\Lambda_{\rm
  QCD}$.  In Refs.~\cite{Becher:2003qh,Becher:2003kh}, factorization in \SCETb
was further investigated, and it was argued that soft-collinear messenger
contributions spoil the possibility of further factorization of the soft $B\to
\pi$ form factor $J^{(0)}$ terms.  In Ref.~\cite{Beneke:2003pa}, the
implications of endpoint singularities for factorization in SCET were
investigated.  It was shown that the hard-scattering contributions are finite to
all orders in $\alpha_s$. It was also argued that endpoint singularities spoil
possible factorization of the soft form factor terms below the scale
$m_b\Lambda_{\rm QCD}$. In Ref.~\cite{Lange:2003pk}, the complete set of \SCETb
operators for the soft $B\to \pi$ form factor contribution were determined,
including three body operators. In Refs.~\cite{Bauer:2004tj} a factorization
formula for $B\to\pi\pi$ was investigated in SCET and it was pointed out that
the same universal jet function occurs as in the $B\to\pi\ell\bar\nu$ process
(for a precursor see~\cite{Chay:2003ju}).

The above results can be summarized by the following formula for the relevant
$B\to\pi$ form factor, derived from a systematic expansion in $E\gg\Lambda_{\rm
  QCD}$:
\begin{align} \label{fplus}
 f_+(E)  &=  T^{(+)}(E)\, \zeta^{B\pi}(E)
  +  \int_0^1\!\! dz\  C_J^{(+)}(z,E)\: \zeta_J^{B\pi}(z,E)
 \\
&=  T^{(+)}(E)\, \zeta^{B\pi}(E)
  + N_0 \int_0^1\!\!\!\! dz\! \int_0^1\!\!\!\! dx\! 
   \int_0^\infty\!\!\!\!\! dk_+ \,
   C_J^{(+)}(z,E) J(z,x,k_+,E) \phi_{\pi}(x) \phi_B^+(k_+)
   \,.   \nonumber
\end{align}
Our notation follows Refs.~\cite{Pirjol:2002km,Bauer:2004tj}, so $N_0=f_B f_\pi
m_B/(4E^2)$. The hard functions $T^{(+)}$ and $C_J^{(+)}$ are perturbative at
$\mu^2\sim m_b^2$, the jet function $J$ is perturbative at $\mu^2\sim
E\Lambda_{\rm QCD}$, and $\phi_\pi$ and $\phi_B^+$ are leading twist
distributions, which are non-perturbative. Finally, $\zeta^{B\pi}(E) =
(m_b^{1/2}/E^2) \hat \zeta^{B\pi}(E)$ is the unfactorized soft form factor with
$\hat \zeta^{B\pi}(E)$ containing dependence on the scales $E\Lambda_{\rm QCD}$
and $\Lambda_{\rm QCD}^2$.  At lowest order in $\alpha_s(m_b)$ at the hard scale
$T^{(+)}=1$, $C_J^{(+)}=1$, and $f_+(E) = \zeta^{B\pi}(E) + \zeta_J^{B\pi}(E)$,
where $\zeta_J^{B\pi}(E) = \int dz\, \zeta_J^{B\pi}(z,E)$. If we also work to
order $\alpha_s(\mu)$ at the intermediate jet scale, $\mu^2\sim E\Lambda_{\rm
  QCD}$, we can expand further, and Eq.~(\ref{fplus}) gives
\begin{align} \label{f+tree}
  f_+(E) = \zeta^{B\pi}(E) + \frac{f_B f_\pi m_B}{4E^2}\
    \frac{4\pi\alpha_s(\mu)}{9} \Big( \frac{2E}{m_B}+\frac{2E}{m_b} -1 \Big)
    \int_0^1\!\!\! dx \: \frac{\phi_\pi(x)}{x}\: 
    \int_0^\infty\!\!\! dk^+\: \frac{\phi_B(k^+)}{k^+} \,.
\end{align}
In this result the $\zeta^{B\pi}(E)$ term is left unexpanded since its
factorization at scales $\mu^2 \lesssim E\Lambda_{\rm QCD}$ is (so far) unknown.
In the remainder of this section we will use the zero-bin in \SCETb to
demonstrate that $\zeta^{B\pi}(E)$ can be factorized further into products of
twist-two and twist-three $\pi$ and $B$ distribution functions.

In \SCETa the $\zeta^{B\pi}(E)$ term is defined by 
\begin{align}
  \big\langle \pi_n(p_\pi^-) \big| T_0+T_3+T_4+T_5+T_6 \big| B_v \rangle 
   &= \bn\mcdot p_\pi \, C(p_\pi^-)\: \zeta^{B\pi}(p_\pi^-)\nn
   & = 2 E\: C(E)\: \zeta^{B\pi}(E)\,.
\end{align}
The $T_i$ are time-ordered products of subleading Lagrangians with the leading
scalar heavy-to-light current $J^{(0)}= \sum_{p^-} C(p^-)(\bar\xi_n W)_{p^-}
(Y^\dagger h_v)$. Their definitions are~\cite{Bauer:2004tj,Pirjol:2002km}
\begin{align}
  T_0 &=  i \int\!\! d^4y\: T J^{(0)}(0)\: {\cal L}_{\xi q}^{(1)}(y) \,,
 & T_3 &=  i \int\!\! d^4y\: T J^{(0)}(0)\: {\cal L}_{\xi q}^{(2b)}(y) \,,
  \nn   
  T_4 &=  i \int\!\! d^4y\: T J^{(0)}(0)\: {\cal L}_{\xi q}^{(2a)}(y) \,,
 & T_5 &= i^2 \int\!\! d^4y d^4z\: T J^{(0)}(0)\: {\cal L}_{\xi \xi}^{(1)}(y)
    {\cal L}_{\xi \xi}^{(1)}(z), 
  \nn
  T_6 &= i^2 \int\!\! d^4y d^4z\: T J^{(0)}(0)\: {\cal L}_{\xi q}^{(1)}(y)
    {\cal L}_{cg}^{(1)}(z) \,,
\end{align}
and the presence of only $J^{(0)}$ guarantees that $\zeta^{B\pi}(E)$ satisfies
the symmetry relations of Ref.~\cite{Charles:1998dr}.  The momentum conserving
$\delta$-function collapse the matrix element of the $T_i$'s into a simple
product. The momentum $p_\pi^-\sim m_b$ is large, and $p_\pi^- =2 E$ up to small
power suppressed $m_\pi$ dependent terms.

The tree level matching onto \SCETb for $J^{(0)}$ currents was carried out in
Ref.~\cite{Lange:2003pk}.  They found the position space operators
\begin{align}
  O_1^{(P)} &= g^2 \Big[ \bar\chi(0) \frac{\bnslash\gamma_5}{2} \chi(s\bn) \Big] 
  \Big[ \bar {\cal Q}_s(tn)\frac{\bnslash\nslash\gamma_5}{4} {\cal H}(0) \Big]
   \,,\nn
  O_2^{(P)} &= g^2 \Big[ \bar\chi(0) \frac{\bnslash\gamma_5}{2}
  i\slash\!\!\!\partial_\perp  \chi(s\bn) \Big] 
  \Big[ \bar {\cal Q}_s(tn)\frac{\nslash\gamma_5}{2} {\cal H}(0) \Big]
    \,,\nn
  O_3^{(P)} &= g^2 \Big[ \bar\chi(0) \frac{\bnslash\gamma_5}{2}
  \: \slash\!\!\!\!\!{\cal A}_{c\perp}\!(r\bn)  \chi(s\bn) \Big] 
  \Big[ \bar {\cal Q}_s(tn)\frac{\nslash\gamma_5}{2} {\cal H}(0) \Big]
    \,,\nn
  O_4^{(P)} &=  g^2 \Big[ \bar\chi(0) \frac{\bnslash\gamma_5}{2}
    \chi(s\bn) \Big] 
  \Big[ \bar {\cal Q}_s(tn) \: \slash\!\!\!\!\!{\cal A}_{s\perp}\!(u n)
   \frac{\nslash\gamma_5}{2} {\cal H}(0) \Big] \,,
\end{align}
where the notation is defined in Ref.~\cite{Lange:2003pk}. The time-ordered
product $T_0$ contributes to the matching onto operators with a gluon field
strength.  They also had a fifth operator, $O_5$, involving a time-ordered
product with soft-collinear messenger fields. In our setup the infrared regions
associated to the messenger fields are covered by the soft and collinear fields
so this fifth operator should not be included.  With labeled fields the operators
are
\begin{align} \label{Oscet2}
  O_1 &= \Big[ (\bar\xi_n W)_{u} \frac{\bnslash\gamma_5}{2} (W^\dagger
  \xi_n)_{-v} \Big] \Big[ (\bar q_s S)_{-k_1} \frac{\bnslash\nslash\gamma_5}{4}
  (S^\dagger h_v)_{k_2} \Big] \,,\\
  O_2 & = \Big[ (\bar\xi_n W)_{u} \frac{\bnslash\gamma_5}{2} 
  \:\slash\!\!\!\!\cP_\perp (W^\dagger \xi_n)_{-v} \Big] 
  \Big[ (\bar q_s S)_{-k_1} \frac{\nslash\gamma_5}{2}
  (S^\dagger h_v)_{k_2} \Big] \,, \nonumber \\
  O_3 &= \Big[ (\bar\xi_n W)_{u} \frac{\bnslash\gamma_5}{2} 
   (ig\,\slash\!\!\!\! {\cal B}_{n}^\perp)_{-w}
    (W^\dagger \xi_n)_{-v}
  \Big]  \Big[ (\bar q_s S)_{-k_1} \frac{\nslash\gamma_5}{2} (S^\dagger h_v)_{k_2}
  \Big] \,,\nn
  O_4 &= \Big[ (\bar\xi_n W)_{u} \frac{\bnslash\gamma_5}{2}
   (W^\dagger \xi_n)_{-v}
  \Big]  \Big[ (\bar q_s S)_{-k_1} (i g\, \slash\!\!\!\! {\cal B}_{s}^\perp)_{k_3}
  \frac{\nslash\gamma_5}{2} 
   (S^\dagger h_v)_{k_2} \Big] \,. \nonumber
\end{align}
Tree level matching gives the Wilson coefficients~\cite{Lange:2003pk}
\begin{align}\label{J134}
  J_1^\pi &= -\frac{4\pi\alpha_s(\mu)}{9E^2} \bigg[ \frac{1}{v^2 k_1} +
  \frac{1}{v\,k_1} \bigg] \,, \\
  J_2^\pi &= - \frac{4\pi\alpha_s(\mu)}{9E^2}\ \frac{1}{u\, v^2\, k_1^2}
   \,,\nn
  J_3^\pi &= \frac{\pi\alpha_s(\mu)}{2E^2} \bigg[ 
  \frac{w}{(v\!+\!w)^2\,  v\, k_1^2} - \frac{7}{9(v\!+\!w)^2\, k_1^2} 
   - \frac{8}{9\, v\, (u\!+\!w)\, k_1^2} \bigg] \,,\nn
  J_4^\pi &= \frac{\pi\alpha_s(\mu)}{2E^2} \bigg[ 
  \frac{k_3}{v^2\, (k_1\!+\!k_3)^2\,  k_1} 
   - \frac{8k_3}{9\,v\,(k_1\!+\!k_3)^2\, k_1} 
   + \frac{1}{9\, v^2\, (k_1\!+\!k_3)^2} \bigg] \,.\nonumber
\end{align}
In Eq.~(\ref{Oscet2}), the $(ig {\cal B}_n^\perp)$ was defined below
Eq.~(\ref{pirhoOPS}), and $ig {\cal B}_{s\perp}^\mu = \big[1/\cP\, S^\dagger [i
n\mcdot D_s, iD_{s\perp}^\mu ] S \big]$ with $S[n\mcdot A_s]$ soft Wilson lines.
As in the previous section, we rescaled the delta functions acting on the
collinear fields by $\bn\mcdot p_\pi$ so that the subscripts involve the
momentum fractions $u,v,w$, and the delta functions set either $u+v=1$ or
$u+v+w=1$.  For the soft fields we left the delta functions dimension-full,
e.g.\ $(\bar q_s S)_{-k_1} = (\bar q_s S)\delta(k_1+\cP^\dagger)$. Note that the
vacuum to pion matrix element of the operators $O_2^{(P)}$ and $O_2$ will
contribute.\footnote{We thank B.~Lange for pointing this out.}

The factorization in rapidity in \SCETb discussed in section~\ref{sec:scetb}
indicates that the jet function matching which determines Eq.~(\ref{J134}) will
give factors of $\alpha_s(\mu)$ that are perturbative at the jet scale
$\mu^2\sim E\Lambda_{\rm QCD}$, similar to the $\alpha_s(\mu)$ in
Eq.~(\ref{f+tree}).  Therefore we included the $\alpha_s(\mu)$ in the $J_i^\pi$
coefficients, rather than grouping a $g^2$ with the operators in
Eq.~(\ref{Oscet2}). To evaluate the $\langle \pi | \cdots | B\rangle$ matrix
elements of the $O_{1,2,3,4}$, we will need Eqs.~(\ref{twist2}) and
(\ref{3bdy}), the pion matrix element
\begin{align}
 & \big\langle \pi_n(p) \big|  (\bar\xi_n W)_{u} \frac{\bnslash\gamma_5}{2} 
  \:\slash\!\!\!\!\cP_\perp (W^\dagger \xi_n)_{-v} \big| 0 \big\rangle
  \\[3pt]
  &\quad 
    = \Big\langle \pi_n(p) \Big|   \frac{(-u\: n\mcdot p)}{2}\, (P_+ \!+\! P_-)(u,v) 
   - i\, \epsilon_{\mu\nu}^\perp \int\!\! dw\: T_3^{\mu\nu}(u\!-\!w,v,w)
    \Big| 0 \Big\rangle \nonumber \\[3pt]
  &\quad
    =   \frac{i}{2}\, { \bn\mcdot p\, f_\pi \mu_\pi}\: 
    \delta_{uv}\: u\Big[\phi_\pi^p(\mu,u,v) 
     + \frac{1}{6}\, \phi_\pi^{\prime\,\sigma}(\mu,u,v) \Big] 
     + i\, \bn\mcdot p\, f_{3\pi}
     \!\! \int\!\! dw\: 
     \delta_{uv}\,  \frac{\phi_{3\pi}(u\!-\!w,v,w) }{w}
      \,,
  \nonumber
\end{align}
and the $B$ matrix elements
\begin{align}
   \big\langle 0 \big| (\bar q_s S)_{-k1} \frac{\bnslash\nslash\gamma_5}{4}
    (S^\dagger h_v)_{k2}
    \big| B\rangle &= -i\, \frac{f_B\, m_B}{2} \:
     \delta(k_1\!+\!k_2\!-\! \overline\Lambda)\:  \phi_B^-(k_1,k_2) \,,
   \\[4pt]
   \big\langle 0 \big| (\bar q_s S)_{-k1} \frac{\nslash\gamma_5}{2}
    (S^\dagger h_v)_{k2}
    \big| B\rangle &= i\, \frac{f_B\, m_B}{2} \:
     \delta(k_1\!+\!k_2\!-\! \overline\Lambda)\:  \phi_B^+(k_1,k_2) \,,
  \nonumber\\[4pt]
  \big\langle 0 \big| (\bar q_s S)_{-k1} (i g \slash\!\!\!\! B_{s}^\perp)_{k_3}
    \frac{\nslash\gamma_5}{2} (S^\dagger h_v)_{k2}
    \big| B\rangle &= i\, {f_{3B}\, m_B} \:
     \delta(k_1\!+\!k_2\!+\!k_3\!-\! \overline\Lambda)\:  \phi_{3B}(k_1,k_2,k_3)
  \,.\nonumber
\end{align}
Here $f_B\sim m_b^{-1/2}$ can be taken to be the decay constant in the heavy
quark limit, and our definition for $\phi_B^\pm$ is that of
Ref.~\cite{Grozin:1996pq}, while $\phi_{3B}$ has been studied in
Ref.~\cite{Lange:2004yh} and is proportional to $\tilde \Psi_A-\tilde\Psi_V$ from
Ref.~\cite{Kawamura:2001jm}.  The momentum conserving delta functions involve
the HQET mass of the $B$-state, $\overline \Lambda = \lim_{m_b\to\infty}
(m_B-m_b)$. For the lowest order factorization formula for the soft form factor
$\zeta^{B\pi}(E)$, we find
\begin{align} \label{zetafinite}
  \zeta^{B\pi}(E) &= \frac{f_\pi f_B m_B}{4 E^2} \: {\pi\alpha_s(\mu)} \!
  \int_0^1\!\! du\, dv\, dw \int\!\! dk_1\, dk_2
  \Bigg\{ 
  \frac{4}{9}\: \delta_{k_1k_2}\, \delta_{uv}
  \frac{ (1\!+\!v)\phi_\pi(u,v)}{(v^2)_{\mbox{\o}}  } 
  \: \frac{\phi_B^-(k_1,k_2)}{ (k_1)_{\mbox{\o}} }  \nonumber \\[5pt]
  &\hspace{-1.2cm}
  + \frac{4\mu_\pi}{9}  \, \delta_{k_1k_2}\, \delta_{uv} 
   \frac{(\phi_\pi^p\!+\!\frac16\,
     \phi_\pi^{\prime\,\sigma})(u,v)}{(v^2)_{\mbox{\o}} }
   \: \frac{\phi_B^+(k_1,k_2)}{ (k_1^2)_{\mbox{\o}} } 
  + \frac{f_{3B}}{f_B} \!\! \int\!\! dk_3\, \delta_{k_1k_2k_3} \,
   \delta_{uv}  \bigg[ \frac{\phi_\pi(u,v)}{(v^2)_{\mbox{\o}}  }
 \nonumber \\[5pt]
  &\hspace{-1.2cm}
   \qquad \times  \phi_{3B}(k_1,k_2,k_3)
  \frac{9k_3\!+\! k_1}{9[(k_1\!+\!k_3)^2 k_1]_{\mbox{\o}} } 
   - \frac{\phi_\pi(u,v)}{v_{\mbox{\o}}  } 
    \frac{8k_3\, \phi_{3B}(k_1,k_2,k_3)}{9 [(k_1\!+\!k_3)^2 k_1]_{\mbox{\o}} } 
   \bigg]
 \nonumber \\[5pt]
  &\hspace{-1.2cm}
   + \frac{f_{3\pi}}{f_\pi}\, \delta_{k_1k_2}\, \delta_{uvw} \bigg[ 
  \frac{\phi_{3\pi}(u,v,w) }{[(v\!+\!w)^2\,  v]_{\mbox{\o}}\, } 
   - \frac{7\, \phi_{3\pi}(u,v,w) }{9[w(v\!+\!w)^2]_{\mbox{\o}} \,  } 
   + \frac{8\,\bar v\, \phi_{3\pi}(u,v,w) }{9[v^2 w (u\!+\!w)]_{\mbox{\o}}\,  } 
   \bigg] \frac{\phi_B^+(k_1,k_2)}{(k_1^2)_{\mbox{\o}} }  \nn
 & \hspace{-1.2cm} + \mbox{D-terms} \Bigg\}
%
   \,.
\end{align}
Here $\bar v=1\!-\!v$, the $k_i$-limits are $-\infty<k_2<\overline\Lambda$, $0<
k_{1,3} < \infty$ and
\begin{align}
 \delta_{k_1k_2} &= \delta(\overline\Lambda -\!k_1\!-\!k_2),
 & \delta_{k_1k_2k_3} & = \delta(\overline\Lambda -\!k_1\!-\!k_2\!-\!k_3), \nn
\delta_{uv} & =\delta(1\!-\!u\!-\! v),
&\delta_{uvw} & =\delta(1\!-\!u\!-\!v\!-\!w) \,.
\end{align}
The ``+ D-terms'' factor in Eq.~(\ref{zetafinite}) indicates that at this order
we must also include the renormalized coefficients $D_i(\mu_+,\mu_-)$ for
operators like the ones in Eq.~(\ref{ct2}).  In Eq.~(\ref{zetafinite}), we have
obtained a finite factorization formula for the soft $B\to\pi$ form factor. The
structure of this contribution is very similar to the hard-scattering term in
Eq.~(\ref{f+tree}), except for the necessity of the zero-bin subtractions.

To study the $\muplusminus$ dependence in Eq.~(\ref{zetafinite}), we can adopt
the standard endpoint behavior for the distributions.  For simplicity, we can
adopt a behavior for $\phi_{3B}(k_1,k_2,k_3)$ so that its tree-level integrals
also all converge without applying the zero-bin and renormalization. This leaves
$\phi_\pi(u)\sim u\bar u$, $\phi_B^+(k_1)\sim k_1$, $\phi_B^-(k_1)\sim 1$, and
$\phi_{3\pi}(u,v,w)\sim uvw^2$.  The zero-bin is then required for
$\phi_\pi(u,v)/(v^2)_{\mbox{\o}}$ given in Eq.~(\ref{Apif}) and for
$\phi_{3\pi}(u,v,w)/[v^2 w (u\!+\! w)]_{\mbox{\o}}$ which generate
$\ln(\muminus/p_\pi^-)=\ln(\muminus/2E)$ terms, and for
$\phi_B^+(k_1,k_2)/(k_1^2)_{\mbox{\o}}$ and
$\phi_B^-(k_1,k_2)/(k_1)_{\mbox{\o}}$. Following the usual procedure, we find
that these last two terms each generate a $\ln[\muplus/(n\mcdot v\overline
\Lambda)]$

At higher orders in $\alpha_s$, we anticipate that the factorization formula
will take the form
\begin{align} \label{zetaall}
 \zeta^{B\pi} &= N_0 \int\!\! du \int\!\! dk_1\:
    \hat J_1(u,k_1,E,\mu,\mu^\pm) \phi_\pi(u,\mu,\muminus)
    \phi_B^-(k_1,\mu,\muplus)\\
 &+ N_0 \int\!\! du \int\!\! dk_1\:
    \hat J_2(u,k_1,E,\mu,\mu^\pm) 
    (\phi_\pi^p\!+\! \frac16\, \phi_\pi^{\prime\,\sigma})(u,\mu,\muminus)
    \phi_B^+(k_1,\mu,\muplus) \nn
 &+ N_0 \int\!\! du\,dv \int\!\! dk_1\:
    \hat J_3(u,v,k_1,E,\mu,\mu^\pm) \phi_{3\pi}(u,v,\mu,\muminus)
    \phi_B^+(k_1,\mu,\muplus)\nn
 &+ N_0 \int\!\! du \int\!\! dk_1 dk_2\:
    \hat J_4(u,k_1,k_2,E,\mu,\mu^\pm) \phi_{\pi}(u,\mu,\muminus)
    \phi_{3B}(k_1,k_2,\mu,\muplus) . \nonumber
\end{align}
where we have used the $\delta_{k_1k_2}$ etc. functions to reduce the number of
integrations, and we note that the zero-bin subtractions are present on the
remaining variables.  Here the $\muplusminus$ dependence in the $\phi_\pi$'s and
$\phi_B$'s is a short hand for the $\ln(\muminus)$ and $\ln(\muplus)$ terms that
are generated from the \o-distribution depending on the structure of the $\hat
J_i$'s and the endpoint behavior of the distributions. The completeness of the
mixed soft-collinear basis of operators $O_{1,2,3,4}$ found in
Ref.~\cite{Lange:2003pk} should guarantee that it is the non-perturbative
functions shown in Eq.~(\ref{zetaall}) which will show up at any order in
$\alpha_s$ in the matching.\footnote{The results in Eq.~(77,78) of
  Ref.~\cite{Hardmeier:2003ig}, allow for the elimination of $\phi_\pi^p$ and
  $\phi_\pi^{\prime\,\sigma}$ in terms of $\phi_{3\pi}$ if so desired.}
One-loop corrections to $T^{(+)}$, $C_J^{(+)}$, and to $J$ are
known~\cite{Bauer:2000yr,Beneke:2004rc,Hill:2004if,Becher:2004kk}.  One-loop
corrections for the $\hat J_i$ can be computed following the method outlined
here.

Since the factorization formula for $B\to\pi\pi$ involves the same
$\zeta^{B\pi}$ and $\zeta_J^{B\pi}$ functions,
\begin{align} \label{Apipi}
%
A(\bar B\to \pi\pi) &= 
\lambda_c^{(f)} A^{M_1M_2}_{c\bar c} +
\frac{G_F m_B^2}{\sqrt2}  \bigg\{ 
  f_{\pi}\, \zeta^{B\pi}  \int_0^1\!\!\!\! du\: 
  \big[T_{1\zeta}(u)\!+\!T_{2\zeta}(u)\big]\, \phi^{\pi}(u)
  \nn
 & 
  +{f_{\pi}} \int_0^1\!\!\!\!du 
  \int_0^1\!\!\!\! dz \:
  \big[T_{1J}(u,z) 
  + T_{2J}(u,z) 
  \big] \phi^{\pi}(u) \zeta_J^{B\pi}(z) \bigg\}   \,, 
\end{align}
the results in this section immediately carry over to that process.  In
Eq.~(\ref{Apipi}) $\zeta_{(J)}^{B\pi}=\zeta^{B\pi}_{(J)}(E=m_b/2)$ and we quoted
the result from Ref.~\cite{Bauer:2004tj}. Our result in Eq.~(\ref{zetafinite})
lends support to the power counting for $B\to\pi\pi$ used in
Ref.~\cite{Bauer:2004tj}, where $\zeta^{B\pi}$ and $\zeta_J^{B\pi}$ were treated
as being parametrically and numerically similar in size. It also indicates that
the entire non-leptonic tree amplitude for $B\to\pi\pi$ can likely be written in
terms of individual $B$ and $\pi$ distribution functions. This differs from the
BBNS~\cite{Beneke:1999br,Beneke:2000fw} type factorization, where the soft-form
factor is taken to be non-perturbative. Our results appear to indicate that the
soft-form factor factorizes, but with the \SCETb factorization in rapidity space
that is different from the standard type of factorization formula.  A
factorization of the form factor in terms of individual $B$ and $\pi$ objects is
similar to the result found in the pQCD analysis in
Ref.~\cite{Keum:2000ph,Keum:2000wi}, however we do not find $k_\perp$ dependent
functions. A more detailed study of Eqs.~(\ref{zetafinite}) will be reported on
in a future publication.


\section{Conclusion} \label{sec:conclusion}

This paper discusses how to formulate effective field theories with multiple
fields for the same physical particle, where each field represents a region of
momentum space with a different power counting. This allowed for a tiling of
all the infrared momentum regions of the theory. The technique was applied to a
non-relativistic field theory (NRQCD) and to field theories for energetic
particles, including \SCETa with collinear and ultrasoft quarks and gluons, and
\SCETb with soft and collinear quarks and gluons.  Effective theories with
multiple fields for the same particle can be formulated using fields with labels
that distinguish the different momentum modes. In converting the sum over
momentum labels into an integral, one must be careful with the zero-bin, where
the momentum label is zero, since this corresponds to a different degree of
freedom. Here the proper treatment of the zero-bin was investigated in detail.
Sums which do not include the zero-bin are converted to integrals over all space
with a zero-bin subtraction. The zero-bin subtraction removes the support of the
integrand in the infrared region of momentum space where overlap between
different modes could occur, and thus avoids double-counting between different
modes in the effective theory.

The zero-bin subtractions give a definition of NRQCD and SCET independent of the
choice of UV and IR regulators, allowing the use of regulators other than
dimensional regularization if desired. They also solve a number of puzzles
encountered in the literature on collinear factorization in QCD and
non-relativistic field theories, which are associated with unphysical pinch and
endpoint singularities, as well as puzzles in the NRQCD and SCET literature with
UV and IR divergences that occur at intermediate length scales.

In NRQCD, the zero-bin subtractions eliminate pinch singularities in box-type
graphs by properly distinguishing between potential and soft heavy fermions.
This result applies to any non-relativistic effective theory (for example, it
also simplifies the resolution of the puzzle discussed in
Ref.~\cite{Cohen:2002im} involving soft pions interacting with nucleons).
Zero-bin subtractions also distinguish between soft and ultrasoft gluons in
NRQCD. In graphs with soft loops, they convert infrared divergences into
ultraviolet divergences.  Converting an infrared into an ultraviolet divergences
in the soft sector corresponds to a rearrangement of degrees of freedom that was
known to be necessary in NRQCD, and had been implemented by hand as the
``pull-up'' procedure~\cite{Manohar:2000kr,Hoang:2001rr}.  Here a proper
treatment of the zero-bin enabled us to derive the pull-up mechanism directly
from the effective Lagrangian.

In \SCETa zero-bin subtractions occur in collinear loops and avoid the overlap
with the ultrasoft momentum region. These subtractions remove infrared
divergences and induce new divergences in the ultraviolet. Previous results in
the SCET literature used anomalous dimensions computed by including infrared
divergences in collinear graphs as though they were ultraviolet divergences.
This gives (by hand) the correct result for the anomalous dimensions.  The
zero-bin subtractions allows the computation of anomalous dimensions in the
effective theory to be carried out by the standard renormalization procedure, in
terms of the counterterms used to cancel ultraviolet divergences, and also
ensures that IR divergences of the full theory are properly reproduced
independent of the regulator choice.  In inclusive decays at large energy, the
zero-bin subtraction applies to both virtual loop integrals as well as phase
space integrals for real emission as described in section~\ref{sect_sceta4}.  We
believe that zero-bin subtractions will play a role in the SCET matching
calculations for parton showering carried out in Ref.~\cite{Bauer:2006qp}, and
perhaps also to the subtractions carried out in Ref.~\cite{Collins:2000gd}. They
are likely to have implications for the singularities encountered with
$k_\perp$-distributions with light-like Wilson lines which were discussed in
Refs.~\cite{Collins:2003fm,Ji:2004wu}.

In exclusive decays, the zero-bin subtraction gives factorized decay rates and
cross-sections with finite convolution integrals. Convolution integrals of the
perturbatively calculable kernel with hadron wavefunctions are sometimes naively
divergent at the endpoints, as is the case for the $\rho$--$\pi$ form factor,
and in factorization of the soft form factor in $B \to \pi \ell \nu$ decays. The
zero-bin subtraction avoids double counting between soft and collinear modes in
\SCETb, and this converts the unphysical infrared divergence in convolution
integrals into an ultraviolet divergence.  These ultraviolet divergences are
canceled by operator renormalization. The final convolution integral is finite,
determined by a distribution we called \o.  The \o--distribution is a plus-type
distribution augmented by additional non-analytic $\ln(E)$ dependence induced by
the renormalization.  Independent of the choice for the UV and IR regulators,
our formulation of \SCETb has only soft and collinear degrees of freedom and
does not have soft-collinear messenger modes~\cite{Becher:2003qh} as
explained in section~\ref{sect:cutoffscalar} and Appendix~\ref{AppB}.\footnote{In
  Refs.~\cite{Beneke:2002ni,Beneke:2002ph,Hill:2002vw} SCET was reformulated in
  position space to avoid the sums over label momenta.  To formulate the
  zero-bin subtractions in position space one can take the Fourier transform of
  Eq.~(\ref{1.02}). }

Thus one gets finite unambiguous formulas for amplitudes in \SCETb, free of
endpoint singularities in the convolution integrals.  In cases where there would
have been a divergent convolution we get a separation of modes in rapidity space
with variables $\mu_\pm$. This preserves the power counting and naive
factorization in the kinematic variables that would be present if the
convolutions were finite.  To illustrate this, we derived finite amplitudes for
the $\rho$--$\pi$ form factor at large $Q^2$, and for the soft form factor
function $\zeta^{B\pi}(E_\pi)$ that appears in $B\to\pi\ell\bar\nu$ and
$B\to\pi\pi$ decays. Without a proper treatment of the zero-bin the convolutions
in these factorization formulas would suffer from endpoint singularities.
Applying the technique to the soft form factor for $B\to\pi\pi$ decays allowed
us to derive an amplitude for non-leptonic decays that is entirely in terms of
$B$ and $\pi$ distribution functions at lowest order.

In previous literature, the field theories, \SCETa and \SCETb, have provided an
algebraic means of deriving factorization formulas at any order in $\Lambda_{\rm
  QCD}/Q$, but not the means to guarantee that the manipulations would result in
finite factorization formulas.  Our work suggests that this will indeed be the
case, so that the power expansion of observables can be carried out to any
desired order without encountering singularities in convolution integrals. It
still remains to carry out the full derivation of an \SCETb factorization
formula and explore the renormalization group properties of the resulting
amplitudes -- tasks which we leave for future work.  The zero-bin procedure
provides the freedom to tile the infrared of an EFT with suitable degrees of
freedom, and makes the connection between the choice of degrees of freedom and
the power counting expansion clear. Exploiting this, we converted the question
of finding a complete set of degrees of freedom for an EFT at any order in the
expansion, to a question that is easier to answer physically, that of
identifying the relevant operators that occur at leading order for an
observable, and give a proper formulation of the power counting.

%
\medskip
{\bf Acknowledgments}
\medskip

  We give a special thanks to A.~Hoang for discussions on this topic over the
  years. We also acknowledge conversations with C.~Bauer, D.~Pirjol, and
  I.~Rothstein.  We thank B.~Lange, Z.~Ligeti, and S.~Mantry for comments on the
  manuscript.
  This work was supported in part by the U.S.  Department of Energy (DOE) under
  grant DE-FG03-97ER40546, the cooperative research agreement DF-FC02-94ER40818,
  and by the Office of Nuclear Science.  AM would like to thank the Alexander
  von Humboldt foundation and the Max-Planck-Institut f\"ur Physik
  (Werner-Heisenberg-Institut) for support. I.S. was also supported in part by
  the DOE Outstanding Junior Investigator program and the Sloan foundation.


\begin{appendix}

\section{Rapidity Cutoff Loop Integrals in \SCETb} \label{AppA}

\setlength\baselineskip{16pt}
\footnotesize

In this section we give some details on the calculation of the integrals
required for the \SCETb diagrams in section~\ref{sect:scalars}. This
calculations uses dimensional regularization for IR divergences, and cutoffs on
a Wick rotated rapidity variable, $\zeta'_k$, to regulate rapidity effects in
the UV as in Eq.~(\ref{harda}). The integrals we wish to compute are
\begin{align} \label{IsIcstart2}
I_s &= \int \frac{{\rm  d}^Dk }{ (2\pi)^D} \frac{1}{-p^- k^++i0^+} 
 \frac{1}{k^+ k^--\mathbf{k}_\perp^2+i0^+}
 \frac{1}{k^+ k^--k^-\ell^+-\mathbf{k}_\perp^2+i0^+} 
  \,, \nn 
I_c &= \int \frac{{\rm  d}^Dk }{(2\pi)^D} \frac{1}{-k^- \ell^++i0^+} 
 \frac{1}{k^+ k^--\mathbf{k}_\perp^2+i0^+}
 \frac{1}{k^+ k^--k^+ p^- -\mathbf{k}_\perp^2+i0^+} \,.
\end{align}
with $\ell^+>0$ and $p^->0$. We use variables $\{k^+,\zeta_k\}$ for the soft
integral with Wick rotation $\zeta_k=i\zeta'_k$, and $\{k^-,\zeta_k\}$ for the
collinear integral with $\zeta_k=-i\zeta'_k$, as discussed in
section~\ref{sect:scalars}. It is easy to verify for $I_s$ and $I_c$ that these
Wick rotations about the origin do not encounter any poles.

\medskip
\noindent\underline{\bf Soft Integral in ${\rm SCET}_{\rm II}$ }
\medskip

Consider the soft integral in Eq.~(\ref{IsIcstart2}) and perform the
$k_\perp$ integral:
\begin{align} \label{A1}
I_s 
 &= \int_0^1 {\rm d}x \int \frac{{\rm  d}^Dk }{ (2\pi)^D} \frac{1}{-p^-
   k^++i0^+} \frac{1}{\left[ \mathbf{k}_\perp^2+k^-\ell^+ x
     -k^+k^--i0^+\right]^2} 
  \nn
 &= \frac{\Gamma(1+\epsilon)} {8\pi} \int_0^1  {\rm d}x 
    \int \frac{{\rm  d}k^+{\rm d}k^- }{ (2\pi)^2} 
   \frac{\left[k^-\ell^+ x -k^+k^--i0^+\right]^{-1-\epsilon}}{-p^- k^++i0^+} 
   \,.
\end{align}
Let $k^-=\zeta k^+$, and note that $dk^- = |k^+| d\zeta $ once we integrate
$-\infty < k^+ < \infty$ and $-\infty < \zeta < \infty$. Thus
\begin{align} \label{A2}
I_s  &= \frac{\Gamma(1+\epsilon)} {8\pi p^-} \int_0^1  {\rm d}x \int \frac{{\rm
    d}k^+{\rm d}\zeta }{ (2\pi)^2} \frac{ |k^+| \left[k^+\ell^+ \zeta x -(k^+)^2
    \zeta -i0^+\right]^{-1-\epsilon}}{- k^++i0^+}  \,.
\end{align}
Rescale by a positive constant $k^+ = (\ell^+ x) k^{\prime +}$ and for
 simplicity rename $k^{\prime +}=k^+$. Then rotate $\zeta=i\zeta'$, so
 \begin{eqnarray}
I_s &=&  \frac{\Gamma(1+\epsilon)} {8\pi p^-}  \int_0^1 \!\!  {\rm d}x \:
(\ell^+ x)^{-1-2 \epsilon} \int \frac{{\rm  d}k^+{\rm d}\zeta }
 { (2\pi)^2} \: \frac{|k^+| \left[ \zeta k^+ (1-k^+)  -i0^+\right]^{-1-\epsilon}}
  {- k^++i0^+} \nn
 &=&  \frac{-\Gamma(1+\epsilon)  (\ell^+ )^{-2 \epsilon} } 
  {16\pi\epsilon (p^-  \ell^+) (2\pi)^2} \int {{\rm  d}k^+ (i\,{\rm d}\zeta')} \:
  \frac{ |k^+| \left[i\zeta' k^+ (1-k^+)  -i0^+\right]^{-1-\epsilon}}{- k^++i0^+} \,.
 \end{eqnarray}
 We divide the $k^+$ integral into $k^+>1$, $0<k^+<1$, and $k^+<0$.  This gives
\begin{eqnarray}  \label{Jint}
J(\zeta') & \equiv & \int_{-\infty}^{\infty}  \!\! 
   {{\rm  d}k^+ }\, \frac{|k^+|  \left[ i\zeta' k^+ (1-k^+) 
    \right]^{-1-\epsilon}}{- k^++i0^+} \nn
   & = & - (i\zeta')^{-1-\epsilon} \int_{0}^{1}  \!\! 
    {{\rm  d}k^+ }  [k^+(1\!-\! k^+)]^{-1-\epsilon} 
       - (-i\zeta')^{-1-\epsilon} \int_{1}^{\infty}  \!\! 
   {{\rm  d}k^+ }   [k^+(k^+\!-\!1)]^{-1-\epsilon} 
 \nn
   && 
   + (-i\zeta')^{-1-\epsilon} \int_{-\infty}^0   \!\!
     {{\rm  d}k^+ }  [(-k^+) (1\!-\!k^+ )]^{-1-\epsilon}
     \nn
    & = &  - (i\zeta')^{-1-\epsilon} \int_{0}^{1}  \!\! 
    {{\rm  d}k^+ }  [k^+(1\!-\! k^+)]^{-1-\epsilon} 
   \nn
   & = &
      -  (i\zeta')^{-1-\epsilon} 
     \frac{  [\Gamma(-\epsilon) ]^2 }{\Gamma(-2\epsilon)} 
   = (i\zeta')^{-1-\epsilon}  \left[  \frac{2}{\epsilon}   
     - \frac{\pi^2\epsilon}{3}  -4 \zeta_3 \epsilon^2 + \ldots \right] \,.
\end{eqnarray}   
In the second equality note that the change of variables $k^+\to 1-k^+$ for
$k^+<0$ causes the second and third terms to cancel.  Putting the pieces back
together and multiplying by the $\mu^{2\epsilon}$ this gives
\begin{eqnarray}
  I_s &=&  \frac{- \Gamma(1+\epsilon)  } {64\pi^3\epsilon\: (p^-\ell^+) }
   \frac{ (\ell^+ )^{-2 \epsilon}}{(\mu)^{-2\epsilon}}  \left[
     \frac{2}{\epsilon}   - \frac{\pi^2\epsilon}{3}  -4 \zeta_3 \epsilon^2 
     + \ldots \right] i \int_{\zeta'_{\rm min}}^{\zeta'_{\rm max}}\!
      d\zeta'\: (i\zeta')^{-1-\epsilon} \nn
    &=& \frac{\Gamma(1+\epsilon)  } {64\pi^3\epsilon^2\: (p^-\ell^+) }
   \frac{ (\ell^+ )^{-2 \epsilon}}{(\mu)^{-2\epsilon}}  \left[
     \frac{2}{\epsilon}   - \frac{\pi^2\epsilon}{3}  -4 \zeta_3 \epsilon^2 
     + \ldots \right] \left[ (i\zeta'_{\text{max}})^{-\epsilon}
     -(i\zeta'_{\text{min}})^{-\epsilon} \right] . 
\end{eqnarray}
Now set $\zeta'_{\rm min}= -a^2$ and $\zeta'_{\rm max}=a^2$ and note that
$(\pm i)^{-\epsilon}=\exp(\mp i\pi\epsilon/2)$, so that we have
\begin{eqnarray} \label{answerIstot}
   I_s   &=&  \left[ \frac{1}{64\pi^3 (p^-\ell^+) } \right]
   \frac{\Gamma(1+\epsilon)  } {\epsilon } 
    \frac{[e^{-i\pi\epsilon/2}-e^{i\pi\epsilon/2}]}{\epsilon}
    \left[  \frac{2}{\epsilon}   - \frac{\pi^2\epsilon}{3} 
        -4 \zeta_3 \epsilon^2 + \ldots \right]
   \frac{ (a\, \ell^+ )^{-2 \epsilon}}{(\mu)^{-2\epsilon}} 
  \nn
  &=&  \left[ \frac{-i\,\Gamma(1+\epsilon) }{32\pi^2 (p^-\ell^+) } \right]  
   \frac{1} {\epsilon^2 }
   \left[ 1 -\frac{5\pi^2\epsilon^2}{24} 
      + \ldots \right]
   \frac{ (a\, \ell^+ )^{-2 \epsilon}}{(\mu)^{-2\epsilon}}  \nn
   &=&  \left[ \frac{-i}{16\pi^2 (p^-\ell^+) } \right]  
   \left[   \frac{1} {2\epsilon^2 }  - \frac{1}{\epsilon} \, \ln\Big( \frac{a\,\ell^+ }{\mu} \Big) + \ln^2\Big( \frac{a\,\ell^+ }{\mu } \Big)  -\frac{\pi^2}{16}
      + \ldots \right].
\end{eqnarray}
(In the last line when expanding in $\epsilon$ we have multiplied by the
$\exp(\epsilon \gamma_E)$ factor to put $\mu$ into the ${\overline {\rm MS}}$
scheme, while the necessary $(4\pi)^{-\epsilon}$ factor was removed already in
Eq.~(\ref{A1}).)  Eq.(\ref{answerIstot}) is the result quoted for the soft
computation in Eq.~(\ref{ssrslt}) in the text.

\medskip
\noindent\underline{\bf Collinear integral in ${\rm SCET}_{\rm II}$  }
\medskip

Lets repeat the computation in the previous section for the collinear integral
in Eq.~(\ref{startsoftcollin}). The integral is
\begin{eqnarray}
I_c 
 &=& \int_0^1 {\rm d}x \int \frac{{\rm  d}^Dk }{ (2\pi)^D} \frac{1}{-k^- \ell^++i0^+} \frac{1}{\left[ \mathbf{k}_\perp^2+p^- k^+ x - k^+k^--i0^+\right]^2}\nn
 &=& \frac{\Gamma(1+\epsilon)} {8\pi} \int_0^1  {\rm d}x \int \frac{{\rm  d}k^+{\rm d}k^- }{ (2\pi)^2} \frac{\left[ k^+p^- x -k^+k^--i0^+\right]^{-1-\epsilon}}{-k^- \ell^++i0^+} 
\end{eqnarray}
Let $k^+=k^-/\zeta$ with $dk^+ = |k^-| d\zeta/\zeta^2 $ and $-\infty <
\zeta<\infty$, so
\begin{eqnarray} \label{Ic}
I_c  &=& \frac{\Gamma(1+\epsilon)} {8\pi \ell^+} \int_0^1  
  {\rm d}x \int \frac{{\rm  d}k^+{\rm d}\zeta }{ (2\pi)^2}
    \frac{ |k^-| \left[ k^-( p^-  x -  k^-)/\zeta -i0^+\right]^{-1-\epsilon}}
   {\zeta^2(- k^- +i0^+)} 
\end{eqnarray}
Rescale by a positive constant $k^+ = (p^- x) k^{\prime +}$ (for simplicity
renaming $k^{\prime +} \to k^+$), and then rotate $\zeta=-i\zeta'$,
\begin{eqnarray}
I_c &=& \frac{\Gamma(1+\epsilon)} {8\pi \ell^+ } 
   \int_0^1\!\!  {\rm d}x \: (p^- x)^{-1-2 \epsilon} 
   \int \frac{{\rm  d}k^-{\rm d}\zeta  }{ (2\pi)^2} 
   \frac{ |k^-| \left[ k^-(1 -k^-)/\zeta  -i0^+\right]^{-1-\epsilon}}
   {\zeta^2(- k^- +i0^+)} \nn
 &=& \frac{i\Gamma(1+\epsilon) (p^-)^{-2\epsilon}} {64\pi^3\epsilon (p^-\ell^+)}
   \int d\zeta'   \int_{-\infty}^{\infty}\!\! {{\rm  d}k^-}\: 
    \frac{ |k^-| \left[ i k^-(1-k^-)/\zeta'   -i0^+\right]^{-1-\epsilon}}
   {\zeta^{\prime 2}(- k^-+i0^+)} \nn
  &=& \frac{\Gamma(1+\epsilon) (p^-)^{-2\epsilon}} {64\pi^3\epsilon
    (p^-\ell^+)}\:  i\! \int_{\zeta'_{\rm min}}^{\zeta'_{\rm max}}
    \frac{d\zeta'}{\zeta^{\prime\, 2}}\ J(1/\zeta') 
\end{eqnarray}
where in the last line we noted that the $k^-$ integral is identical
$J(1/\zeta')$ defined via Eq.~(\ref{Jint}). Multiplying by $\mu^{2\epsilon}$
this gives
\begin{eqnarray}
  I_c &=&  \frac{\Gamma(1+\epsilon)  } {64\pi^3\epsilon\: (p^-\ell^+) }
   \frac{ (p^- )^{-2 \epsilon}}{(\mu)^{-2\epsilon}}  \left[
     \frac{2}{\epsilon}   - \frac{\pi^2\epsilon}{3}  -4 \zeta_3 \epsilon^2 
     + \ldots \right] i \int_{\zeta'_{\rm min}}^{\zeta'_{\rm max}}
      \frac{d\zeta'}{\zeta^{\prime\,2}}
   \: \Big(\frac{i}{\zeta'}\Big)^{-1-\epsilon} 
    \nn
    &=& \frac{\Gamma(1+\epsilon)  } {64\pi^3\epsilon^2\: (p^-\ell^+) }
   \frac{ (p^- )^{-2 \epsilon}}{(\mu)^{-2\epsilon}}  \left[
     \frac{2}{\epsilon}   - \frac{\pi^2\epsilon}{3}  -4 \zeta_3 \epsilon^2 
     + \ldots \right] \left[ \Big(\frac{i}{\zeta'_{\text{max}}}\Big)^{-\epsilon}
     -\Big(\frac{i}{\zeta'_{\text{min}}}\Big)^{-\epsilon} \right] . 
\end{eqnarray}
Now we add the contributions from the regions $\zeta'\in [a^2,\infty]$ and
 $\zeta'\in[-\infty,-a^2]$ to give
\begin{eqnarray} \label{answer}
  I_c   &=&  
  \left[ \frac{1}{64\pi^3 (p^-\ell^+) } \right] 
   \frac{\Gamma(1+\epsilon)} {\epsilon } 
    \frac{[-e^{-i\pi\epsilon/2} + e^{i\pi\epsilon/2} ]}{\epsilon}
   \left[ \frac{2}{\epsilon} - \frac{\pi^2\epsilon}{6} - 2\zeta_3 \epsilon^2 
      + \ldots \right]
   \frac{ (p^-/a )^{-2 \epsilon}}{(\mu)^{-2\epsilon}}  \nn
   &=&  \left[ \frac{-i \Gamma(1+\epsilon)}{32\pi^2 (p^-\ell^+) } \right] 
    \frac{1} {\epsilon^2 }
   \left[  1 - \frac{5\pi^2\epsilon^2}{24}
      + \ldots \right]
   \frac{ (  p^-/a  )^{-2 \epsilon}}{(\mu)^{-2\epsilon}}  \nn
   &=&  \left[ \frac{-i}{16\pi^2 (p^-\ell^+) } \right]  
   \left[   \frac{1} {2\epsilon^2 }  - \frac{1}{\epsilon} \, \ln\Big( \frac{p^- }{a\mu} \Big) + \ln^2\Big( \frac{p^- }{a\mu } \Big)  -\frac{\pi^2}{16}
      + \ldots \right].
\end{eqnarray}
This is the collinear integral result quoted in Eq.~(\ref{scrslt}). A simple way
to get this answer is to note that the original collinear integral is identical
to the soft integral with the replacements $k^+ \leftrightarrow k^-$,
$p^-\leftrightarrow \ell^+$, and $a\to 1/a$. The answers in
Eqs.~(\ref{answerIstot}) and (\ref{answer}) agree with this.
 

\section{\SCETb Loops in Dim. Reg. with Different IR Regulators}
\label{AppB}

In this appendix we repeat the matching computation done in
section~\ref{sect:dimregscalar} of a scalar loop integral in \SCETb.  We use
dimensional regularization, but modify the treatment of the IR regulator. The
structure of the full theory and effective theory diagrams changes, but again
the IR divergences are properly reproduced and the same contribution to the
matching coefficient is obtained. The calculation is done for two classes of IR
regulators: i) taking three different IR masses, $m_1$, $m_2$, and $m_3$ rather
than just the single mass used in section~\ref{sect:dimregscalar}, and ii) with
$m_1\ne 0$, $m_2=m_3=0$, and external momenta offshell, $\ell^2\ne 0$ and
$p^2\ne 0$.  Finally, in a part iii) we discuss subtleties related to the
$m_1=0$ limit of these two cases. For simplicity we leave off the diagram
prefactor $i e g^2 G/(p^-\ell^+)$ and just quote results for the integrals in
this appendix. In all cases $p^->0$ and $\ell^+>0$, and $(p^-\ell^+)$ is the
perturbative scale.

\subsection*{i) Three IR Masses, $\mathbf{m_1}$, $\mathbf{m_2}$,
  $\mathbf{m_3}$}

The full theory loop is the generalization of Eq.~(\ref{sect:dimregscalar}) with
three IR masses,
\begin{align} \label{scetIIfullp}
 I_{\rm full}^{\rm scalar} &= \int  
  \!\!\frac{d^Dk}{(2\pi)^D}  \frac{1}
{[(k-\ell)^2- m_2^2+i0^+][k^2 -m_1^2 + i0^+][(k- p)^2 - m_3^2 + i0^+]} 
 \nn
   &= \frac{-i}{16\pi^2 (p^-\ell^+)} 
   \bigg[ \frac{1}{2} 
   \ln^2\Big(\frac{m_1^2}{p^-\ell^+} \Big) 
     + {\rm Li}_2\Big(1-\frac{m_2^2}{m_1^2}\Big)
     + {\rm Li}_2\Big(1-\frac{m_3^2}{m_1^2}\Big)
  \bigg] .
\end{align}
In Eq.~(\ref{scetIIfullp}) factors of the IR regulators, $m_1^2$, $m_2^2$, and
$m_3^2$ appear in all propagators, external momenta are taken onshell,
$p^2=\ell^2=0$, and we have expanded in $m_i^2/(p^-\ell^+)$. The result is valid
as long as $(p^-\ell^+) \gg m_i^2$, $(p^-\ell^+)m_1^2 \gg m_2^2 m_3^2$, and can
not be used for the case $m_1=0$ since it blows up. The result which is valid
for $m_1\to 0$ and also reproduces Eq.~(\ref{scetIIfullp}) is given below in
Eq.~(\ref{scetIIfullpboth}). The $m_1=0$ result is in Eq.~(\ref{m1=0}).

The LO \SCETb currents for dimensional regularization are given in
Eq.~(\ref{OIIdr}). Using the $m_{1,2,3}$ IR regulators for the scalar and
collinear loops in Figs.~\ref{fig:scalargamma}b,c we have
\begin{align}
 I_{\rm soft}^{\rm scalar} &= \sum_{k^+\ne 0} \int  
  \!\!\frac{d^Dk_r}{(2\pi)^D}  \frac{\mu^{2\epsilon}}
 {[k^2\!-\!\ell^+\,k^- \!-\!m_2^2 \!+\! i 0^+]
 [k^2 \!-\!m_1^2\!+\! i 0^+][-p^-\,  k^+ \!+\! i0^+]}
\ \frac{|k^+|^\epsilon |k^+\!-\!\ell^+|^\epsilon}{\muplusee} \,, \\
 I_{\rm cn}^{\rm scalar} &= \sum_{k^-\ne 0} \int  
  \!\!\frac{d^Dk_r'}{(2\pi)^D}  \frac{\mu^{2\epsilon}}
 {[-\ell^+\,k^- \!+\! i 0^+]
 [k^2  \!-\!m_1^2\!+\! i 0^+][k^2\!-\! p^- \, k^+  \!-\!m_3^2\!+\! i 0^+]} 
 \ \frac{|k^-|^\epsilon |k^-\!-\!p^-|^\epsilon}{\muminusee} \,.
  \nonumber
\end{align}
Note that we keep $m_2^2$ and $m_3^2$ only in the propagators that are allowed
to become small by the power counting in \hbox{\SCETb\!.}  Eq.~(\ref{1.02})
tells us that we have zero-bin subtractions for the soft and collinear diagrams
which avoid the IR singularities from the $[-p^-k^+]$ and $[-\ell^+k^-]$
propagators. The naive integrals and subtraction integrals are
\begin{align}
 \tilde I_{\rm soft}^{\rm scalar} &=  \int  
  \!\!\frac{d^Dk}{(2\pi)^D}  \frac{\mu^{2\epsilon}}
 {[k^2\!-\!\ell^+\,k^-\!-\!m_2^2 \!+\! i 0^+]
 [k^2  \!-\!m_1^2\!+\! i 0^+][-p^-\,  k^+ \!+\! i0^+]}
\ \frac{|k^+|^\epsilon |k^+\!-\!\ell^+|^\epsilon}{\muplusee} \,, \\
  I_{\rm 0soft}^{\rm scalar} &=  \int  
  \!\!\frac{d^Dk}{(2\pi)^D}  \frac{\mu^{2\epsilon}}
 {[-\!\ell^+\,k^- \!+\! i 0^+]
 [k^2  \!-\!m_1^2\!+\! i 0^+][-p^-\,  k^+ \!+\! i0^+]}
\ \frac{|k^+|^\epsilon |k^+\!-\!\ell^+|^\epsilon}{\muplusee} \,, \nn
 \tilde I_{\rm cn}^{\rm scalar} &=  \int  
  \!\!\frac{d^Dk}{(2\pi)^D}  \frac{\mu^{2\epsilon}}
 {[-\ell^+\,k^- \!+\! i 0^+]
 [k^2  \!-\!m_1^2\!+\! i 0^+][k^2\!-\! p^- \, k^+\!-\!m_3^2 \!+\! i 0^+]} 
 \ \frac{|k^-|^\epsilon |k^-\!-\!p^-|^\epsilon}{\muminusee} \,,\nn
 I_{\rm 0cn}^{\rm scalar} &=  \int  
  \!\!\frac{d^Dk}{(2\pi)^D}  \frac{\mu^{2\epsilon}}
 {[-\ell^+\,k^- \!+\! i 0^+]
 [k^2  \!-\!m_1^2\!+\! i 0^+][-\! p^- \, k^+\!+\! i 0^+]} 
 \ \frac{|k^-|^\epsilon |k^-\!-\!p^-|^\epsilon}{\muminusee} \,.\nonumber
\end{align}
To determine the form of the subtraction integrals, we considered the collinear
limit of the soft loop momentum in $\tilde I_{\rm soft}^{\rm scalar}$, and the
soft limit of the collinear loop momentum in $\tilde I_{\rm cn}^{\rm scalar}$.
Note that the $m_2^2$ and $m_3^2$ dependence is dropped in the subtraction
integrals because in these limits $\ell^+k^-\gg m_2^2$ and $p^-k^+\gg m_3^2$.
The UV rapidity regulator factors, $|\cdots |^\epsilon$ are not affected by the
subtractions (despite the way we are organizing the computation here, one really has
an integrand defined with subtractions and then multiplies it by these factors). From
Eq.~(\ref{1.02}) the differences $\tilde I_{\rm soft}^{\rm scalar}- I_{\rm
  0soft}^{\rm scalar}$ and $\tilde I_{\rm cn}^{\rm scalar}- I_{\rm 0cn}^{\rm
  scalar}$ will give the result for $I_{\rm soft}^{\rm scalar}$ and $I_{\rm
  cn}^{\rm scalar}$ respectively.

For the soft graph we do the $k^-$ integral by contours. Due to the pole
structure this restricts the $k^+$-integration to the region $0< k^+<\ell^+$.
The $k_\perp$ integral is then done. For the soft subtraction integral we follow
the same procedure which this time leaves the integration region $0<
k^+<\infty$. We find
\begin{align}  \label{Isoftm1m2m3}
  \tilde I_{\rm soft}^{\rm scalar} 
  &= \frac{-i\: \Gamma(\epsilon)\, \mu^{2\epsilon} }{16\pi^2(p^-\ell^+)}  
   \int_0^{\ell^+} \! \frac{dk^+}{k^+} \, \left[ m_1^2
     \Big(1-\frac{k^+}{\ell^+}\Big)
      + m_2^2\, \frac{k^+}{\ell^+}\right]^{-\epsilon} 
   \ \ \bigg|  \frac{k^+ (k^+\!-\!\ell^+)}
   {\muplus{}^2} \bigg|^\epsilon \,,
   \nonumber \\
  I_{\rm 0soft}^{\rm scalar} 
  &= \frac{-i\: \Gamma(\epsilon)\, \mu^{2\epsilon} }{16\pi^2(p^-\ell^+)}  
   \int_0^{\infty} \! \frac{dk^+}{k^+} \, \Big[ m_1^2 \Big]^{-\epsilon} 
   \ \ \bigg|  \frac{k^+ (k^+\!-\!\ell^+)}
   {\muplus{}^2} \bigg|^\epsilon \,.
\end{align}
The double counting with the collinear integral comes from the $k^+\to 0$ part
of the integral, but the divergence from this limit exactly cancels in $I_{\rm
  soft}^{\rm scalar} = \tilde I_{\rm soft}^{\rm scalar} - I_{\rm 0soft}^{\rm
  scalar}$ as long as $m_1\ne 0$.  Computing the integrals we find
\begin{align} \label{ssrslt2p}
 I_{\rm soft}^{\rm scalar} 
  &= \frac{-i\: \Gamma(\epsilon)\,  }{16\pi^2(p^-\ell^+)}  \, 
   \Big(\frac{\ell^+}{\muplus}\Big)^{2\epsilon} 
  \bigg\{ \int_0^1\!\!\! dx\: \frac{|1\!-\!x|^\epsilon |x|^\epsilon}{x} \:
    \Big[ \frac{m_1^2(1\!-\!x)+m_2^2\, x}{\mu^2} \Big]^{-\epsilon}
  - \int_0^\infty\!\!\!\! dx  \frac{|1\!-\!x|^\epsilon |x|^\epsilon}{x} \: 
   \Big[ \frac{m_1^2}{\mu^2} \Big]^{-\epsilon}
  \bigg\}
   \nonumber \\[5pt]
&= \frac{-i\: \Gamma(\epsilon)\,  }{16\pi^2(p^-\ell^+)}  
   \Big(\frac{\mu^2}{m_1^2}\Big)^{\epsilon}\!
   \Big(\frac{\ell^+}{\muplus}\Big)^{2\epsilon} 
  \bigg\{ \frac{\Gamma(\epsilon)\Gamma(1\!+\!\epsilon)}{\Gamma(1\!+\!2\epsilon)}
    {}_2F_1\Big(\epsilon,\epsilon,1\!+\!2\epsilon,\frac{m_1^2\!-\!m_2^2}{m_1^2}\Big)
  \!-\!  \frac{\Gamma(\epsilon)\Gamma(1\!+\!\epsilon)}{\Gamma(1\!+\!2\epsilon)}
   \!-\!  \frac{\Gamma(1\!+\!\epsilon)\Gamma(-2\epsilon)}{\Gamma(1\!-\!\epsilon)}
  \bigg\}
   \nonumber \\[5pt]
 &= \frac{-i\: }{16\pi^2(p^-\ell^+)}  
   \bigg[ \frac{1}{2\epsilon_{\rm
      UV}^2} 
   \!+\! \frac{1}{\epsilon_{\rm UV}} \ln\Big(\frac{\ell^+  }{\muplus} \Big)
   \!-\! \frac{1}{2\epsilon_{\rm UV}} \ln\Big(\frac{m_1^2 }{\mu^2} \Big)
  + \ln^2\Big(\frac{\ell^+}{\muplus} \Big) + \frac{5\pi^2}{24} \nn
 & \qquad\qquad\qquad\qquad
  \!+\! \frac{1}{4} \ln^2\Big(\frac{m_1^2}{\mu^2} \Big) 
   - \ln\Big(\frac{m_1^2}{\mu^2} \Big) \ln\Big(\frac{\ell^+}{\muplus}\Big)
  +  {\rm Li}_2\Big(1-\frac{m_2^2}{m_1^2}\Big)\bigg] \,.
\end{align}
For the collinear integrals we do the contour integration in $k^+$ which
restricts the remaining integration region in $k^-$. For the naive and
subtraction integrals we find
\begin{align}  \label{Icnsp}
  \tilde I_{\rm cn}^{\rm scalar} 
  &= \frac{-i\: \Gamma(\epsilon)\, \mu^{2\epsilon} }{16\pi^2(p^-\ell^+)}  
   \int_0^{p^-} \! \frac{dk^-}{k^-} \, 
    \left[ m_1^2
     \Big(1-\frac{k^-}{p^-}\Big)
      + m_3^2\, \frac{k^-}{p^-}\right]^{-\epsilon} 
   \ \ \bigg|  \frac{k^- (k^-\!-\!p^-)}
   {\muminus{}^2} \bigg|^\epsilon \,,
   \nonumber \\[5pt]
  I_{\rm 0cn}^{\rm scalar} 
  &= \frac{-i\: \Gamma(\epsilon)\, \mu^{2\epsilon} }{16\pi^2(p^-\ell^+)}  
   \int_0^{\infty} \! \frac{dk^-}{k^-} \, \Big[ m_1^2
     \Big]^{-\epsilon} 
   \ \ \bigg[  \frac{k^- (k^-\!-\!p^-)}
   {\muminus{}^2} \bigg]^\epsilon \,.
\end{align}
The subtraction integral cancels the singularity in $\tilde I_{\rm cn}^{\rm
  scalar}$ as $k^-\to 0$ as long as $m_1\ne 0$.  The complete collinear result,
$I_{\rm cn}^{\rm scalar}= \tilde I_{\rm cn}^{\rm scalar}-I_{\rm 0cn}^{\rm
  scalar}$, is very similar to the soft result
\begin{align}  \label{scrslt2p}
  I_{\rm cn}^{\rm scalar} &=
 \frac{-i\: \Gamma(\epsilon)\,  }{16\pi^2(p^-\ell^+)}  \, 
   \Big(\frac{p^-}{\muminus}\Big)^{2\epsilon} 
  \bigg\{ \int_0^1\!\!\! dx\: \frac{|1\!-\!x|^\epsilon |x|^\epsilon}{x} \:
    \Big[ \frac{m_1^2(1\!-\!x)+m_3^2\, x}{\mu^2} \Big]^{-\epsilon}
  - \int_0^\infty\!\!\!\! dx  \frac{|1\!-\!x|^\epsilon |x|^\epsilon}{x} \: 
   \Big[ \frac{m_1^2}{\mu^2} \Big]^{-\epsilon}
  \bigg\}
   \nonumber \\[5pt]
 &= \frac{-i\: }{16\pi^2(p^-\ell^+)}  
   \bigg[ \frac{1}{2\epsilon_{\rm
      UV}^2} 
   \!+\! \frac{1}{\epsilon_{\rm UV}} \ln\Big(\frac{p^-  }{\muminus} \Big)
   \!-\! \frac{1}{2\epsilon_{\rm UV}} \ln\Big(\frac{m_1^2 }{\mu^2} \Big)
  +  \ln^2\Big(\frac{p^-}{\muminus} \Big) + \frac{5\pi^2}{24} \nn
 & \qquad\qquad\qquad\qquad
  \!+\! \frac{1}{4} \ln^2\Big(\frac{m_1^2}{\mu^2} \Big) 
   -  \ln\Big(\frac{m_1^2}{\mu^2} \Big) \ln\Big(\frac{p^-}{\muminus}\Big)
  +   {\rm Li}_2\Big(1-\frac{m_3^2}{m_1^2}\Big)\bigg] \,.
\end{align}

The results in Eqs.~(\ref{ssrslt2p}) and (\ref{scrslt2p}) have $1/\epsilon_{\rm
  UV} \ln(m^2)$ divergences, which are canceled by the $\phi(0,\mu)/\epsilon$
type counterterms. For this scalar calculation these divergences are canceled by
a graph containing the insertion of the renormalized currents in Eq.~(\ref{J0d})
with additional counterterm coefficients for the convolution integral as given
in Eq.~(\ref{C0d0e}).  Contracting the scalar gluon as in
Fig.~\ref{fig:scalargamma}b), using the same IR mass regulator, and pulling out
the same prefactor as the other diagrams gives
\begin{align} \label{Ictscalar}
 I_{\rm ct}^{\rm scalar} = 2 \Big(\frac{1}{2\epsilon_{\rm UV}} \Big)
   \frac{i\: }{16\pi^2(p^-\ell^+)}  
   \bigg[ - \ln\Big(\frac{m_1^2}{\mu^2} \Big) 
    \bigg]\,.
\end{align}
Note that dependence on $m_{2,3}$ drops out of the answer for $I_{\rm ct}^{\rm scalar}$.
Due to our choice of $\delta C^{(0d,0e)}$ this exactly cancels the
$1/\epsilon_{\rm UV} \ln(m_1^2)$ terms in the collinear and soft loops.  Adding
the soft, collinear, and counterterm graphs we find the full \SCETb result
\begin{align} \label{scetIIDRrslt0p}
 I_{\rm soft+cn}^{\rm scalar} 
 &= \frac{-i\: }{16\pi^2(p^-\ell^+)}  
   \bigg[  \frac12\,\ln^2\Big(\frac{m_1^2}{\mu^2} \Big) 
   \!-  \ln\Big(\frac{m_1^2}{\mu^2} \Big) \ln\Big(\frac{p^-}{\muminus}\Big)
  \!-  \ln\Big(\frac{m_1^2}{\mu^2} \Big) \ln\Big(\frac{\ell^+}{\muplus}\Big)
  \!+   {\rm Li}_2\Big(1-\frac{m_2^2}{m_1^2}\Big)
  \!+    {\rm Li}_2\Big(1-\frac{m_3^2}{m_1^2}\Big)
   \nn 
 & \qquad
  +\frac{1}{\epsilon^2_{\rm UV}}  
  \!+\! \frac{1}{\epsilon_{\rm UV}} \ln\Big(\frac{p^- \ell^+ }{\muminus\muplus} \Big)
   + \ln^2\Big(\frac{p^-}{\muminus} \Big) + \ln^2\Big(\frac{\ell^+}{\muplus} \Big)
  + \frac{5\pi^2}{12} 
  \bigg] \nonumber \\[5pt]
&= \frac{-i\: }{16\pi^2(p^-\ell^+)}  
   \bigg[ \frac12\, \ln^2\Big(\frac{m_1^2}{p^-\ell^+} \Big)
   - \ln\Big(\frac{m_1^2}{\mu^2} \Big) \ln\Big(\frac{\mu^2}{\muminus\muplus}\Big)
  +  {\rm Li}_2\Big(1-\frac{m_2^2}{m_1^2}\Big)
  +  {\rm Li}_2\Big(1-\frac{m_3^2}{m_1^2}\Big)
  \nonumber \\
 & \qquad 
   +\frac{1}{\epsilon^2_{\rm UV}}
  \!+\! \frac{1}{\epsilon_{\rm UV}} \ln\Big(\frac{p^- \ell^+ }{\muminus\muplus} \Big)
  +  \ln^2\Big(\frac{p^-}{\muminus} \Big) +  \ln^2\Big(\frac{\ell^+}{\muplus} \Big)
    - \frac12\,\ln^2\Big(\frac{p^-\ell^+}{\mu^2} \Big) 
   + \frac{5\pi^2}{12}  \bigg]
 \,.
\end{align}
The effective theory result in Eq.~(\ref{scetIIDRrslt0p}) has UV divergences
which are the same as in Eq.~(\ref{scetIIDRrslt0}), and are canceled by a
counterterm for the jet function coefficient $J^{(0a)}$, as given in
Eq.~(\ref{delJrslt}). The renormalized EFT result is
\begin{align}
  \label{scetIIDRrsltp}
 I_{\rm soft+cn}^{\rm scalar} 
 &=  \frac{-i\: }{16\pi^2(p^-\ell^+)}  
   \bigg[ \frac12\: \ln^2\Big(\frac{m_1^2}{p^-\ell^+} \Big)
   +   {\rm Li}_2\Big(1-\frac{m_2^2}{m_1^2}\Big)
   +   {\rm Li}_2\Big(1-\frac{m_3^2}{m_1^2}\Big)
   -  \ln\Big(\frac{m_1^2}{\mu^2} \Big) \ln\Big(\frac{\mu^2}{\muminus\muplus}\Big)
   \nn
 & \qquad \qquad\qquad
  +  \ln^2\Big(\frac{p^-}{\muminus} \Big) + \ln^2\Big(\frac{\ell^+}{\muplus} \Big)
    - \frac12\, \ln^2\Big(\frac{p^-\ell^+}{\mu^2} \Big) 
   + \frac{5\pi^2}{12} \bigg]
 \,.
\end{align}
The first three terms exactly reproduces the IR divergences in the full theory
result in Eq.~(\ref{scetIIfullp}), including the entire functional dependence on
the ratios of $m_i^2$, and the fourth term vanishes since $\mu^2 = \muplus\:
\muminus$.  The difference of the remaining finite terms gives a contribution to
the one-loop matching coefficient
\begin{align} \label{matchrsltp}
   I_{\rm match}^{\rm scalar} 
  &= \frac{-i\: }{16\pi^2(p^-\ell^+)}  
   \bigg[ \frac12\,\ln^2\Big(\frac{p^-\ell^+}{\muminus\muplus}\Big) 
  -  \ln^2\Big(\frac{p^-}{\muminus} \Big) -  \ln^2\Big(\frac{\ell^+}{\muplus} \Big)
  - \frac{5\pi^2}{12}
  \bigg] \nn
&= \frac{-i\: }{16\pi^2(p^-\ell^+)}  
   \bigg[ - \frac12\,\ln^2\Big(\frac{p^-\muplus}{\muminus\ell^+} \Big)
  - \frac{5\pi^2}{12}
  \bigg]
 \,.
\end{align}
This result exactly reproduces the matching coefficient in
Eq.~(\ref{matchrslt}), as anticipated.  In the limit $m_{2,3}\to 0$ all results
go smoothly over to those in section~\ref{sect:dimregscalar}.


\subsection*{ii) Offshellness $\mathbf{p^2 =-P^2\ne 0}$, 
  $\mathbf{\ell^2=-L^2\ne 0}$, with $\mathbf{m_1\ne 0}$ and $\mathbf{m_{2,3}=0}$}

The full theory loop integral is now
\begin{align} \label{scetIIfullp2}
 I_{\rm full}^{\rm scalar} &= \int  
  \!\!\frac{d^Dk}{(2\pi)^D}  \frac{1}
{[(k-\ell)^2+i0^+][k^2 -m_1^2 + i0^+][(k- p)^2  + i0^+]} 
 \nonumber\\[5pt]
   &= \frac{-i}{16\pi^2 (p^-\ell^+)} 
   \bigg[ \frac{1}{2} 
   \ln^2\Big(\frac{m_1^2}{p^-\ell^+} \Big) 
     + {\rm Li}_2\Big(\frac{-L^2}{m_1^2}\Big)
     + {\rm Li}_2\Big(\frac{-P^2}{m_1^2}\Big)
     + \frac{\pi^2}{3}
  \bigg] .
\end{align}
where $\ell^2 = -L^2$, $p^2=-P^2$ and we have expanded in $P^2/(p^-\ell^+)$,
$L^2/(p^-\ell^+)$, and $m_1^2/(p^-\ell^+)$.  The result is valid as long as
$(p^-\ell^+) m_1^2 \gg P^2 L^2$, and so can not be used for the special case
$m_1=0$. The result for $m_1\to 0$ is discussed below in case iii).

Using the same IR regulators for the scalar and
collinear \SCETb loops in Figs.~\ref{fig:scalargamma}b,c we have
\begin{align}
 I_{\rm soft}^{\rm scalar} &= \sum_{k^+\ne 0} \int  
  \!\!\frac{d^Dk_r}{(2\pi)^D}  \frac{\mu^{2\epsilon}}
 {[(k-\ell)^2\!+\! i 0^+]
 [k^2 \!-\!m_1^2\!+\! i 0^+][-p^-\,  k^+ \!+\! i0^+]}
\ \frac{|k^+|^\epsilon |k^+\!-\!\ell^+|^\epsilon}{\muplusee} \,, \\
 I_{\rm cn}^{\rm scalar} &= \sum_{k^-\ne 0} \int  
  \!\!\frac{d^Dk_r'}{(2\pi)^D}  \frac{\mu^{2\epsilon}}
 {[-\ell^+\,k^- \!+\! i 0^+]
 [k^2  \!-\!m_1^2\!+\! i 0^+][(k-p)^2\!+\! i 0^+]} 
 \ \frac{|k^-|^\epsilon |k^-\!-\!p^-|^\epsilon}{\muminusee} \,.
  \nonumber
\end{align}
Note that we keep $\ell^2$ and $p^2$ only in the propagators that are allowed to
become small by the power counting in \hbox{\SCETb\!.}  Eq.~(\ref{1.02}) tells
us that we have zero-bin subtractions for the soft and collinear diagrams and the
naive integrals and subtraction integrals are
\begin{align}
 \tilde I_{\rm soft}^{\rm scalar} &=  \int  
  \!\!\frac{d^Dk}{(2\pi)^D}  \frac{\mu^{2\epsilon}}
 {[(k-\ell)^2 \!+\! i 0^+]
 [k^2  \!-\!m_1^2\!+\! i 0^+][-p^-\,  k^+ \!+\! i0^+]}
\ \frac{|k^+|^\epsilon |k^+\!-\!\ell^+|^\epsilon}{\muplusee} \,, \\
  I_{\rm 0soft}^{\rm scalar} &=  \int  
  \!\!\frac{d^Dk}{(2\pi)^D}  \frac{\mu^{2\epsilon}}
 {[-\!\ell^+\,k^- \!+\! i 0^+]
 [k^2  \!-\!m_1^2\!+\! i 0^+][-p^-\,  k^+ \!+\! i0^+]}
\ \frac{|k^+|^\epsilon |k^+\!-\!\ell^+|^\epsilon}{\muplusee} \,, \nn
 \tilde I_{\rm cn}^{\rm scalar} &=  \int  
  \!\!\frac{d^Dk}{(2\pi)^D}  \frac{\mu^{2\epsilon}}
 {[-\ell^+\,k^- \!+\! i 0^+]
 [k^2  \!-\!m_1^2\!+\! i 0^+][(k-p)^2   \!+\! i 0^+]} 
 \ \frac{|k^-|^\epsilon |k^-\!-\!p^-|^\epsilon}{\muminusee} \,,\nn
 I_{\rm 0cn}^{\rm scalar} &=  \int  
  \!\!\frac{d^Dk}{(2\pi)^D}  \frac{\mu^{2\epsilon}}
 {[-\ell^+\,k^- \!+\! i 0^+]
 [k^2  \!-\!m_1^2\!+\! i 0^+][-\! p^- \, k^+\!+\! i 0^+]} 
 \ \frac{|k^-|^\epsilon |k^-\!-\!p^-|^\epsilon}{\muminusee} \,.\nonumber
\end{align}
To determine the form of the subtraction integrals, we considered the collinear
limit of the soft loop momentum in $\tilde I_{\rm soft}^{\rm scalar}$, and the
soft limit of the collinear loop momentum in $\tilde I_{\rm cn}^{\rm scalar}$.
The $p^2$ and $\ell^2$ dependence is dropped in the subtraction integrals
because in these limits $\ell^+k^-\gg -\ell^2$ and $p^-k^+\gg -p^2$. Note that
the subtraction integrals are identical to the case with $m_{1,2,3}\ne 0$. The
final results $I_{\rm soft}^{\rm scalar}$ and $I_{\rm cn}^{\rm scalar}$ are
defined by the differences $\tilde I_{\rm soft}^{\rm scalar}- I_{\rm 0soft}^{\rm
  scalar}$ and $\tilde I_{\rm cn}^{\rm scalar}- I_{\rm 0cn}^{\rm
  scalar}$ respectively.

To compute the soft graph we work in the frame where $\ell^\perp=0$, thus 
$\ell^+\ell^-=-L^2<0$ with $\ell^+>0$, so the offshell momentum is
$\ell^-=-L^-<0$. We do the $k^-$ integral by contours. Due to the pole structure
this restricts the $k^+$-integration to the region $0< k^+<\ell^+$.  The
$k_\perp$ integral is then done. For the soft subtraction integral we follow the
same procedure which gives the integration region $0< k^+<\infty$. Thus
\begin{align}  \label{Isoftm1L2P2}
  \tilde I_{\rm soft}^{\rm scalar} 
  &= \frac{-i\: \Gamma(\epsilon)\, \mu^{2\epsilon} }{16\pi^2(p^-\ell^+)}  
   \int_0^{\ell^+} \! \frac{dk^+}{k^+} \, \left[
     \Big(1-\frac{k^+}{\ell^+}\Big)\big(  m_1^2 + k^+ L^-\big)
     \right]^{-\epsilon} 
   \ \ \bigg|  \frac{k^+ (k^+\!-\!\ell^+)}
   {\muplus{}^2} \bigg|^\epsilon \,,
   \nonumber \\
  I_{\rm 0soft}^{\rm scalar} 
  &= \frac{-i\: \Gamma(\epsilon)\, \mu^{2\epsilon} }{16\pi^2(p^-\ell^+)}  
   \int_0^{\infty} \! \frac{dk^+}{k^+} \, \Big[ m_1^2 \Big]^{-\epsilon} 
   \ \ \bigg|  \frac{k^+ (k^+\!-\!\ell^+)}
   {\muplus{}^2} \bigg|^\epsilon \,.
\end{align}
The double counting with the collinear integral comes from the $k^+\to 0$ part
of the integral, but the divergence from this limit exactly cancels in $I_{\rm
  soft}^{\rm scalar} = \tilde I_{\rm soft}^{\rm scalar} - I_{\rm 0soft}^{\rm
  scalar}$ as long as $m_1\ne 0$.  Computing the integrals we find
\begin{align} \label{ssrsltL2P2}
 I_{\rm soft}^{\rm scalar} 
  &= \frac{-i\: \Gamma(\epsilon)\,  }{16\pi^2(p^-\ell^+)}  \, 
   \Big(\frac{\ell^+}{\muplus}\Big)^{2\epsilon} 
  \bigg\{ \int_0^1\!\!\! dx\: \frac{|1\!-\!x|^\epsilon |x|^\epsilon}{x} \:
    \Big[ \frac{(1\!-\!x)(m_1^2+x L^2)}{\mu^2} \Big]^{-\epsilon}
  - \int_0^\infty\!\!\!\! dx  \frac{|1\!-\!x|^\epsilon |x|^\epsilon}{x} \: 
   \Big[ \frac{m_1^2}{\mu^2} \Big]^{-\epsilon}
  \bigg\}
   \nonumber \\[5pt]
&= \frac{-i\: \Gamma(\epsilon)\,  }{16\pi^2(p^-\ell^+)}  
   \Big(\frac{\mu^2}{m_1^2}\Big)^{\epsilon}\!
   \Big(\frac{\ell^+}{\muplus}\Big)^{2\epsilon} 
  \bigg\{ \frac{\Gamma(\epsilon)}{\Gamma(1\!+\!\epsilon)}
    {}_2F_1\Big(\epsilon,\epsilon,1\!+\!\epsilon,\frac{-L^2}{m_1^2}\Big)
  \!-\!  \frac{\Gamma(\epsilon)\Gamma(1\!+\!\epsilon)}{\Gamma(1\!+\!2\epsilon)}
   \!-\!  \frac{\Gamma(1\!+\!\epsilon)\Gamma(-2\epsilon)}{\Gamma(1\!-\!\epsilon)}
  \bigg\}
   \nonumber \\[5pt]
 &= \frac{-i\: }{16\pi^2(p^-\ell^+)}  
   \bigg[ \frac{1}{2\epsilon_{\rm
      UV}^2} 
   \!+\! \frac{1}{\epsilon_{\rm UV}} \ln\Big(\frac{\ell^+  }{\muplus} \Big)
   \!-\! \frac{1}{2\epsilon_{\rm UV}} \ln\Big(\frac{m_1^2 }{\mu^2} \Big)
  + \ln^2\Big(\frac{\ell^+}{\muplus} \Big) + \frac{3\pi^2}{8} \nn
 & \qquad\qquad\qquad\qquad
  \!+\! \frac{1}{4} \ln^2\Big(\frac{m_1^2}{\mu^2} \Big) 
   - \ln\Big(\frac{m_1^2}{\mu^2} \Big) \ln\Big(\frac{\ell^+}{\muplus}\Big)
  +  {\rm Li}_2\Big(\frac{-L^2}{m_1^2}\Big)\bigg] \,.
\end{align}
For the collinear integrals we take $p^\perp=0$, so $p^-p^+=-P^2 <0$ with
$p^->0$, and it is $p^+=-P^+ <0$ that takes the $p^2$ offshell. We do the contour
integration in $k^+$ which restricts the remaining integration region in $k^-$.
For the naive and subtraction integrals we find
\begin{align}  \label{IcnL2P2}
  \tilde I_{\rm cn}^{\rm scalar} 
  &= \frac{-i\: \Gamma(\epsilon)\, \mu^{2\epsilon} }{16\pi^2(p^-\ell^+)}  
   \int_0^{p^-} \! \frac{dk^-}{k^-} \, 
    \left[
     \Big(1-\frac{k^-}{p^-}\Big) \big( m_1^2 + k^- P^+\big)
      \right]^{-\epsilon} 
   \ \ \bigg|  \frac{k^- (k^-\!-\!p^-)}
   {\muminus{}^2} \bigg|^\epsilon \,,
   \nonumber \\[5pt]
  I_{\rm 0cn}^{\rm scalar} 
  &= \frac{-i\: \Gamma(\epsilon)\, \mu^{2\epsilon} }{16\pi^2(p^-\ell^+)}  
   \int_0^{\infty} \! \frac{dk^-}{k^-} \, \Big[ m_1^2
     \Big]^{-\epsilon} 
   \ \ \bigg[  \frac{k^- (k^-\!-\!p^-)}
   {\muminus{}^2} \bigg]^\epsilon \,.
\end{align}
The complete collinear result, $I_{\rm cn}^{\rm scalar}= \tilde I_{\rm
  cn}^{\rm scalar}-I_{\rm 0cn}^{\rm scalar}$, is very similar to the soft result
\begin{align}  \label{scrsltL2P2}
  I_{\rm cn}^{\rm scalar} &=
 \frac{-i\: \Gamma(\epsilon)\,  }{16\pi^2(p^-\ell^+)}  \, 
   \Big(\frac{p^-}{\muminus}\Big)^{2\epsilon} 
  \bigg\{ \int_0^1\!\!\! dx\: \frac{|1\!-\!x|^\epsilon |x|^\epsilon}{x} \:
    \Big[ \frac{(1\!-\!x)(m_1^2+x P^2)}{\mu^2} \Big]^{-\epsilon}
  - \int_0^\infty\!\!\!\! dx  \frac{|1\!-\!x|^\epsilon |x|^\epsilon}{x} \: 
   \Big[ \frac{m_1^2}{\mu^2} \Big]^{-\epsilon}
  \bigg\}
   \nonumber \\[5pt]
 &= \frac{-i\: }{16\pi^2(p^-\ell^+)}  
   \bigg[ \frac{1}{2\epsilon_{\rm
      UV}^2} 
   \!+\! \frac{1}{\epsilon_{\rm UV}} \ln\Big(\frac{p^-  }{\muminus} \Big)
   \!-\! \frac{1}{2\epsilon_{\rm UV}} \ln\Big(\frac{m_1^2 }{\mu^2} \Big)
  +  \ln^2\Big(\frac{p^-}{\muminus} \Big) + \frac{3\pi^2}{8} \nn
 & \qquad\qquad\qquad\qquad
  \!+\! \frac{1}{4} \ln^2\Big(\frac{m_1^2}{\mu^2} \Big) 
   -  \ln\Big(\frac{m_1^2}{\mu^2} \Big) \ln\Big(\frac{p^-}{\muminus}\Big)
  +   {\rm Li}_2\Big(\frac{-P^2}{m_1^2}\Big)\bigg] \,.
\end{align}

The results in Eqs.~(\ref{ssrsltL2P2}) and (\ref{scrsltL2P2}) have
$1/\epsilon_{\rm UV} \ln(m^2)$ divergences, which are canceled by the
$\phi(0,\mu)/\epsilon$ type counterterms just as in our $m_{1,2,3}$ case. The
counterterms and result to be added are the same as in Eq.~(\ref{Ictscalar}) and
exactly cancels the $1/\epsilon_{\rm UV} \ln(m^2)$ terms in the collinear and
soft loops.  Adding the soft, collinear, and counterterm graphs and simplifying
we find the full \SCETb result
\begin{align} \label{scetIIDRrslt0L2P2}
 I_{\rm soft+cn}^{\rm scalar} 
&= \frac{-i\: }{16\pi^2(p^-\ell^+)}  
   \bigg[ \frac12\, \ln^2\Big(\frac{m_1^2}{p^-\ell^+} \Big)
   - \ln\Big(\frac{m_1^2}{\mu^2} \Big) \ln\Big(\frac{\mu^2}{\muminus\muplus}\Big)
  +  {\rm Li}_2\Big(\frac{-P^2}{m_1^2}\Big)
  +  {\rm Li}_2\Big(\frac{-L^2}{m_1^2}\Big)
  \nonumber \\
 & \qquad 
   +\frac{1}{\epsilon^2_{\rm UV}} -\frac{1}{\epsilon_{\rm UV}} 
  \!+\! \frac{1}{\epsilon_{\rm UV}} \ln\Big(\frac{p^- \ell^+ }{\muminus\muplus} \Big)
  +  \ln^2\Big(\frac{p^-}{\muminus} \Big) +  \ln^2\Big(\frac{\ell^+}{\muplus} \Big)
    - \frac12\,\ln^2\Big(\frac{p^-\ell^+}{\mu^2} \Big) 
   + \frac{3\pi^2}{4}  \bigg]
 \,.
\end{align}
The effective theory result in Eq.~(\ref{scetIIDRrslt0L2P2}) has UV divergences
which are canceled by the counterterm for the jet function coefficient $J^{(0a)}$,
as already given in Eq.~(\ref{delJrslt}). The renormalized EFT result is
\begin{align}
  \label{scetIIDRrsltL2P2}
 I_{\rm soft+cn}^{\rm scalar} 
 &=  \frac{-i\: }{16\pi^2(p^-\ell^+)}  
   \bigg[ \frac12\: \ln^2\Big(\frac{m_1^2}{p^-\ell^+} \Big)
   +   {\rm Li}_2\Big(\frac{-P^2}{m_1^2}\Big)
   +   {\rm Li}_2\Big(\frac{-L^2}{m_1^2}\Big)
   -  \ln\Big(\frac{m_1^2}{\mu^2} \Big) \ln\Big(\frac{\mu^2}{\muminus\muplus}\Big)
   \nn
 & \qquad \qquad\qquad
  +  \ln^2\Big(\frac{p^-}{\muminus} \Big) + \ln^2\Big(\frac{\ell^+}{\muplus} \Big)
    - \frac12\, \ln^2\Big(\frac{p^-\ell^+}{\mu^2} \Big) 
   + \frac{3\pi^2}{4} \bigg]
 \,.
\end{align}
The first three terms exactly reproduces the IR divergences in the full theory
result in Eq.~(\ref{scetIIfullp2}), including the entire functional dependence
on $P^2/m_1^2$ and $L^2/m_1^2$. The fourth term vanishes since $\mu^2 =
\muplus\: \muminus$.  The difference of the remaining finite terms gives a
contribution to the one-loop matching coefficient
\begin{align} \label{matchrsltL2P2}
   I_{\rm match}^{\rm scalar} 
  &= \frac{-i\: }{16\pi^2(p^-\ell^+)}  
   \bigg[ \frac12\,\ln^2\Big(\frac{p^-\ell^+}{\muminus\muplus}\Big) 
  -  \ln^2\Big(\frac{p^-}{\muminus} \Big) -  \ln^2\Big(\frac{\ell^+}{\muplus} \Big)
  - \frac{5\pi^2}{12}
  \bigg] \nn
&= \frac{-i\: }{16\pi^2(p^-\ell^+)}  
   \bigg[ - \frac12\,\ln^2\Big(\frac{p^-\muplus}{\muminus\ell^+} \Big)
  - \frac{5\pi^2}{12}
  \bigg]
 \,.
\end{align}
This result exactly reproduces the matching coefficient in
Eqs.~(\ref{matchrslt}) and (\ref{matchrsltp}) using a different IR regulator. This is
as expected since the full theory was UV finite, and the same UV regulator was
used in the \SCETb calculation. In the limit $L^2,P^2\to 0$ all the results go
smoothly over to the results presented in section~\ref{sect:dimregscalar}.


\subsection*{iii) The limit $\mathbf{m_1\to 0}$ of cases i) and ii)}

Finally, we discuss the limit $m_1\to 0$ of the IR regulators considered above
in cases i) and ii). This is not a smooth limit in either the full or effective
theories. In the following we use the notation $Q^2\equiv (p^-\ell^+)$ as a
shorthand for our large perturbative scale.

We first consider the full theory loop integrals, but in expanding out the IR
regulators we keep the first subleading terms in the expansions in cases where
the leading term vanishes as $m_1^2\to 0$. So for the expansion in
$m_{1,2,3}^2/Q^2$ in case i), we keep subleading $m_{2,3}^2$ terms if the
leading term is proportional to $m_1^2$. For the expansion in $m_1^2/Q^2$,
$L^2/Q^2$, and $P^2/Q^2$ in case ii) we keep subleading $L^2$ and $P^2$ terms
when the leading term is proportional to $m_1^2$. For the $m_{1,2,3}\ne 0$
regulator this gives
\begin{eqnarray} \label{scetIIfullpboth}
  I_{\rm full}^{\rm scalar} &=&
    \frac{-i}{16\pi^2(p^-\ell^+)} \bigg\{
        \frac{1}{2} \ln^2\left[ \frac{\xi -i0^+}{Q^4} \! \right]
  +{\rm Li}_2\left[ \frac{Q^2(m_1^2\!-\!m_2^2)}{\xi}
  -i0^+  \right] \nn
 && \qquad\qquad\qquad
   +{\rm Li}_2\left[ \frac{Q^2(m_1^2\!-\!m_3^2)}{\xi}
  -i0^+  \right]
    - {\rm Li}_2\left[ \frac{-(m_1^2\!-\!m_2^2)(m_1^2\!-\!m_3^2)}
      {\xi} \right] \bigg\} \,,\nn
\xi &\equiv& Q^2 m_1^2-m_2^2m_3^2 \,,
\end{eqnarray}
while for the $m_1^2\ne 0$, $p^2=-P^2\ne 0$, and $\ell^2=-L^2\ne 0$ we have
\begin{eqnarray} \label{scetIIfullbothL2P2}
  I_{\rm full}^{\rm scalar} &=&
    \frac{-i}{16\pi^2(p^-\ell^+)} \bigg\{
        \frac{1}{2} \ln^2\left[\frac{\xi -i0^+}{Q^4} \! \right]
  \! +{\rm Li}_2\left[ \frac{Q^2 P^2}{-\xi +i0^+}
   \right] \! +{\rm Li}_2\left[ \frac{Q^2 L^2}{-\xi+i0^+}  \right]
   + \frac{\pi^2}{3} - {\rm Li}_2\left[ \frac{L^2P^2}{-\xi +i0^+}  \right]
     \bigg\} \,,\nonumber \\[5pt]
\xi &\equiv& Q^2 m_1^2-L^2 P^2 
  .
\end{eqnarray}

From Eq.~(\ref{scetIIfullpboth}) we see that as long as $Q^2 m_1^2 \gg m_2^2
m_3^2$, as is the case if all the IR masses are the same order in the power
counting, then $\xi \to Q^2 m_1^2$, and expanding Eq.~(\ref{scetIIfullpboth})
reproduces the result quoted in Eq.~(\ref{scetIIfullp}). If we set $m_1=0$, then
the leading $m_i^2/Q^2$ terms in the double log and di-logs vanish, and
subleading terms regulate the IR divergences. In this case $\xi\to -m_2^2m_3^2$,
and we obtain
\begin{eqnarray} \label{m1=0}
 I_{\rm full}^{\rm scalar}(m_1\!=\!0) &=&
    \frac{-i}{16\pi^2(p^-\ell^+)} \bigg[ \ln\Big(\frac{m_2^2}{Q^2}\Big) 
    \ln\Big(\frac{m_3^2}{Q^2}\Big) \bigg] \,,
\end{eqnarray}
We have checked that Eq.~(\ref{m1=0}) agrees with the result obtained by setting
$m_1=0$ before evaluating the integral. Since subleading terms are regulating
the IR divergences the result depends on the product $\ln(m_2^2)\ln(m_3^2)$ and
thus no longer has a form that can be factorized in a straightforward manner
into soft and collinear parts. The situation is very similar for case ii).
Expanding Eq.~(\ref{scetIIfullbothL2P2}) when $Q^2 m_1^2 \gg L^2 P^2$ gives
$\xi\to Q^2 m_1^2$ and reproduces the result quoted in
Eq.~(\ref{scetIIfullp2}). If we set $m_1=0$ then the leading term in $\xi$
vanishes and subleading $L^2$ and $P^2$ terms regulate divergences in the double
log and di-logs, with $\xi \to -L^2 P^2$. In this case 
\begin{eqnarray} \label{m1=02}
 I_{\rm full}^{\rm scalar}(m_1\!=\!0) &=&
    \frac{-i}{16\pi^2(p^-\ell^+)} \bigg[ \ln\Big(\frac{L^2}{Q^2}\Big) 
    \ln\Big(\frac{P^2}{Q^2}\Big) + \frac{\pi^2}{3} \bigg] \,,
\end{eqnarray}

\begin{figure}[t!]
 \begin{center}
\includegraphics[height=8cm]{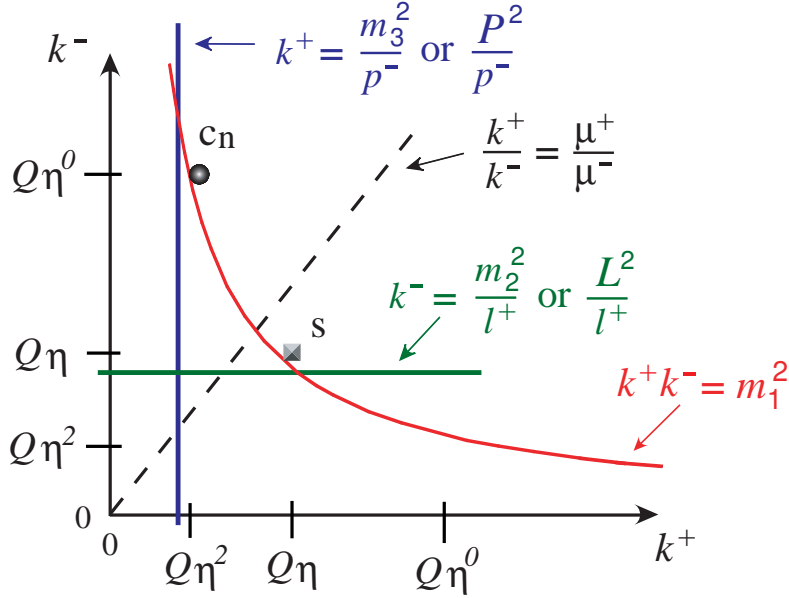}
 \end{center}
\vskip-0.7cm
 \caption{
   Regulation of IR divergences in \SCETb in perturbation theory in the
   $k^+$--$k^-$ plane. The $c_n$ and $s$ denote the collinear and soft modes
   respectively, and the dashed $k^+/k^-=\mu^+/\mu^-$ line indicates how these
   modes are distinguished in perturbation theory.  For $k^\mu\to 0$ the solid
   lines indicate at what scales the IR divergences are cutoff by the $m_1^2$,
   $m_2^2$, and $m_3^2$ regulators (and also for the $m_1^2$, $L^2$, $P^2$
   regulator choice). Since $p^-\gg m_i$ the intersection of the $k^+=m_3^2/p^-$
   and $k^-=m_2^2/\ell^+$ lines is always below the curve $k^+k^-=m_1^2$ (and
   same for $k^+=P^2/p^-$ and $k^-=L^2/\ell^+$). This intersection occurs at the
   messenger scale.  }
  \label{fig:scet2pert}
\end{figure}

To see why this happens we can examine the IR divergences in the full theory
integral. Divergences occur for $k\to p^-$, $k\to \ell^+$, and $k\to 0$, and the
issue with $m_1\to 0$ arises from the $k\to 0$ case where the propagators
carrying the soft and collinear momenta are both singular in opposite light-like
directions. The $k\to 0$ limit of the denominator of the integrand in
Eq.~(\ref{scetIIfullp}) is $[-\ell^+k^- - m_2^2][k^2-m_1^2][-p^-k^+-m_3^2]$ and
for Eq.~(\ref{scetIIfullp2}) is $[-\ell^+k^- - L^2][k^2-m_1^2][-p^-k^+-P^2]$.
The effect of the IR regulators is pictured by the solid curves in
Fig.~\ref{fig:scet2pert}, where we show how they shield the integrand from
blowing up when we go towards the $k^-=0$ or $k^+=0$ lines. Without the $m_1$
regulator the intersection of the $k^+=m_3^2/p^-$ and $k^-=m_2^2/\ell^+$ lines
generates unphysical sensitivity to a very small scale $\sim m_2^2 m_3^2/Q^2$.
For the case with offshellness it is the intersection of the lines $k^+=P^2/p^-$
and $k^-=L^2/\ell^+$ generating sensitivity to the very small scale $\sim L^2
P^2/Q^2$.  The sensitivity to this new small scale was first pointed out in
Ref.~\cite{Becher:2003qh} where the IR regulator $L^2\ne 0$, $P^2\ne 0$ was used.
The result in Eq.~(\ref{m1=02}) agrees with the one studied there.  In
Ref.~\cite{Becher:2003qh} messenger modes with very small invariant mass were
added to the effective theory to account for the dependence on this ``messenger
scale''.  However, in QCD the sensitivity to this small messenger scale is
unphysical because IR divergences are cutoff at an earlier stage by
$\Lambda_{\rm QCD}$~\cite{Manohar:2005az}.  In perturbation theory with 
$m_1\ne 0$, sensitivity to the messenger scale also never appears.
Comparing Fig.~\ref{fig:scet2} with Fig.~\ref{fig:scet2pert} we see that the
$m_1^2\ne 0$ regulator behaves in a similar manner to $\Lambda_{\rm QCD}^2$.
Other IR regulators are also known which remove the unphysical sensitivity to
the messenger scale, including analytic regulators~\cite{Beneke:2003pa}, and an
energy dependent gluon mass~\cite{Bauer:2003td}.

It is possible to choose IR regulators that complicate the choice of the
matching coefficient; $m_1=0$ is such a choice.  We have seen the complication
in the full theory result in Eq.~(\ref{scetIIfullpboth}). In the effective
theory the choice $m_1=0$ also causes problems for $\tilde I_{\rm soft}^{\rm
  scalar}-I_{\rm 0soft}^{\rm scalar}$ and $\tilde I_{\rm cn}^{\rm scalar}-I_{\rm
  0cn}^{\rm scalar}$.  In this situation the $k_\perp$ integration for the
subtraction integral is scaleless, and the subtraction terms do not cancel the
problematic $k^+\to 0$ and $p^-\to 0$ regions in the naive integrals. For
example the naive and subtraction integrals in Eq.~(\ref{Isoftm1m2m3}) scale
with a different power of $k^+$ as $k^+\to 0$. The same is true in for
Eq.~(\ref{Isoftm1L2P2}), and for the collinear integrals.  If one chooses
$m_1=0$ some of the IR divergences are regulated at the scale $m_2^2m_3^2/Q^2$
or $L^2 P^2/Q^2$, and so part of the IR regulator is of subleading order in the
power counting with our definitions of the soft and collinear modes. This makes
it difficult to compute the IR behavior in the effective theory, since one has
to sum up a class of subleading terms in the power counting to all orders. We
expect a resummation procedure in \SCETb could be developed to reproduce the
matching result in Eq.~(\ref{matchrslt}) with $m_1=0$ for the IR regulators
considered in i) and ii), but we will not attempt it here.

It is worth emphasizing that the matching coefficient, Eqs.~(\ref{matchrslt},
\ref{matchrsltp}), which depends on the difference between the full and
effective theory results, does not depend on the IR regulator, as we have
demonstrated with several different calculations. The matching coefficient is
only sensitive to ultraviolet physics. It can be computed by using an IR
regulator that is homogeneous in the power counting.  We have seen this
explicitly by computing the matching for arbitrary $m_{1,2,3}$, and for
$P^2,L^2,m_1^2$ with $m_1\not=0$.  The purpose of the perturbation theory
Feynman graph computations in the full and effective theories is to compute the
matching coefficient Eq.~(\ref{Isoftm1m2m3}), which is a short distance
quantity. \emph{The effective theory is not being used to reproduce the IR
  behavior of the full theory in perturbation theory for arbitrary IR
  regulators}, especially regulators that only become effective at a small 
  scale that is subleading order in the power counting.  The effective theory is
being applied to QCD, where the infrared dynamics is nonperturbative, and cutoff
at $\lqcd$.

\end{appendix}

\newpage



\bibliography{0bin}

\end{document}